\documentclass[twocolumn,longauth]{aa}

\usepackage{graphicx}
\usepackage{amsmath,amsfonts,amssymb}
\usepackage{txfonts}
\usepackage{color}
\usepackage{natbib}
\usepackage{float}
\usepackage{dblfloatfix}
\usepackage{afterpage}
\usepackage{ifthen}
\usepackage[morefloats=12]{morefloats}
\usepackage{placeins}
\usepackage{multicol}
\bibpunct{(}{)}{;}{a}{}{,}
\usepackage[switch]{lineno}
\definecolor{linkcolor}{rgb}{0.6,0,0}
\definecolor{citecolor}{rgb}{0,0,0.75}
\definecolor{urlcolor}{rgb}{0.12,0.46,0.7}
\usepackage[breaklinks, colorlinks, urlcolor=urlcolor,
    linkcolor=linkcolor,citecolor=citecolor,pdfencoding=auto]{hyperref}
\hypersetup{linktocpage}

\def\setsymbol#1#2{\expandafter\def\csname #1\endcsname{#2}}
\def\getsymbol#1{\csname #1\endcsname}

\def\Planck{\textit{Planck}}

\newbox\tablebox    \newdimen\tablewidth
\def\leaderfil{\leaders\hbox to 5pt{\hss.\hss}\hfil}
\def\endPlancktable{\tablewidth=\columnwidth 
    $$\hss\copy\tablebox\hss$$
    \vskip-\lastskip\vskip -2pt}
\def\endPlancktablewide{\tablewidth=\textwidth 
    $$\hss\copy\tablebox\hss$$
    \vskip-\lastskip\vskip -2pt}
\def\tablenote#1 #2\par{\begingroup \parindent=0.8em
    \abovedisplayshortskip=0pt\belowdisplayshortskip=0pt
    \noindent
    $$\hss\vbox{\hsize\tablewidth \hangindent=\parindent \hangafter=1 \noindent
    \hbox to \parindent{$^#1$\hss}\strut#2\strut\par}\hss$$
    \endgroup}
\def\doubleline{\vskip 3pt\hrule \vskip 1.5pt \hrule \vskip 5pt}

\def\L2{\ifmmode L_2\else $L_2$\fi}

\def\DeltaT{\ifmmode \Delta T\else $\Delta T$\fi}
\def\deltat{\ifmmode \Delta t\else $\Delta t$\fi}
\def\fknee{\ifmmode f_{\rm knee}\else $f_{\rm knee}$\fi}
\def\Fmax{\ifmmode F_{\rm max}\else $F_{\rm max}$\fi}
\def\solar{\ifmmode{\rm M}_{\mathord\odot}\else${\rm M}_{\mathord\odot}$\fi}
\def\Msolar{\ifmmode{\rm M}_{\mathord\odot}\else${\rm M}_{\mathord\odot}$\fi}
\def\Lsolar{\ifmmode{\rm L}_{\mathord\odot}\else${\rm L}_{\mathord\odot}$\fi}
\def\inv{\ifmmode^{-1}\else$^{-1}$\fi}
\def\mo{\ifmmode^{-1}\else$^{-1}$\fi}
\def\sup#1{\ifmmode ^{\rm #1}\else $^{\rm #1}$\fi}
\def\expo#1{\ifmmode \times 10^{#1}\else $\times 10^{#1}$\fi}
\def\,{\thinspace}
\def\lsim{\mathrel{\raise .4ex\hbox{\rlap{$<$}\lower 1.2ex\hbox{$\sim$}}}}
\def\gsim{\mathrel{\raise .4ex\hbox{\rlap{$>$}\lower 1.2ex\hbox{$\sim$}}}}

\def\simprop{\mathrel{\raise .4ex\hbox{\rlap{$\propto$}\lower 1.2ex\hbox{$\sim$}}}}
\def\deg{\ifmmode^\circ\else$^\circ$\fi}
\def\pdeg{\ifmmode $\setbox0=\hbox{$^{\circ}$}\rlap{\hskip.11\wd0 .}$^{\circ}
          \else \setbox0=\hbox{$^{\circ}$}\rlap{\hskip.11\wd0 .}$^{\circ}$\fi}
\def\arcs{\ifmmode {^{\scriptstyle\prime\prime}}
          \else $^{\scriptstyle\prime\prime}$\fi}
\def\arcm{\ifmmode {^{\scriptstyle\prime}}
          \else $^{\scriptstyle\prime}$\fi}
\newdimen\sa  \newdimen\sb
\def\parcs{\sa=.07em \sb=.03em
     \ifmmode \hbox{\rlap{.}}^{\scriptstyle\prime\kern -\sb\prime}\hbox{\kern -\sa}
     \else \rlap{.}$^{\scriptstyle\prime\kern -\sb\prime}$\kern -\sa\fi}
\def\parcm{\sa=.08em \sb=.03em
     \ifmmode \hbox{\rlap{.}\kern\sa}^{\scriptstyle\prime}\hbox{\kern-\sb}
     \else \rlap{.}\kern\sa$^{\scriptstyle\prime}$\kern-\sb\fi}
\def\ra[#1 #2 #3.#4]{#1\sup{h}#2\sup{m}#3\sup{s}\llap.#4}
\def\dec[#1 #2 #3.#4]{#1\deg#2\arcm#3\arcs\llap.#4}
\def\deco[#1 #2 #3]{#1\deg#2\arcm#3\arcs}
\def\rra[#1 #2]{#1\sup{h}#2\sup{m}}

\def\dots{\relax\ifmmode \ldots\else $\ldots$\fi}
\def\WHzsr{\ifmmode $W\,Hz\mo\,sr\mo$\else W\,Hz\mo\,sr\mo\fi}
\def\mHz{\ifmmode $\,mHz$\else \,mHz\fi}
\def\GHz{\ifmmode $\,GHz$\else \,GHz\fi}
\def\mKs{\ifmmode $\,mK\,s$^{1/2}\else \,mK\,s$^{1/2}$\fi}
\def\muKs{\ifmmode \,\mu$K\,s$^{1/2}\else \,$\mu$K\,s$^{1/2}$\fi}
\def\muKRJs{\ifmmode \,\mu$K$_{\rm RJ}$\,s$^{1/2}\else \,$\mu$K$_{\rm RJ}$\,s$^{1/2}$\fi}
\def\muKHz{\ifmmode \,\mu$K\,Hz$^{-1/2}\else \,$\mu$K\,Hz$^{-1/2}$\fi}
\def\MJysr{\ifmmode \,$MJy\,sr\mo$\else \,MJy\,sr\mo\fi}
\def\MJysrmK{\ifmmode \,$MJy\,sr\mo$\,mK$_{\rm CMB}\mo\else \,MJy\,sr\mo\,mK$_{\rm CMB}\mo$\fi}
\def\microns{\ifmmode \,\mu$m$\else \,$\mu$m\fi}

\def\muK{\ifmmode \,\mu$K$\else \,$\mu$\hbox{K}\fi}
\def\microK{\ifmmode \,\mu$K$\else \,$\mu$\hbox{K}\fi}
\def\muW{\ifmmode \,\mu$W$\else \,$\mu$\hbox{W}\fi}
\def\kms{\ifmmode $\,km\,s$^{-1}\else \,km\,s$^{-1}$\fi}
\def\kmsMpc{\ifmmode $\,\kms\,Mpc\mo$\else \,\kms\,Mpc\mo\fi}
\providecommand{\sorthelp}[1]{}

\def\WMAP{WMAP}

\def\nside{N_{\mathrm{side}}}

\def\healpix{\texttt{HEALPix}}
\def\commander{\texttt{Commander}}
\def\commanderone{\texttt{Commander1}}
\def\commandertwo{\texttt{Commander2}}

\def\nilc{\texttt{NILC}}
\def\gnilc{\texttt{GNILC}}
\def\sevem{\texttt{SEVEM}}
\def\smica{\texttt{SMICA}}

\def\Plik{\texttt{Plik}}

\renewcommand{\d}[0]{\vec{d}}
\renewcommand{\t}[0]{\vec{t}}

\newcommand{\Y}[0]{\tens{Y}}
\newcommand{\n}[0]{\vec{n}}

\newcommand{\s}[0]{\vec{s}}
\renewcommand{\a}[0]{\vec{a}}
\newcommand{\m}[0]{\vec{m}}
\newcommand{\f}[0]{\vec{f}}
\newcommand{\F}[0]{\tens{F}}
\newcommand{\B}[0]{\tens{B}}
\newcommand{\T}[0]{\tens{T}}
\newcommand{\Cp}[0]{\tens{C}}
\renewcommand{\L}[0]{\tens{L}}

\newcommand{\N}[0]{\tens{N}}
\newcommand{\M}[0]{\tens{M}}

\renewcommand{\S}[0]{\tens{S}}

\renewcommand{\P}[0]{\tens{P}}

\newcommand{\mathsc}[1]{{\normalfont\textsc{#1}}}
\newcommand{\hi}{\ensuremath{\mathsc {Hi}}}

\def\bC{\tens{C}}
\def\ba{\vec{a}}
\def\ncha{N_\mathrm{cha}}
\def\nfg{N_\mathrm{fg}}

\def\adj{^{\dagger}}

\def\inv{^{-1}}
\def\lm{{\ell m}}

\begin{document}

\title{\textit{Planck} 2018 results. IV. Diffuse component separation}
\author{\small
Planck Collaboration: Y.~Akrami\inst{14, 51, 53}
\and
M.~Ashdown\inst{60, 5}
\and
J.~Aumont\inst{85}
\and
C.~Baccigalupi\inst{71}
\and
M.~Ballardini\inst{20, 37}
\and
A.~J.~Banday\inst{85, 8}
\and
R.~B.~Barreiro\inst{55}
\and
N.~Bartolo\inst{26, 56}
\and
S.~Basak\inst{77}
\and
K.~Benabed\inst{50, 84}
% ...
}
\institute{\small
AIM, CEA, CNRS, Universit\'{e} Paris-Saclay, Universit\'{e} Paris-Diderot, Sorbonne Paris Cit\'{e}, F-91191 Gif-sur-Yvette, France\goodbreak
\and
APC, AstroParticule et Cosmologie, Universit\'{e} Paris Diderot, CNRS/IN2P3, CEA/lrfu, Observatoire de Paris, Sorbonne Paris Cit\'{e}, 10, rue Alice Domon et L\'{e}onie Duquet, 75205 Paris Cedex 13, France\goodbreak
\and
African Institute for Mathematical Sciences, 6-8 Melrose Road, Muizenberg, Cape Town, South Africa\goodbreak
\and
Aix Marseille Univ, CNRS, CNES, LAM, Marseille, France\goodbreak
\and
Astrophysics Group, Cavendish Laboratory, University of Cambridge, J J Thomson Avenue, Cambridge CB3 0HE, U.K.\goodbreak
\and
Astrophysics \& Cosmology Research Unit, School of Mathematics, Statistics \& Computer Science, University of KwaZulu-Natal, Westville Campus, Private Bag X54001, Durban 4000, South Africa\goodbreak
\and
CITA, University of Toronto, 60 St. George St., Toronto, ON M5S 3H8, Canada\goodbreak
\and
CNRS, IRAP, 9 Av. colonel Roche, BP 44346, F-31028 Toulouse cedex 4, France\goodbreak
\and
Cahill Center for Astronomy and Astrophysics, California Institute of Technology, Pasadena CA,  91125, USA\goodbreak
\and
California Institute of Technology, Pasadena, California, U.S.A.\goodbreak
\and
Centre for Theoretical Cosmology, DAMTP, University of Cambridge, Wilberforce Road, Cambridge CB3 0WA, U.K.\goodbreak
\and
Computational Cosmology Center, Lawrence Berkeley National Laboratory, Berkeley, California, U.S.A.\goodbreak
\and
D\'{e}partement de Physique Th\'{e}orique, Universit\'{e} de Gen\`{e}ve, 24, Quai E. Ansermet,1211 Gen\`{e}ve 4, Switzerland\goodbreak
\and
D\'{e}partement de Physique, \'{E}cole normale sup\'{e}rieure, PSL Research University, CNRS, 24 rue Lhomond, 75005 Paris, France\goodbreak
\and
Departamento de Astrof\'{i}sica, Universidad de La Laguna (ULL), E-38206 La Laguna, Tenerife, Spain\goodbreak
\and
Departamento de F\'{\i}sica, Universidad de Oviedo, C/ Federico Garc\'{\i}a Lorca, 18 , Oviedo, Spain\goodbreak
\and
Departamento de F\'{i}sica Matematica, Instituto de F\'{i}sica, Universidade de S\~{a}o Paulo, Rua do Mat\~{a}o 1371, S\~{a}o Paulo, Brazil\goodbreak
\and
Department of Astrophysics/IMAPP, Radboud University, P.O. Box 9010, 6500 GL Nijmegen, The Netherlands\goodbreak
\and
Department of Physics \& Astronomy, University of British Columbia, 6224 Agricultural Road, Vancouver, British Columbia, Canada\goodbreak
\and
Department of Physics \& Astronomy, University of the Western Cape, Cape Town 7535, South Africa\goodbreak
\and
Department of Physics and Astronomy, University of Sussex, Brighton BN1 9QH, U.K.\goodbreak
\and
Department of Physics, Gustaf H\"{a}llstr\"{o}min katu 2a, University of Helsinki, Helsinki, Finland\goodbreak
\and
Department of Physics, Princeton University, Princeton, New Jersey, U.S.A.\goodbreak
\and
Department of Physics, University of California, Santa Barbara, California, U.S.A.\goodbreak
\and
Department of Physics, University of Illinois at Urbana-Champaign, 1110 West Green Street, Urbana, Illinois, U.S.A.\goodbreak
\and
Dipartimento di Fisica e Astronomia G. Galilei, Universit\`{a} degli Studi di Padova, via Marzolo 8, 35131 Padova, Italy\goodbreak
\and
Dipartimento di Fisica e Scienze della Terra, Universit\`{a} di Ferrara, Via Saragat 1, 44122 Ferrara, Italy\goodbreak
\and
Dipartimento di Fisica, Universit\`{a} La Sapienza, P. le A. Moro 2, Roma, Italy\goodbreak
\and
Dipartimento di Fisica, Universit\`{a} degli Studi di Milano, Via Celoria, 16, Milano, Italy\goodbreak
\and
Dipartimento di Fisica, Universit\`{a} degli Studi di Trieste, via A. Valerio 2, Trieste, Italy\goodbreak
\and
Dipartimento di Fisica, Universit\`{a} di Roma Tor Vergata, Via della Ricerca Scientifica, 1, Roma, Italy\goodbreak
\and
European Space Agency, ESAC, Planck Science Office, Camino bajo del Castillo, s/n, Urbanizaci\'{o}n Villafranca del Castillo, Villanueva de la Ca\~{n}ada, Madrid, Spain\goodbreak
\and
European Space Agency, ESTEC, Keplerlaan 1, 2201 AZ Noordwijk, The Netherlands\goodbreak
\and
Gran Sasso Science Institute, INFN, viale F. Crispi 7, 67100 L'Aquila, Italy\goodbreak
\and
Haverford College Astronomy Department, 370 Lancaster Avenue, Haverford, Pennsylvania, U.S.A.\goodbreak
\and
Helsinki Institute of Physics, Gustaf H\"{a}llstr\"{o}min katu 2, University of Helsinki, Helsinki, Finland\goodbreak
\and
INAF - OAS Bologna, Istituto Nazionale di Astrofisica - Osservatorio di Astrofisica e Scienza dello Spazio di Bologna, Area della Ricerca del CNR, Via Gobetti 101, 40129, Bologna, Italy\goodbreak
\and
INAF - Osservatorio Astronomico di Padova, Vicolo dell'Osservatorio 5, Padova, Italy\goodbreak
\and
INAF - Osservatorio Astronomico di Trieste, Via G.B. Tiepolo 11, Trieste, Italy\goodbreak
\and
INAF, Istituto di Radioastronomia, Via Piero Gobetti 101, I-40129 Bologna, Italy\goodbreak
\and
INAF/IASF Milano, Via E. Bassini 15, Milano, Italy\goodbreak
\and
INFN - CNAF, viale Berti Pichat 6/2, 40127 Bologna, Italy\goodbreak
\and
INFN, Sezione di Bologna, viale Berti Pichat 6/2, 40127 Bologna, Italy\goodbreak
\and
INFN, Sezione di Ferrara, Via Saragat 1, 44122 Ferrara, Italy\goodbreak
\and
INFN, Sezione di Milano, Via Celoria 16, Milano, Italy\goodbreak
\and
INFN, Sezione di Roma 1, Universit\`{a} di Roma Sapienza, Piazzale Aldo Moro 2, 00185, Roma, Italy\goodbreak
\and
INFN, Sezione di Roma 2, Universit\`{a} di Roma Tor Vergata, Via della Ricerca Scientifica, 1, Roma, Italy\goodbreak
\and
Imperial College London, Astrophysics group, Blackett Laboratory, Prince Consort Road, London, SW7 2AZ, U.K.\goodbreak
\and
Institut d'Astrophysique Spatiale, CNRS, Univ. Paris-Sud, Universit\'{e} Paris-Saclay, B\^{a}t. 121, 91405 Orsay cedex, France\goodbreak
\and
Institut d'Astrophysique de Paris, CNRS (UMR7095), 98 bis Boulevard Arago, F-75014, Paris, France\goodbreak
\and
Institute Lorentz, Leiden University, PO Box 9506, Leiden 2300 RA, The Netherlands\goodbreak
\and
Institute of Astronomy, University of Cambridge, Madingley Road, Cambridge CB3 0HA, U.K.\goodbreak
\and
Institute of Theoretical Astrophysics, University of Oslo, Blindern, Oslo, Norway\goodbreak
\and
Instituto de Astrof\'{\i}sica de Canarias, C/V\'{\i}a L\'{a}ctea s/n, La Laguna, Tenerife, Spain\goodbreak
\and
Instituto de F\'{\i}sica de Cantabria (CSIC-Universidad de Cantabria), Avda. de los Castros s/n, Santander, Spain\goodbreak
\and
Istituto Nazionale di Fisica Nucleare, Sezione di Padova, via Marzolo 8, I-35131 Padova, Italy\goodbreak
\and
Jet Propulsion Laboratory, California Institute of Technology, 4800 Oak Grove Drive, Pasadena, California, U.S.A.\goodbreak
\and
Jodrell Bank Centre for Astrophysics, Alan Turing Building, School of Physics and Astronomy, The University of Manchester, Oxford Road, Manchester, M13 9PL, U.K.\goodbreak
\and
Kavli Institute for Cosmological Physics, University of Chicago, Chicago, IL 60637, USA\goodbreak
\and
Kavli Institute for Cosmology Cambridge, Madingley Road, Cambridge, CB3 0HA, U.K.\goodbreak
\and
Kavli Institute for the Physics and Mathematics of the Universe (Kavli IPMU, WPI), UTIAS, The University of Tokyo, Chiba, 277- 8583, Japan\goodbreak
\and
Laboratoire de Physique Subatomique et Cosmologie, Universit\'{e} Grenoble-Alpes, CNRS/IN2P3, 53, rue des Martyrs, 38026 Grenoble Cedex, France\goodbreak
\and
Laboratoire de Physique Th\'{e}orique, Universit\'{e} Paris-Sud 11 \& CNRS, B\^{a}timent 210, 91405 Orsay, France\goodbreak
\and
Lawrence Berkeley National Laboratory, Berkeley, California, U.S.A.\goodbreak
\and
Low Temperature Laboratory, Department of Applied Physics, Aalto University, Espoo, FI-00076 AALTO, Finland\goodbreak
\and
Max-Planck-Institut f\"{u}r Astrophysik, Karl-Schwarzschild-Str. 1, 85741 Garching, Germany\goodbreak
\and
Mullard Space Science Laboratory, University College London, Surrey RH5 6NT, U.K.\goodbreak
\and
NAOC-UKZN Computational Astrophysics Centre (NUCAC), University of KwaZulu-Natal, Durban 4000, South Africa\goodbreak
\and
National Centre for Nuclear Research, ul. A. Soltana 7, 05-400 Otwock, Poland\goodbreak
\and
Nicolaus Copernicus Astronomical Center, Polish Academy of Sciences, Bartycka 18, 00-716 Warsaw, Poland\goodbreak
\and
SISSA, Astrophysics Sector, via Bonomea 265, 34136, Trieste, Italy\goodbreak
\and
San Diego Supercomputer Center, University of California, San Diego, 9500 Gilman Drive, La Jolla, CA 92093, USA\goodbreak
\and
School of Chemistry and Physics, University of KwaZulu-Natal, Westville Campus, Private Bag X54001, Durban, 4000, South Africa\goodbreak
\and
School of Physical Sciences, National Institute of Science Education and Research, HBNI, Jatni-752050, Odissa, India\goodbreak
\and
School of Physics and Astronomy, Cardiff University, Queens Buildings, The Parade, Cardiff, CF24 3AA, U.K.\goodbreak
\and
School of Physics and Astronomy, Sun Yat-sen University, 2 Daxue Rd, Tangjia, Zhuhai, China\goodbreak
\and
School of Physics, Indian Institute of Science Education and Research Thiruvananthapuram, Maruthamala PO, Vithura, Thiruvananthapuram 695551, Kerala, India\goodbreak
\and
Simon Fraser University, Department of Physics, 8888 University Drive, Burnaby BC, Canada\goodbreak
\and
Sorbonne Universit\'{e}, Observatoire de Paris, Universit\'{e} PSL, \'{E}cole normale sup\'{e}rieure, CNRS, LERMA, F-75005, Paris, France\goodbreak
\and
Sorbonne Universit\'{e}-UPMC, UMR7095, Institut d'Astrophysique de Paris, 98 bis Boulevard Arago, F-75014, Paris, France\goodbreak
\and
Space Science Data Center - Agenzia Spaziale Italiana, Via del Politecnico snc, 00133, Roma, Italy\goodbreak
\and
Space Sciences Laboratory, University of California, Berkeley, California, U.S.A.\goodbreak
\and
The Oskar Klein Centre for Cosmoparticle Physics, Department of Physics, Stockholm University, AlbaNova, SE-106 91 Stockholm, Sweden\goodbreak
\and
UPMC Univ Paris 06, UMR7095, 98 bis Boulevard Arago, F-75014, Paris, France\goodbreak
\and
Universit\'{e} de Toulouse, UPS-OMP, IRAP, F-31028 Toulouse cedex 4, France\goodbreak
% ...
}

\authorrunning{Planck Collaboration}
\titlerunning{Diffuse component separation}

\abstract{We present full-sky maps of the cosmic microwave background  (CMB) and polarized synchrotron and thermal dust emission, derived  from the third set of \Planck\ frequency maps. These products have  significantly lower contamination from instrumental systematic  effects than previous versions. The methodologies used to derive  these maps follow closely those described in earlier papers,  adopting four methods (\commander, \nilc,  \sevem, and \smica) to extract the CMB component, as well as three  methods (\commander, \gnilc, and \smica) to extract astrophysical components. Our revised CMB temperature maps  agree with corresponding products in the \Planck\ 2015 delivery,  whereas the polarization maps exhibit significantly lower  large-scale power, reflecting the improved data processing described  in companion papers; however,  the noise properties  of the resulting data products are complicated, and the best  available end-to-end simulations exhibit relative biases with  respect to the data at the few percent level.  Using these maps, we are for the first time able to fit the spectral index of  thermal dust independently over $3^{\circ}$ regions. We derive a  conservative estimate of the mean spectral index of polarized  thermal dust emission of $\beta_{\mathrm{d}}=1.55\pm0.05$, where the  uncertainty marginalizes both over all known systematic  uncertainties and different estimation techniques. For polarized  synchrotron emission, we find a mean spectral index of  $\beta_{\mathrm{s}}=-3.1\pm0.1$, consistent with previously reported  measurements.  We note that the current data processing does not  allow for construction of unbiased single-bolometer maps, and this  limits our ability to extract CO emission and correlated  components. The foreground results for intensity derived in this  paper therefore do not supersede corresponding \Planck\ 2015  products.  For polarization the new results supersede the  corresponding 2015 products in all respects.}

\maketitle


\hypersetup{linkcolor=black}
\tableofcontents
\hypersetup{linkcolor=red}

\section{Introduction}
\label{sec:introduction}

This paper, one of a set associated with the 2018 release of data from the \Planck\footnote{ \Planck\ (\url{http://www.esa.int/Planck}) is a  project of the European Space Agency (ESA) with instruments provided  by two scientific consortia funded by ESA member states and led by  Principal Investigators from France and Italy, telescope reflectors  provided through a collaboration between ESA and a scientific  consortium led and funded by Denmark, and additional contributions  from NASA (USA).  } mission \citep{planck2014-a01}, describes the cosmological and astrophysical component maps derived from the full set of \Planck\ observations \citep{planck2016-l01}, and compares these to earlier versions of the corresponding products. \Planck\  was launched on 14 May 2009, and observed the sky nearly without interruption for four years. The raw, time-ordered observations were released to the public in their entirety in February 2015 as part of the second \Planck\ data release (PR2), together with associated frequency and component sky maps and higher-level science data products, including cosmic microwave background (CMB) power spectra and cosmological parameters. These observations represent a cornerstone of modern cosmology, and they severely constrain the history of the early Universe.

The time-ordered data selection adopted for the current (third, PR3) release is similar to that used in the second release (\citealt{planck2016-l02}; \citealt{planck2016-l03}); the second and third \Planck\ product deliveries therefore have nearly identical scientific constraining power, as measured in terms of raw integration time and instrumental noise levels. The difference between the two releases lies in their overall levels of instrumental systematic uncertainties \textcolor{black}{and calibration}.  A substantial fraction of the second-release papers was dedicated to identifying, quantifying, and characterizing residual uncertainties due to a wide range of instrumental effects, including effective gain variations, analogue-to-digital converter (ADC) nonlinearities, residual temporal transfer functions, and foreground bandpass leakage. Indeed, these residuals were sufficiently large to prohibit extraction of a robust polarization signal on large angular scales from the \Planck\ High Frequency Instrument (HFI) observations, significantly limiting the science scope of the \Planck\ polarization observations as a whole. Fortunately, as discussed extensively in \citet{planck2016-l03}, these residuals are now not only better understood and modelled, but also greatly reduced in the final dataset, particularly through the use of improved end-to-end processing techniques.

In this paper, we present updated full-sky CMB maps in both temperature and polarization, as well as new synchrotron and thermal dust emission maps in polarization, and compare these to previous versions \citep{planck2013-p06, planck2014-a11, planck2014-a12}. In terms of temperature foreground products, we provide an update of the Generalized Needlet Internal Linear Combination (\gnilc;\citealp{Remazeilles2011b}) thermal dust model, to be used in conjunction with the updated 2018 \gnilc\ polarization map, but no new \commander\ \citep{eriksen2008} foreground products. The reason for this is one of necessity: as described in \citet{planck2016-l03}, the latest HFI processing exploits the full information content of each frequency in order to suppress large-scale polarization systematics, and the processing has thus been tuned to optimize the polarization solution. The cost of this choice, however, is that individual single-bolometer maps are no longer available; see section~3.1.2 of \citealp{planck2016-l03} for details. Specifically, some of the single-bolometer maps only contain part of the sky signal and thus cannot be used for component separation. This, in turn, has an impact on the ability of the \commander\ algorithm to resolve individual foreground components in temperature. The single most important effect is on our ability to constrain CO line emission, which benefits particularly strongly from intra-frequency measurements.  \textcolor{black}{Because each unfiltered bolometer in principle has a different bandpass amplitude} at the CO-line centre frequency of 115.27\,GHz (and multiples thereof), each bolometer observes the true CO signal with different effective responses, and these differences provide a strong handle on the true intensity of the CO signal. Furthermore, both thermal dust and free-free emission correlate strongly with CO emission, and are therefore also negatively affected by the lack of single-bolometer maps. In turn, free-free emission is strongly correlated with both synchrotron and anomalous microwave emission. In summary, we believe that the \Planck\ 2015 Commander-based temperature (i.e., Stokes $I$) foreground model represents a more accurate description of the true temperature sky than what can be extracted from the current (2018) data set.  To avoid confusion, we therefore do not release the latest version publicly, although we compare the two models in Sect.~\ref{sec:foregrounds}. For the CMB component, we find that the latest processing produces results that are fully consistent with the previous incarnation, while for polarization the new results represent a major improvement, both in terms of CMB and foregrounds.

The methodologies adopted in this paper mirror those used in earlier \Planck\ releases, with only minor algorithmic updates and improvements.  In particular, for CMB extraction we adopt the same four component-separation implementations used in earlier releases, namely \commander, \nilc, \sevem, and \smica, each of which was initially selected as a representative of a particular class of algorithms(blind versus non-blind methods and pixel-based versus harmonic-based methods).  In combination, they represent most approaches proposed in the literature. In the current release, all four CMB methods adopt the same data selection, based only on full-frequency \Planck\ maps, in order to facilitate a direct comparison of the results. As in previous releases, we strongly suggest considering all four CMB maps in any higher-level map-based CMB analysis, in order to assess robustness with respect to algorithmic choices. We also provide again cleaned CMB maps at individual frequencies constructed by \sevem. More specifically, in this release, intensity and polarization CMB maps are produced at four different frequencies from 70 to 217 GHz. These maps are particularly useful to test, for example, the robustness of results versus the presence of foregrounds and/or systematics. In addition, one fundamentally new data product is delivered in this release, namely a CMB temperature map generated by \smica\ from which Sunyaev-Zeldovich(SZ) sources have been projected out.  This can be used, for instance, in lensing studies \citep{planck2016-l08}.

For astrophysical component separation, which depends inherently on explicit parametric modelling, we adopt \commander\ as our primary computational engine, mirroring the processing adopted in the two previous \Planck\ releases. However, since the last release the internal mechanics of this code have been significantly re-written.  \commander\ now allows for analysis of data sets with different angular resolutions at each frequency, and thereby allows for production of frequency maps at the full angular resolution of the data \citep{seljebotn2017}. In addition, we employ both \gnilc\ and \smica\ for foreground reconstruction in the new release.

The rest of the paper is organized as follows. Section~\ref{sec:methods} reviews the algorithms and methods used in the analysis, focusing primarily on updates and improvements made since the 2015 release.  Section~\ref{sec:inputs} describes the data selection and pre-processing steps applied to the data before analysis.  Section~\ref{sec:cmb} presents the \Planck\ 2018 CMB maps in both temperature and polarization, and characterizes their properties in terms of residuals with respect to earlier versions, along with angular power spectra, cosmological parameters, and simple higher-order statistics.  Section~\ref{sec:foregrounds} discusses the updated polarization foreground products.  Section~\ref{sec:conclusions} gives conclusions. The various algorithms are reviewed in Appendices~\ref{app:commander}--\ref{app:gnilc}.  A brief summary of temperature foregrounds derived from the \Planck\ 2018 frequency maps is provided in Appendix~\ref{app:foregrounds} and, finally, additional CMB figures are provided in Appendices~\ref{app:splits} and \ref{app:npt_functions}.

\section{Component-separation methods}
\label{sec:methods}

Earlier publications give detailed descriptions of the four main component-separation methods used in this paper \citep{planck2013-p06,planck2014-a11,planck2014-a12}. For some methods, notable improvements have been implemented since the last release, and these are described below. Further technical details may be found in the Appendices.

We also employ the \gnilc\ algorithm for thermal dust extraction. This method and corresponding results are described in detail in \citet{Remazeilles2011b}, \citet{planck2016-XLVIII}, and \citet{planck2016-l11B}. A detailed comparison of the foreground
products derived with \commander\ and \gnilc\ is presented in the current paper.

\subsection{\commander}
\label{sec:commander}

\commander\ \citep{eriksen2004,eriksen2008,planck2014-a12} has undergone the most significant changes since the previous release. \commander\ is a Bayesian approach employing a Monte Carlo method called Gibbs sampling as its central computational engine. Within this Bayesian framework, a parametric model is fitted to the data set in question with standard posterior sampling or maximization techniques, including cosmological, astrophysical, and instrumental parameters.

We start by writing down a generic model on the form,
\begin{equation}
  \d_{\nu}(p) = g_{\nu} \sum_{c=1}^{N_{\rm c}} \F_{\nu}(\beta_c) \T(p) \a_{c} + \n_{\nu}(p).
\end{equation}
Here $\d_{\nu}(p)$ denotes the observed data at frequency $\nu$ and pixel $p$.  The sum runs over $N_{\rm c}$ components, each with an amplitude vector $\a_c$, a map projection operator $\T(p)$, and frequency scaling operator $\F_{\nu}(\beta_c)$ that depends on astrophysical spectral parameters $\beta_c$.  The quantity $g(\nu)$ denotes an overall instrumental calibration factor per frequency channel, and $n_{\nu}(p)$ indicates instrumental noise. With this notation, the component sum runs over both astrophysical components (CMB, synchrotron, CO, thermal dust emission etc.) and possible spurious monopole and dipole terms. The projection operator $\T$ indicates any step required in going from a general amplitude vector
(such as a pixelized sky map, a set of spherical harmonic coefficients, or a template amplitude) to a map as observed by the
current detector. Thus, this matrix encodes both the choice of basis vectors (pixels, spherical harmonics, templates) and higher-level operations such as beam convolution. Given this data model, samples are drawn from the full posterior as described in \citet{eriksen2004,eriksen2008} and \citet{seljebotn2017}.

In previous releases the above model was fitted to the combination of \Planck\ and external data using the \commander\ implementation described by \citet{eriksen2008}. This implementation adopted map-space pixels as its basis set for astrophysical foregrounds, for coding efficiency reasons. Although computationally fast, this approach has a significant limitation in that it requires all data sets under consideration to have the same angular resolution. Specifically, this implies that the angular resolution of the final output maps are limited to that of the lowest resolution frequency channel under consideration, which typically is $1^{\circ}$ FWHM for the combination of \Planck, \WMAP, and Haslam 408\,MHz, which
formed the basis of the previous astrophysically oriented foreground analysis. Higher-resolution products could then only be derived by dropping lower-resolution channels, which in turn carried  a significant cost in terms of model fidelity.

In the current release, we implement the \commander\ algorithm described by \citet{seljebotn2017}, which we refer to as \commandertwo. This approach, which models the foreground amplitude maps in terms of spherical harmonics instead of pixels, offers three important improvements over the pixel-based approach.

First, since amplitudes are modelled in harmonic space, it is computationally trivial to convolve with a separate instrumental beam transfer function at separate frequencies, so that for the first time we can solve for full-resolution signal models with
multi-resolution data sets. \commandertwo\ is thus able to produce a foreground model at native \Planck\ resolution, limited only by the effective signal-to-noise ratio of each component. The computational cost is greater; however, as shown by \citet{seljebotn2017}, this is manageable with modern computers, even for \Planck-sized data sets.

Second, the new approach offers the option of imposing a prior on the foreground signal amplitudes in the form of an angular power spectrum. This can be used to regularize the foreground solution at small angular scales, and thereby reduce degeneracies between different components at high multipoles.

Third, the improvements allow for joint fitting of compact or unresolved objects and diffuse components. This improves the reconstruction of the diffuse components themselves, including the CMB, and also allows production of a new catalogue of compact objects. The details of this procedure are described in Appendix~\ref{app:commander}. 

Overall, from an algorithmic point of view the \commandertwo\ implementation used in the current data release is more
powerful than in previous releases. At the same time, there is also one important aspect of the \Planck\ 2018 data release that limits our ability to perform a component separation as detailed as that in the 2015 analysis. As mentioned in Sect.~\ref{sec:introduction}, the \Planck\ 2018 data set includes only full-frequency maps, not single-bolometer maps. For the
\commander\ temperature analysis, this implies that a simpler foreground model must be employed than in the corresponding 2015 analysis. In the previous analysis we considered seven different physical components, namely CMB, synchrotron, free-free, spinning and thermal dust emission, a general line emission component at 95 and 100\,GHz, and CO with individual components at 100, 217, and 353\,GHz. Single-detector maps played a central part in constraining this rich model, in particular with respect to CO line emission. With the new and more limited data set, we instead adopt a similar model as employed in the 2013 analysis, which includes only four diffuse signal components in temperature, namely CMB, a single general
low-frequency power-law component, thermal dust, and a single CO component with spatially constant line ratios between 100, 217, and 353\,GHz. For polarization the model remains the same as in 2015, and includes only CMB, synchrotron, and thermal dust emission. The latter two components are as usual modelled in terms of simple power-law and modified blackbody SEDs, respectively.

The above general specification provides a basic summary of the framework used for parametric fitting. However, there are still some free choices that must be made, the two most important of which are: (1)~the angular resolution of the foreground spectral indices; and (2)~the spatial priors imposed on the foreground amplitudes. For the spectral indices, we are primarily driven by signal-to-noise considerations, as adopting too high resolution for such parameters leads to an undesirable increase in noise in all components. In the temperature case, we adopt a smoothing scale of $40\arcm$ FWHM for low-frequency
foregrounds, slightly larger than the 30-GHz instrumental beam. For the dust spectral index, we adopt $10\arcm$ FWHM, which is slightly larger than the 100-GHz beam. The dust temperature is fitted at the full \Planck\ resolution of $5\arcm$ FWHM \textcolor{black}{of the frequencies between 217 and 857\,GHz}. For polarization, we fit only a spatially-constant spectral index for synchrotron,\footnote{Note that the numerical value derived for the  spectral index of polarized synchrotron emission is not directly  comparable to the mean of the low-frequency component spectral index  map derived in intensity, since the latter also includes free-free and spinning dust emission.} while for thermal dust emission, we fit the dust spectral index at $3\deg$ FWHM. The dust temperature for the polarization model is fixed at the values derived in the intensity
analysis, as the \Planck\ 545- and 857-GHz frequency channels are unpolarized, and the \Planck\ observations therefore do not constrain the thermal dust temperature in polarization.

Finally, for spatial priors, we adopt minimally informative power-spectrum priors, defined simply as flat spectra in units of $C_\ell \ell(\ell+1)/2\pi$ for all components, with an amplitude that is larger than that observed in the high signal-to-noise regime. In addition, this flat spectrum is smoothly apodized at high multipoles in order to suppress ringing around bright compact objects. For the low-frequency temperature foreground and the CO line-emission components, the apodization is performed with a Gaussian beam with a FWHM roughly matching the dominant frequency for the respective component, while for thermal dust only a mild apodization is applied in the form of an exponentially-falling cut-off between $\ell=5000$ and 6000. For polarization, we apodize with Gaussian smoothing kernels, as in the low-frequency foreground and CO case.\footnote{The   decision on whether to use a Gaussian kernel or a mild exponential high-$\ell$ cut-off for prior apodization is determined by the effective signal-to-noise ratio of the component in question.} Full details regarding these choices are summarized in
Appendix~\ref{app:commander}.

\subsection{\nilc} 
\label{sec:nilc}

\nilc\ (Needlet Internal Linear Combination) is described by \citet{2012MNRAS.419.1163B,   2013MNRAS.435...18B}. The overall goal of \nilc\ is to extract the CMB component from multi-frequency observations while minimizing the contamination from Galactic and extragalactic foregrounds and instrumental noise. This is done by computing the linear combination of input maps that minimizes the variance in a basis spanned by a particular class of spherical wavelets called needlets
\citep{narcowich06localizedtight}.  Needlets allow localized filtering in both pixel space and harmonic space.  Localization in pixel space allows the weights of the linear combination to adapt to local conditions of foreground contamination and noise, whereas localization in harmonic space allows the method to favour foreground rejection on large scales and noise rejection on
small scales.  Needlets permit the weights to vary smoothly on large scales and rapidly on small scales, which is not possible by cutting the sky into zones prior to processing \citep{2009A&A...493..835D}. The \nilc\ pipeline is applicable to scalar fields on the sphere, hence we work separately on maps of temperature and the $E$ and $B$ modes of polarization.  The decomposition of input polarization maps into $E$ and $B$ is done on the full sky.  At the end, the CMB $Q$ and $U$ maps
are reconstructed from the $E$ and $B$ maps. Further details of the method are provided in Appendix~\ref{app:nilc}.

The \nilc\ pipeline employed in the \Planck\ 2018 analysis is essentially unchanged from that employed in the 2015 analysis; we therefore refer to \citet{planck2014-a11} and references therein for full details.

\subsection{\sevem} 
\label{sec:sevem}

\sevem\ \citep{leach2008,fernandez-cobos2012} is an implementation of an internal template-cleaning approach in real space. It has been used in the previous \Planck\ releases to produce clean CMB maps in both intensity and polarization, and has been demonstrated to provide robust results. A detailed description of the \sevem\ pipeline can be found in Appendix~\ref{app:sevem}.

The starting point for \sevem\ is a set of internal templates typically constructed as difference maps between two neighboring
\Planck\ channels convolved to the same resolution, ensuring that the CMB signal vanishes. These templates trace the foreground contaminants at the corresponding frequency ranges. Next, a linear combination of such templates is then subtracted from some set of CMB-dominated frequency maps, typically 70 to 217\,GHz for \Planck. The coefficients of the linear fit are derived by minimizing the variance of the clean map outside a given mask. A final, co-added CMB map is obtained
by combining individually-cleaned frequency maps in harmonic space. 

\sevem\ is also able to produce cleaned CMB maps at specific channels. Individually-cleaned frequency CMB maps are  useful to test the robustness of results versus the presence of foregrounds and/or systematics, for instance for isotropy and statistics estimators \citep{planck2014-a16} or the integrated Sachs-Wolfe effect stacking analysis \citep{planck2014-a26}. They are also valuable to construct cross-frequency estimators, which allow one to minimize the impact of certain types of systematic effects (e.g., possible correlated noise in data splits). In addition, they can be used to search for frequency-dependent effects in the CMB itself, such as those arising from relativistic boosting \citep{planck2013-pipaberration} or the Sunyaev-Zeldovich effect \citep{sunyaev:1970}, although for this type of analysis the contribution of the templates (which would contain a certain level of any effect that is not constant with frequency) to the cleaned maps should be taken into account.
 
Since the 2015 release, we have introduced two significant improvements to the \sevem\ pipeline for polarization. First, in the previous release we produced cleaned maps at three frequencies, 70, 100, and 143\,GHz, and the final map was produced by combining the cleaned 100 and 143-GHz maps. However, given the improvements in the new \Planck\ polarization data, we are now also able to robustly clean the 217-GHz channel map, and this is now included in the final combination. As a result, the signal-to-noise ratio of the cleaned \sevem\ CMB polarization map is significantly improved with respect to the previous version. Second, in the updated pipeline, we now produce polarization maps at full resolution ($N_{\rm side}=2048$), whereas in the last release all polarization maps were constructed at N$_{\rm side}$=1024. However, recognizing the fact that the 217-GHz channel is likely to be somewhat more susceptible to large-scale systematic residuals and calibration uncertainties due its
higher foreground levels than the two lower frequencies \citep{planck2016-l03}, we introduce at the same time a relative
down-weighting of the 217-GHz channel on the largest scales. In summary, these modifications yield significantly improved
\sevem\ polarization maps, both in terms of the combined CMB map and individually cleaned frequency maps. Regarding intensity, the \sevem\ pipeline is essentially identical to that used in the previous release; however, we now also provide a cleaned 70-GHz map in intensity. In addition to the final CMB map, \sevem\ therefore now provides the complete set of $\{T,Q,U\}$ CMB maps for each of the four frequency channels between 70 and 217\,GHz.

\subsection{\smica} 
\label{sec:smica}

\smica\ (Spectral Matching Independent Component Analysis) is described in \citet{cardoso2008}, and details regarding
the actual implementation used in the following analysis (pre-processing, masking and mask correction, beam correction,
binning, possible re-calibration, etc.) are provided in Appendix~\ref{app:smica}.

\smica\ synthesizes CMB $\{T, E, B\}$ maps from spherical harmonic coefficients $\hat s_\lm$ obtained by combining the coefficients of $\ncha$ frequency maps with an $\ell$-dependent $\ncha\times 1$ vector of weights $\vec w_\ell$,
\begin{equation}\label{eq:smica:ilc}
  \hat s_\lm = \vec{w}_\ell \adj \vec{x}_\lm
  \quad\text{where }\quad
  \vec{w}_\ell = \frac{\bC_\ell\inv \ba}{\ba\adj \bC_\ell\inv \ba}.
\end{equation}
Here the $\ncha\times 1 $ vector $\vec a$ describes the emission law of the CMB, and the $\ncha\times \ncha$ spectral covariance matrix $\bC_\ell$ contains (estimates of) all auto- and cross-spectra of the $\ncha$ input maps.
On small angular scales, where a large number of harmonic coefficients are available, $\bC_\ell$ may be accurately estimated as
\begin{equation}\label{eq:smica:scm}
  \widehat\bC_\ell = \frac1{2\ell+1}\sum_m \vec x_\lm^{\vphantom{\dagger}}\vec x_\lm\adj ,
\end{equation}
which is used ``as is,'' in Eq.~(\ref{eq:smica:ilc}). 
On large angular scales, we resort to a parametric model $\bC_\ell(\theta)$ of the spectral covariance matrices in order to reduce the estimation variance and mitigate the  effects of chance correlation between the CMB field and the foregrounds.
The model is adjusted to the data by selecting best-fit parameters $\theta$ obtained as
\begin{equation}
  \label{eq:smica:mmcrit}
  \hat\theta
  =
  \arg\min_\theta \sum_\ell (2\ell+1) 
  \left[ 
    \mathrm{Tr}\left(\widehat\bC_\ell \bC_\ell (\theta)\inv \right)
    +
    \log\det \bC_\ell (\theta) 
  \right]
  .
\end{equation}
The minimization in Eq.~(\ref{eq:smica:mmcrit}) is equivalent to maximizing the joint likelihood of the $\ncha$ input maps assuming that they follow a Gaussian isotropic distribution characterized by the spectra and cross-spectra collected in the spectral covariance matrices $\bC_\ell(\theta)$.  For a motivation of this likelihood, see~\citet{LVAICA2017}.

\begin{figure}
  \begin{center}
    \begin{tabular}{ccc}
      \includegraphics[width=0.8\linewidth]{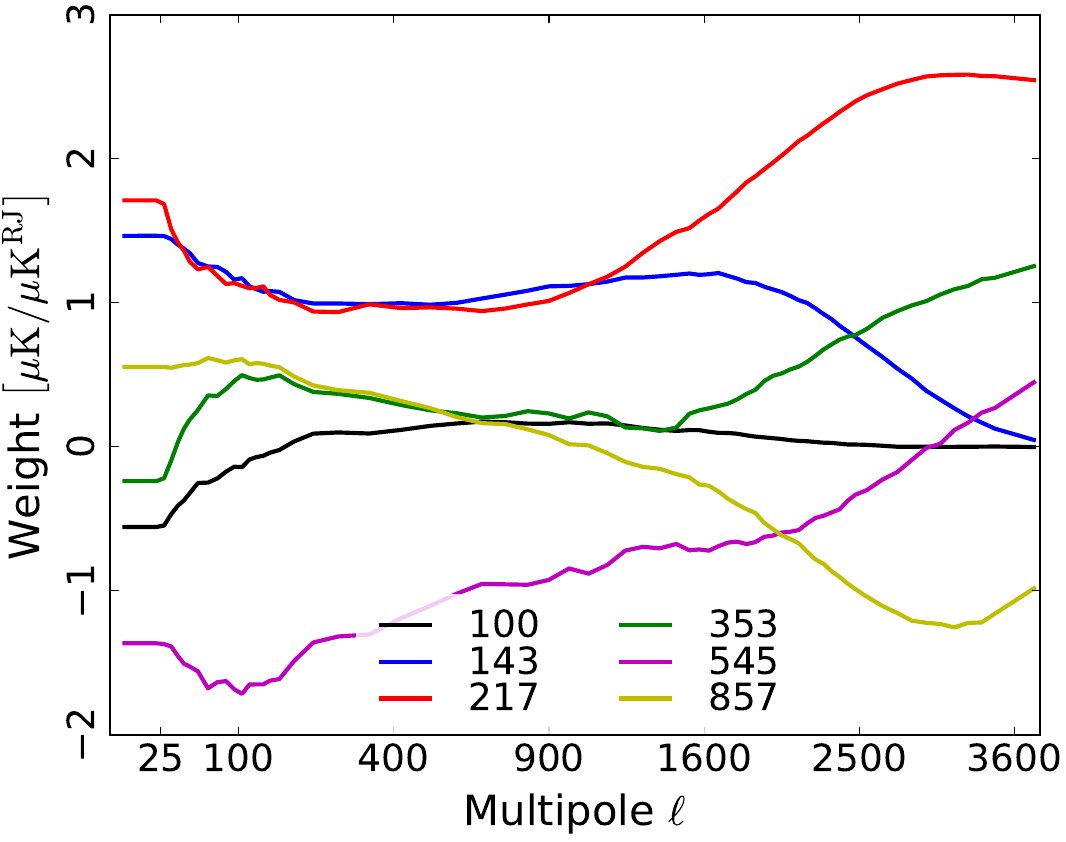}\\
      \includegraphics[width=0.8\linewidth]{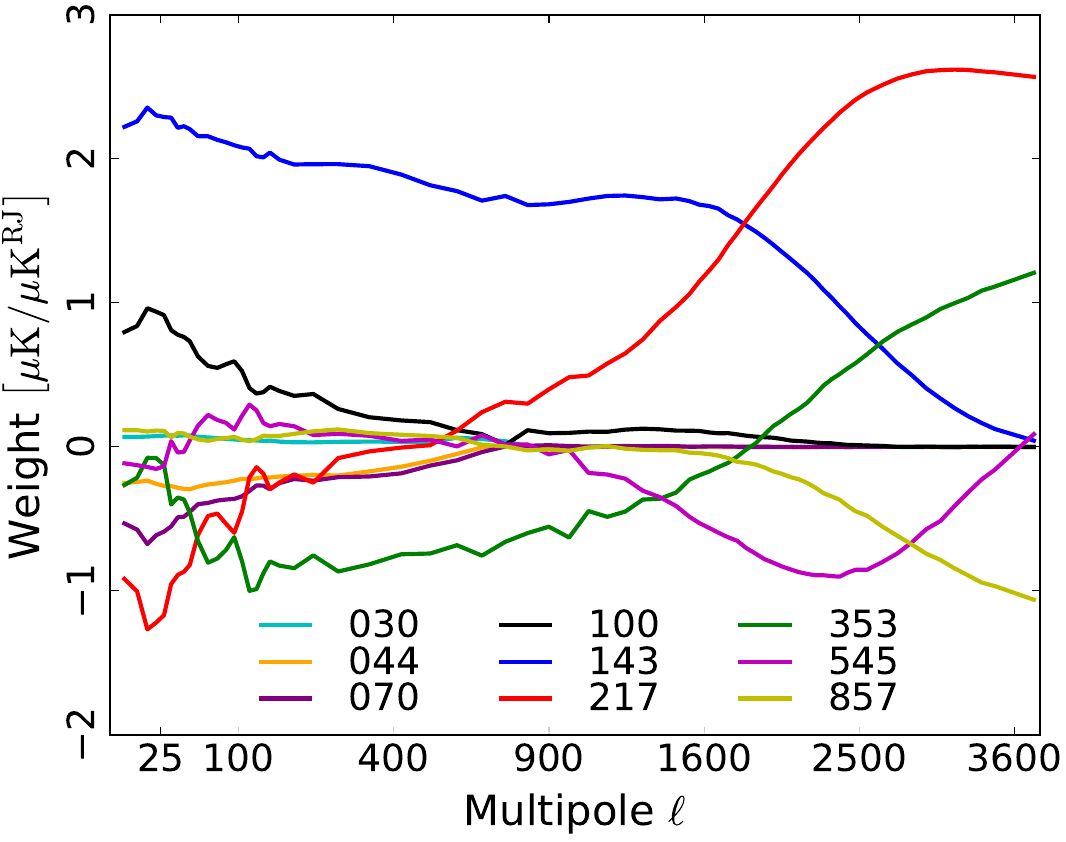}\\
      \includegraphics[width=0.8\linewidth]{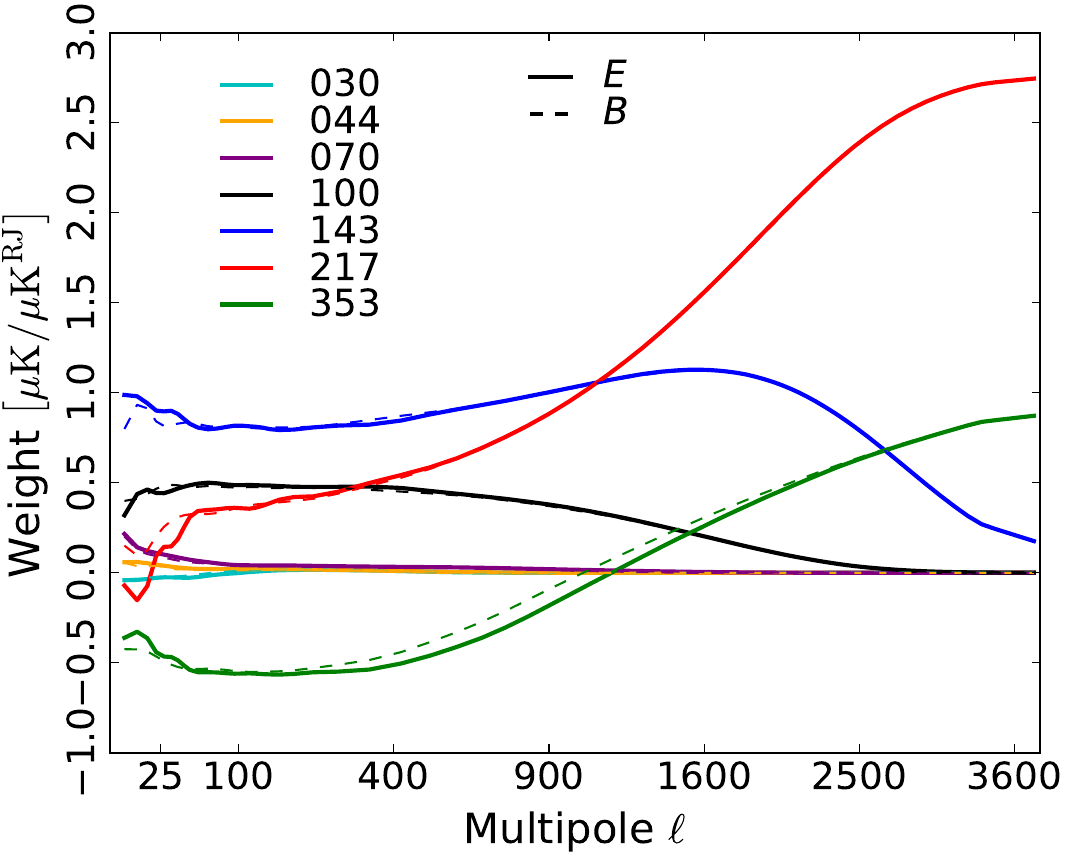}\\
    \end{tabular}
  \end{center}
  \caption{\smica\ harmonic weights used to obtain the temperature $X_\textrm{high}$ map (top), $X_\textrm{full}$ map (middle), and polarization map (bottom).}
  \label{fig:smica_T_weights}
\end{figure}

The spectral model fitted by \smica, $\Cp_\ell(\theta)$, is agnostic, as it assumes only that the foreground emission can be
described by an unconstrained $\nfg$-dimensional component with a covariance matrix of the form
\begin{equation}
  \label{eq:smica:model}
  \bC_\ell(\theta)
  = 
  \left[\begin{array}{cc} \vec a  & \tens F \end{array}  \right]
  \left[\begin{array}{cc} C_\ell^\mathrm{cmb} & 0  \\ 0  & \tens P_\ell \end{array}  \right]
  \left[\begin{array}{cc} \vec a  & \tens F \end{array}  \right]\adj
  +
  \tens N_\ell.
\end{equation}
Here the $\ncha\times\nfg$ matrix $\tens F$ represents the foreground emissivities, which are $\ell$-independent, and the $\nfg\times\nfg$ matrix $\tens P_\ell$ contains the foreground auto- and cross-spectra.  The diagonal matrix $\tens N_\ell$ represents the noise contribution, and $\theta$ contains whatever parameters are needed to determine the quantities $C_\ell^\mathrm{cmb}$, $\vec a$, $\tens F$, $\tens P_\ell$, and $\mathrm{diag}(\tens N_\ell)$.  In most cases, a \smica\ fit
is conducted with $\vec a$ fixed to assumed known values (i.e., assuming perfect calibration) and leaving all other parameters free.  $\tens P_\ell$ is only constrained to be positive.  In other words, foreground spectra (emissivities and angular spectral behaviour) and their correlations are freely fitted by \smica. 

In this release, however, we also consider two variations that include constraints on foreground emissions. The first of these is used to produce an SZ-free CMB map in intensity (see Appendix~\ref{app:smica}) \textcolor{black}{used in~\cite{planck2016-l08}}, and the second results in thermal dust and synchrotron maps in polarization (see Sect.~\ref{sec:foregrounds}). No attempt is made to reconstruct temperature foregrounds, since the combination of synchrotron, free-free, spinning and thermal dust, and CO emission is intrinsically much more tightly coupled and difficult to disentangle than synchrotron and thermal dust emission in polarization.

Since the last release, changes have been introduced for both intensity and polarization maps.  Starting with the temperature case, the most important change in this release is the introduction of hybrid CMB rendering, merging two different CMB maps produced independently by the \smica\ pipeline.  The first CMB map, $X_\text{high}$, is designed to describe the cleanest region of the sky and intermediate-to-small angular scales.  It is obtained from all six HFI channels using a foreground dimension of $\nfg=4$.  The second CMB map, $X_\text{full}$, is designed to describe the full sky and all harmonic scales.  It includes all nine \Planck\ frequency channels using a maximal foreground dimension of $\nfg=8$.  The final hybrid CMB map $X$ is then computed by merging $X_\text{high}$ and $X_\text{full}$ according to
\begin{equation}
  \label{eq:smica:merge}
  X 
  = \mathcal{P} X_\text{high} + (\mathcal{I}-\mathcal{P})\, X_\text{full}
  \ =\ 
  X_\text{full} + \mathcal{P} (X_\text{high}-X_\text{full}),
\end{equation}
where $\mathcal{P}$ is a linear operator that smoothly removes large harmonic scales, and masks out an area close to the Galactic plane.  Hence, in the resulting hybridized map, the multipoles of highest degree and the areas of highest Galactic latitude are provided by $X_\text{high}$, while the remaining information is provided by $X_\text{full}$.  In practice, the hybridization operator $\mathcal{P}$ is implemented by high-pass filtering in the harmonic domain (with a transfer function
that smoothly transitions from $0$ to $1$ according to an arc-cosine function over the multipole range $50\leq \ell\leq150$), followed by multiplication by an apodized Galactic mask that is similar to the mask used at 100\,GHz in the \Planck\ 2018 likelihood ({\tt \Plik})  \citep[see][for details]{planck2016-l05}. 

Hybridization of two CMB renderings has several benefits compared to using a single set of harmonic weights over all areas of the sky.  First, the data suggest it: the \smica\ weights are quite different if they are based on spectral statistics computed over the full sky rather than over a region with much lower foreground contamination.  This is the rationale behind \nilc, which  extends the idea to many more than the two sky regions considered regions considered by \smica.  Second, the reason for leaving out the LFI channels in producing $X_\text{high}$ except at large angular scales is that \smica\ would put very small weights on those channels (this is not the case when the weights are based on statistics computed for $X_\text{full}$, as
seen on Fig.~\ref{fig:smica_T_weights} which shows a significant contribution from the 70-GHz channel).  We could still include those channels and let \smica\ automatically down-weight them, but by excluding the channels with the lowest resolution, we  avoid large, `low-resolution' holes in the common point source mask, and therefore in the final CMB map.  Finally, hybridization matches well the high-$\ell$ TT likelihood function in \Plik, uses a \textcolor{black}{low-foreground-contaminated} fraction of the sky, does not include LFI channels, and involves only high frequency foregrounds. \textcolor{black}{The spectral weights, $\vec{w}_\ell$, for temperature (both full-sky and high latitudes) and polarization are shown in Fig.~\ref{fig:smica_T_weights}}.   

\smica\ adopts its own relative calibration between frequency channels.  In 2015, this process was applied to frequency
channels from 44 to 353\,GHz; however, since then we have found that the uncertainty in the 44-GHz channel was larger than
expected, and that the previously reported value was inaccurate (see Fig.~\ref{fig:smica_diff2015}).  In the new release, we adopt a more conservative approach, and limit re-calibration to 70, 100, and 217\,GHz, taking the 143-GHz channel as a reference; see Appendix~\ref{app:smica:temp} for further details.

For polarization, we have introduced two changes since the previous release.  First, the CMB polarization maps are now generated by independently processing $E$ and $B$ modes, while in 2015 they were jointly fitted and filtered.  Second, we run two independent \smica\ fits, one targeted at CMB extraction, the other at foreground separation.

For CMB extraction, we conduct a fit using a maximal foreground dimension of $\nfg=7-1=6$, which makes $[\vec a \ \tens F ]$ a square matrix.  This is the largest dimension supported blindly (i.e., without any constraint on the foreground contribution)
by \smica, given the number of available polarized channels. 

For foreground separation, we conduct a separate fit using a foreground model of dimension $\nfg=2$, implicitly targeting
synchrotron and dust emissions.  The degeneracy of the \smica\ foreground model (Eq.~\ref{eq:smica:model}) can then be  fixed by requesting that synchrotron (thermal dust) emission should be negligible at 353\,GHz (30\,GHz);
Appendix~\ref{app:smica} describes the implementation details.  This analysis yields, without any other prior information, the angular spectra and emissivities of both foreground components and the corresponding synchrotron and dust maps.  The results are summarized in Sect.~\ref{sec:foregrounds}.  Note that in 2015, a foreground model at $\nfg=2$ dimensions for
capturing synchrotron and thermal dust emissions was already explored, but no maps were released (although a dust comparison appeared in \citealp{planck2014-a12}) because additional ``foreground dimensions'' were clearly needed to accommodate the systematic errors.  In 2018, we use the same dimensions as in 2015 (a \smica\ fit with maximal dimension for CMB cleaning and a \smica\ fit with $\nfg=2$ for dust and synchrotron maps); however, contrary to 2015, the $\nfg=2$
fit yields a clean CMB reconstruction, almost as clean as when using the maximal foreground dimension.  For that reason, this \Planck\ release includes \smica-derived synchrotron and dust polarized maps.

\subsection{\gnilc}
\label{sec:gnilc}

The above four methods were the standard CMB extraction algorithms in each of the three \Planck\ data releases.  In this release, we also consider the Generalized Needlet Internal Linear Combination (\gnilc; \citealt {Remazeilles2011b}) method as a foreground extraction algorithm.  \gnilc\ is not designed to extract CMB information from the data.\footnote{I\gnilc\ should not be confused with the   "Constrained ILC" method \citep{Remazeilles2011a}, which was designed  to extract SZ-free CMB temperature anisotropies.}  \gnilc\  is a wavelet-based component-separation method that generalizes the \nilc\ method by exploiting not only the \emph{spectral} information (SED) but also the \emph{spatial} information (angular power spectra) from non-astrophysical components (cosmic infrared background, CIB, CMB, and instrumental noise) to extract clean estimates of the correlated emission from Galactic foregrounds, with reduced contamination from CIB, CMB, and noise. This additional spatial discriminator adopted by \gnilc\ enables in particular disentanglement of emission components that suffer from spectral degeneracies, such as modified blackbody emissions like the CIB and Galactic dust.  \gnilc\ has been successfully applied to \Planck\ 2015 intensity data to disentangle Galactic thermal dust emission and CIB anisotropies over the entire sky \citep{planck2016-XLVIII}. In this paper, CMB and instrumental noise were also filtered out from the \Planck\ \gnilc\ dust intensity map by using the same strategy as for CIB removal.

In this work, we apply \gnilc\ to the \Planck\ 2018 polarization data in order to extract the Stokes parameters $Q$ and $U$ of Galactic thermal dust polarization, while removing the contamination from CMB polarization and instrumental noise over the entire sky.   $I$, $Q$, and $U$ dust maps have been produced in a self-consistent way by processing the seven \Planck\ polarized channels (30 to 353\,GHz).  The reason for discarding the 545- and 857-GHz channels is as follows. The main characteristic of the \gnilc\ method is to estimate the local number of independent foreground degrees of freedom over the sky and over angular scales. The estimated dimension of the foreground subspace depends on the local signal-to-noise ratio in the $9\times 9$ (intensity) or $7\times 7$ (polarization) observation space of the frequency-by-frequency data covariance matrix. In some parts of sky where the data are found by \gnilc\ to be fully compatible with CIB, CMB, and noise at small angular scales, the dimension of the Galactic foreground subspace can go down to zero. The result of this is that the \gnilc\ dust products have a variable resolution over the sky, with the local FWHM fully determined and publicly released \citep{planck2016-XLVIII}. However, because of decorrelation effects, the local dimension of the foreground subspace found by \gnilc\ will be larger in a 9-dimensional space of observations (30--857\,GHz) than in a 7-dimensional space of observations (30--353\,GHz), so that the effective local resolution of the \gnilc\ dust products will be different over the sky for intensity and polarization. For the purpose of polarization fraction studies in the 2018 release \citep{planck2016-l11B}, we prefer to have the same local resolution over the sky both for intensity and polarization, hence our choice of processing with \gnilc\ the same data set for $I$, $Q$, and $U$, namely the seven \Planck\ polarized channels (30--353\,GHz).

Omission of the 545- and 857-GHz channels limits the ability of \gnilc\ to clean CIB anisotropies in the \Planck\ 2018 dust intensity map compared to the \Planck\ 2015 dust intensity map \citep{planck2016-XLVIII}, for which the full set of unpolarized
channels (30--857\,GHz) and the IRAS map were used in the component-separation pipeline. For analyses of dust intensity (e.g., dust optical depth, emissivity, and temperature), we recommend use of the \Planck\ 2015 \gnilc\ dust intensity map, which has reduced CIB contamination. Conversely, for analysis of dust polarization (e.g., polarization fraction) we recommend use of
\gnilc\ 2018 $I$, $Q$, and $U$ maps.

\section{Data selection, preprocessing, splits, and simulations}
\label{sec:inputs}

\subsection{Frequency maps}

The low-level data processing and mapmaking algorithms adopted for the current release are described in detail in
\citet{planck2016-l02} and \citet{planck2016-l03}. For the LFI maps at 30, 44, and 70\,GHz, there are only minor changes compared to the previous release, the most important of which is a better calibration procedure that explicitly accounts for polarized foregrounds in the calibration sources. For HFI, more significant changes have been implemented, all designed to suppress instrumental systematics at various scales. These include better ADC and transfer-function corrections, and explicit bandpass corrections employing a detailed foreground model.

A particularly important problem for both LFI and HFI with respect to polarization reconstruction is bandpass mismatch between multiple detectors within a single frequency channel. The issue may be summarized as follows.  In order to solve for both temperature and linear polarization in each pixel on the sky, a total of three parameters per pixel, it is necessary to include information from at least three polarization-sensitive detectors in any given mapmaking operation.  The polarization signal is estimated by taking pairwise differences between the signals observed by these detectors, while accounting for the relative orientation of their polarization detector angles at any given time. However, there are other effects in addition to true sky polarization signals that may induce effective signal differences between detectors. The largest of these is different effective bandpasses.  Since each detector has a slightly different frequency response function, each detector observes a slightly different foreground signal.  Unless explicitly accounted for during mapmaking, these differences create a spurious polarization signal in the maps.

In the LFI mapmaking procedure, this effect is accounted for in two different ways, as described in \citet{planck2016-l02}. First, for gain estimation, an iterative scheme is established, in which a proper foreground model is derived jointly with the sky maps using \commander.  Each iteration of this procedure consists of three individual steps. First, a gain model is established for each radiometer, accounting for the orbital and Solar dipoles as well as astrophysical foregrounds as estimated by \commander. Second, frequency maps are derived based on this gain model using {\tt MADAM}  \citep{keihanen2005, planck2014-a07}, a well-established destriper. Third, these frequency maps are used by \commander\ to derive a new foreground model.  A total of four such iterations are used to derive the final LFI maps; however, even after these iterations there may be non-negligible large-scale residuals present in the 70-GHz sky map, as described by \citet{planck2016-l02}. To account for this, a gain correction template, based on differences between consecutive iterations, is subtracted from the final LFI 70-GHz map, with an amplitude derived from a low-resolution likelihood fit \citep{planck2016-l05}.  \textcolor{black}{These procedures account for biases in the time-variable gain solutions; however, they do} not remove direct temperature-to-polarization leakage from bandpass mismatch. That effect, which is stationary on the sky, is corrected through use of static templates, as described in detail in \citet{planck2014-a03}.  The same procedure is applied to the LFI sky maps in the current release with an updated foreground model \citep{planck2016-l02}.

For HFI a different but related approach is adopted.  The 2015 \commander\ temperature model is used to explicitly adjust the effective bandpass response of all bolometers within a frequency channel, by subtracting a small fraction of each foreground signal (thermal dust, free-free, and CO emission, but not synchrotron or spinning dust emission) from the individual bolometer timestreams. These ``foreground-equalized'' timestreams are then combined into a single frequency map by standard destriping.  Since only a spin-0 temperature signal is subtracted in this procedure, the resulting polarization maps are unbiased with respect to foreground leakage, to the extent that the foreground model is accurate. However, the resulting temperature maps will be very slightly biased, in the sense that the predicted bandpass response of a given map does not perfectly match the observed signal, and this causes complications for any method that explicitly employs such information. In the current paper, this applies to \commander\ and \gnilc. The three remaining methods (\nilc, \sevem\ and \smica) do not explicitly use such information.

An additional complication arises from the updated 2018 HFI mapmaking procedure, due to the fact that the single-bolometer maps produced by the latest processing are not reliable for component-separation purposes \citep{planck2016-l03}. Since the CO emission lines are very narrow, their measured amplitudes are very sensitive to small variations in bandpass shape between individual detectors. In 2015, this sensitivity was exploited to extract line-emission maps at each of the affected frequencies. However, since single-bolometer maps are not available in 2018, this is no longer possible. The new processing  represents a conscious choice of optimizing the polarization extraction at non-negligible expense in terms of our ability to perform high-fidelity astrophysical foreground reconstruction with temperature maps. For individual foreground components in temperature, we therefore recommend continued usage of the \Planck\ 2015 data products.

To summarize the overall data selection, all diffuse component separation codes \textcolor{black}{except \gnilc} employ all nine \Planck\ frequency maps between 30 and 857\,GHz in temperature, and all seven frequency maps between 30 and 353\,GHz in polarization, for the 2018 analysis.  \textcolor{black}{\gnilc\  uses only the seven lowest frequencies in temperature in order to match the polarization analysis}. For the LFI polarization maps, we apply a set of template corrections that account for bandpass mismatch and gain corrections, as described in \citet{planck2016-l02}, while no additional corrections are applied to the HFI maps. All maps are defined by the \healpix\footnote{\url{http://healpix.jpl.nasa.gov}} pixelization \citep{gorski2005}.

\begin{figure*}
\begin{center}
  \includegraphics[width=0.33\textwidth]{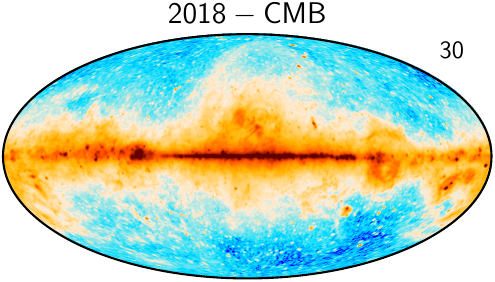}
  \includegraphics[width=0.33\textwidth]{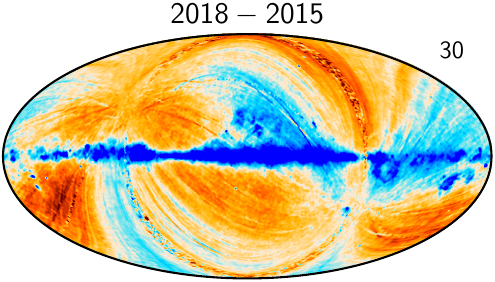}
  \includegraphics[width=0.33\textwidth]{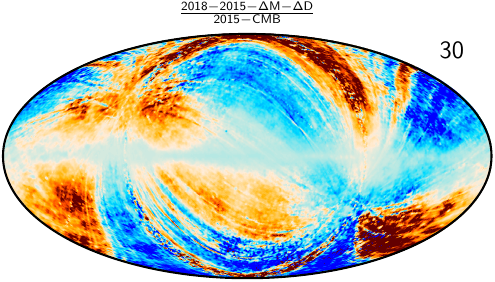}\\
  \includegraphics[width=0.33\textwidth]{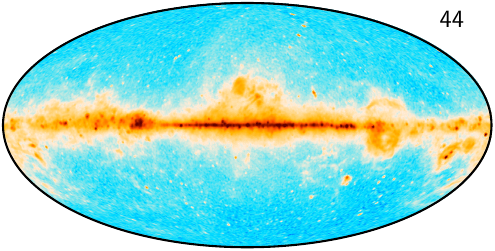}
  \includegraphics[width=0.33\textwidth]{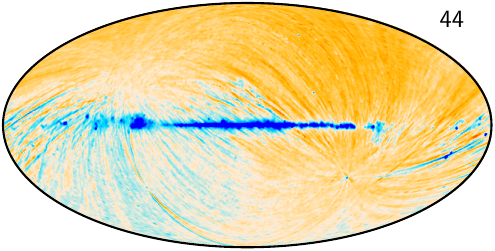}
  \includegraphics[width=0.33\textwidth]{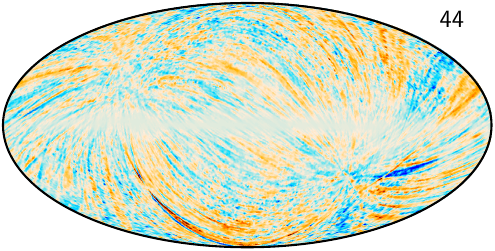}\\
  \includegraphics[width=0.33\textwidth]{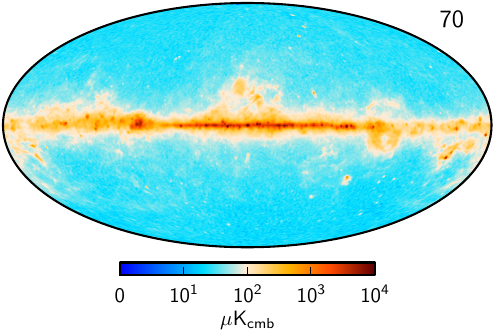}
  \includegraphics[width=0.33\textwidth]{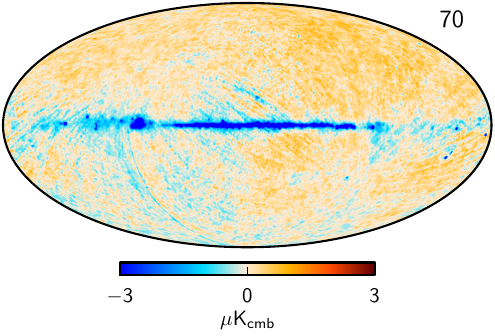}
  \includegraphics[width=0.33\textwidth]{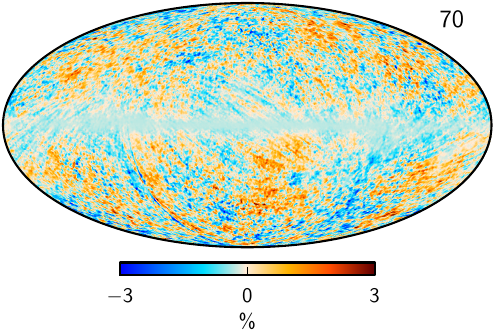}
\end{center}
\caption{Comparison of 2018 and 2015 LFI temperature maps. Columns show, from left to right: (1)~the difference between the 2018 intensity maps and the 2018 \commander\ CMB map; (2)~the  difference between the 2018 and 2015 frequency maps; and (3)~the fractional difference between the 2018 and 2015 frequency maps. Note the different temperature scales.  In the third column, $\Delta M$ and $\Delta D$ denote the relative monopole and dipole differences between the 2018 and 2015 sky maps. Rows indicate results for each of the three LFI frequency channels. All maps are smoothed to a common resolution of $1^{\circ}$ FWHM. }
\label{fig:dx11_vs_dx12_lfi}
\end{figure*}

\begin{figure*}
\begin{center}
  \includegraphics[width=0.33\textwidth]{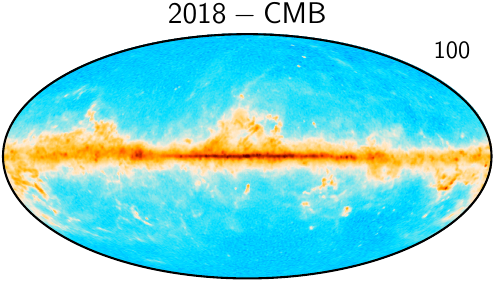}
  \includegraphics[width=0.33\textwidth]{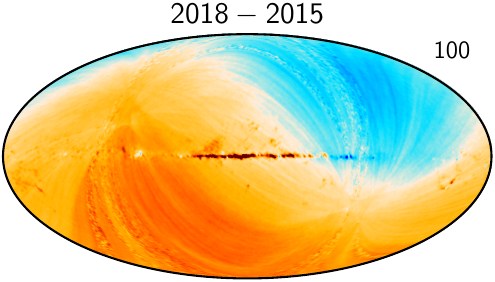}
  \includegraphics[width=0.33\textwidth]{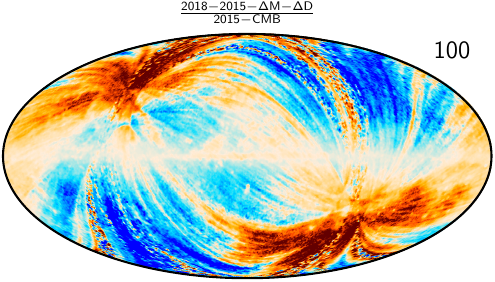}\\
  \includegraphics[width=0.33\textwidth]{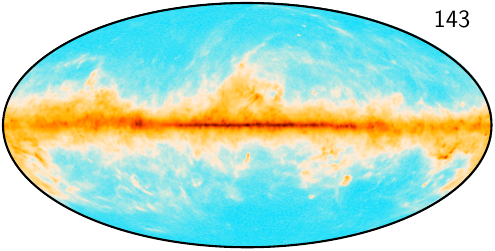}
  \includegraphics[width=0.33\textwidth]{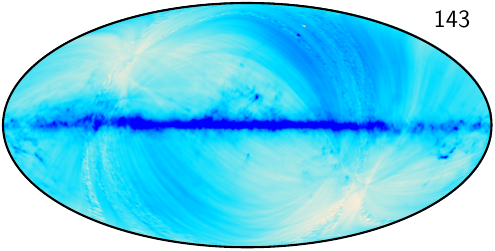}
  \includegraphics[width=0.33\textwidth]{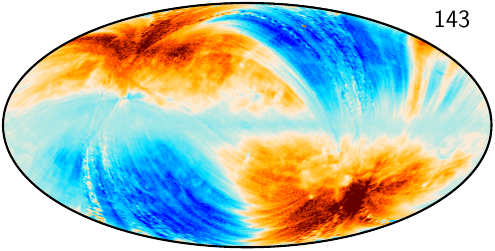}\\
  \includegraphics[width=0.33\textwidth]{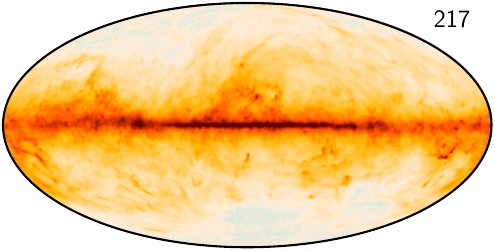}
  \includegraphics[width=0.33\textwidth]{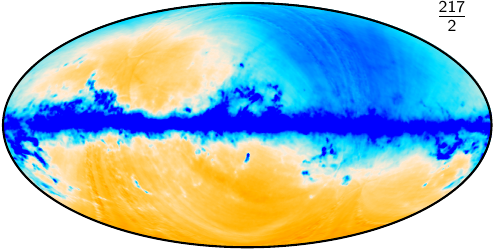}
  \includegraphics[width=0.33\textwidth]{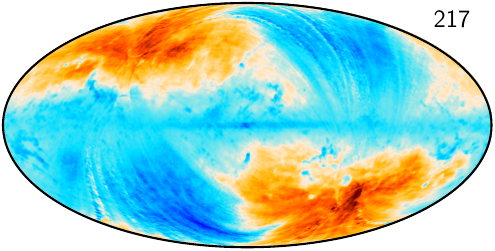}\\
  \includegraphics[width=0.33\textwidth]{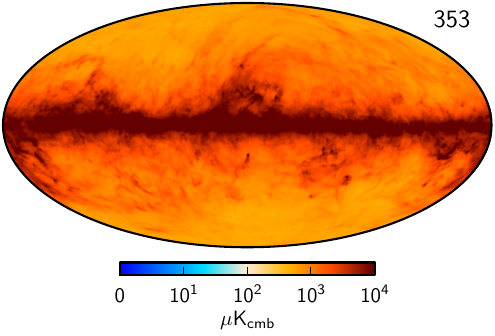}
  \includegraphics[width=0.33\textwidth]{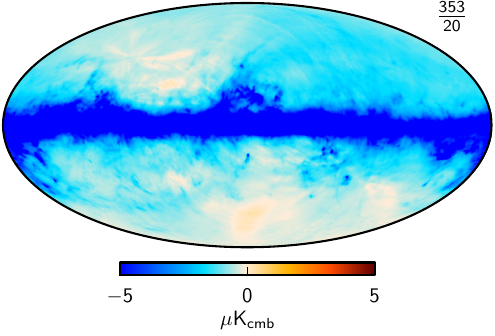}
  \includegraphics[width=0.33\textwidth]{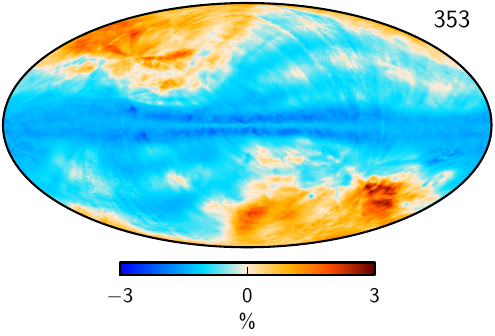}\\
  \includegraphics[width=0.33\textwidth]{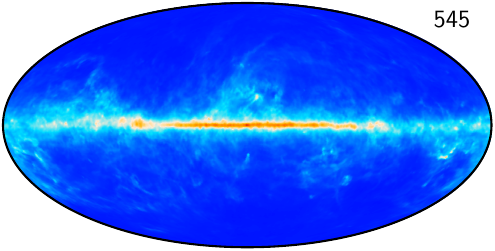}
  \includegraphics[width=0.33\textwidth]{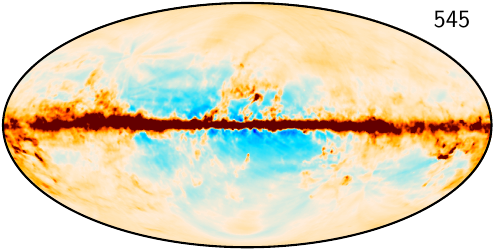}
  \includegraphics[width=0.33\textwidth]{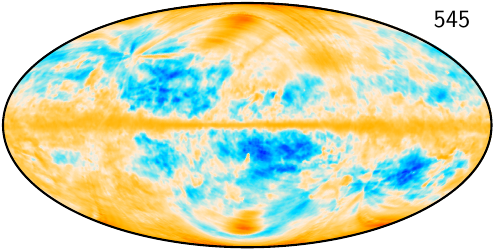}\\
  \includegraphics[width=0.33\textwidth]{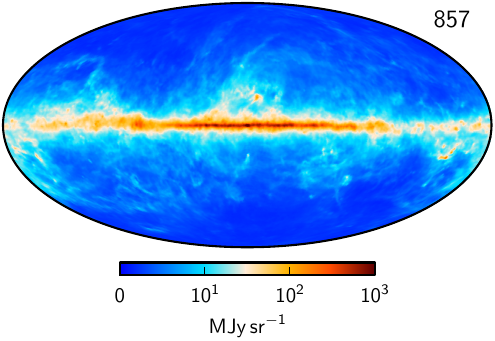}
  \includegraphics[width=0.33\textwidth]{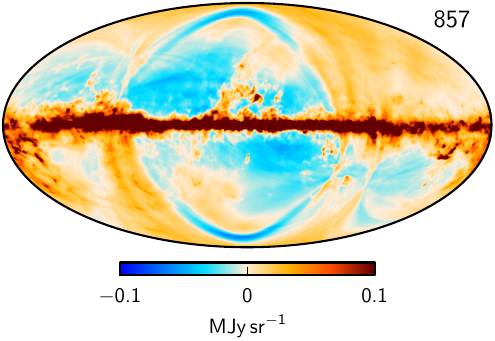}
  \includegraphics[width=0.33\textwidth]{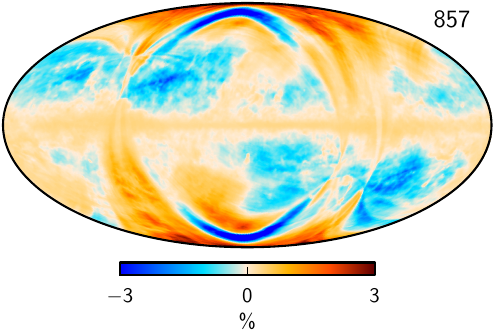}
\end{center}
\caption{Comparison of 2018 and 2015 HFI temperature maps, similar to Fig.~\ref{fig:dx11_vs_dx12_lfi} for LFI. Columns
show, from left to right: (1)~the difference between the 2018 intensity maps and the 2018 \commander\ CMB map; (2)~the 
difference between the 2018 and 2015 frequency maps; and (3)~the fractional difference between the 2018 and 2015 frequency maps. Note the different temperature scales.  Rows indicate results for each of the six HFI frequency channels. All maps are smoothed to a common resolution of $1^{\circ}$ FWHM.  Note that the 217- and 353-GHz difference maps have been scaled by factors of $1/2$ and $1/20$, respectively, to conform numerically to the same range as the 100- and 143-GHz maps.}
\label{fig:dx11_vs_dx12_hfi}
\end{figure*}

\begin{figure*}
\begin{center}
  \includegraphics[width=0.24\textwidth]{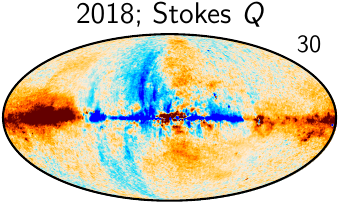}
  \includegraphics[width=0.24\textwidth]{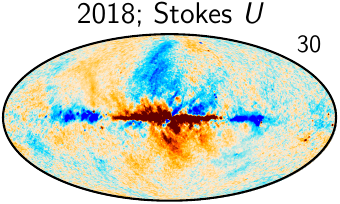}
  \includegraphics[width=0.24\textwidth]{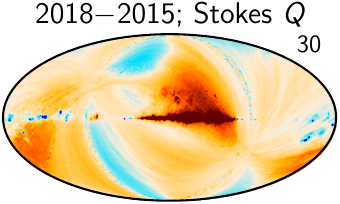}
  \includegraphics[width=0.24\textwidth]{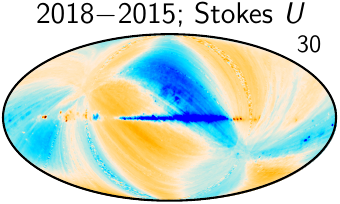}\\
  \includegraphics[width=0.24\textwidth]{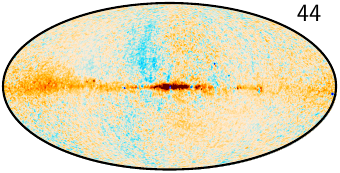}
  \includegraphics[width=0.24\textwidth]{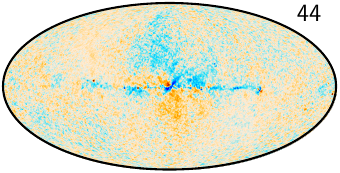}
  \includegraphics[width=0.24\textwidth]{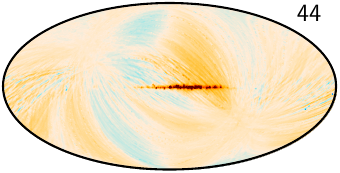}
  \includegraphics[width=0.24\textwidth]{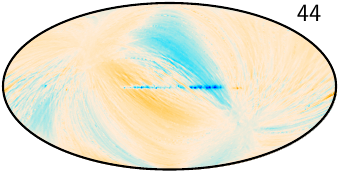}\\
  \includegraphics[width=0.24\textwidth]{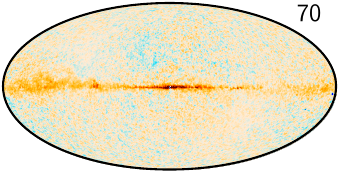}
  \includegraphics[width=0.24\textwidth]{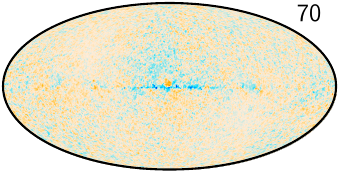}
  \includegraphics[width=0.24\textwidth]{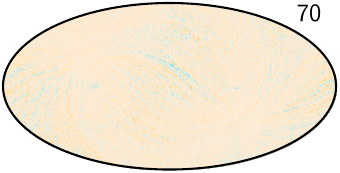}
  \includegraphics[width=0.24\textwidth]{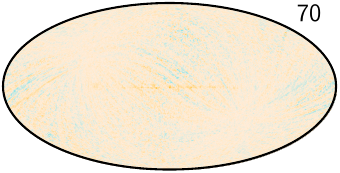}\\
  \includegraphics[width=0.24\textwidth]{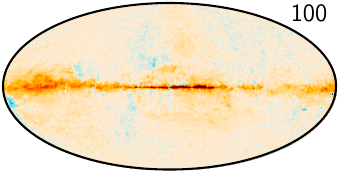}
  \includegraphics[width=0.24\textwidth]{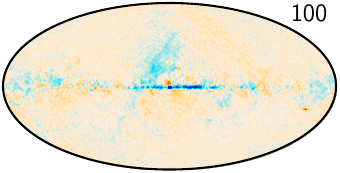}
  \includegraphics[width=0.24\textwidth]{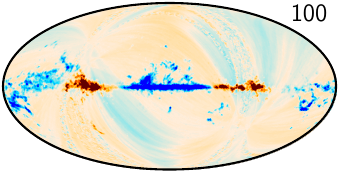}
  \includegraphics[width=0.24\textwidth]{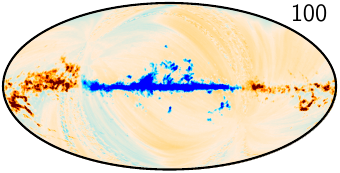}\\
  \includegraphics[width=0.24\textwidth]{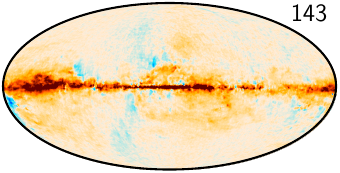}
  \includegraphics[width=0.24\textwidth]{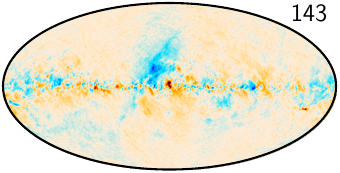}
  \includegraphics[width=0.24\textwidth]{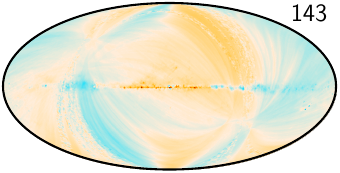}
  \includegraphics[width=0.24\textwidth]{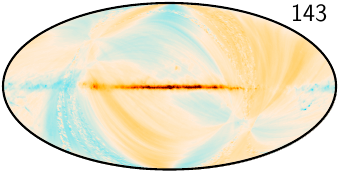}\\
  \includegraphics[width=0.24\textwidth]{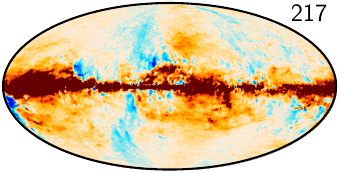}
  \includegraphics[width=0.24\textwidth]{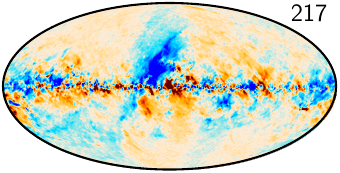}
  \includegraphics[width=0.24\textwidth]{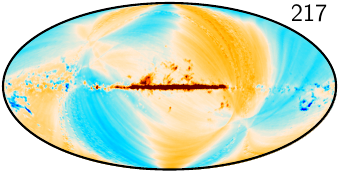}
  \includegraphics[width=0.24\textwidth]{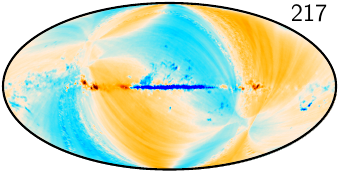}\\
  \includegraphics[width=0.24\textwidth]{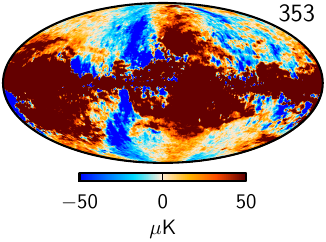}
  \includegraphics[width=0.24\textwidth]{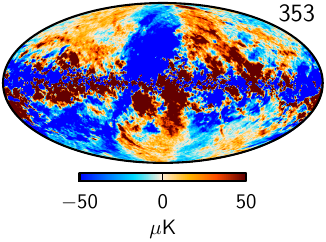}
  \includegraphics[width=0.24\textwidth]{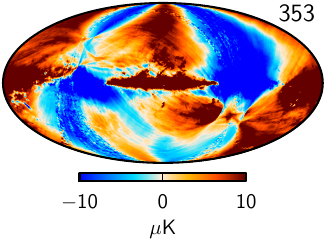}
  \includegraphics[width=0.24\textwidth]{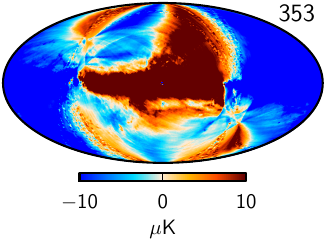}
\end{center}
\caption{Comparison of 2015 and 2018 polarization frequency maps. Columns show, from left to right: (1)~2018 Stokes $Q$ maps; (2)~2018 Stokes $U$ maps; (3)~Stokes $Q$ difference map between 2018 and 2015; and (4)~the Stokes $U$ difference map between 2018 and 2015. Note the different temperature scales.  Rows indicated results for each of the seven polarized \Planck\ frequency channels. All maps are smoothed to a common resolution of $1^{\circ}$ FWHM. }
\label{fig:dx11_vs_dx12_pol}
\end{figure*}

\subsection{Instrument characterization}

In addition to the raw frequency maps, each method requires various degrees of knowledge about the \Planck\ instrument itself. The most important characterization is the beam response of the individual frequency channels. These have been updated to reflect the latest changes in the data processing pipelines, and are described in \citet{planck2016-l02} and \citet{ planck2016-l03}.  We note that in the 2015 data release, CMB polarization maps for two of the methods (\commander\ and \sevem) were given at 10\arcm\ FWHM,  compared to 5\arcm\ FWHM  for the temperature maps; however, in this release all CMB maps in both polarization and temperature are provided at the maximum angular resolution of 5\arcm\ FWHM.

Each CMB map must also be associated with a statistical characterization of the instrumental noise. For this purpose, we
compute and analyse null maps derived from subsets of the full data set, as done in earlier releases. In the previous release, we focused on half-mission splits, yearly splits, and half-ring splits \citep{planck2014-a11}. In the current release, we drop the yearly split, since this behaves similarly to the half-mission split, and we replace the half-ring split with a so-called ``odd-even'' split, in which scanning rings from HFI are grouped according to odd or even pointing IDs. The odd-even split nullifies long-time-correlated signals, similarly to the half-ring split, but suffers less from inter-ring correlations. For LFI, we still adopt the same half-ring split as in 2015, but nevertheless refer to this split as ``odd-even,'' recognizing the different signal-to-noise ratios of the LFI and HFI maps. We consider this to be our best instrumental noise tracer among the splits, whereas the half-mission split represents the best instrumental systematics tracer. Simulations including either pure CMB signal or the sum of instrumental noise and residual systematics are individually propagated through each analysis pipeline, and these simulations form the basis of all subsequent goodness-of-fit tests.

As described in \citet{planck2016-l03}, the HFI polarization frequency maps are associated with a significant uncertainty regarding polarization efficiencies, corresponding in effect to an uncertainty in the overall calibration of the Stokes $Q$ and $U$ maps. Ideally, such polarization efficiencies would be perfectly accounted for during mapmaking. However, as reported by \citet{planck2016-l05}, a cosmological analysis of power spectra of the individual frequency maps suggests that small but notable residual calibration uncertainties may remain in a few channels. The reported best-fit correction values are $+0.7 \pm 1.0$\,\% (100\,GHz), $-1.7 \pm 1.0$\,\% (143\,GHz), and $+1.9 \pm 1.0$\,\% (217\,GHz).  For 353\,GHz, the foreground contribution is too large to allow a robust CMB-based measurement.  These corrections are only marginally statistically significant, therefore we do not apply them by default in this paper.  Instead, we compute results with and without the corrections, and report the difference between the two solutions as a known systematic error.  For the CMB, we find that the differences due to polarization efficiency uncertainties are small, while for polarized foregrounds, we find that the inclusion of polarization efficiencies changes the spectral index of thermal dust by $\Delta\beta_{\mathrm{d}}=-0.03$.  See Sect.~\ref{sec:foregrounds} for details.

\subsection{Treatment of unobserved pixels}
\label{sec:misspix}

As described in \citet{planck2016-l03}, the HFI split maps contain a non-negligible number of unobserved pixels at the full
$N_{\textrm{side}}=2048$ \healpix\ resolution. These are pixels that were either never seen by any bolometer at a given frequency, or for which the polarization angle coverage is too poor to support a reliable decomposition into the three Stokes parameters. For most methods considered in this paper,\footnote{\commander\ behaves differently from the other codes with respect to unobserved pixels.  It applies per-pixel inverse noise weighting per frequency channel, and unobserved pixels in a given channel are simply given zero weight in the parametric fits.} such unobserved pixels represent a notable algorithmic problem, and must be treated before analysis. For these methods, we simply replace all unobserved pixels in a given frequency map by the same pixels in a corresponding map downgraded to a \healpix\ resolution of $\nside=64$, corresponding to a pixel size of 55\arcm. Of course, this procedure introduces correlations between neighbouring unobserved pixels, and we therefore mask all high-resolution pixels after the analysis; separate masks for each data split are provided to account for this effect. The details of how the unobserved pixel mask has been generated are described in Sect.~\ref{sec:masks}. Finally, to account for possible leakage from unobserved to observed pixels during inter-analysis smoothing operations, we apply the same procedure to the reference simulations described below.

\subsection{Comparison between 2015 and 2018 frequency maps}

It is useful to compare the new 2018 frequency maps to the previous 2015 frequency maps. Structures seen in these difference maps should be expected to propagate into the corresponding CMB differences at some level. Starting with the temperature case, the left columns of Figs.~\ref{fig:dx11_vs_dx12_lfi} and \ref{fig:dx11_vs_dx12_hfi} show the differences between each 2018 frequency map and the 2018 \commander\ CMB solution.\footnote{We remove a common estimate of the CMB signal in order to highlight the foreground and residual monopole and dipole contents of each map. Visually identical results would be obtained by adopting any of the other solutions as a reference instead of \commander.} Overall, the behaviour is consistent with what has been found in earlier releases, with: an absolute foreground minimum around 70\,GHz; LFI monopoles of 10--20\muK; increasing HFI monopoles with frequency, corresponding to the expected offset due to the cosmic infrared background (CIB), which is manually introduced into the HFI frequency maps \citep{planck2016-l03}; and overall morphologies consistent with some combination of synchrotron, free-free, CO, and \textcolor{black}{thermal and spinning} dust emission.

More interesting are the second and third columns in each figure, which show the raw and the fractional differences between the 2018 and 2015 frequency maps, respectively. In the latter we have removed the best-fit offset and dipole outside a Galactic mask, defining the fractional difference, $f$, as
\begin{equation}
\f = \frac{\m^{2018} - \m^{2015} - \Delta M - \Delta D }{\m^{2015}-\m^{\mathrm{CMB}}},
\end{equation}
where $\m^{2018}$ is the new \Planck\ 2018 frequency map, $\m^{2015}$ is the \Planck\ 2015 map, $\Delta M$ and $\Delta D$ are the monopole and dipole differences between these maps, and $\m^{\mathrm{CMB}}$ is the \commander\ 2015 CMB temperature map.

Starting with the LFI 30-GHz difference maps, two effects stand out. At high latitudes, we see broad stripes following the
\Planck\ scanning pattern. These are due to an improved time-varying gain calibration procedure in the 2018 analysis that takes into account astrophysical foregrounds as computed by \commander, in an iterative gain-estimation$\rightarrow$mapmaking$\rightarrow$component-separation procedure. This new iterative scheme is one of the main new features of the LFI 2018 processing pipeline \citep{planck2016-l02}. A second effect is seen in the Galactic plane, where the 2018 amplitude is lower by about 0.2\,\% compared to 2015. This is due to re-estimation of the overall absolute calibration, due to a new estimate of the Solar CMB dipole \citep{planck2016-l01}.

Similar considerations hold for the 44-GHz channel, although with a significantly lower striping level. In fact, in this case the striping is sufficiently low to reveal a small residual dipole of about 1\muK\ in the raw difference map, directly showing the effect of the new Solar dipole estimate. Even smaller differences are seen in the 70-GHz channel, but in this case the iterative
foreground estimation process was not used, because the foreground level of this channel near the foreground minimum is too low to allow robust foreground estimation \citep{planck2016-l02}.

The HFI frequencies (Fig.~\ref{fig:dx11_vs_dx12_hfi}) show many qualitatively similar structures, in addition to a few unique HFI-type features. First, in the 100-GHz channel we see a fairly large dipole of 2--3\muK. In the new HFI processing, thermal dust emission is explicitly included in the dipole estimation model, resulting in improved consistency in the dipole estimates among the various frequency channels. As a result of this process, the best-fit 2018 dipole estimate changed by 2.4\muK\ relative to 2015, and this is visually apparent in the 100-GHz raw difference map. In addition, we see significant striping in the fractional difference map, with an amplitude of more than 3\,\% of the foreground level at high latitudes. As is the case for LFI, these stripes are due to improved time-variable gain estimation, which in turn is responsible for the overall improvement in the large-scale polarization reconstruction. Of course, for this channel the absolute foreground levels are low at high Galactic latitudes, and a 3\,\% relative difference corresponds only to 1--2\muK\ in absolute value. For temperature this is small, while for polarization it is highly relevant, as we discuss below.

Qualitatively speaking, similar considerations hold for the 143 and 217-GHz channels as well. However, in these cases we see an additional effect, namely a significantly blue Galactic plane in the fractional difference map, indicating relative absolute differences of about 1\,\% in the high signal-to-noise regime. At first sight, this may appear puzzling, since the absolute CMB calibration between the 2018 and 2015 has changed by less than 0.1\,\% \citep{planck2016-l05}. The explanation is the new HFI treatment of bandpass differences among individual bolometers. As described in Sect.~\ref{fig:dx11_vs_dx12_hfi}, each frequency map is now generated as the sum over all bolometer timestreams within that frequency channel, each of which has been \emph{bandpass equalized} prior to co-addition. This equalization is implemented by fitting \commander\ foreground templates of thermal dust, CO, and free-free emission jointly with other instrumental parameters, with the goal of minimizing inter-bolometer bandpass differences that otherwise generate spurious polarization contamination.

For component-separation purposes, this implies that the overall bandpass profile of each HFI frequency channel has changed. Furthermore, this process also leads to a complicated bandpass definition overall, in which the bandpass in principle is component dependent. While thermal dust, free-free, and CO emission are associated with bandpasses given as straight averages of the individual bolometer bandpasses (due to their inclusion in the bandpass equalization procedure), synchrotron, spinning dust, and thermal Sunyaev-Zeldovich signals are associated with inverse noise-variance-weighted bandpasses as in earlier releases. In practice, though, we adopt the straight averaged bandpasses for all HFI channels in the current release, since the affected non-equalized components are sub-dominant at HFI frequencies, and implementing multi-bandpass integration would require significant algorithm re-structuring. However, this is also one of the reasons why we do not release new individual synchrotron and spinning dust products in temperature in the current release.

Turning to the 353-GHz frequency channel, two additional effects are seen. First, at high latitudes one can see a weak imprint of zodiacal light emission \citep{planck2013-pip88} in the fractional difference map, taking the form of a blue band along the Ecliptic plane with an amplitude of 1\,\%. Second, we also see two deep blue bands on either side of the Galactic plane with amplitudes of 2\,\%; these are due to changes in the 353-GHz transfer function. From such difference maps alone, it is of course impossible to conclude whether the additional residuals are due to defects in the 2015 or 2018 maps. On the other
hand, such structures tend to stand out quite prominently in maps of foreground spectral indices, which in essence measure small differentials between frequencies. Thus, through subsequent \commander-type astrophysical analyses, we find that these two 353-GHz effects are indeed present in the 2018 maps, and not in the corresponding 2015 maps. These residual effects, along with the lack of single-bolometer maps, are thus part of the cost of producing as clean polarization maps as possible, which is the primary goal of the current data release. 

\begin{figure*}
  \begin{center}
    \begin{tabular}{ccc}
      \includegraphics[width=0.3\linewidth]{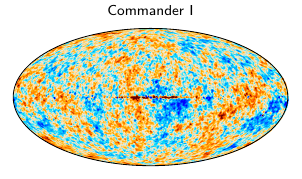}&
      \includegraphics[width=0.3\linewidth]{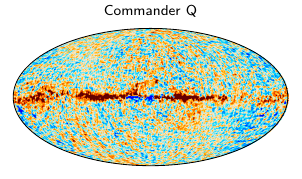}&
      \includegraphics[width=0.3\linewidth]{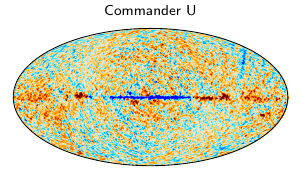}\\
      \includegraphics[width=0.3\linewidth]{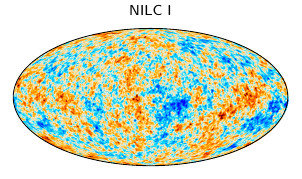}&
      \includegraphics[width=0.3\linewidth]{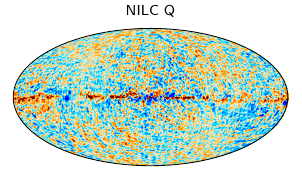}&
      \includegraphics[width=0.3\linewidth]{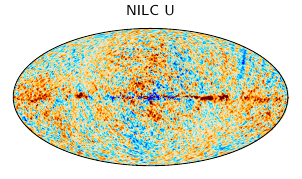}\\
      \includegraphics[width=0.3\linewidth]{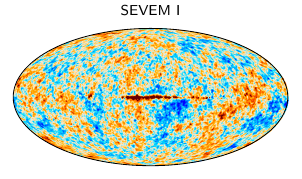}&
      \includegraphics[width=0.3\linewidth]{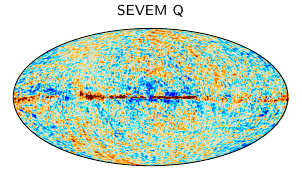}&
      \includegraphics[width=0.3\linewidth]{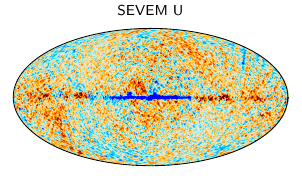}\\
      \includegraphics[width=0.3\linewidth]{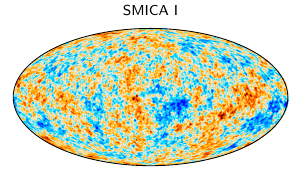}&
      \includegraphics[width=0.3\linewidth]{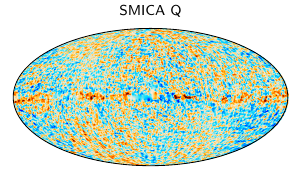}&
      \includegraphics[width=0.3\linewidth]{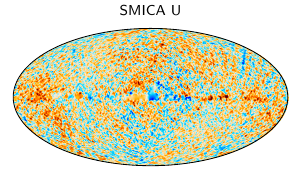}\\
      \includegraphics[width=0.25\linewidth]{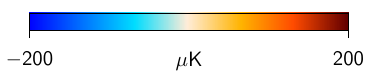}&
      \multicolumn{2}{c}{
        \includegraphics[width=0.25\linewidth]{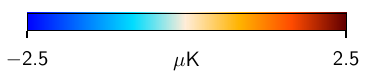}
      }
    \end{tabular}
  \end{center}
  \caption{Component-separated CMB maps at 80\arcmin\ resolution. Columns show Stokes $I$, $Q$, and $U$, respectively,
while rows show results derived with different component-separation methods. The Galactic plane region in the \smica\ maps results from a pre-processing step (masking and diffusive inpainting of a narrow Galactic region in all frequency channels), while no masks are applied to the other maps. In this plot, monopoles and dipoles have been subtracted with parameters fitted outside a $|b|<30^{\circ}$ mask.}
\label{fig:cmb_maps}
\end{figure*}

At 545 and 857\,GHz, most of the effects are similar to those described above, with one additional effect for the 857-GHz channel, where residual sidelobe contamination dominates the high-latitude residuals, with amplitudes of 2--3\,\% of the full foreground signal. In this case, the 2018 processing represents an absolute improvement over the 2015 processing, in the sense that the full 2018 frequency map has lower sidelobe contamination than the corresponding 2015 frequency map. At the same time, it is worth noting that single-bolometer maps are available in the 2015 release, and the 857-2 bolometer map has significantly lower sidelobe contamination than any of the other three \citep{planck2014-a12}. Thus, if a given scientific analysis does not require the signal-to-noise ratio of the full 857-GHz channel, the \Planck\ 2015 857-2 bolometer channel may be an even better choice than the full 857-GHz 2018 frequency map. However, in the current paper, which is dedicated to the 2018 release itself, we adopt the 2018 full-frequency map in all analyses.

Figure~\ref{fig:dx11_vs_dx12_pol} shows the corresponding plots for polarization.  Here we do not subtract any CMB component (since it is small), and we also do not show fractional difference maps (since polarized foreground amplitudes can go both positive and negative). The two leftmost columns show the raw 2018 frequency maps in Stokes $Q$ and $U$, and the two rightmost columns show the straight differences between the 2018 and 2015 frequency maps. 

As expected, the various features seen in the polarization difference maps trace those observed in the corresponding temperature differences. For 30 and 44\,GHz, the main features at high latitudes are due to bandpass mismatch and time-variable gain corrections, achieved by iterating between gain estimation, mapmaking and component separation. For 70\,GHz, only very small differences are seen, since the gain estimation procedure is unchanged from 2015; however, it is important to note that a separate residual gain template has been produced for this channel, and this is applied in the scientific processing (see \citealt{planck2016-l02}).

For the HFI channels, we see similar effects of improved effective gain estimation at high latitudes, as well as improved bandpass corrections at low latitudes, in particular for 100, 217, and 353\,GHz, which are strongly affected by CO emission. \textcolor{black}{Most strikingly, the low-latitude structures seen in the 100-GHz map are typical examples of temperature-to-polarization leakage, where the CO morphology is modulated by the scanning orientation of the \Planck\ satellite.}   At 353\,GHz, we additionally see the residual effect of transfer-function convolution near the Galactic plane in Stokes $U$. Thus, caution is warranted when studying polarized thermal dust emission near the Galactic plane with this frequency map.

\begin{figure}[htbp]
  \begin{center}
    \includegraphics[width=\columnwidth]{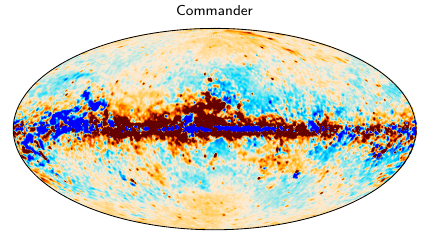}\\
    \includegraphics[width=\columnwidth]{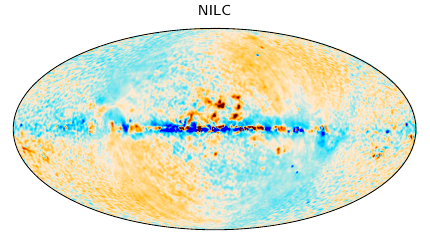}\\
    \includegraphics[width=\columnwidth]{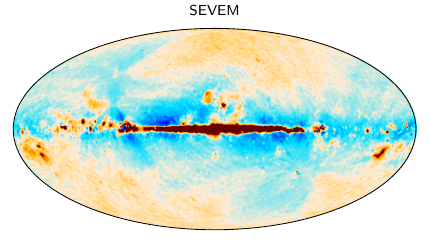}\\
    \includegraphics[width=\columnwidth]{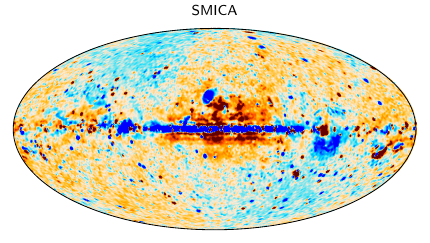}\\
    \includegraphics[width=0.7\columnwidth]{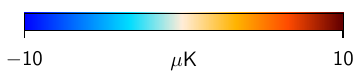}
  \end{center}
  \caption{Differences between 2015 and 2018 CMB $I$ maps at 80\arcmin\ resolution. From top to bottom, results \textcolor{black}{are shown} for \commander,  \nilc, \sevem, and \smica. Monopoles and dipoles have been subtracted with parameters fitted outside a $|b|<30^{\circ}$ mask.}
  \label{fig:cmb_diff_release_maps}
\end{figure}

To summarize, we observe typically (at most) 2--3\,\% differences between the 2015 and 2018 frequency maps at high latitudes, as measured in units of foreground signal. In most cases, these differences are directly due to improvements in the updated processing \textcolor{black}{\citep{planck2016-l02,planck2016-l03}}, although with a few notable exceptions, in particular for the 353-GHz channel. It is important to note, however, that the design philosophy of the 2018 release has been to optimize the quality of the polarization products, which in some cases comes at the expense of temperature analysis. In particular, the non-availability of single-bolometer maps represents a limiting factor for astrophysical component separation in temperature. For this reason, we expect both 2015 and 2018 temperature products to be in common use in the future, depending on the needs of a particular application, whereas for polarization we strongly recommend usage of the 2018 products.

\subsection{Simulations}
\label{sec:simulations}

The instrumental noise characteristics of the \Planck\ observations are complex, and a simple white-noise approximation is inadequate for high-precision analyses of these data. The only realistic approach to handling both instrumental noise and residual systematics is through end-to-end simulations. As part of the \Planck\ 2018 data release, we therefore provide a set of 300 independent noise-plus-systematics simulations for each frequency band and for each of the data splits described above, as well as 999 CMB-only simulations that include the effects of satellite scanning and asymmetric beams; see \citealt{planck2016-l02} and \citealt{planck2016-l03} for full details. These simulations are available through the Planck Legacy
  Archive.\footnote{\url{http://pla.esac.esa.int}}

These simulations are propagated through each of the pipelines; we adopt the same frequency weights (mixing matrices, spectral indices etc.) as for the real data. The two main advantages of fixing the weights are, first, that the noise properties actually correspond to the real final maps; and, second, that the system becomes linear, and CMB and noise may be propagated independently through each pipeline. In the following, we will employ CMB-only, noise-only, and CMB-plus-noise combinations for various applications.

\subsection{Standardization of simulations and data}
\label{sec:standardization}

Each of the four pipelines processes both the data and simulations somewhat differently with respect to harmonic space truncation ($\ell_{\mathrm{max}}$) and high-$\ell$ regularization. In order to facilitate meaningful direct comparisons between the various maps, we convolve all four data sets to a common effective resolution prior to analysis, as described below. We emphasize, however, that the released data products are provided at their native resolution, in order to allow external users to exploit the full resolution of each data set, if so desired.

For temperature, the most aggressive smoothing applied by any of the four pipelines is defined by \nilc, for which the effective
high-$\ell$ apodization kernel reads
\begin{equation}
  B(\ell) =
\begin{cases}
    1, & \ell\leq \ell_{\mathrm{peak}}, \\
    \cos^2\left[(\pi/2)(\ell-\ell_{\mathrm{peak}})/(\ell_{\mathrm{max}}-\ell_{\mathrm{peak}})\right],
    & \ell > \ell_{\mathrm{peak}},
\end{cases}
\end{equation}
where $\ell_{\mathrm{peak}} = 3400$ and $\ell_{\mathrm{max}}=3999$. We therefore apply this kernel to each of the three other pipelines, on top of their intrinsic $5\arcm$ FWHM smoothing kernels. For \smica\ we additionally apply the \healpix\ pixel window for $N_{\mathrm{side}}=2048$, which is not by default applied for this code.

For polarization, the most aggressive high-$\ell$ truncation is applied by \sevem, which enforces a hard harmonic space truncation at $\ell_{\mathrm{max}}=3000$. This same truncation is applied to each of the three other codes in polarization as a post-processing step.

\section{CMB maps}
\label{sec:cmb}

\begin{figure*}
  \begin{center}
      \includegraphics[width=0.33\linewidth]{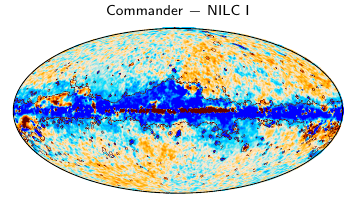}
      \includegraphics[width=0.33\linewidth]{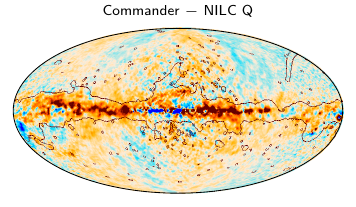}
      \includegraphics[width=0.33\linewidth]{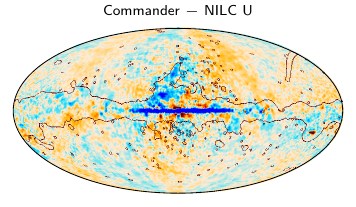}\\
      \includegraphics[width=0.33\linewidth]{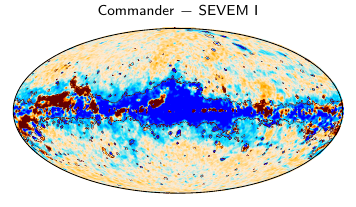}
      \includegraphics[width=0.33\linewidth]{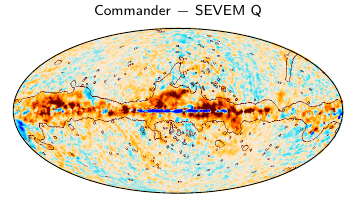}
      \includegraphics[width=0.33\linewidth]{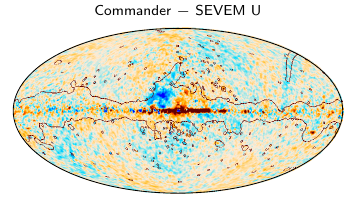}\\
      \includegraphics[width=0.33\linewidth]{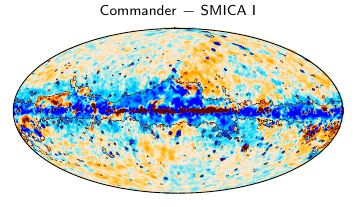}
      \includegraphics[width=0.33\linewidth]{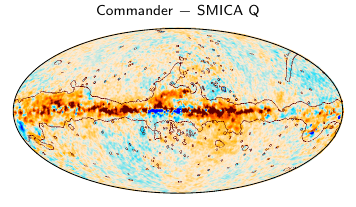}
      \includegraphics[width=0.33\linewidth]{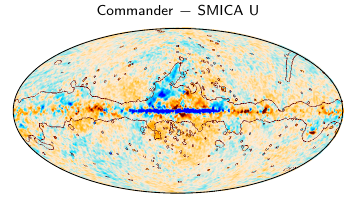}\\
      \includegraphics[width=0.33\linewidth]{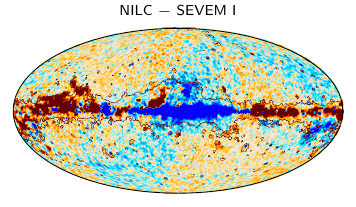}
      \includegraphics[width=0.33\linewidth]{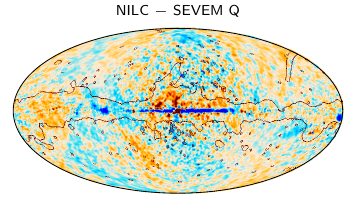}
      \includegraphics[width=0.33\linewidth]{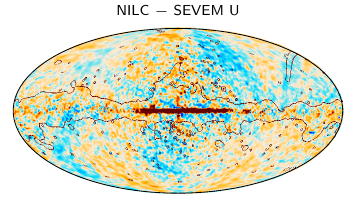}\\
      \includegraphics[width=0.33\linewidth]{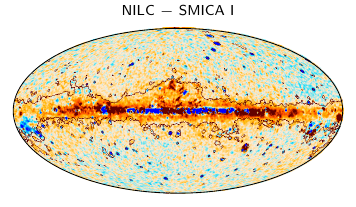}
      \includegraphics[width=0.33\linewidth]{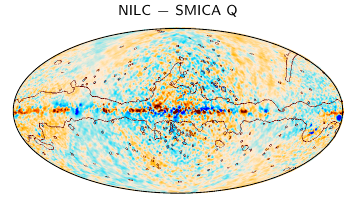}
      \includegraphics[width=0.33\linewidth]{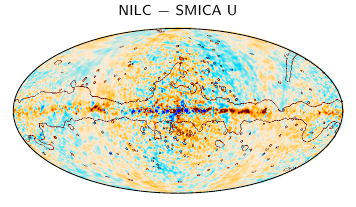}\\
      \includegraphics[width=0.33\linewidth]{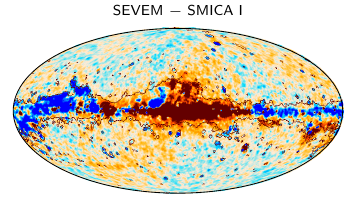}
      \includegraphics[width=0.33\linewidth]{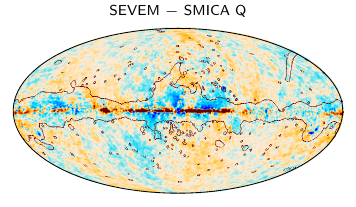}
      \includegraphics[width=0.33\linewidth]{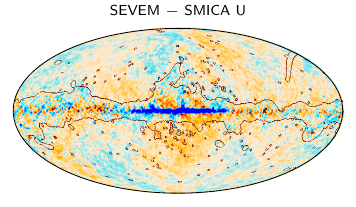}\\
  \end{center}
  \vspace*{-0.5cm}
  \begin{tabular}{ccc}
    \hspace*{0.45cm}
      \includegraphics[width=0.25\linewidth]{figs/colourbar_10uK}\hspace*{4cm}&
      \multicolumn{2}{c}{
        \includegraphics[width=0.25\linewidth]{figs/colourbar_2p5uK}
      }
    \end{tabular}
  \caption{Pairwise differences between maps from the four CMB component separation pipelines, smoothed to 80\arcmin\ resolution. Columns show Stokes $I$, $Q$, and $U$, respectively, while rows show results for different pipeline combinations.  The lines show the regions masked in component separation. Monopoles and dipoles have been subtracted with parameters fitted outside a $|b|<30^{\circ}$ mask.}
\label{fig:cmb_diff_pipe_maps}
\end{figure*}

The CMB maps and associated products obtained by the various pipelines as applied to the \Planck\ 2018 data are presented in this section; astrophysical foreground results are presented in the next section. For a detailed analysis of the higher-order statistical properties of these maps, see \citet{planck2016-l07}.

\subsection{Full-mission maps and comparison with 2015 release}
\label{sec:fullmission}

\begin{figure}[htbp]
  \begin{center}
    \includegraphics[width=\linewidth]{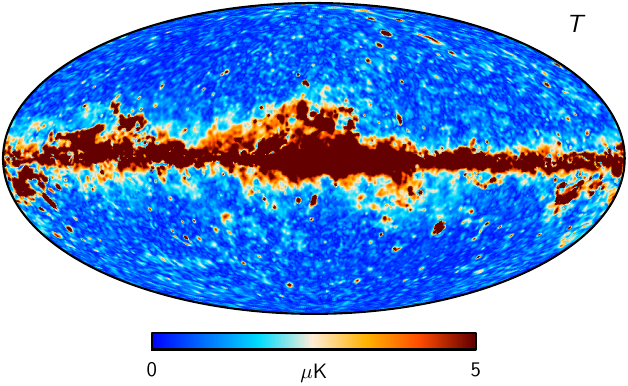}\\
    \includegraphics[width=\linewidth]{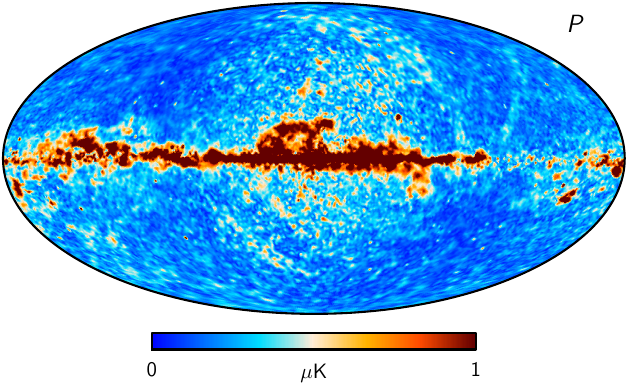}
  \end{center}
  \caption{Standard deviation of the CMB maps between the four
    component-separation methods, at 80\arcmin\ resolution.
    Temperature is shown in the top panel and polarization in the
    bottom panel. The polarization standard deviation is defined as
    $\sqrt{\mathrm{var}(Q) + \mathrm{var}(U)}$.}
  \label{fig:cmb_maps_stddev}
\end{figure}

Figure~\ref{fig:cmb_maps} shows the final full-mission \Planck\ 2018 CMB component-separated maps derived by each of the four pipelines,\footnote{The four cleaned frequency maps (from 70 to 217\,GHz)  provided by \sevem\ are also shown in Fig.~\ref{fig:sevem_freqmaps}.} both in intensity (left column) and polarization (middle and right columns). Only {\tt SMICA} has been inpainted within a Galactic mask (see Appendix~\ref{app:smica}). All maps are smoothed to a common resolution of $80\arcm$ FWHM for visualization purposes.

At first sight, the consistency among the various pipeline maps appears to be reasonable outside the central Galactic plane, and, as expected, more so in temperature than in polarization.

\begin{figure*}
\begin{center}
  \includegraphics[width=\columnwidth]{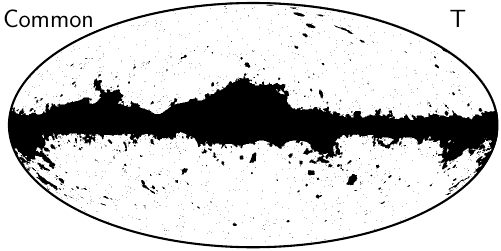}
  \includegraphics[width=\columnwidth]{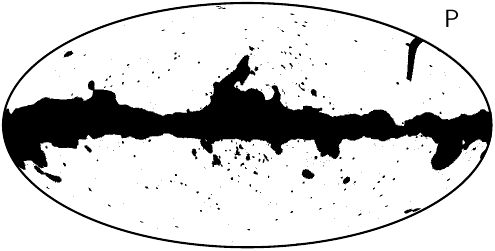}\\
  \includegraphics[width=\columnwidth]{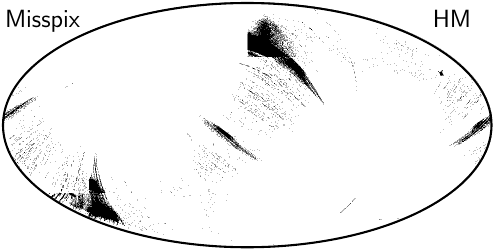}
  \includegraphics[width=\columnwidth]{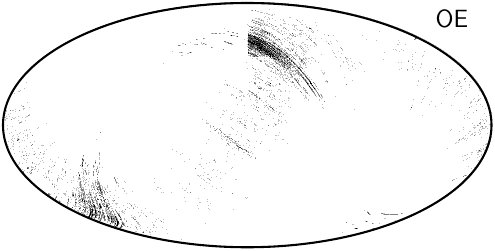}
\end{center}
\caption{Masks recommended for analysis of the cleaned CMB maps. \emph{Top}:~Common confidence masks for temperature (left) and polarization (right). These masks should always be applied to any scientific analysis of the maps presented in this paper. \emph{Bottom}:~Unobserved pixel masks for half-mission (left) and odd-even (right) data splits. These panels show the products of the individual unobserved pixel masks for temperature and polarization, whereas separate (but very similar) temperature and polarization masks are applied during analysis. }
\label{fig:commonmask}
\end{figure*}

\begin{figure}[htbp]
\begin{center}
  \includegraphics[width=\columnwidth]{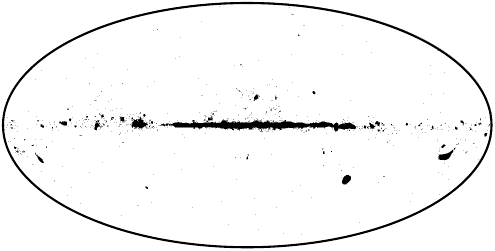}
\end{center}
\caption{Mask used for inpainting the cleaned CMB temperature maps.}
\label{fig:inpaint}
\end{figure}

In the polarization maps, however, we can identify several notable artefacts already at this stage, which prospective future users of these maps need to be aware of. The visually most striking features are of course residual foreground contamination in the Galactic plane. In particular, the alternating sign along the plane is a classic signature of temperature-to-polarization leakage, and the \Planck\ data set is particularly sensitive to residual CO emission in this respect. These features are extremely difficult to suppress to the level of the CMB fluctuations during processing, and must in practice be removed by standard Galactic masking.

The second most striking feature in the polarization maps is a blue stripe in the upper right quadrant of the Stokes $U$ map. This stripe corresponds to a few bad scanning rings that ideally should have been removed by flagging during mapmaking. Unfortunately, this issue was not caught at a sufficiently early stage of the processing, and remains in the final maps. We therefore mask this stripe in the same way that we mask Galactic residuals.

Third, and somewhat less obvious, we observe broad large-scale structures in both Stokes $Q$ and $U$ that are aligned with the \Planck\ scanning strategy. These structures are effectively due to gain-modelling uncertainties coupled to monopole and dipole leakage, and corresponding features are present in the associated simulations. In principle, therefore, these need not be removed prior to subsequent analyses, as long as the appropriate simulations are used to quantify all relevant uncertainties. In practice, however, we note that these modes are associated with significant additional systematic uncertainties, and we therefore caution against over-interpretation of the very largest scales in these maps. In particular, we warn against employing these maps for auto-correlation type analysis, unless the statistic of choice is explicitly shown to be robust against this type of systematic effect, based on end-to-end simulations.

Figure~\ref{fig:cmb_diff_release_maps} shows maps of temperature differences between each of the 2018 pipeline maps and the corresponding 2015 pipeline maps. Corresponding maps of polarization differences are not shown, since the high level of large-scale systematics in the 2015 maps renders a direct difference-map comparison non-informative. In Fig.~\ref{fig:cmb_diff_release_maps}, we recognize many of the features seen in the raw input frequency difference maps shown in Figs.~\ref{fig:dx11_vs_dx12_lfi} and \ref{fig:dx11_vs_dx12_hfi}, and
discussed in Sect.~\ref{sec:inputs}.

Starting with \commander, the most striking difference is a dark blue Galactic plane residual with a clear CO-like morphology. This reflects the fact that it is more difficult for the parametric \commander\ pipeline to estimate CO emission from co-added frequency maps (as in the 2018 processing) than with individual bolometer maps (as in the 2015 processing). For this reason, the \commander\ map adopts a larger Galactic mask in the new release than in the previous one, specifically targeting CO emission; see Appendix~\ref{app:commander} for further details. The second most notable feature in the \commander\ difference map is a $\lesssim2\muK$ blue signal at intermediate latitude with a thermal dust imprint, and this is due to
the changes in bandpass modelling in the high-frequency channels.

Only small differences are observed for \nilc, for which very few pipeline modifications have been introduced since 2015. 
\nilc\ already used full-frequency maps in the previous release. The most significant change is a large-scale quadrupolar
structure at high latitudes, which directly reflects the effective gain changes at 100, 143, and 217\,GHz seen in
Fig.~\ref{fig:dx11_vs_dx12_hfi}. Likewise, \sevem\ also used full-frequency maps in 2015, and only minor pipeline modifications
have been introduced, and consequently, only minor differences are observed in temperature from 2015 to 2018.

For \smica, three qualitatively different types of differences are seen. First, the weak large-scale background pattern is similar to that observed in \nilc, and simply reflects the slight changes in input data discussed above. In addition, we see significant changes in the compact sources that can be explained by the change of masking strategy described in Sect.~\ref{sec:methods}. Third and finally, the strong near-Galactic-plane differences that include free-free sources (e.g., the Gum Nebula and Rho Ophucius) are explained by the miscalibration of the 44-GHz channel in the 2015 released map (as recalled in Sect.~\ref{sec:methods}). The impact of this issue is assessed in Appendix~\ref{app:smica}.

Figure~\ref{fig:cmb_diff_pipe_maps} shows all pairwise difference maps between each of the pipeline CMB maps. The structures seen in these plots correspond closely to those already discussed above.  Finally, Fig.~\ref{fig:cmb_maps_stddev} shows the standard deviation evaluated from the four cleaned CMB maps, smoothed to $80\arcm$ FWHM angular scales; the polarization standard deviation is here defined as
$\sqrt{\mathrm{var}{Q} + \mathrm{var}{U}}$.

\subsection{Confidence masks}
\label{sec:masks}

From the above discussion, it is clear that significant residuals are present in the CMB maps, in particular close to the Galactic
plane. Therefore, appropriate masking is required for scientific exploration of the \Planck\ 2018 maps in both temperature and
polarization, as in earlier releases. For this purpose, we adopt a conservative strategy similar to that of 2015, and we construct a common confidence mask for all maps, even if the various maps may have different levels of residuals.

In previous releases, a common mask was generated simply as the product of the individual confidence masks derived for each pipeline. However, the pipeline masks were established using qualitatively different criteria in each case, and a direct comparison was therefore non-trivial. In the current analysis, we adopt a more direct route, starting with the inter-pipeline standard deviation maps shown in Fig.~\ref{fig:cmb_maps_stddev}. \textcolor{black}{The single most striking feature in these maps is the Galactic plane. At high latitudes, one can additionally see point sources and a few rings in the \Planck\ scanning strategy that were observed fewer times than average, resulting in coherent stripes of excess variance.}

Specifically, for temperature we first threshold at 3\muK\ the standard deviation map evaluated at 80\arcm\ FWHM smoothing scale, and adopt this as our primary mask. The specific smoothing scale of 80\arcm\ represents a compromise between suppressing noise while still retaining small features, while the threshold of 3\muK\ is defined by the region at high Galactic latitude in the top panel of Fig.~\ref{fig:cmb_maps_stddev}. Second, we smooth this binary mask, consisting of 0 and 1s, with a $10^\circ$ FWHM Gaussian beam, and remove any pixels with a value lower than 0.5; this is to remove isolated small ``islands'' within the main Galactic plane. Third, we threshold at 10\muK\ a corresponding standard deviation map evaluated at 10\arcm\  FWHM smoothing scale in order to remove compact objects.

The resulting mask ensures that only pixels for which the four pipelines agree in their CMB solutions to better than 3\muK\ in standard deviation on large scales (10\muK\ on small scales)  are allowed in the final analysis. However, quantitative agreement among codes is only a necessary criterion; it is not sufficient. We therefore augment this mask by the absolute individual confidence masks of \commander\ and \sevem\ (see Appendices~\ref{app:commander} and \ref{app:sevem} for
details), by the point-source masks used for inpainting by \sevem\ and \smica, and by the processing mask employed by \smica. The first two of these employ $\chi^2$ and difference maps to define their acceptable regions, and thereby correspond to standard absolute goodness-of-fit statistics, while the latter two correspond to basic processing masks. The \sevem\ inpainted point-source mask is constructed from point sources detected in the 143- and 217-GHz channels, and is described in detail in Appendix~\ref{app:sevem} (see also Fig.~\ref{fig:dx12_masks_inpainting_sevem}). We find no evidence for significant artefacts in the \nilc\ and \smica\ maps outside the \commander\ and \sevem\ masks defined above, and we therefore do not apply any special measure for these maps.

We adopt a similar procedure for polarization, but with a few notable additions. First, the inter-pipeline standard deviation map evaluated at 80\arcm\ FWHM is thresholded at 1\muK. The resulting mask is smoothed to 5\deg\ FWHM, and thresholded at a value of 0.9, effectively expanding the original mask by a few degrees in all directions. This mask is then multiplied with a corresponding mask derived by thresholding at 0.6\muK\ the original standard deviation map at 80\arcm\ FWHM, to remove sharper features. We then exclude all pixels flagged by the \commander\ and \sevem\ confidence masks. Next, we remove the region contaminated by cosmic rays discussed in Sect.~\ref{sec:fullmission}, as defined in Ecliptic
coordinates following \Planck's scanning path. Third, as an additional guard against temperature-to-polarization leakage from CO emission, we exclude any pixels for which the CO emission at 100\,GHz (see Sect.~\ref{sec:foregrounds}) is brighter than $20\muK$, evaluated at a smoothing scale of $5^{\circ}$ FWHM. Isolated ``islands'' in the main Galactic plane are then removed with the same procedure as for temperature. Finally, we also add the point-source masks used for inpainting by \sevem\ and \smica. For polarization, the \sevem\ inpainted point-source mask is constructed from point sources detected in the 100-, 143-, and 217-GHz channels and is shown in Fig.~\ref{fig:dx12_masks_inpainting_sevem}. 

The resulting common masks are shown in the top row of Fig.~\ref{fig:commonmask} for temperature (left panel) and
polarization (right panel). The final accepted sky fractions are $f_{\mathrm{T}} = 0.780$ and $f_{\mathrm{P}} = 0.782$. These sky fractions are similar to those reported in 2015, namely $f_{\mathrm{T}}^{2015} = 0.77$ and $f_{\mathrm{P}}^{2015}
= 0.78$.

As discussed in Sect.~\ref{sec:misspix}, the half-mission and odd-even split maps contain a number of unobserved or poorly conditioned pixels. For split-map analysis, we therefore recommend additional unobserved pixel masks. These are produced by thresholding the $3\times3$ Stokes parameter condition number hit-count maps produced during mapmaking \citep{planck2016-l02,planck2016-l03}. The resulting unobserved pixel mask is further extended in a three-step iterative process in which the neighbours of each unobserved pixel have been masked. The bottom row in Fig.~\ref{fig:commonmask} shows the products of the temperature and polarization unobserved pixel masks for both the half-mission (left panel) and odd-even (right panel) splits.

As a final mask-related issue, we note that the \Planck\ 2018 product delivery includes Wiener-filtered versions of each pipeline map, in which all high-foreground regions are replaced with a Gaussian constrained realization. For temperature, these regions are defined simply by thresholding the maximum difference between any of the four cleaned CMB maps at $100\muK$, and additionally removing all pixels excluded by the \smica\ processing mask. This mask is shown in Fig.~\ref{fig:inpaint}, and excludes 2\,\% of the sky. For polarization inpainting we conservatively adopt the common confidence mask defined above. In either case, we note that the inpainted CMB maps are primarily intended for publication and presentation purposes, rather than scientific analysis. For full scientific analysis \textcolor{black}{of the high-foreground-contaminated regions}, we recommend corresponding processing of end-to-end simulations. \textcolor{black}{These are, however, not provided due to large data volume and processing costs, although they may be generated by applying a Wiener filter code such as \commander\ \citep{seljebotn2017} to the cleaned CMB simulations that are provided}.

\subsection{Effective transfer functions}

As noted in Sect.~\ref{sec:methods}, all \Planck\ 2018 CMB maps have a common nominal target resolution of $5\arcm$ FWHM, as output by each of the respective pipelines. However, this resolution is not exact, as it does not take into account the effect of spatially-varying asymmetric beams on the sky. The nominal $5\arcm$ beam kernel must therefore be corrected by an effective transfer function for each pipeline prior to any harmonic space analysis of these maps, including cosmological power spectrum and parameter estimation.

\begin{figure}[htbp]
\begin{center}
  \includegraphics[width=\columnwidth]{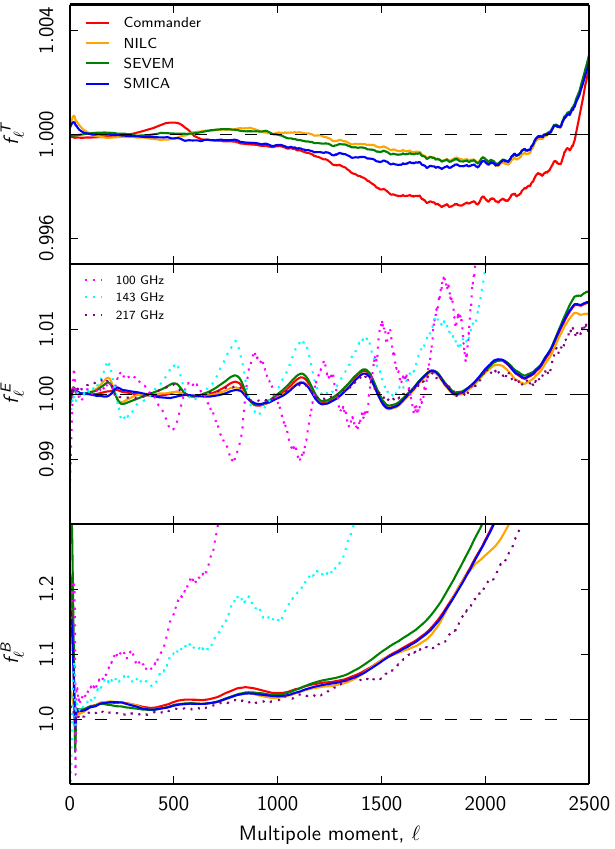}
\end{center}
\caption{Effective transfer functions $f_{\ell}$ for each of the four pipeline CMB maps, after deconvolving a $5\arcm$ FWHM Gaussian beam and $\nside=2048$ \healpix\ pixel window. From top to bottom, the panels show results for $T$, $E$, and $B$. In the two bottom panels, the dotted lines show the effective residual transfer functions for the three CMB-dominated HFI frequencies between 100 and 217\,GHz, after deconvolving the azimuthally-symmetric {\tt QuickBeam}-based transfer function and \healpix\ pixel window in each case.}
\label{fig:transferfunctions}
\end{figure}

We estimate the effective transfer functions from the CMB signal-only simulations discussed in Sect.~\ref{sec:simulations} through the following expression,
\begin{equation}
f_{\ell} = \frac{1}{b_{\ell}^{5\arcm} p_{\ell}^{2048}}
\sqrt{\left<\frac{C_{\ell}^{\mathrm{out}}}{C_{\ell}^{\mathrm{in}}}\right>},
\label{eq:transfunc}
\end{equation}
where $C_{\ell}^{\mathrm{out}}$ and $C_{\ell}^{\mathrm{in}}$ denote the simulated output and input power spectra of each CMB signal realization. The former includes both instrumental beam and pixel window convolution, while the latter includes
neither.  The quantity $b_{\ell}^{5\arcm}$ denotes the beam transfer function of a 5\arcm\ FWHM Gaussian beam, $p_{\ell}^{2048}$ is the $N_{\mathrm{side}}=2048$ pixel window \citep{gorski2005}, and brackets indicate an average over 50 simulations. Equation~\ref{eq:transfunc} is evaluated independently for temperature and both $E$- and $B$-mode polarization. Finally, each transfer function is smoothed with a third-order Savitzky-Golay filter with a window size of $\Delta\ell=51$ to reduce residual uncertainty from the finite number of Monte Carlo simulations. The examples shown in this paper correspond to full-sky transfer functions; these functions should in principle be re-evaluated for each sky fraction used in a given analysis. \footnote{ For the particular case of  \sevem, the evaluation of the transfer function is in principle  affected by the pixels inpainted in the cleaned frequency maps.   Since those pixels are all excluded in the common confidence mask,
we have evaluated this function from full-sky CMB simulations  without applying this inpainting. Therefore, this effective transfer function should be a good approximation for the regions passed by the common mask. However, if a transfer function is needed for a region of the sky that contains inpainted pixels, it is recommended to re-evaluate this function for that particular sky coverage, taking into account the inpainting. Although the effect is very small, it can be noticeable, especially for agressive masks that remove only a small fraction of the Galaxy, since in the Galactic regions a relatively large number of sources are inpainted.}

The resulting transfer functions are shown in Fig.~\ref{fig:transferfunctions}. Starting with the temperature case, we first note that the range spanned by the four curves is well within $\pm0.5\,$\%, and, therefore, these effects are quite minor for all the considered multipoles.  Overall, qualitatively similar behaviour is observed for the four codes, with \commander\ showing a slightly larger deviation. In particular, for \commander, we see that the effective residual transfer function is very close to unity up to $\ell\approx700$, after which it starts to fall off, eventually reaching an amplitude of about 0.3\,\% at $\ell\approx2000$, before it begins to rise sharply. The small excess of $\lesssim0.1\%$ around $\ell=500$ is associated with the effective cut-off of the LFI-dominated low-frequency signal component employed by \commander. These general trends are due to small mismatches between the full asymmetric beams, as implemented through pixel-space {\tt FEBeCoP} \citep{mitra2010} convolutions, and the azimuthally symmetric effective beam transfer functions, as implemented with {\tt QuickBeam} \citep{hivon2017}. For instance, a fall-off of 0.3\,\% at $\ell\approx2000$ corresponds to a mismatch of about $0.05'$ FWHM in
the two models. \textcolor{black}{Further, we note that \commander\ is the only code that applies per-pixel inverse variance weighting of the raw frequency maps, and as such it has a different response to small angular scales than the other codes.}

Turning our attention to the $E$-mode transfer functions, the most striking new feature is a pattern of systematic wiggles. These are seen both in transfer functions derived from each frequency alone (shown as dotted lines) and in the component-separated maps. These wiggles are due to temperature-to-polarization leakage through the asymmetric beam shapes, and the positions of the peaks coincide with the peaks in the CMB temperature power spectrum. \textcolor{black}{Note that most of the total weight below $\ell\approx300$ is determined by the 143-GHz channel, while above $\ell\approx300$ it is dominated by the 217-GHz channel. The 100-GHz channel does not dominate at any angular scale, due to its lower angular resolution and higher noise as compared to the 143-GHz channel.}

\begin{figure*}[t]  
\begin{center}
  \includegraphics[width=0.97\textwidth]{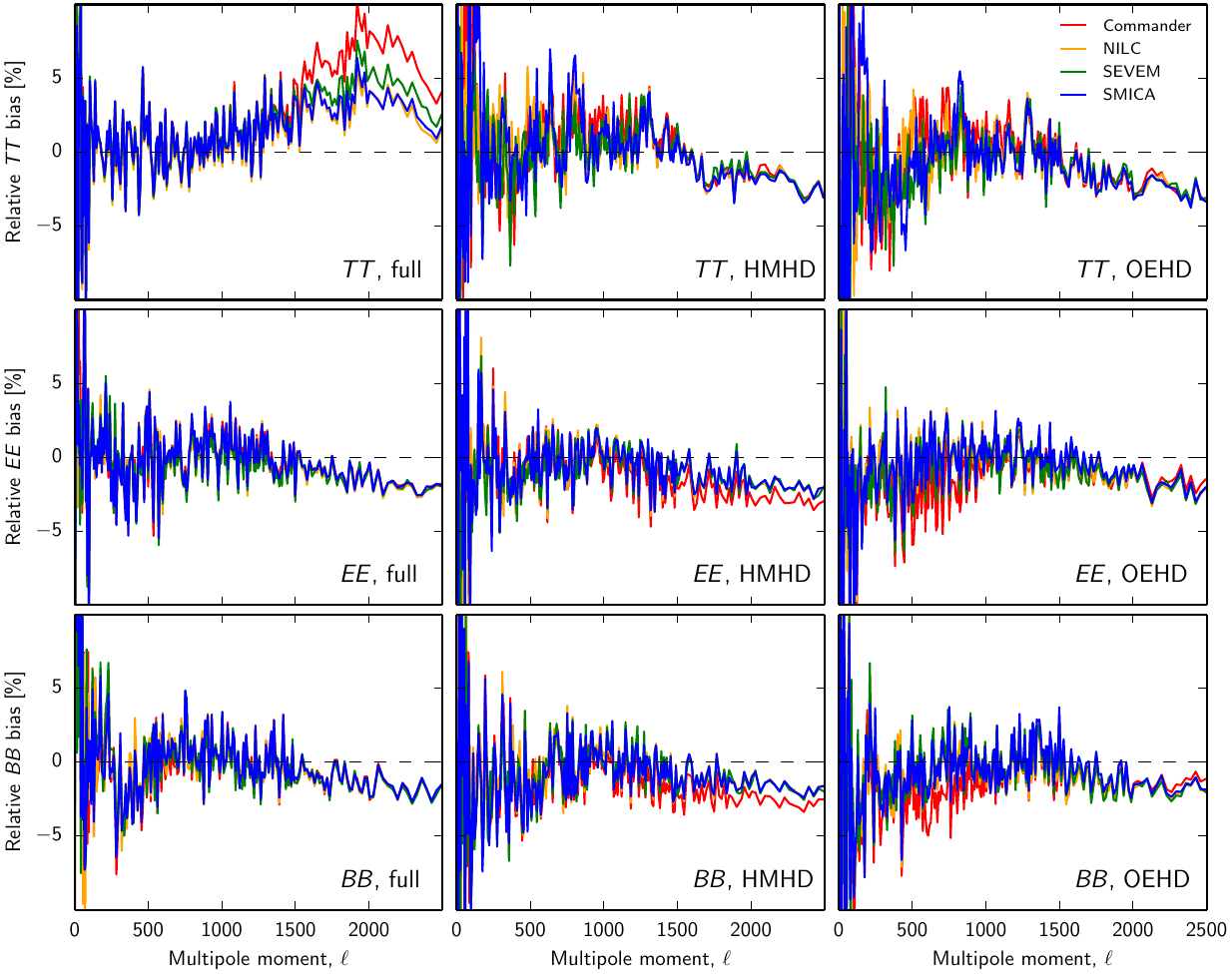}
\end{center}
\caption{Power spectrum consistency between cleaned CMB maps and end-to-end simulations. Each panel shows the fractional difference between the angular power spectrum computed from the observed data and the mean of the simulations. Rows show different polarization spectra ($TT$, $EE$, and $BB$), while columns show different data splits (full, HMHD, and OEHD).}
\label{fig:simspec}
\end{figure*}

Similar considerations apply to the $B$-mode transfer functions, although in this case the wiggles are largely dominated by an increasing trend caused by a wide range of both temperature-to-polarization and polarization-to-polarization leakage effects. Overall, however, the net sum of all these effects is smaller than 10\,\% of the underlying (lensing-induced) $B$-mode signal up to $\ell\lesssim1600$. We also see that the component-separated map is strongly dominated by the 217-GHz channel for $\ell\gtrsim500$.

\subsection{Noise characterization and consistency with simulations}
\label{sec:noise_consistency}

We now characterize the statistical properties of the component-separated CMB map, and we start with a description of instrumental noise and residual systematic effects. We adopt three main measures for this purpose, each designed to highlight different aspects of the effective noise properties; these are designed for different applications.

Our first noise measure is defined in terms of the so-called odd-even half-difference (OEHD) maps, in which the full time-ordered data volume is divided according to odd and even ring numbers. This is a fine-grained time split, and as such, the OEHD map tends to cancel most systematic effects. This noise measure is thus our cleanest probe of pure instrumental (white and correlated) noise. OEHD maps are plotted in Appendix~\ref{fig:cmb_oehd_maps} for each pipeline and for each of the three Stokes parameters. Overall, we see that these difference maps exhibit very few visually-apparent systematic effects at high latitudes, and the only significant residuals occur in the Galactic centre, where the overall signal amplitude is very larget.

Our second noise measure is defined in terms of the half-mission half-difference (HMHD) maps, in which the time-ordered data are split according to long time periods, defined by years \citep{planck2016-l02,planck2016-l03}. This measure is thus a coarse-grained time split, and more sensitive to systematic effects that vary on long time scales, such as gain variations or sidelobe contamination. This is our preferred estimate for the combined impact of instrumental noise and systematic effects. HMHD maps are shown in Appendix~\ref{fig:cmb_hmhd_maps} for each pipeline and for each of the three Stokes parameters. In these maps, we clearly see the imprint of the Galactic plane, which is largely caused by calibration uncertainties, as well as more pronounced scan-aligned structures at high latitudes.

The third noise measure comprises the full-blown end-to-end simulations, in which all known systematics have been modelled to the best of our ability (see \citet{planck2016-l02} and \citet{planck2016-l03} for full details). These simulations are generated as raw time-ordered data, and processed through each step of the analysis pipeline, including map making and component separation. Unfortunately, this process is computationally very expensive, and only a limited set of 300~realizations has been produced for the current release. However, for each realization a full set of results are produced, including full mission maps, half-mission and odd-even splits. Combined, these form the basis of most goodness-of-fit statistics presented in the following sections.

\subsubsection{Power spectrum analysis}

In Fig.~\ref{fig:simspec} we compare the power spectra of the cleaned CMB maps with the simulations. All spectra are evaluated outside the common mask described in Sect.~\ref{sec:masks} using {\tt PolSpice} \citep{chon2004}. Furthermore, all spectra have been normalized relative to the mean of the simulated ensemble, and plotted in terms of the fractional deviation,
\begin{equation}
  \eta_{\ell} \equiv \frac{D_{\ell}^{\mathrm{data}} - \left< D_{\ell}^{\mathrm{sim}}\right>}{\left< D_{\ell}^{\mathrm{sim}}\right>}.
\end{equation}
This function thus measures the fractional difference of the observed power spectrum from the mean of the simulations, plotted  as a percentage in Fig.~\ref{fig:simspec}. This function is evaluated both for temperature and polarization ($TT$, $EE$, and $BB$), as well as for full-mission, HMHD, and OEHD data splits. For clarity, each function has been binned with $\Delta\ell=25$ after computing the above single-$\ell$ quantity.

For full-mission temperature data, we find that the CMB-plus-noise simulations agree well with the data in terms of angular power up to $\ell\lesssim 750$. At higher multipoles, we see a slow increase in power up to $\ell\approx2000$, corresponding to a positive contribution from point sources not included in the simulations. The level of point-source residuals is highest in \commander\ and lowest in \smica. At high multipoles, $\ell \gtrsim 2000$, the spectra turn over. As described in \citet{planck2016-l03}, the power in the HFI simulations for the 100--217\,GHz channels underestimates the true noise in the real data by a few percent (with variations depending both on angular scale and frequency), and this translates into a negative bias at high multipoles in the cleaned CMB maps presented in this paper.

Similar features are seen even more clearly in the polarization $EE$ and $BB$ full-mission spectra, for which the signal-to-noise ratio is lower. In these cases, the simulations agree well with the data up to $\ell\lesssim200$, after which a negative bias of a few percent is observed in the range $200 \lesssim \ell \lesssim 500$. Then, in the range $500\lesssim\ell\lesssim1500$ the agreement is good, before we see the same negative high-$\ell$ bias as in the temperature case. The same trends are even more prominent in the HMHD and OEHD spectra, which by construction are entirely noise-dominated.

\begin{figure*}     
\begin{center}
  \includegraphics[width=\textwidth]{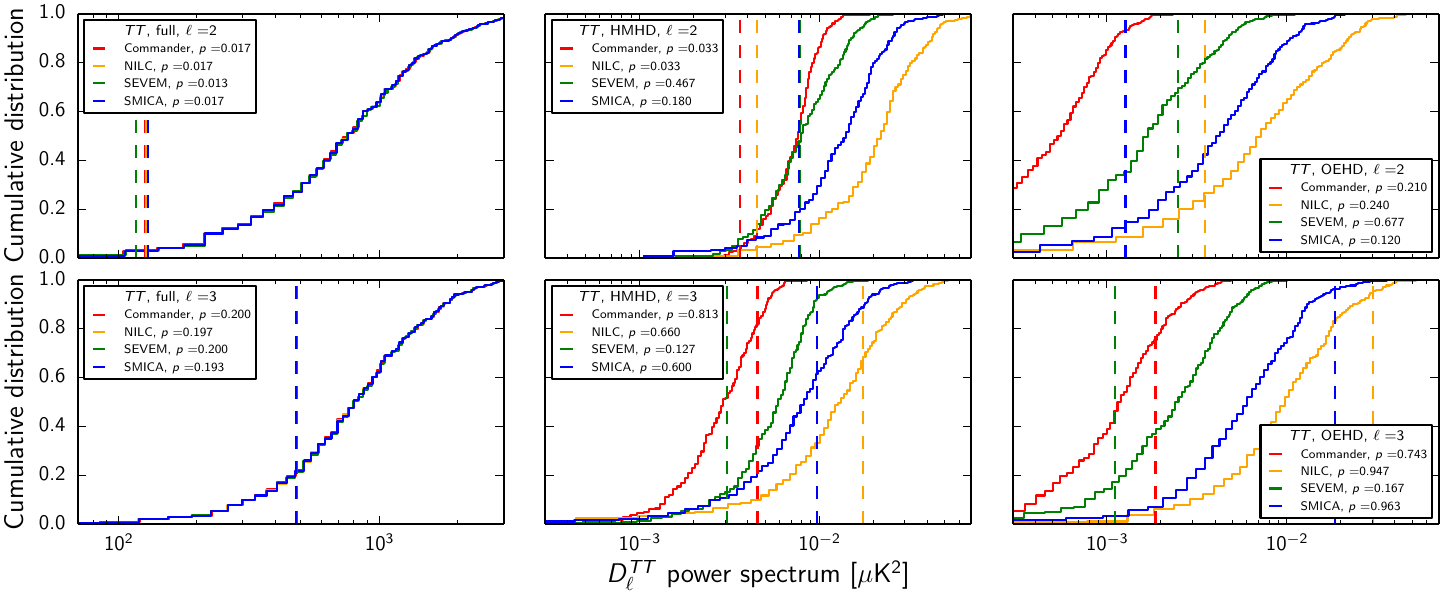}\\
  \includegraphics[width=\textwidth]{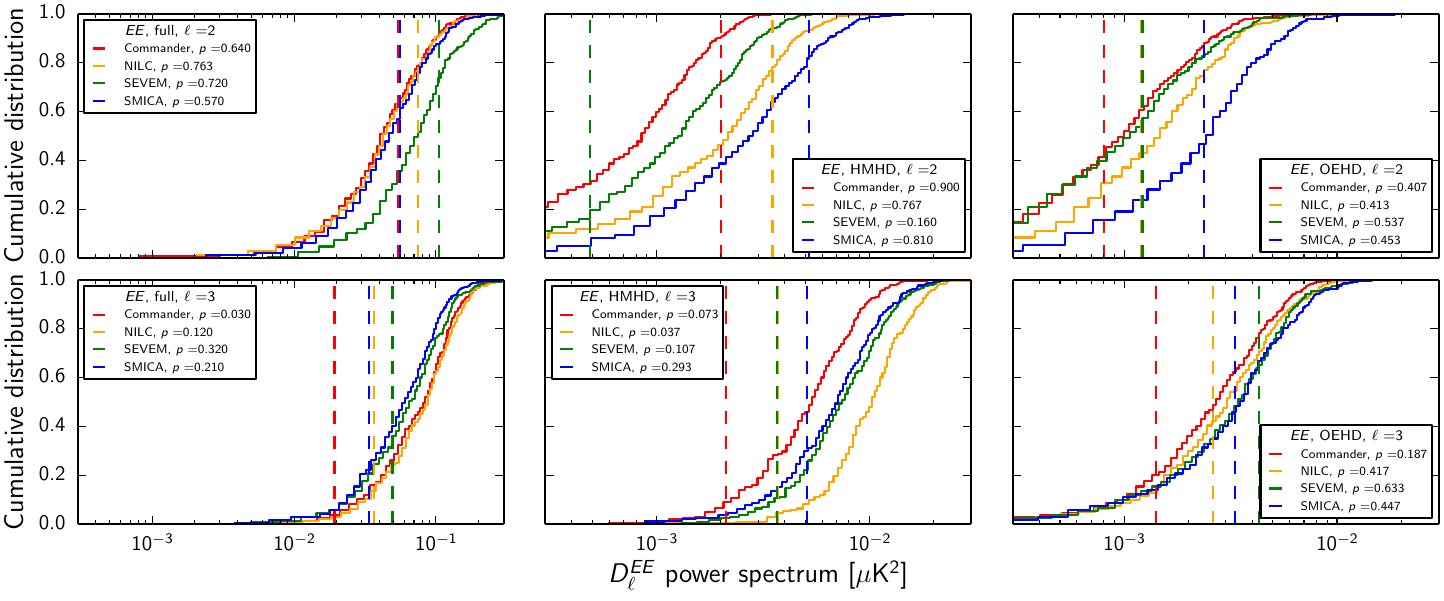}\\
  \includegraphics[width=\textwidth]{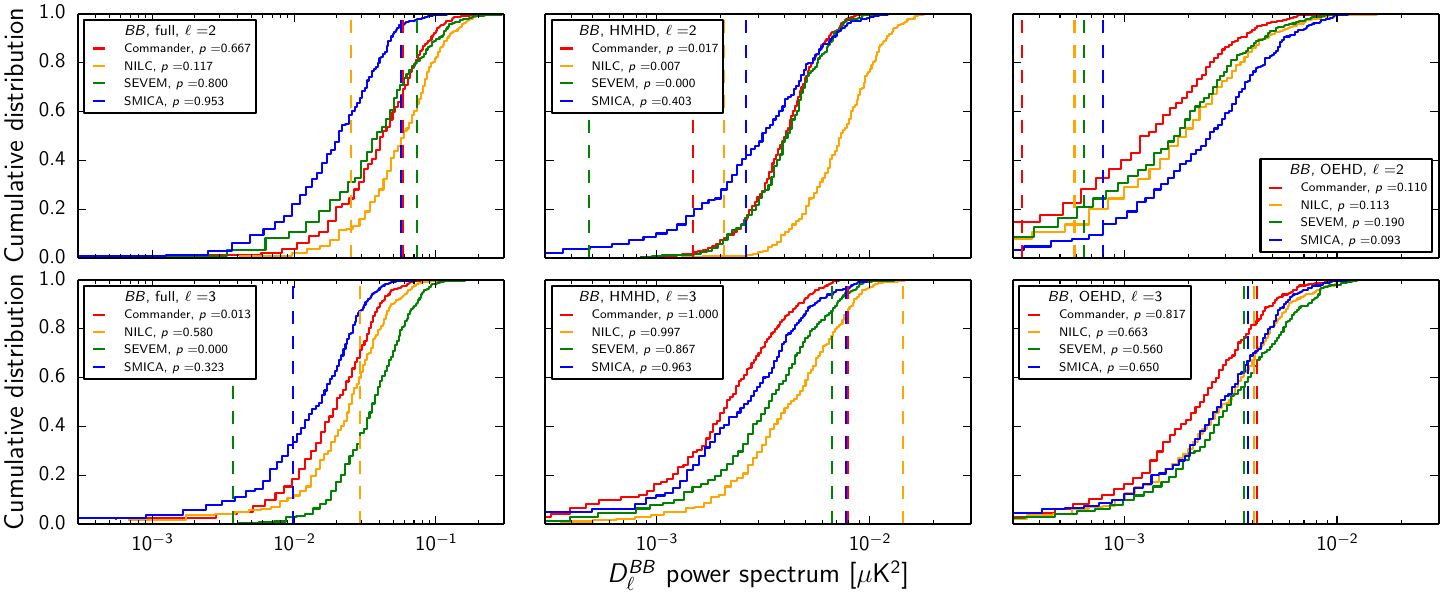}
\end{center}
\caption{Power spectrum consistency between cleaned CMB maps and end-to-end simulations for $\ell=2$ and 3. Solid lines show cumulative distributions computed from 300 simulations, and dashed lines show the value derived from the \Planck\ data. From top to bottom, the three sections show $TT$, $EE$, and $BB$, and within each section the top and bottom rows show distributions for $\ell=2$ and 3. Columns from left to right show full, HMHD, and OEHD splits. The fraction of simulations with a lower power amplitude is given in the legends of each panel for each code.}
\label{fig:simspec_l23}
\end{figure*}

In Fig.~\ref{fig:simspec_l23} we focus on the first two multipoles, and compare the observed power to the full simulated distributions in terms of cumulative distribution functions. Overall, we observe acceptable statistical agreement between the data and simulations at these largest scales, with only a few points showing extreme values of 0 or 1; however, even in these cases the observed values lie just at the edge of the simulated histogram. No large outliers are observed. Nevertheless, it is important to note that the effective noise varies greatly between the various analysis pipelines, and it is therefore essential to compare any given data set with its corresponding simulations.

To summarize, the end-to-end simulations presented and employed in this paper exhibit power biases of several percent with respect to the true observations on intermediate and small scales, while reasonable agreement is observed on large angular scales. These biases originate from corresponding discrepancies at the level of individual frequency bands, as reported in \citealt{planck2016-l02} and \citealt{planck2016-l03}. When employing these simulations for scientific analysis, it is important
to verify that the statistic of choice is not sensitive to such percentage-level differences. This will usually be the case for linear
or cross-correlation type analyses, but not necessarily for quadratic or auto-correlation type analyses.

\subsubsection{Pixel-space variance analysis}

A complementary consistency measure is given by the total variance as measured in pixel space at different pixel resolutions (see, e.g., \citealt{Monteserin2008};  \citealt{Cruz2011};  \citealt{planck2014-a18}). This method normalizes the map with respect to the total variance of the signal plus noise, where the noise variance is estimated through the simulations described above, and the variance of the signal is determined as the value that gives a normalized map variance equal to unity. For the HMHD and OEHD maps, the method simplifies, since the CMB signal is cancelled through the half-difference calculation.

Recognizing the fact that \Planck\ polarization maps are generally noise dominated, we apply the methods described in \citet{planck2016-l08} and Molinari et al.\ (in preparation) for polarization. These methods include both auto- and cross-estimates for the variance, which is the result of the subtraction between the variance of the $\left<Q^2+U^2\right>$ signal-plus-noise map and the variance of the $\left<Q_N^2+U_N^2\right>$ noise estimated from the MC simulations.  For both temperature and polarization, we employ the respective union masks described above. When dealing with HMHD and OEHD maps, we consider the union mask combined with the corresponding unobserved pixel mask.

\begin{figure} [t]
\begin{center}
  \includegraphics[width=\columnwidth]{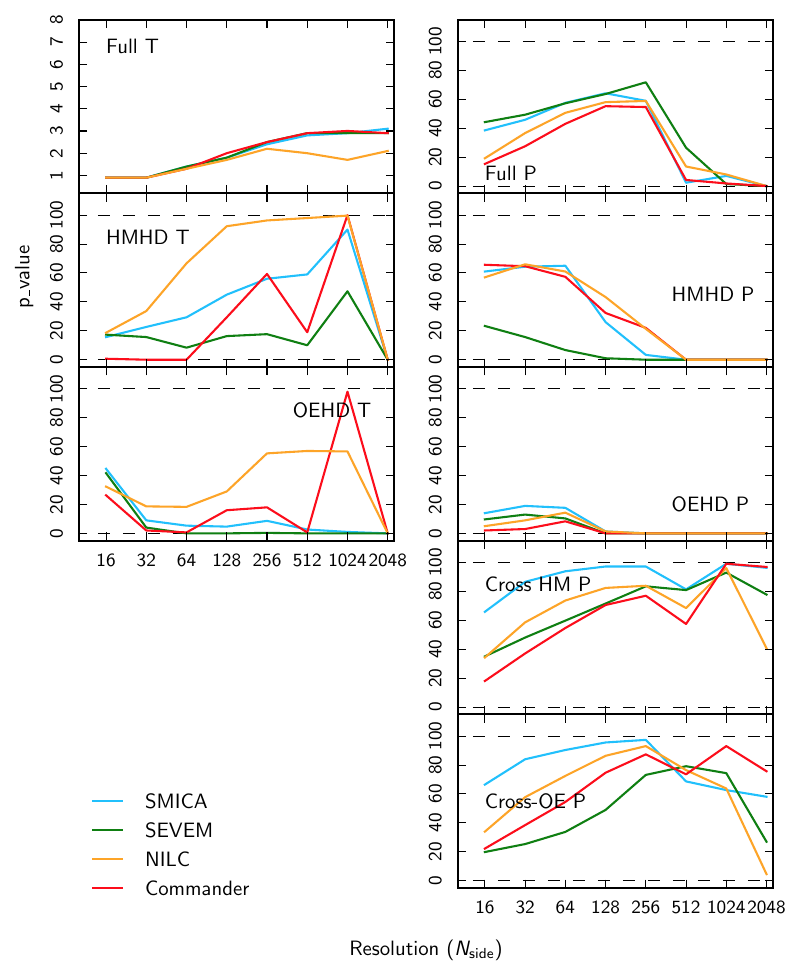}
\end{center}
\caption{Consistency between data and simulations as quantified in terms of pixel-space variance for both temperature (left) and polarization (right), and for both full mission maps (\textcolor{black}{top row), half difference maps (middle two rows), and for half-mission and odd-even cross variances (bottom two} rows). Coloured lines show results for the four different component-separation pipelines\textcolor{black}{, \commander\ (red), \smica\ (cyan), \sevem\ (green), and \nilc\ (orange),} as a function of pixel  resolution, $N_{\textrm{side}}$.}
\label{fig:simspec_variance}
\end{figure}

In Fig.~\ref{fig:simspec_variance} we plot the percentage of simulations with a lower variance than the real data, as a function of pixel resolution. The results from this analysis are in good agreement with those found in the power spectrum analysis. In temperature we find a generally good consistency between real data and half difference noise simulations, with few exceptions. We find that only a few simulations have a lower variance for the HMHD \commander\ map at very large scales and for the OEHD \sevem\ map at intermediate resolutions. At the maximum resolution of $N_{\rm side}=2048$, there is a lack of compatibility with MC simulations for both HMHD and OEHD maps, showing that at high resolution noise in temperature data is poorly described by the simulations. However, given the very high signal-to-noise ratio of the \Planck\ temperature data at all  resolutions, a small noise mismatch is irrelevant compared to the CMB cosmic variance.

For the signal-plus-noise data, we observe satisfactory consistency in temperature at high pixel resolutions. At lower resolutions we observe low probabilities, with p values of about 1.0\,\%, which are associated with the well known lack of power on large angular scales. These results are compatible with results reported in the previous release described in \cite{planck2014-a18}. We have also investigated the higher order moments, skewness and kurtosis in temperature as shown in \citet{planck2016-l07}, and find good consistency with Monte Carlo simulations at all resolutions.

\begin{figure}[htbp]  
\begin{center}
  \includegraphics[width=\columnwidth]{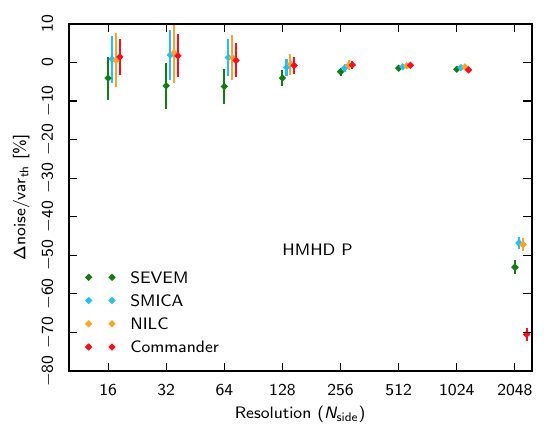}
  \includegraphics[width=\columnwidth]{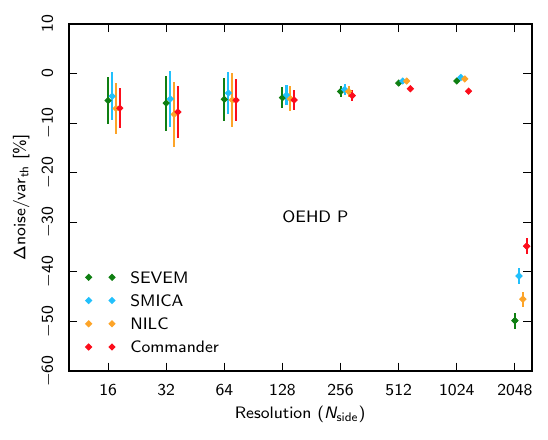}
\end{center}
\caption{Amplitude of the noise mismatch in terms of expected signal amplitude, $\textrm{var}_{\mathrm{th}}$, in percentage as a function of the pixel resolution,     $N_{\textrm{side}}$ for HMHD polarization data (top plot) and OEHD polarization data (bottom plot). Coloured lines show results for the four different component-separation pipelines\textcolor{black}{, \commander\ (red), \smica\ (cyan), \sevem\ (green), and \nilc\ (orange)}. Error bars show the amplitude of the MC dispersion at $\pm$ 1 $\sigma$, showing that where the noise is well characterized the bias is embedded in the uncertainty in the variance extraction and hence it is not significant.}
\label{fig:noise_mismatch}
\end{figure}

In polarization at high resolutions, results are not as robust, due to the noise mismatch. We observe an incompatibility for all the component-separated HMHD and OEHD maps between the MC distribution and real data at intermediate and high resolutions. This suggests that noise in the data (including systematic effects) is not fully characterized by the simulations. At lower resolutions ( $\nside=256$ for HMHD and 128 for OEHD), however, data are compatible with simulations, showing that  the noise properties are better represented. In Fig. \ref{fig:noise_mismatch} we show the amplitude of the noise mismatch with respect to the amplitude of the expected CMB variance as a function of pixel resolution. These results give an estimation of
the bias due to the noise mismatch in the extraction of the variance from signal plus noise data. The bias is very important at the highest resolution ($N_{\rm side}=2048$), with values of about 40--50\,\% for all the methods. At intermediate resolutions it is of the order of few percent. At large scales the bias is not significant, since half-difference data are compatible with the MC dispersion.

In spite of the presence of a noise mismatch, we find that cross- and auto-analyses of the signal-plus-noise maps are in agreement with MC simulations. At intermediate and large scales, auto- and cross-analyses are in good agreement with each other, showing the robustness of the analysis, although with some differences among the component-separation methods due to the presence of residual foregrounds, or systematic effects, or a different impact of the noise mismatch. At high resolution the differences between auto and cross results are mainly due to the noise mismatch, whose impact is more important for the auto analysis. On the other side, the cross-analyses may be biased by a poor description of the correlated noise that we cannot investigate with the above analyses.

In \cite{planck2016-l08} we consider more detailed analyses of this kind, including also the analysis of the \sevem\ frequency maps, in a way that minimizes the impact of the correlated noise.

\subsubsection{Assessing the impact of simulation noise bias}

In order to understand whether these percent-level noise discrepancies \textcolor{red}{in polarisation} are relevant for a given analysis, we strongly recommend considering the following questions while assessing the results.
\begin{enumerate}
\item \emph{Which angular scales are relevant for the statistic of  choice?} If the statistic is sensitive only to large angular scales  ($\ell \lesssim 50$), then the simulations are likely to be adequate. If not, see next question.

\item \emph{Is the statistic of choice sensitive to signal-plus-noise or noise alone?} If the former, then the simulations are likely to be adequate for $\ell \lesssim 1500$ for temperature, and $\ell \lesssim 250$ for polarization; if the latter, then see next
question.

\item \emph{Is the statistic of choice sensitive to $\lesssim$5\,\% errors in the noise model?} To quantify this, we recommend applying the statistic of choice to simulations for which the noise contribution is artificially re-scaled either up or down by 5\,\% (see Fig.~\ref{fig:simspec}), while the signal contribution is unchanged.  If the statistic of choice is unable to distinguish between the scaled and the unscaled ensembles, then the statistic is likely robust against the uncertainties in the current simulations. If not, caution is warranted. Typically, linear, cubic, or cross-spectrum type statistics are only marginally sensitive to this type of error in the noise model, whereas quadratic and auto-spectrum type statistics are typically highly sensitive.
\end{enumerate}

Clearly, no general prescription can be given for all analyses, and we therefore stress that caution is warranted when using the end-to-end simulations. That being said, they do provide the most complete description of the uncertainties in the data set currently available, and with an appropriate level of care, they should form the basis of most goodness-of-fit tests with the current data set. For several worked examples of applications of these simulations, see \citet{planck2016-l07}.

\subsection{Foreground template fits}

Next, we consider residual foreground contamination as measured by correlation between known foreground templates and the
cleaned CMB maps. Specifically, for a given cleaned temperature CMB map $\d$, a foreground template $\t$, and the common confidence mask $\m$ (consisting of 0's and 1's), we compute the correlation coefficient
\begin{equation}
r = \frac{1}{N_{\mathrm{pix}}-1} \sum_{i\in\m} \frac{\d_i -
  \left<\d\right>}{\sigma_{\d}} \frac{\t_i - \left<\t\right>}{\sigma_{\t}},
\end{equation}
where the sum runs over the $N_{\mathrm{pix}}$ pixels not excluded by the mask, $\left<\d\right> = 1/N_{\mathrm{pix}} \sum_i \d_i$, $\sigma_{\d} =  \left[1/N_{\mathrm{pix}-1} \sum_i   (\d_i-\left<\d\right>)^2\right]^{1/2}$, and similarly for $\t$. All maps are smoothed to a common resolution of $80\arcm$ FWHM, and pixelized at $\nside=128$. 

We consider four foreground templates in intensity, namely, the 408\,MHz \citet{haslam1982} map as processed by \citet{remazeilles2014} for synchrotron emission, the \Planck\ 2018 857-GHz map for thermal dust emission, the \citet{dame2001} map for CO line emission, and the \citet{finkbeiner2003} $H_\alpha$ map for free-free emission.  For polarization, we consider \textcolor{black}{the difference between the \WMAP\ 23-GHz and 33-GHz maps} as a synchrotron tracer. Uncertainties are evaluated from 300 end-to-end
simulations.  Corresponding results were reported in \citet{planck2014-a11} for the \Planck\ 2015 CMB sky maps.

The results from these calculations are summarized in Table~\ref{tab:tempfit}. In nearly all cases, we see that the correlation coefficients are lower for the 2018 maps than the corresponding 2015 maps, and most are within the $1\sigma$ confidence
limits. The only notable issue is a marginally significant polarization correlation with the \WMAP\ synchrotron tracer, ranging in statistical significance between $2.8\sigma$ for \commander\ to $3.5\sigma$ for \sevem. The absolute level of the correlation is low, however, ranging between 3 and 4\,\%.  

\begin{table*}[tb]       
\begingroup                                                                            
\newdimen\tblskip \tblskip=5pt
\caption{Correlation coefficients between known diffuse foreground templates and each of the four component-separated CMB
maps. The intensity templates are: (1)~a 408-MHz map for synchrotron emission from \citet{haslam1982}; (2)~an H$\alpha$ template for free-free emission from \citet{finkbeiner2003}; (3)~a tracer of CO emission in the Galactic plane from \citet{dame2001}; and (4)~the \Planck\ 857-GHz map for thermal dust  emission. For polarization, we only include the difference between the \emph{WMAP} K (23\,GHz) and Ka (33\,GHz) frequency maps as a synchrotron tracer. All maps have been smoothed to a common resolution of $80\arcm$ FWHM prior to the fitting process, and the correlation coefficients are evaluated outside the confidence     masks described in Sect.~\ref{sec:masks}.\label{tab:tempfit}}
\nointerlineskip                                                                                                                                                                                     
\vskip -2mm
\footnotesize                                                                                                                                      
\setbox\tablebox=\vbox{ %
\newdimen\digitwidth                                                                                                                          
\setbox0=\hbox{\rm 0}
\digitwidth=\wd0
\catcode`*=\active
\def*{\kern\digitwidth}
\newdimen\signwidth
\setbox0=\hbox{+}
\signwidth=\wd0
\catcode`!=\active
\def!{\kern\signwidth}
\newdimen\decimalwidth
\setbox0=\hbox{.}
\decimalwidth=\wd0
\catcode`@=\active
\def@{\kern\signwidth}
\halign{ \hbox to 1.2in{#\leaderfil}\tabskip=1.0em&
  \hfil$#$\hfil\tabskip=3em&
  \hfil$#$\hfil\tabskip=3em&
  \hfil$#$\hfil\tabskip=3em&
  \hfil$#$\hfil\tabskip=3em&
  \hfil$#$\hfil\tabskip=0em\cr
\noalign{\doubleline}
\omit&\omit&\multispan4\hfil {\sc Correlation Coefficient}\hfil\cr
\noalign{\vskip -3pt}
\omit&\omit&\multispan4\hrulefill\cr
\noalign{\vskip 3pt}
\omit&\omit\hfil {\sc Data Set}\hfil& \hfil \commander\hfil&\hfil \nilc\hfil&\hfil \sevem\hfil&\hfil \smica\hfil\cr
\noalign{\vskip 3pt\hrule\vskip 5pt}
\omit{\bf Intensity}\hfil\cr
\noalign{\vskip 6pt}
\hglue 1em Haslam&  2018&   -0.037\pm0.134& -0.023\pm0.135& -0.055\pm0.077& -0.027\pm0.077\cr
\omit&              2015&   -0.062\pm0.115& -0.051\pm0.116& -0.065\pm0.115& -0.023\pm0.069\cr
\noalign{\vskip 4pt}
\hglue 1em $H\alpha$& 2018& !0.000\pm0.032& !0.006\pm0.032& !0.003\pm0.028& !0.004\pm0.028\cr
\omit               & 2015& !0.010\pm0.071& !0.011\pm0.071& !0.019\pm0.071& !0.003\pm0.057\cr
\noalign{\vskip 4pt}
\hglue 1em CO&      2018&   -0.002\pm0.019& !0.000\pm0.019& !0.001\pm0.020& !0.001\pm0.020\cr
\omit&              2015&   -0.004\pm0.027& !0.003\pm0.027& !0.003\pm0.027& -0.007\pm0.022\cr
\noalign{\vskip 4pt}
\hglue 1em 857\,GHz& 2018&  -0.033\pm0.097& -0.019\pm0.097& -0.018\pm0.098& -0.019\pm0.098\cr
\omit              & 2015&  -0.043\pm0.084& -0.032\pm0.084& -0.037\pm0.084& -0.029\pm0.083\cr
\noalign{\vskip 10pt}
\omit{\bf Polarization}\hfil\cr
\noalign{\vskip 6pt}
\hglue 1em \emph{WMAP} K$-$Ka& 2018& -0.031\pm0.011& -0.037\pm0.011& -0.039\pm0.011& -0.033\pm0.011\cr
\omit                 & 2015& -0.057\pm0.026& -0.116\pm0.024& -0.026\pm0.025& -0.027\pm0.026\cr
\noalign{\vskip 5pt\hrule\vskip 3pt}
}}
\endPlancktablewide                                                                                                                                            
\endgroup
\end{table*}

\subsection{Power spectrum comparison}

Next, we characterize the cleaned CMB maps in terms of angular power spectra. As above, we employ the {\tt PolSpice} estimator for these calculations, and all spectra are evaluated outside the common mask defined in Sect.~\ref{sec:masks}.

\begin{figure}[htbp]  
\begin{center}
  \includegraphics[width=\columnwidth]{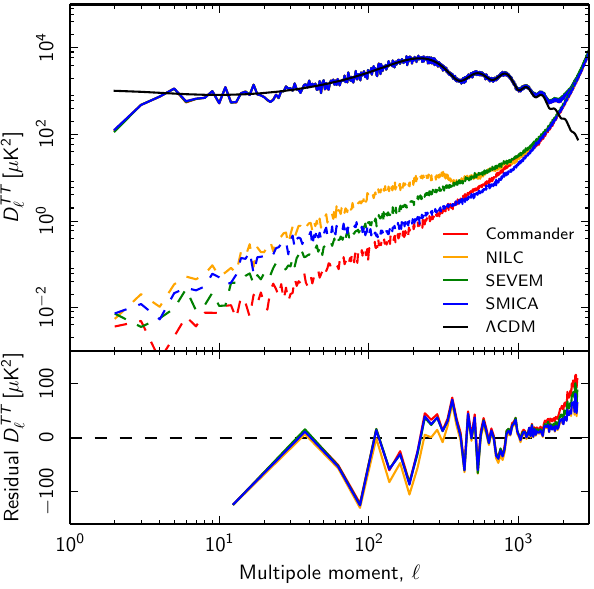}
\end{center}
\caption{Comparison of half-mission temperature power spectra. The top panel shows the half-sum (HMHS; solid lines) and half-difference (HMHD; dashed lines) power spectra, while the bottom panel shows the difference between the half-sum and the best-fit \Planck\ 2018 $\Lambda$CDM and half-difference spectra. The latter residual spectrum is binned with $\Delta\ell=25$.}
\label{fig:powspec_HM_TT}
\end{figure}

Figure~\ref{fig:powspec_HM_TT} shows a comparison of power spectra evaluated from the four cleaned CMB temperature half-mission maps. In the top panel, the solid lines show spectra computed from the half-mission half-sum (HMHS) maps, and thereby contain both signal and noise, while dashed lines show spectra computed from the HMHD, and thereby should contain only instrumental noise and systematic uncertainties. The black solid line shows the best-fit \Planck\ 2018 $\Lambda$CDM model derived from the combination of the low-$\ell$ $TT$, low-$\ell$ $EE$, high-$\ell$ $TT+TE+EE$, and lensing likelihoods \citep{planck2014-a15}. The bottom panel shows the residuals after subtracting both the best-fit $\Lambda$CDM model (as a signal tracer)
and the half-difference spectrum (as a noise tracer) from each of the half-sum spectra.

Overall, we observe good agreement among the four pipelines in terms of half-sum spectra up to $\ell \lesssim 1500$. The main notable feature is a small power deficit of about $10\muK^2$ in \nilc\ between $\ell=100$ and 300, corresponding to a relative deficit of 0.2\,\%. At these multipoles, the \Planck\ data are strongly CMB dominated, and algorithmic variations make little difference in terms of overall power. However, at higher multipoles the noise and compact source contributions become relevant, and in that regime the various approaches show slightly different behaviour, with \commander\ having the largest \textcolor{black}{unresolved} source imprint and \nilc\ the smallest.  

At low multipoles we also see differences among the codes in terms of noise. The lowest noise is achieved by \commander, which also exhibits nearly white noise with a scaling of $\mathcal{O}(\ell^2)$. The highest low-$\ell$ noise -- almost an order of magnitude higher than \commander\ -- is seen for \nilc\ for $\ell\lesssim300$. This is not unexpected given the nature of the
\nilc\ algorithm. On large angular scales, the \nilc\ frequency weights primarily adjust themselves to suppress foregrounds, while on small scales, they converge to inverse-noise-variance weighting. In this respect, \commander\ is different from the other three codes in that it explicitly uses estimates of the noise standard deviation to perform inverse-variance noise
weighting per pixel. Finally, for \smica\ we note that the noise decreases around $\ell\approx100$, which corresponds to the multipole at which the three LFI frequencies are excluded at high latitudes (see Appendix~\ref{app:smica}).

In Fig.~\ref{fig:powspec_HM_EE_BB} we present a similar comparison for the $EE$ and $BB$ HMHD polarization power spectra. As in Fig.~\ref{fig:powspec_HM_TT}, the solid lines includes contributions from both signal and noise. Here we see, at
least at the level of visual inspection, that all four codes perform similarly in terms of polarization power spectrum reconstruction, both for HMHS and HMHD spectra. The only marginal outlier is \nilc, which exhibits slightly higher $BB$ HMHS and HMHD spectra at multipoles lower than $\ell \lesssim 100$. However, this excess disappears in the difference between the HMHS and HMHD spectrum (bottom panels of Fig.~\ref{fig:powspec_HM_EE_BB}), suggesting that it is due to a somewhat higher noise level in the \nilc\ map compared to the others, and not a signal bias.

\begin{figure*} 
\begin{center}
  \includegraphics[width=\columnwidth]{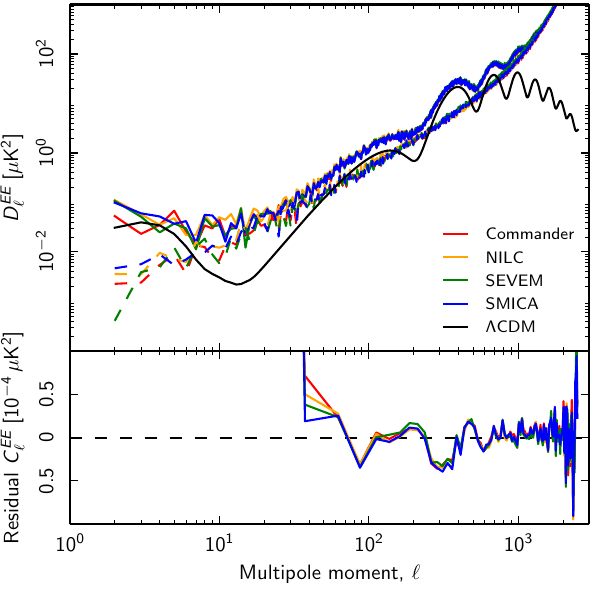}
  \includegraphics[width=\columnwidth]{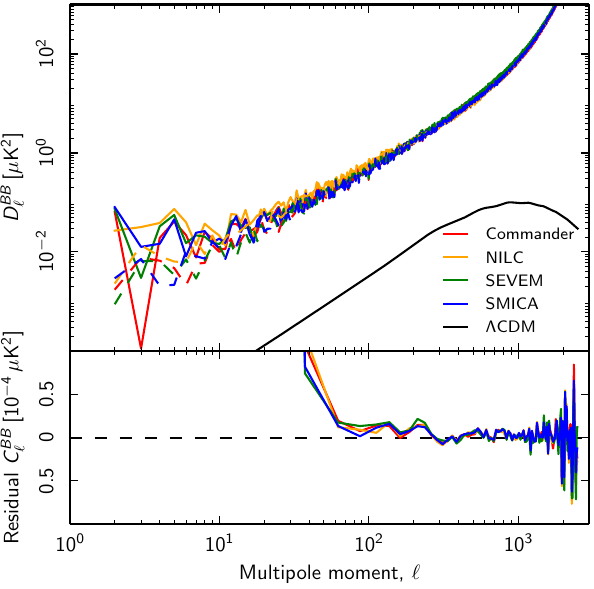}
\end{center}
\caption{Comparison of half-mission polarization power spectra. The top panel shows the half-sum (solid lines) and half-difference (dashed lines) power spectra, while the bottom panel shows the difference between the half-sum and the best-fit \Planck\ 2018 $\Lambda$CDM and half-difference spectra. The latter residual spectrum is binned with $\Delta\ell=25$.}
\label{fig:powspec_HM_EE_BB}
\end{figure*}

\begin{figure*}   
\begin{center}
  \includegraphics[width=\columnwidth]{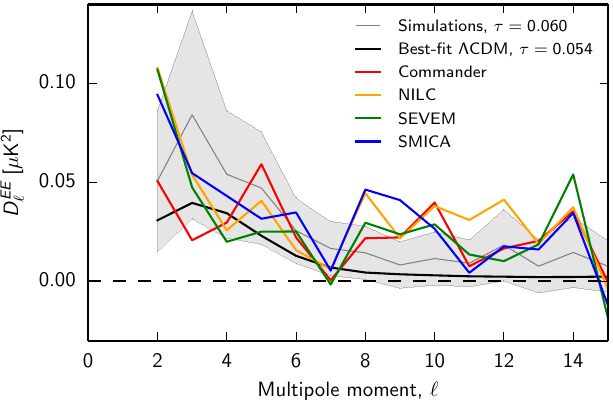}
  \includegraphics[width=\columnwidth]{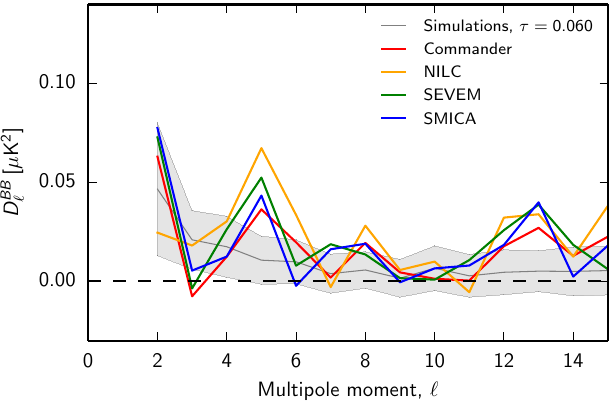}
\end{center}
\caption{Comparison of low-$\ell$ half-mission polarization EE (\emph{left}) and BB (\emph{right}) power spectra. Each spectrum is computed as the difference between the respective half-sum and the half-difference spectrum. Grey bands show $1\sigma$ confidence regions derived from 300 end-to-end simulations as processed by \commander. Formally speaking, these can therefore only be directly compared with the red curve. Note that the value for the optical depth of reionization adopted for these simulations is $\tau=0.060$ (see \citealp{planck2016-l02,planck2016-l03} for details), which is larger than the best-fit \Planck\ 2018 value of $\tau=0.054$. }
\label{fig:powspec_HM_EE_BB_lowl}
\end{figure*}

Finally, in Fig.~\ref{fig:powspec_HM_EE_BB_lowl} we show an expansion of the low multipole part of the polarization power spectra without applying any multipole binning. The black solid line in the left panel shows the best-fit \Planck\ 2018 $\Lambda$CDM model for which the posterior mean optical depth of re-ionization is $\tau=0.054\pm0.019$ \citep{planck2016-l06}. The left and right panels show the $EE$ and $BB$ spectra, respectively. The grey curves indicate corresponding
spectra computed from 300 end-to-end \commander\ simulations. 

Starting with the $BB$ spectrum, we note that there is an overall significant excess compared to zero. This excesss, however, is reproduced in the simulations, as seen by the non-zero mean of simulations, and therefore reflects the presence of understood residual correlations in the data; see \citet{planck2016-l03} for further discussion. Consequently, we re-emphasize the importance of comparing these data with full end-to-end simulations when subjecting them to cosmological analysis, in order to adequately capture this type of residual noise correlation. A similar noise excess is seen in  the $EE$ spectrum for $\ell \gtrsim 8$, with an amplitude similar to the $BB$ spectrum.

The $EE$ spectrum does not not show a clear detection of the reionization peak. As mentioned, the solid black curve shown in
Fig.~\ref{fig:powspec_HM_EE_BB_lowl} indicates a spectrum for which $\tau=0.054$, and this amplitude is too low to be visually observed in an $\ell$-by-$\ell$ spectrum, in particular in the presence of the noise excess mentioned above. In order to detect this peak, a full likelihood analysis is essential, as presented in \citet{planck2016-l05} and \citet{planck2016-l06}.

While the \Planck\ 2018 best-fit value for the optical depth of reionization is $\tau=0.054$, the value used to generate the simulations (which had to be adopted well before the final results were available for computational expense reasons) was
$\tau=0.060$. The effect of this difference can be seen in Fig.~\ref{fig:powspec_HM_EE_BB_lowl} as an excess of power in the simulations relative to the best-fit model at low multipoles in $EE$.  While the difference is within $1\sigma$, it is worth having this issue in mind when studying low-$\ell$ polarization effects with these maps and simulations; see section~3 of \citet{planck2016-l07} for a quantitative analysis of this issue. 

A full cosmological likelihood and parameter analysis of the \Planck\ 2018 data is presented in \citet{planck2016-l05}, based on
cross-spectrum techniques. In this paper, we perform a simple consistency test between the full likelihood analysis and the cleaned CMB maps presented in this paper, by fitting a CMB spectrum (parametrised by an amplitude $A_{\mathrm{CMB}}$ and tilt $n$ relative to the best-fit \Planck\ 2018 $\Lambda$CDM model) and an $\ell^2$ point-source contribution to the difference between the HMHS and HMHD spectra. Explicitly, we adopt the signal model
\begin{equation}
 D_{\ell} = A_{\mathrm{CMB}} \left(\ell/\ell_{0}\right)^n f_{\ell}^2
 D_{\ell}^{\Lambda\mathrm{CDM}} + A_{\mathrm{ps}}
 \ell(\ell+1)/(\ell_{\mathrm{ps}}(\ell_{\mathrm{ps}}+1)),
\end{equation}
where $f_{\ell}$ is the transfer function shown in Fig.~\ref{fig:transferfunctions}, $\ell_0 = 600$ is a pivot multipole for the CMB fit, and $\ell_{\mathrm{ps}}=500$ is a pivot multipole for the point source contribution.  The quantity $D_{\ell} {\Lambda\mathrm{CDM}}$ is the best-fit \Planck\ 2018 $\Lambda$CDM power spectrum. Each analysis includes multipoles between $\ell=2$ and 1500 (for which the simulations agree well with the data; see Fig.~\ref{fig:simspec}), and
all spectra are binned with $\Delta\ell=20$. Uncertainties within each bin are defined as the standard deviation of the observed spectrum within the bin. The number of degrees of freedom for the $\chi^2$ is $n_{\mathrm{dof}}=75$. The fit is performed with a simple Metropolis MCMC sampler.

The results from these calculations are summarized in Table~\ref{tab:params}. Overall, we find good agreement between the
cleaned CMB maps and the likelihood analysis, with most $\Lambda$CDM amplitudes consistent with unity within $2\sigma$ and all tilt parameters consistent with zero within $1.5\sigma$. All $\chi^2$s are also reasonable, ranging between 57.8 and 74.0 for 75 degrees of freedom, corresponding to probabilities-to-exceed (PTEs) ranging between 0.07 and 0.49. Finally, as already noted, \commander\ exhibits the largest point-source contribution, with an amplitude of $A_{\mathrm{ps}}=2.7\pm0.5\muK^2$ at $\ell=500$ in temperature, while \nilc\ and \smica\ show the smallest contribution, with amplitudes of
$A_{\mathrm{ps}}=1.9\muK^2$ at $\ell=500$.

\begin{table}[t]  
\begingroup
\newdimen\tblskip \tblskip=5pt
\caption{Parameter fits to HM power spectra. In each case, the observed spectrum is taken as the difference between the HMHS and HMHD spectra, and the model fitted reads $D_{\ell} = A_{\mathrm{CMB}} \left(\ell/\ell_{0}\right)^n f_{\ell}^2 D_{\ell}^{\Lambda\mathrm{CDM}} + A_{\mathrm{ps}} \ell(\ell+1)/(\ell_{\mathrm{ps}}(\ell_{\mathrm{ps}}+1))$, where $f_{\ell}$ is the transfer function shown in Fig.~\ref{fig:transferfunctions}, $\ell_0 = 600$, and $\ell_{\mathrm{ps}}=500$. $D_{\ell}^{\Lambda\mathrm{CDM}}$ is the best-fit \Planck\ 2018 $\Lambda$CDM power spectrum. Each analysis includes multipoles between $\ell=2$ and 1500.  Spectra are binned with $\Delta\ell=20$. The uncertainties within each bin are defined as the standard deviation of the observed spectrum within the bin. The number of degrees of freedom for the $\chi^2$ is $n_{\mathrm{dof}}=75$. \label{tab:params}}
\vskip -6mm
\footnotesize
\setbox\tablebox=\vbox{
\newdimen\digitwidth
\setbox0=\hbox{\rm 0}
\digitwidth=\wd0
\catcode`*=\active
\def*{\kern\digitwidth}
\newdimen\signwidth
\setbox0=\hbox{+}
\signwidth=\wd0
\catcode`!=\active
\def!{\kern\signwidth}
\newdimen\decimalwidth
\setbox0=\hbox{.}
\decimalwidth=\wd0
\catcode`@=\active
\def@{\kern\signwidth}
\halign{\hbox to 0.9in{#\leaderfil}\tabskip=1.0em&
    \hfil#\hfil\tabskip=0.4em&
    \hfil#\hfil\tabskip=0.2em&
    \hfil#\hfil\tabskip=0.2em&
    \hfil#\hfil\tabskip=0em\cr
\noalign{\doubleline}
\omit\hfil Pipeline\hfil& $A_{\mathrm{CMB}}$& $n$& $A_{\mathrm{ptsrc}}$ [$\mu\mathrm{K}^2$]& $\chi^2$\cr 
\noalign{\vskip 5pt\hrule\vskip 5pt}
\noalign{\vskip 3pt}
\multispan5{\bf \boldmath $TT$}\hfil\cr 
\noalign{\vskip 3pt}
\hglue 0em \commander& $0.997\pm0.002$& $-0.003\pm0.004$& $2.7\pm0.5$& 61.6\cr
\hglue 0em \nilc& $0.994\pm0.002$& $-0.006\pm0.004$& $1.9\pm0.6$& 59.3\cr
\hglue 0em \sevem& $0.996\pm0.002$& $-0.001\pm0.004$& $2.1\pm0.6$& 60.3\cr
\hglue 0em \smica& $0.996\pm0.002$& $-0.001\pm0.002$& $1.9\pm0.5$& 58.5\cr
\noalign{\vskip 3pt}
\multispan5{\bf \boldmath $EE$}\hfil\cr 
\noalign{\vskip 3pt}
\hglue 0em \commander& $0.985\pm0.007$& $-0.002\pm0.009$& $0.35\pm0.06$& 74.0\cr
\hglue 0em \nilc& $0.984\pm0.007$& $-0.007\pm0.010$& $0.26\pm0.08$& 70.3\cr
\hglue 0em \sevem& $0.983\pm0.008$& $-0.001\pm0.010$& $0.27\pm0.08$& 62.7\cr
\hglue 0em \smica& $0.983\pm0.007$& $-0.002\pm0.008$& $0.30\pm0.08$& 66.7\cr
\noalign{\vskip 5pt\hrule\vskip 2pt}
}}
\endPlancktablewide 
\endgroup
\end{table}

\subsection{The real-space N-point correlation functions} 
\label{sec:2point_correlation}

A complementary measure of correlations and non-Gaussianity are given by real-space 2- and 3-point correlation functions. These functions are defined as the average product of $N$ observed fields, measured in a fixed relative distance on the sky.  In the case of the CMB, the fields correspond to temperature anisotropy $\Delta T$ and two Stokes parameters $Q$ and $U$ describing the linear polarization of the radiation in a given direction (see \citealt{planck2016-l07} for more detail).

Because of computational limitations, we restrict our analysis to the pseudo-collapsed and equilateral configurations of the 3-point functions and low resolution CMB maps with a resolution parameter $N_{\rm side}=64$ and smoothed with a $160\arcm$ FWHM Gaussian beam (see \citealt{planck2016-l07} for a description of the procedure for downgrading and smoothing  the maps). For both temperature and polarization, we employ the common masks described above, downgraded in the same way as CMB maps.  Because we analyse half-difference maps, the common mask is combined with the corresponding unobserved pixel mask.  The resulting 2- and 3-point correlation functions for the \commander\ HMHD and OEHD maps are presented in Fig.~\ref{fig:npt_commander} (figures for the remaining component separation maps can be found in the Appendix \ref{app:npt_functions}).  We use a simple $\chi^2$ statistic to quantify the agreement between the observed data and the full-focal-plane noise simulations (FFP10; \citealp{planck2014-a14}).  Table~\ref{tab:prob_npt} lists the significance level in terms of the fraction of simulations with a larger $\chi^2$ value than the observed map. Corresponding
analysis for full CMB maps is provided in \citet{planck2016-l07}.

\begin{figure*}[htp!]  
\begin{center}
\includegraphics[width=0.48\textwidth]{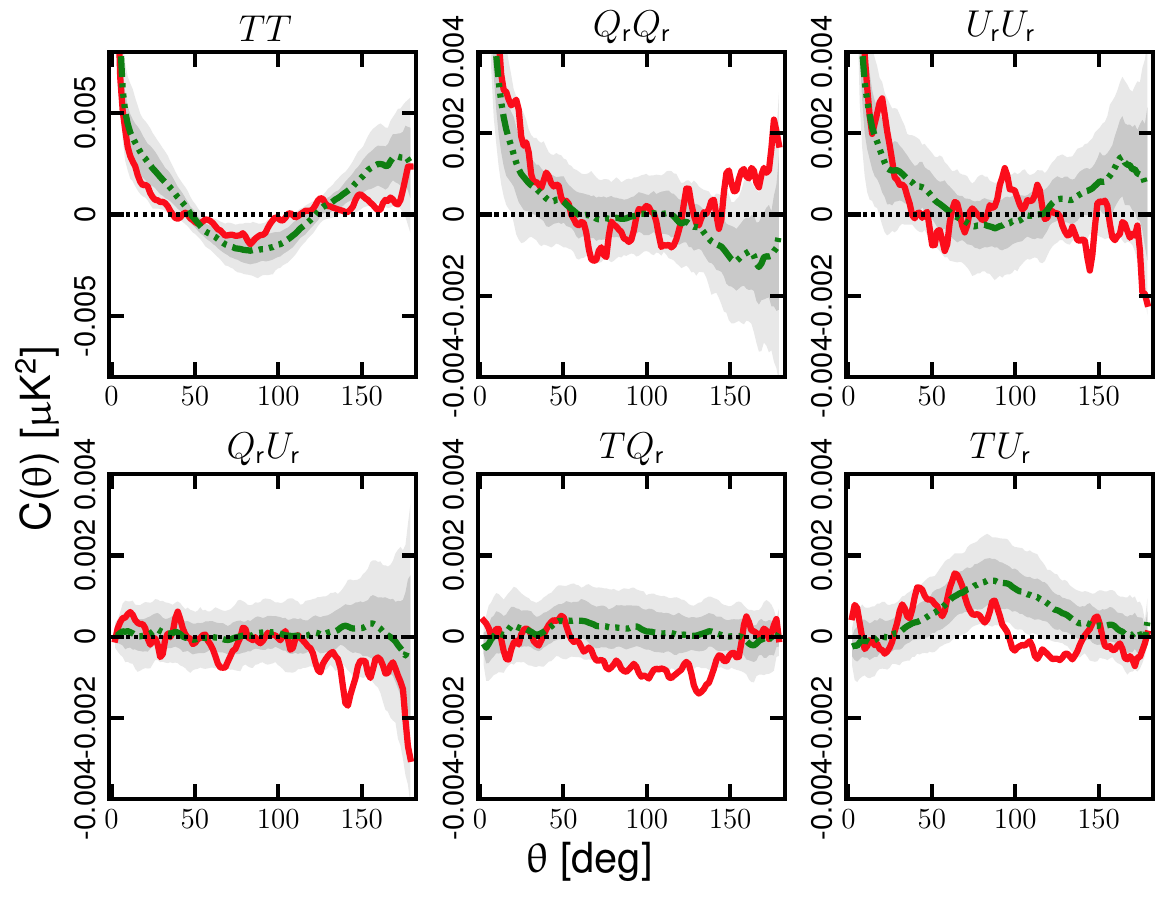}
\includegraphics[width=0.48\textwidth]{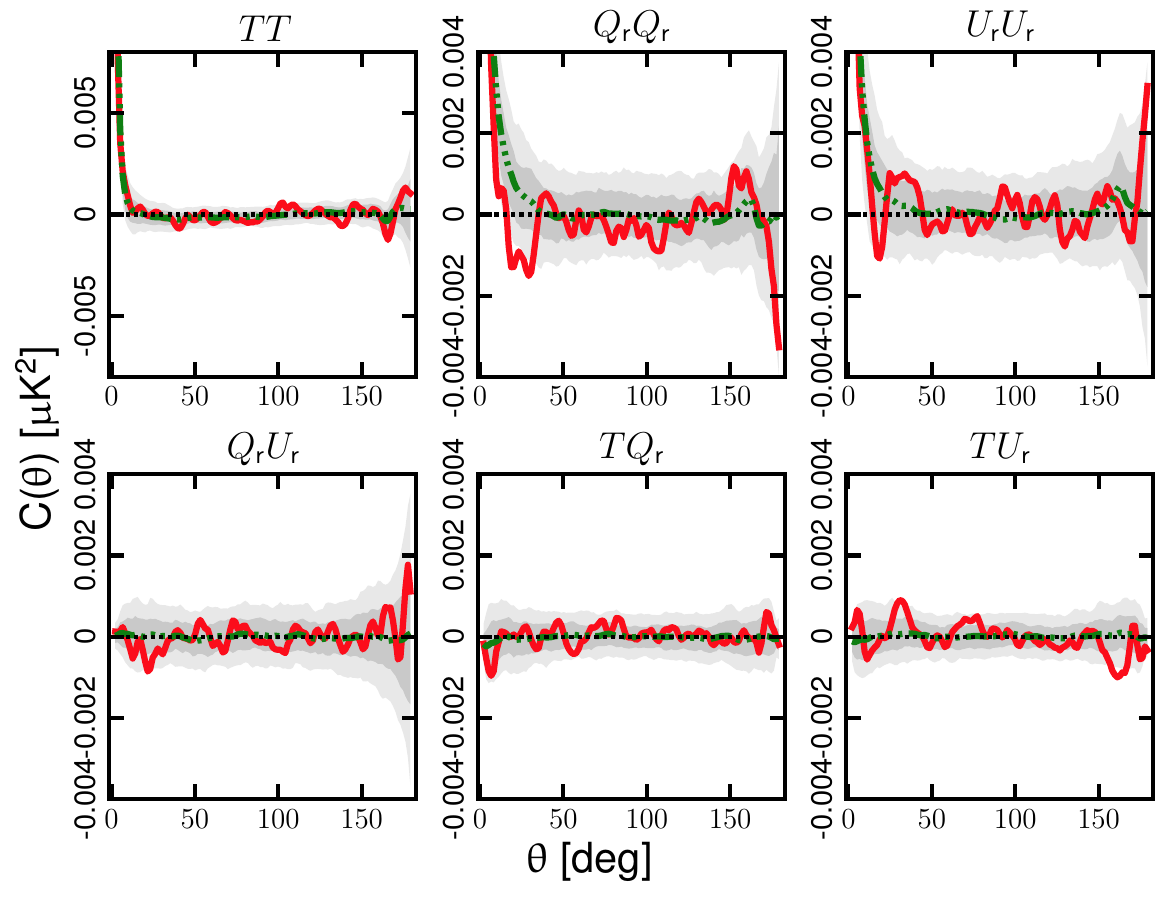} \\
\includegraphics[width=0.48\textwidth]{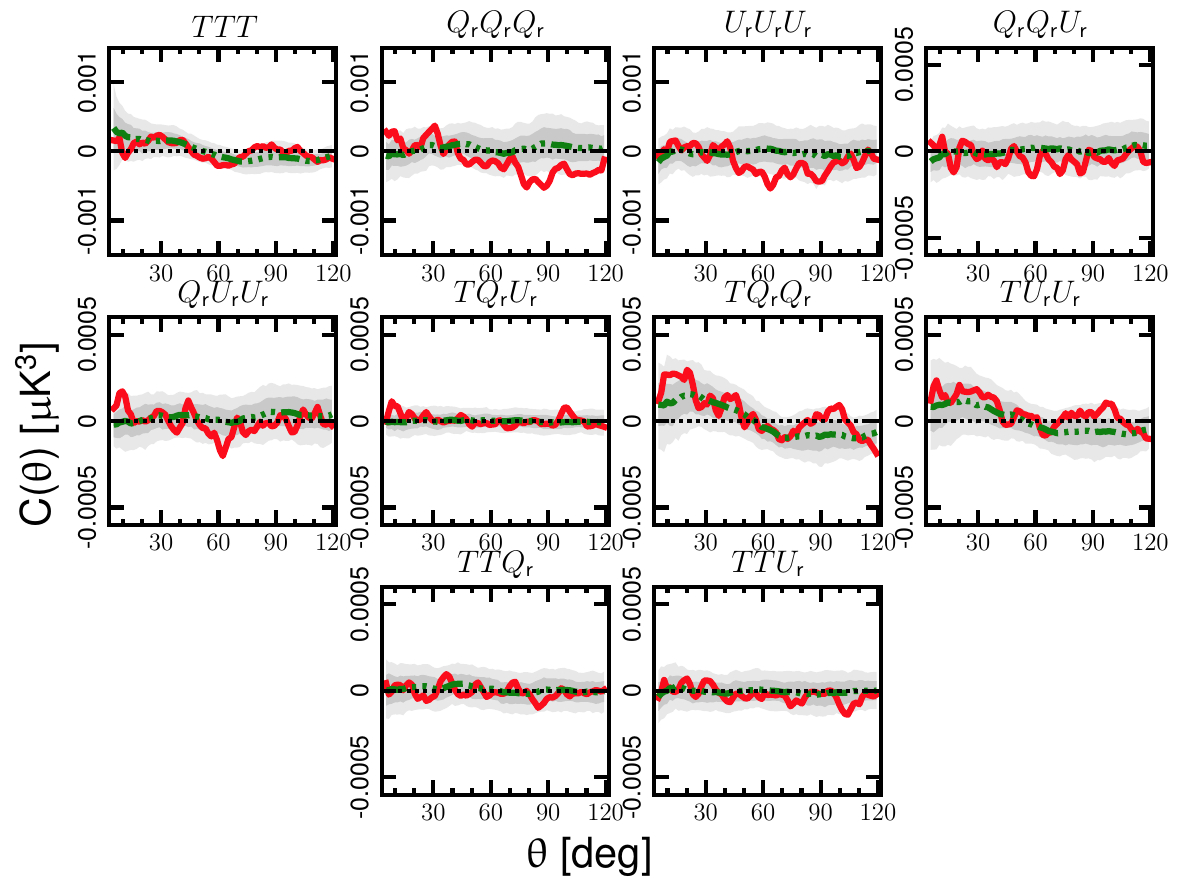}
\includegraphics[width=0.48\textwidth]{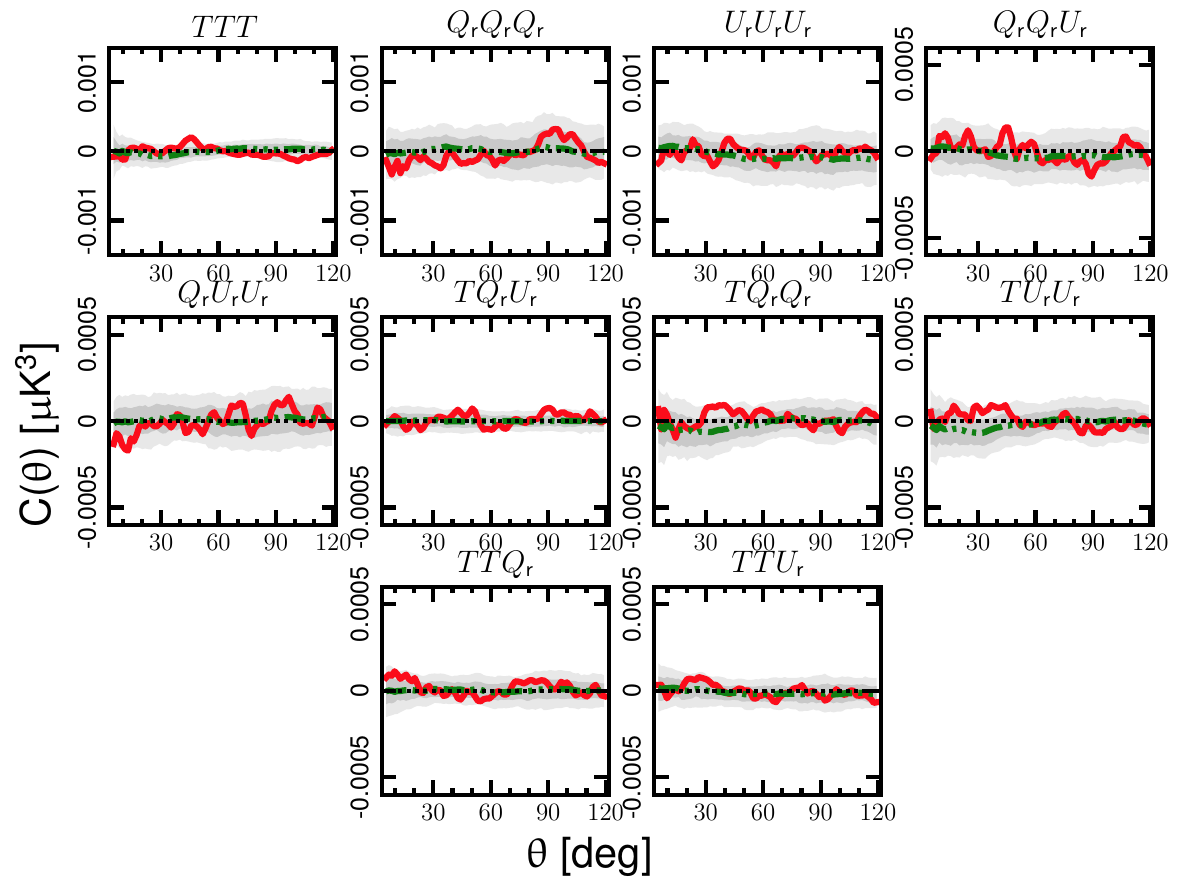}\\
\includegraphics[width=0.48\textwidth]{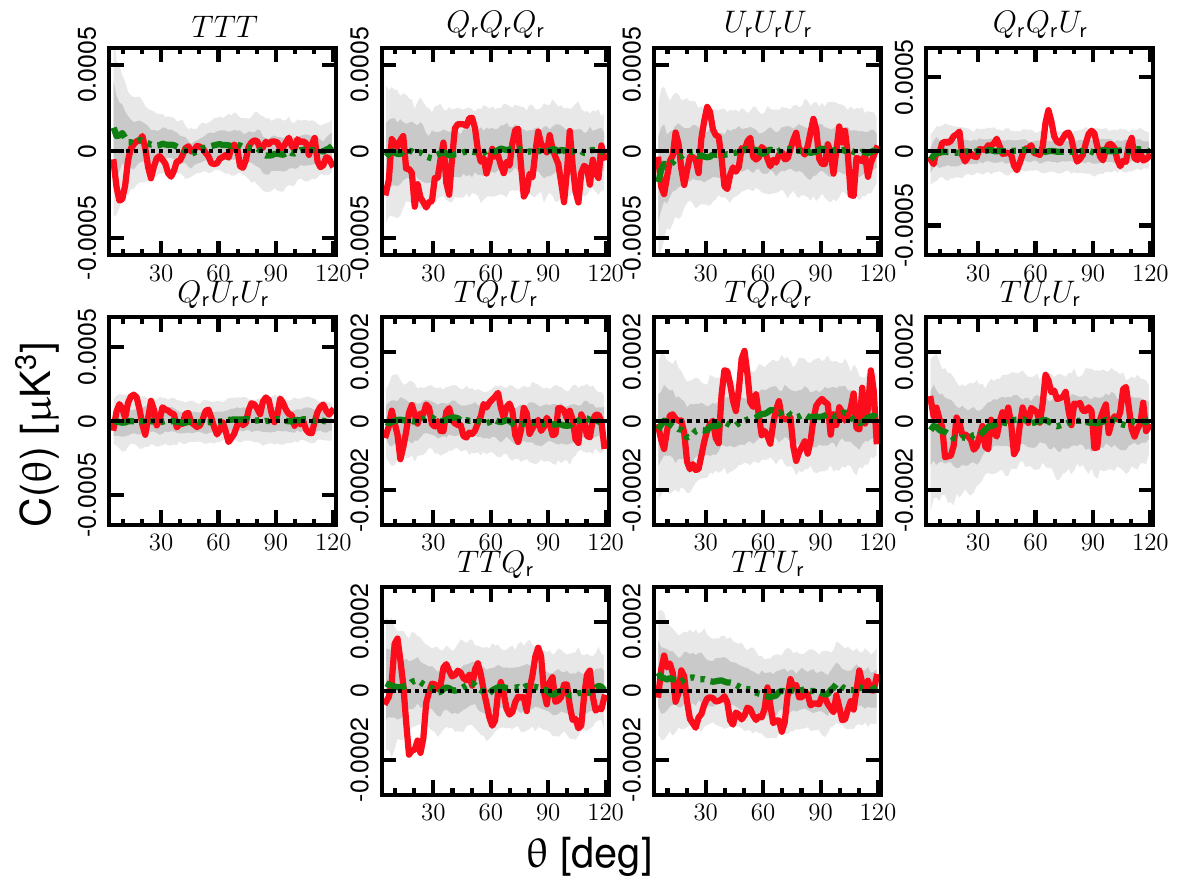}
\includegraphics[width=0.48\textwidth]{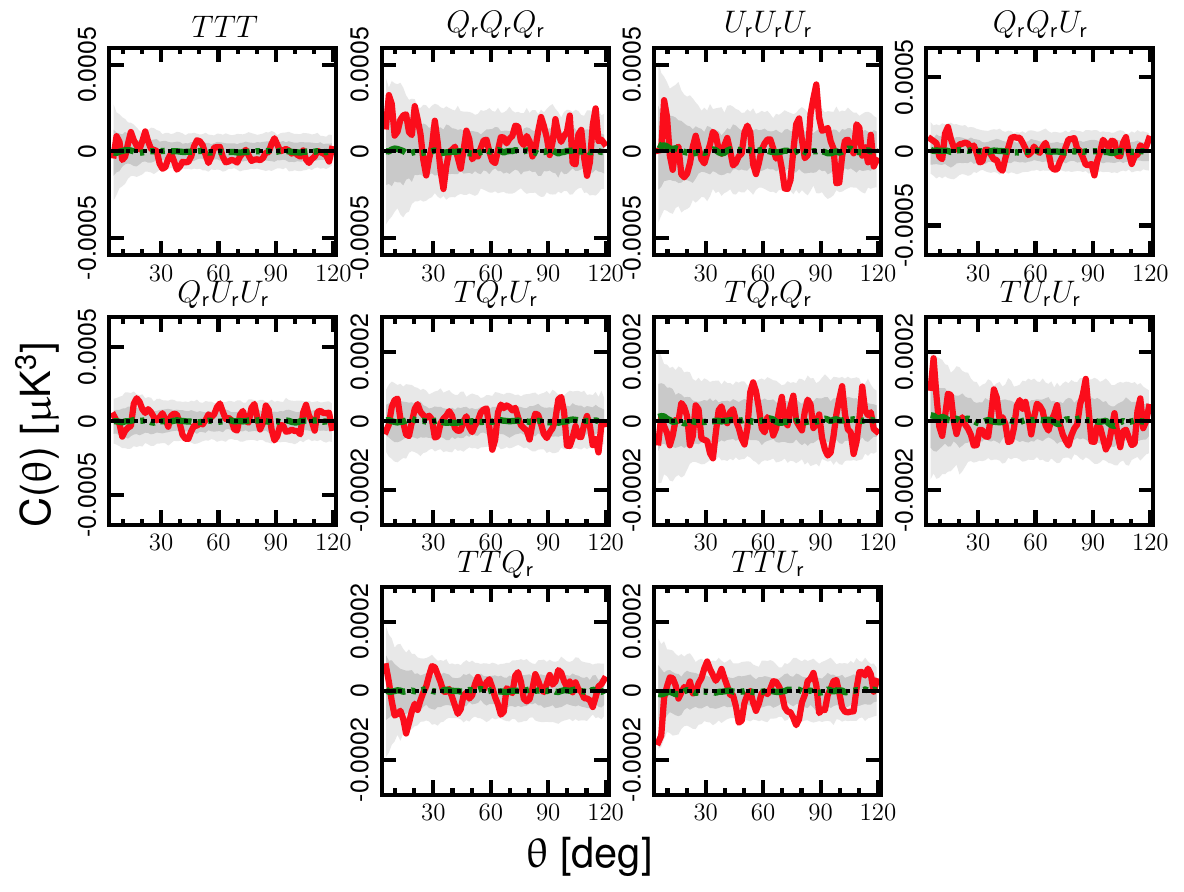}
\caption{The 2-point (upper panels), pseudo-collapsed (middle
  panels) and equilateral (lower panels) 3-point correlation functions determined from the $\nside=64$  \Planck\ \commander\ HMHD (left panels) and OEHD (right panels) temperature and polarization map. The red solid  lines correspond to the half-difference maps (HMHD or OEHD). The green triple-dot-dashed lines indicate the mean  determined from 300 FFP10 noise simulations. The shaded dark and light  grey regions indicate the corresponding 68\,\% and 95\,\% confidence
 regions, respectively.}
\label{fig:npt_commander}
\end{center}
\end{figure*}

\begin{table}[t]  
\begingroup
\newdimen\tblskip \tblskip=5pt
\caption{Probabilities in percentages of obtaining values for the $\chi^2$ statistic of the $N$-point functions for FFP10 simulations at least as large as the those obtained from the observed {\tt   Commander}, {\tt NILC}, {\tt SEVEM}, and {\tt SMICA} temperature and polarization maps with resolution parameter $N_{\rm side}=64$. Results are given for both the HMHD and OEHD data splits.}
\label{tab:prob_npt}
\nointerlineskip
\vskip -3mm
\scriptsize
\setbox\tablebox=\vbox{
   \newdimen\digitwidth 
   \setbox0=\hbox{\rm 0} 
   \digitwidth=\wd0 
   \catcode`*=\active 
   \def*{\kern\digitwidth}
   \newdimen\signwidth 
   \setbox0=\hbox{+} 
   \signwidth=\wd0 
   \catcode`!=\active 
   \def!{\kern\signwidth}
\halign{\hbox to 0.7in{#\leaderfil}\tabskip 4pt&
\hfil#\hfil\tabskip=1em&
\hfil#\hfil\tabskip=1em&
\hfil#\hfil\tabskip=1em&
\hfil#\hfil\tabskip=2em&
\hfil#\hfil\tabskip=1em&
\hfil#\hfil\tabskip=1em&
\hfil#\hfil\tabskip=1em&
\hfil#\hfil\tabskip 0pt\cr
\noalign{\doubleline\vskip -1pt}
\omit&\multispan4 \hfil {\sc HMHD split}\hfil&\multispan4 \hfil {\sc OEHD split}\hfil\cr
\noalign{\vskip -4pt}
\omit&\multispan4\hrulefill&\multispan4\hrulefill\cr
\omit\hfil {\sc Function}\hfil&{\tt Comm.}&{\tt NILC}&{\tt SEVEM}&{\tt SMICA}&{\tt Comm.}&{\tt NILC}&{\tt SEVEM}&{\tt SMICA}\cr
\noalign{\vskip 3pt\hrule\vskip 6pt}
\multispan5{\bf 2-point functions}\hfil\cr
\noalign{\vskip 3pt}
\hglue 1em$TT$ & 50.0& 81.3& 97.0& 87.7& *6.7& 81.7& 75.3& 60.3\cr
\hglue 1em$Q_r Q_r$& 20.7& 42.3& 40.3& 49.3& 52.7& 30.3& 90.0& 54.7\cr
\hglue 1em$U_r U_r$& 12.0& 21.7& 48.3& 51.0& 40.7& 51.7& 75.3& 33.7\cr
\hglue 1em$T Q_r$& 36.0& *2.0& 29.0& 36.0& 98.0& 90.0& 93.3& 83.7\cr
\hglue 1em$T U_r$& *1.3& 21.3& 45.0& 33.7& 49.0& 35.7& 96.7& 44.7\cr
\hglue 1em$Q_r U_r$& 47.7& 51.3& 76.3& 68.7& 76.0& 78.7& 77.3& 96.3\cr
\noalign{\vskip 6pt}
\multispan5{\bf Pseudo-collapsed 3-point functions}\hfil\cr
\noalign{\vskip 3pt}
 \hglue 1em$TTT$& 32.3& 29.3& 22.7& 50.3& 50.0& 28.0& 20.0& 90.7\cr
 \hglue 1em$Q_r Q_r Q_r$& 33.7& 18.3& 25.0& 30.7& 96.0& 27.0& 72.3& 75.7\cr
\hglue 1em $U_r U_r U_r$& *8.0& *9.0& 42.7& 11.0& 84.0& 99.7& 45.3& 94.3\cr
\hglue 1em $TTQ_r$& 63.7& 20.7& 77.0& 67.7& 17.0& 60.3& 99.0& 95.3\cr
 \hglue 1em$TTU_r$& 82.0& 59.0& 42.3& 74.7& 94.3& 18.3& 63.0& 26.3\cr
\hglue 1em $TQ_r Q_r$& 23.0& *2.3& 33.7& *7.0& 99.7& 94.7& 50.7& 78.7\cr
\hglue 1em $TU_r U_r$& 70.0& 64.3& 55.7& 48.3& 98.3& 95.3& 75.3& 90.7\cr
 \hglue 1em$TQ_r U_r$& 82.7& 69.7& 96.0& 39.7& 30.3& 98.0& 29.7& 94.0\cr
 \hglue 1em$Q_r Q_r U_r$& 50.0& 68.3& 73.7& 80.0& 66.3& 72.7& 77.7& 98.3\cr
 \hglue 1em$Q_r U_r U_r$& 64.7& 90.0& 89.0& 69.3& 45.3& 27.3& 52.7& 79.3\cr
 \noalign{\vskip 6pt}
 \multispan5{\bf Equilateral 3-point functions}\hfil\cr
 \noalign{\vskip 3pt}
 \hglue 1em$TTT$& 74.0& 62.0& 70.7& 91.7& 82.7& 84.3& 85.0& 53.3\cr
 \hglue 1em$Q_r Q_r Q_r$& 30.3& 26.0& 95.7& 14.3& 81.7& 94.7& 23.3& 41.0\cr
 \hglue 1em$U_r U_r U_r$& 90.7& 91.0& 91.0& 74.0& 35.3& 93.3& *4.7& 70.0\cr
 \hglue 1em$TTQ_r$& 34.0& 50.3& 50.7& 38.0& 93.0& 56.3& 49.7& 55.3\cr
\hglue 1em $TTU_r$& 94.0& 29.0& 37.0& 40.3& 82.3& 75.3& 96.0& 86.3\cr
\hglue 1em $TQ_r Q_r$& 54.3& 72.7& 51.0& 58.3& 51.7& 73.0& 82.3& 85.7\cr
\hglue 1em $TU_r U_r$& 92.3& 99.0& 96.7& 96.0& 88.7& 70.3& >99.7*& 69.0\cr
 \hglue 1em$TQ_r U_r$& 96.3& 87.3& 27.0& 89.3& 90.7& 52.0& 70.0& 54.0\cr
 \hglue 1em$Q_r Q_r U_r$& 64.0& 82.7& 97.3& 74.0& 68.0& 55.7& 73.3& 72.0\cr
\hglue 1em $Q_r U_r U_r$& 58.7& 69.7& 68.3& 39.0& 38.7& 26.3& 41.3& 29.3\cr
 \noalign{\vskip 3pt\hrule\vskip 3pt}}}
\endPlancktable  
\endgroup
\end{table} 

We can observe quite significant scatter in the results for the half-difference maps estimated using the different component-separation methods. This is not surprising, as different component-separation methods respond to noise and systematic effects in a different way.
No statistically significant deviations between data and simulations are found at these angular scales except a few cases for the 3-point functions with deviation significance around 99\,\%.
Note, however, that the confidence regions derived from the noise simulations do vary between methods, indicating that each method results in different effective statistical properties. To avoid biases, it is therefore essential to analyse each map together with the simulations constructed specifically for that map.

\subsection{Gravitational lensing}\label{sec:lensing}

As an example of science that may be extracted from the cleaned CMB maps presented in this paper, we consider reconstruction of the gravitational lensing potential. For a complete analysis of this topic, we refer the interested reader to \citet{planck2016-l08}, from which the following results are reproduced.

Gravitational lensing of CMB photons by large-scale structures induces slight distortions in the statistics of the CMB. In particular, lensing deflections result in a characteristic acoustic peak smoothing signature in the angular power spectrum, and they induce a non-zero four-point CMB correlation function. We use the methodology described in \cite{planck2016-l08} to reconstruct the lensing power spectrum.  With the sensitivity and sky coverage of \Planck, this approach constrains the lensing deflection power spectrum to a few percent, with most of the signal coming from temperature observations at high multipoles $\ell \sim 1500$. These measurements therefore result in a stringent consistency test between the various component-separation methods at small angular scales.

For masking, we employ the union of the intensity and polarization mask recommended in Sect.~\ref{sec:masks}, combined with a Galactic mask allowing $f_{\rm sky} = 0.70$, as well as a mask removing resolved SZ clusters with $\hbox{S/N} > 5$, as given by the \Planck\ 2015 SZ catalogue (\citealt{planck2014-a35}; for full details, see \citealt{planck2016-l08}). We consider quadratic lensing estimates built from temperature only ($\hat \phi^{\rm TT}$), as well as the full minimum variance combination ($\hat \phi^{\rm MV}$).  The minimum-variance estimator is derived from the full set of quadratic estimators $TT, TE, TB, EE$, and $EB$, which increases the signal-to-noise ratio with respect to $TT$ by roughly 20\,\%. As discussed in  Sect.~\ref{sec:cmb}, there is slight power mismatch between data and simulation power on the scales relevant for lensing. To account for this, we add in each case additional power as an isotropic, Gaussian component either to the simulations or to the data.

\begin{figure}[htbp]  
\begin{center}
  \includegraphics[width=\columnwidth]{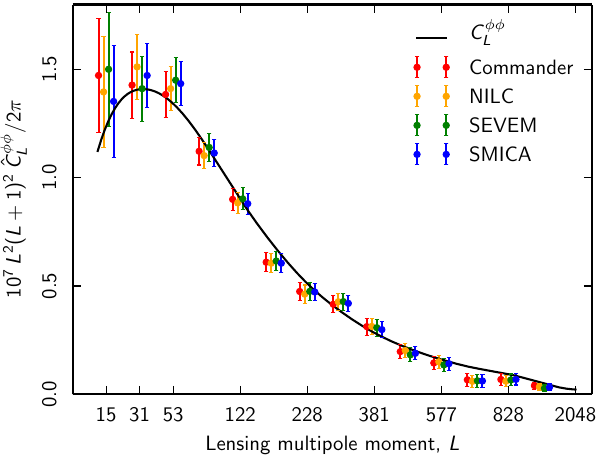}
\end{center}
\caption{Lensing reconstruction power spectrum from the four cleaned CMB maps, including lensing multipoles $8 \leq L \leq 2048$ in the minimum variance estimator. For comparison, the black line shows the lensing potential power spectrum adopted for the FFP10 simulation suite.}
\label{fig:lensing}
\end{figure}

Figure~\ref{fig:lensing} shows the our minimum-variance lensing spectrum estimates evaluated from lensing multipoles $8 \leq L \leq 2048 $. Summary amplitude statistics are listed in Table~\ref{tab:lensing_amplitude}, both on the conservative ($8
\leq L \leq 400$) and high-$L$~($401 \leq L \leq 2048$) ranges. As we see, the four component-separation methods result in almost identical constraining power. No clear band-power outliers are observed in Fig.~\ref{fig:lensing}, and all summary statistics are consistent with each other within uncertainties. However, all four methods show a lensing power that is slightly tilted with respect to the fiducial model, with slightly less power at high multipoles. More detailed analysis and consistency tests are presented in \cite{planck2016-l08}.

\begin{table}[t]
\begingroup                                                                            
\newdimen\tblskip \tblskip=5pt
\caption{Summary of reconstructed gravitational lensing amplitudes. These amplitudes are defined relative to the $\Lambda$CDM spectrum adopted for the FFP10 simulation, which is close, but not identical, to the best-fit \Planck\ 2018 $\Lambda$CDM spectrum. The first two lines show results derived with the minimum variance estimator that includes both temperature and polarization data, while the last two rows show results derived from temperature data alone.\label{tab:lensing_amplitude}}
\nointerlineskip                                                                                                                                                                                     
\vskip -4mm
\scriptsize                                                                                                                                      
\setbox\tablebox=\vbox{ %
\newdimen\digitwidth                                                                                                                          
\setbox0=\hbox{\rm 0}
\digitwidth=\wd0
\catcode`*=\active
\def*{\kern\digitwidth}
\newdimen\signwidth
\setbox0=\hbox{+}
\signwidth=\wd0
\catcode`!=\active
\def!{\kern\signwidth}
\newdimen\decimalwidth
\setbox0=\hbox{.}
\decimalwidth=\wd0
\catcode`@=\active
\def@{\kern\signwidth}
\halign{ \hbox to 1.in{#\leaderfil}\tabskip=1.0em&
  \hfil#\hfil\tabskip=1em&
  \hfil#\hfil\tabskip=1em&
  \hfil#\hfil\tabskip=1em&
  \hfil#\hfil\tabskip=0em\cr
\noalign{\doubleline}
\omit&\multispan{4}{\hfil \sc Lensing Amplitude, $\hat{A}$\hfil}\cr
\noalign{\vskip -3pt}
\omit&\multispan4\hrulefill\cr
\omit\hfil\sc Multipole Range\hfil& \commander& \nilc& \sevem& \smica\cr
\noalign{\vskip 5pt\hrule\vskip 3pt}
MV, $L=8$--400& 0.99 $\pm$ 0.03& 0.98 $\pm$ 0.02& 1.00 $\pm$ 0.03& 0.98 $\pm$ 0.02\cr
MV, $L=401$--2048& 0.87 $\pm$ 0.10& 0.86 $\pm$ 0.10& 0.80 $\pm$ 0.10& 0.83 $\pm$ 0.10\cr
$TT$, $L=8$--400& 1.00 $\pm$ 0.03& 0.99 $\pm$ 0.03& 0.99 $\pm$ 0.03& 0.99 $\pm$ 0.03\cr
$TT$, $L=401$--2048& 0.77 $\pm$ 0.10& 0.78 $\pm$ 0.10& 0.69 $\pm$ 0.11& 0.75 $\pm$ 0.10\cr
\noalign{\vskip 5pt\hrule\vskip 3pt}
}}
\endPlancktablewide                                                                                                                                            
\endgroup
\end{table}

\subsection{Limits on primordial non-Gaussianity}

\textcolor{black}{
The foreground-cleaned CMB maps may also be used to constrain
primordial non-Gaussianity, which is often parameterized in terms of
the amplitude, $f_{\mathrm{NL}}$, of quadratic corrections to the
gravitational potential. This amplitude may be measured through the
harmonic-space 3-point correlation function, evaluated for different
triangle configurations. A detailed $f_{\mathrm{NL}}$ analysis applied
to the current cleaned CMB maps is presented in
\citet{planck2016-l09}.}

\textcolor{black}{
Table~\ref{tab:ng} summarizes some of the main results presented in
that paper, specifically $f_{\mathrm{NL}}$ as evaluated from each map
by the KSW estimator after correcting for gravitational lensing and
the ISW effect. Three sets of results are provided, corresponding to
constraints derived from temperature data alone, from polarization
data alone, and from temperature and polarization
combined.  Corresponding results for the \Planck\ 2015 data were presented in
\citet{planck2014-a09}. However, due to
the presence of systematic effects, the largest angular scales in
polarization were excluded from that analysis. In contrast, the new
results presented for the 2018 data set include all angular scales.
}

\textcolor{black}{
As in previous analyses with \Planck\ measurements
\citep{planck2013-p09a,planck2014-a19}, no statistically significant
detection of primordial non-Gaussianity is found in the \Planck\ 2018
data set, even when including large angular scales in
polarization. Statistically speaking, the most significant excursion
from zero corresponds to a $2.4\sigma$ deviation. With six
statistically independent tests (three in each of temperature and
polarization), this has a probability-to-exceed of about 10\,\% by
chance alone.
}

\begin{table}[t]
\begingroup
\newdimen\tblskip \tblskip=5pt
\caption{\textcolor{black}{Amplitude of primordial non-Gaussianity, $f_{\rm{NL}}$,
  estimated by the KSW estimator after correcting for gravitational
  lensing and the ISW effect. See Table~1 in
  \citet{planck2016-l09} for full details.}}
\label{tab:ng}
\vskip -4mm
\footnotesize
\setbox\tablebox=\vbox{
\newdimen\digitwidth
\setbox0=\hbox{\rm 0}
\digitwidth=\wd0
\catcode`*=\active
\def*{\kern\digitwidth}
\newdimen\signwidth
\setbox0=\hbox{+}
\signwidth=\wd0
\catcode`!=\active
\def!{\kern\signwidth}
\newdimen\decimalwidth
\setbox0=\hbox{.}
\decimalwidth=\wd0
\catcode`@=\active
\def@{\kern\signwidth}
\halign{ \hbox to 0.85in{#\leaderfil}\tabskip=0.5em&
    \hfil$#$\hfil\tabskip=0.5em&
    \hfil$#$\hfil\tabskip=0.5em&
    \hfil$#$\hfil\tabskip=0.5em&
    \hfil$#$\hfil\tabskip=0pt\cr
\noalign{\doubleline}
\omit&\multispan4\hfil $f_{\rm NL}$\hfil\cr 
\noalign{\vskip -3pt}
\omit&\multispan4\hrulefill\cr
\noalign{\vskip 3pt}
\omit\hfil{\sc Type}\hfil&\omit\hfil\commander\hfil&\omit\hfil\nilc\hfil&\omit\hfil\sevem\hfil&\omit\hfil\smica\hfil\cr
\noalign{\vskip 4pt\hrule\vskip 8pt} 
\omit{\boldmath$T$}\hfil\cr
\noalign{\vskip 4pt}
\hglue 0.5em Local       & **-2\pm*6*& !**0\pm *6*& !**0\pm *6*& **-2\pm *6*\cr
\hglue 0.5em Equilateral & !*15\pm66*& *-10\pm 66*& !*17\pm 66*& *!14\pm 66*\cr
\hglue 0.5em Orthogonal  & !*25\pm37*& !**0\pm 36*& !*24\pm 37*& *-15\pm 36*\cr
\noalign{\vskip 6pt}
\omit{\boldmath$E$}\hfil\cr
\noalign{\vskip 4pt}
\hglue 0.5em Local       & !*31\pm*29& !**9\pm *30& !*38\pm *29& !*47\pm *28\cr
\hglue 0.5em Equilateral & !170\pm170& !*39\pm 160& !180\pm 170& !170\pm 160\cr
\hglue 0.5em Orthogonal  & -180\pm*88& -130\pm *88& -180\pm *88& -210\pm *86\cr
\noalign{\vskip 6pt}
\omit{\boldmath$T+E$}\hfil\cr
\noalign{\vskip 4pt}
\hglue 0.5em Local       & **-2\pm*5*& **-1\pm *5*& **-2\pm *5*& **-1\pm *5*\cr
\hglue 0.5em Equilateral & *-10\pm47*& *-31\pm 46*& **-9\pm 47*& *-18\pm 47*\cr
\hglue 0.5em Orthogonal  & *-13\pm23*& *-24\pm 23*& *-15\pm 23*& *-37\pm 23*\cr
\noalign{\vskip 5pt\hrule\vskip 5pt}
}}
\endPlancktablewide
\endgroup
\end{table}

\subsection{Analysis of end-to-end simulations}

We finish this CMB-targeted analysis section with a brief discussion of end-to-end simulations, focusing on polarization extraction from the FFP10 set. For a corresponding analysis of temperature simulations, see \citet{planck2014-a11}.

Unlike the simulations discussed in Sect.~\ref{sec:noise_consistency}, which only included the CMB and instrumental noise, the simulations considered in this section also includes polarized synchrotron and thermal dust emission. These simulations are processed through each pipeline, allowing each code to estimate spectral parameters (i.e., weights for \nilc, \sevem\, and \smica, and spectral indices for \commander) directly from the simulations.

\begin{figure*}[htbp]  
\begin{center}
  \includegraphics[width=0.48\textwidth]{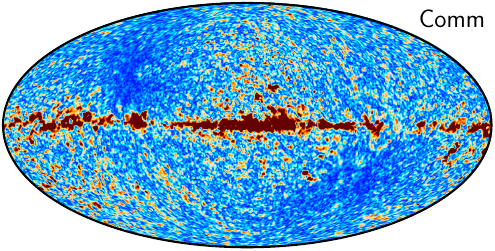}
  \includegraphics[width=0.48\textwidth]{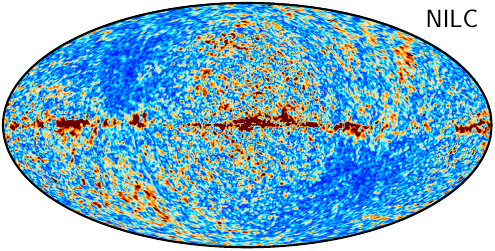}\\
  \includegraphics[width=0.48\textwidth]{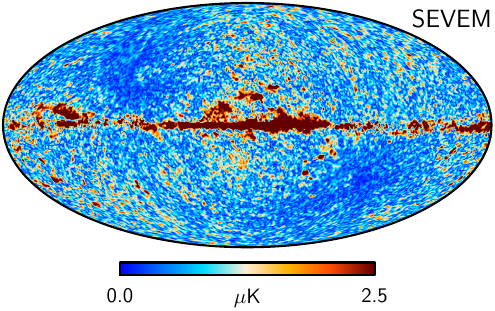}
  \includegraphics[width=0.48\textwidth]{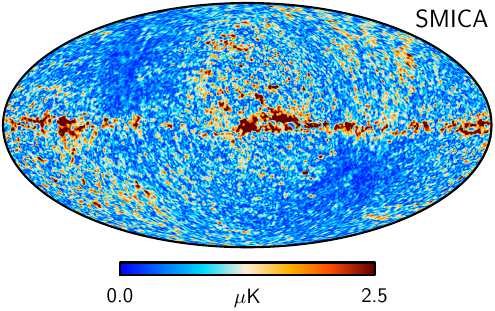}\\
\end{center}
\caption{CMB polarization reconstruction error for each of the four CMB analysis pipelines, as evaluated from the end-to-end FFP10 analysis pipeline. This error is defined as $\sqrt{(Q_{\mathrm{out}}-Q_{\mathrm{in}})^2 + (U_{\mathrm{out}}-U_{\mathrm{in}})^2}$, where $Q_{\mathrm{out}}$  and $U_{\mathrm{out}}$ are the estimated Stokes parameters, and $Q_{\mathrm{in}}$ and $U_{\mathrm{in}}$ are the true Stokes parameters. Each difference map has been smoothed to $80\arcm$ FWHM before computing the polarization amplitude, to reduce the impact of instrumental noise.
}
\label{fig:ffp10_absdiff}
\end{figure*}

Figure~\ref{fig:ffp10_absdiff} shows the CMB polarization reconstruction error for each of the four CMB analysis pipelines, as
evaluated from the end-to-end FFP10 analysis pipeline, defined by
\begin{equation}
  \Delta P = \sqrt{(Q_{\mathrm{out}}-Q_{\mathrm{in}})^2 +
    (U_{\mathrm{out}}-U_{\mathrm{in}})^2},
\end{equation}
where $Q_{\mathrm{out}}$ and $U_{\mathrm{out}}$ are the estimated Stokes parameters, and $Q_{\mathrm{in}}$ and $U_{\mathrm{in}}$ are the true Stokes parameters. All maps have been smoothed to $80\arcm$ FWHM before computing this quantity, to reduce the impact of instrumental noise.

In these plots, one may observe generally similar behaviour between \commander\ and \sevem, and between \nilc\ and \smica. Explicitly, \nilc\ and \smica\ result in slightly lower residuals in the Galactic plane, whereas \commander\ and \sevem\ appear slightly less sensitive to stripes at high Galactic latitues. As evaluated over the common polarization mask, the standard deviations of the four maps (in alphabetical order) are $0.74\muK$, $0.86\muK$, $0.74\muK$, and $0.75\muK$, respectively.

\section{Polarized foregrounds}
\label{sec:foregrounds}

We now turn to the scientific characterization of diffuse microwave foregrounds as derived from the \Planck\ 2018
polarization maps; a corresponding discussion of temperature foreground products is given in Appendix~\ref{app:foregrounds}. Three different algorithms are employed in the following, namely \commander\ \citep{eriksen2004,eriksen2008,planck2014-a12,seljebotn2017}, \gnilc\ \citep{Remazeilles2011b}, and \smica\ \citep{cardoso2008}.

\begin{figure}[htbp]  
\begin{center}
  \includegraphics[width=0.49\columnwidth]{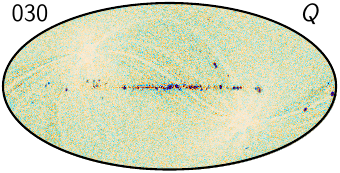}
  \includegraphics[width=0.49\columnwidth]{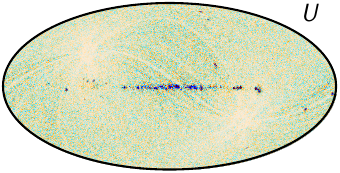}\\
  \includegraphics[width=0.49\columnwidth]{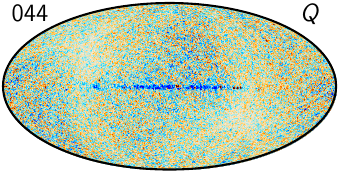}                 
  \includegraphics[width=0.49\columnwidth]{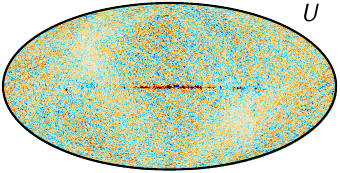}\\
  \includegraphics[width=0.49\columnwidth]{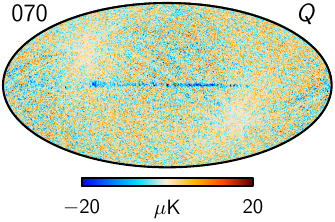}
  \includegraphics[width=0.49\columnwidth]{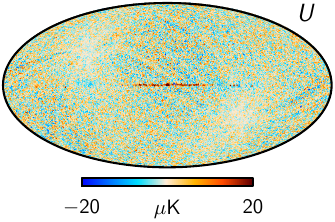}\\
  \includegraphics[width=0.49\columnwidth]{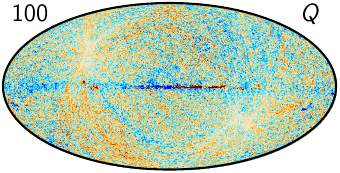} 
  \includegraphics[width=0.49\columnwidth]{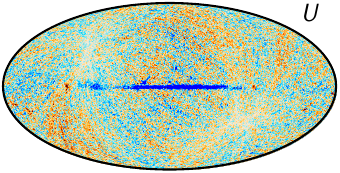}\\
  \includegraphics[width=0.49\columnwidth]{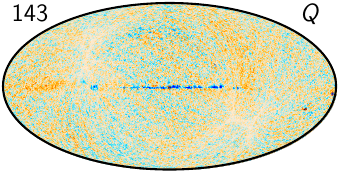}                 
  \includegraphics[width=0.49\columnwidth]{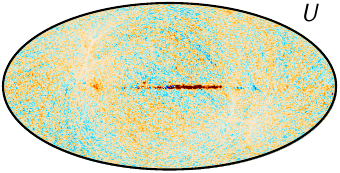}\\
  \includegraphics[width=0.49\columnwidth]{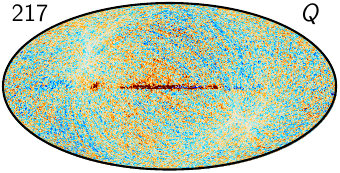}
  \includegraphics[width=0.49\columnwidth]{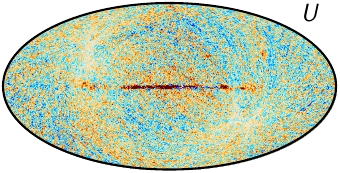}\\
  \includegraphics[width=0.49\columnwidth]{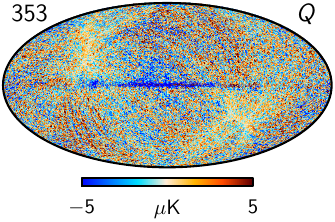}
  \includegraphics[width=0.49\columnwidth]{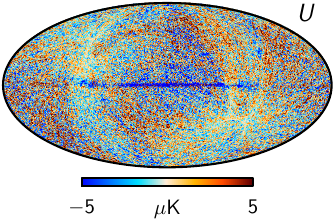}\\
  \includegraphics[width=\columnwidth]{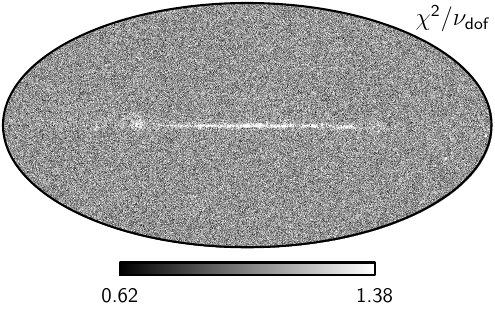}
  \vspace*{-7mm}
\end{center}
\caption{(\emph{Top}:) \commander\ polarization residual maps, $\d_{\nu}-\s_{\nu}$, for each polarized \Planck\ frequency
channel. All maps are smoothed to a common resolution of $40\arcm$ FWHM. (\emph{Bottom}:) \textcolor{black}{Reduced $\chi^2$ map for the
high-resolution polarization analysis. The grayscale range corresponds to $\pm3\sigma$ in terms of expected statistical variation.}  }
\label{fig:comm_pol_residuals}
\end{figure}

\subsection{Internal consistency and goodness-of-fit}

\begin{figure*}[htbp]  
  \begin{center}
    \includegraphics[width=0.95\textwidth]{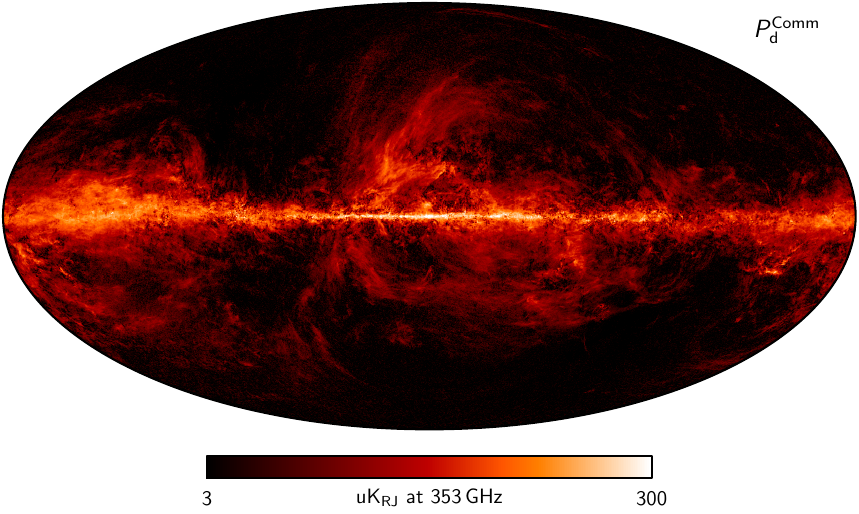}
  \end{center}
\caption{\commander\ 2018 polarized thermal dust amplitude map at $5\arcm$ FWHM resolution, evaluated at a mono-chromatic reference frequency of 353\,GHz.}
\label{fig:comm_pol_dust}
\end{figure*}

\begin{figure*}[htbp]  
\begin{center}
  \includegraphics[width=0.95\textwidth]{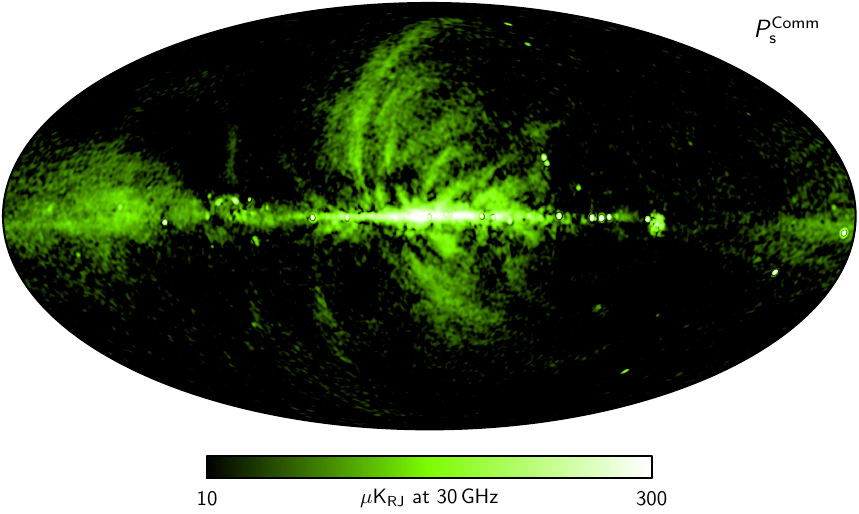}
\end{center}
\caption{\commander\ 2018 polarized synchrotron amplitude map at $40\arcm$ FWHM resolution, evaluated at a mono-chromatic reference frequency of 30\,GHz.}
\label{fig:comm_pol_synch}
\end{figure*}

\begin{figure*}[htbp]  
  \begin{center}
    \includegraphics[width=0.95\textwidth]{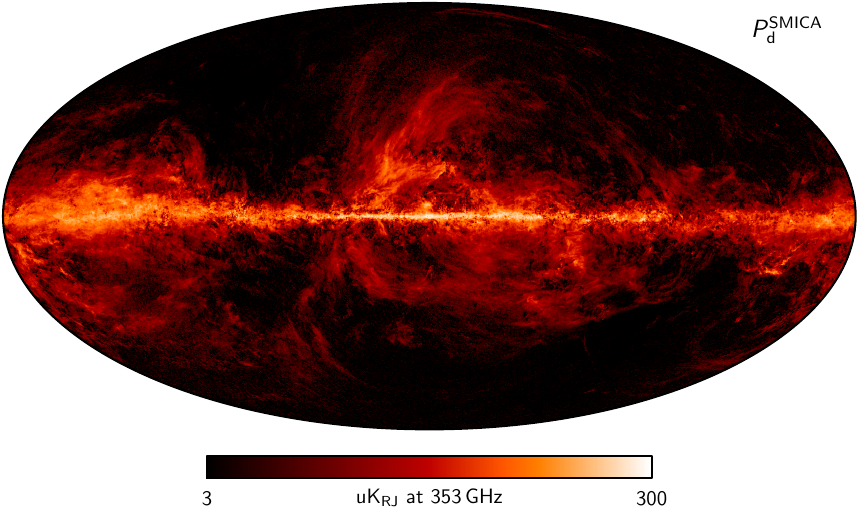}
  \end{center}
  \caption{\smica\ 2018 polarized thermal dust amplitude map at $12\arcm$ FWHM resolution, evaluated at 353\,GHz. No colour corrections have been applied to this map.}
\label{fig:smica_pol_dust}
\end{figure*}

\begin{figure*}[htbp]  
  \vspace*{-0.8mm}
  \begin{center}
    \includegraphics[width=0.95\textwidth]{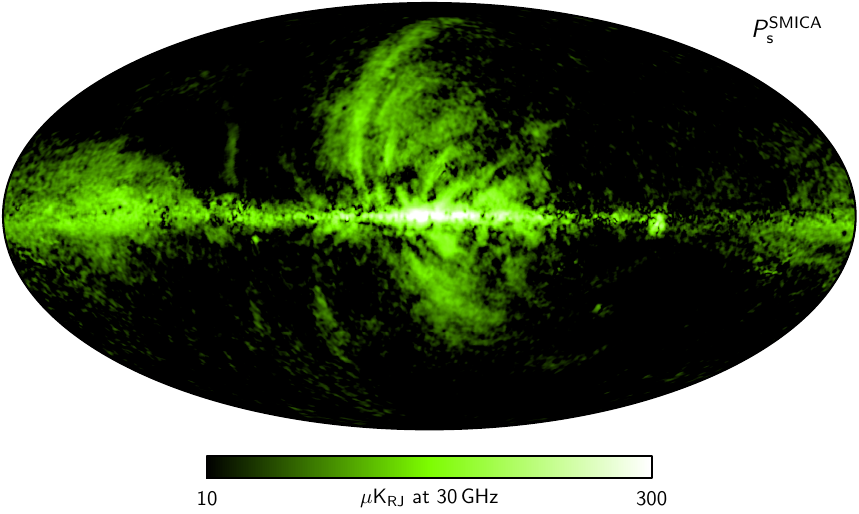}
  \end{center}
  \caption{\smica\ 2018 polarized synchrotron amplitude map at $40\arcm$ FWHM resolution, evaluated at 30\,GHz. No colour corrections have been applied to this map.}
\label{fig:smica_pol_synch}
\end{figure*}

\begin{figure*}[htbp]  
  \begin{center}
    \includegraphics[width=0.95\textwidth]{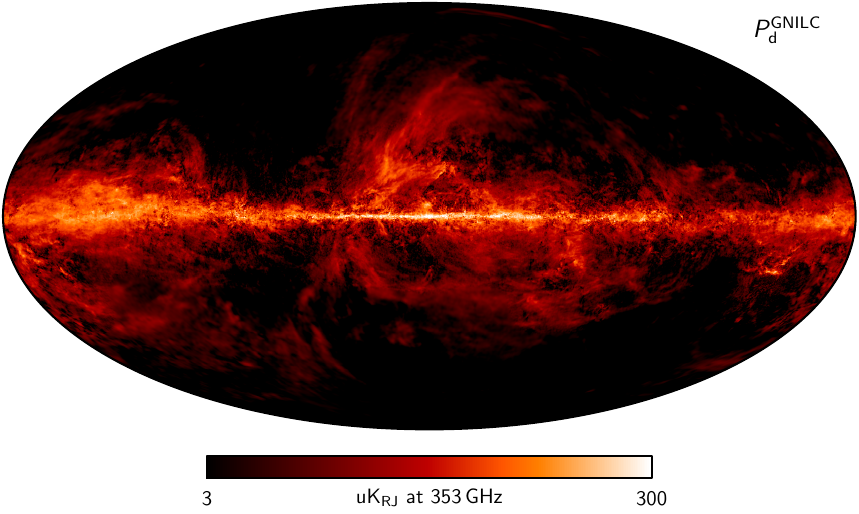}
  \end{center}
  \caption{\gnilc\ 2018 polarized thermal dust amplitude map evaluated at 353~GHz. The angular resolution varies over the sky, as described in \citet{Remazeilles2011b}. No colour corrections have been applied to this map.}
\label{fig:gnilc_pol_dust}
\end{figure*}

\begin{figure}[htbp]  
\begin{center}
  \includegraphics[width=\columnwidth]{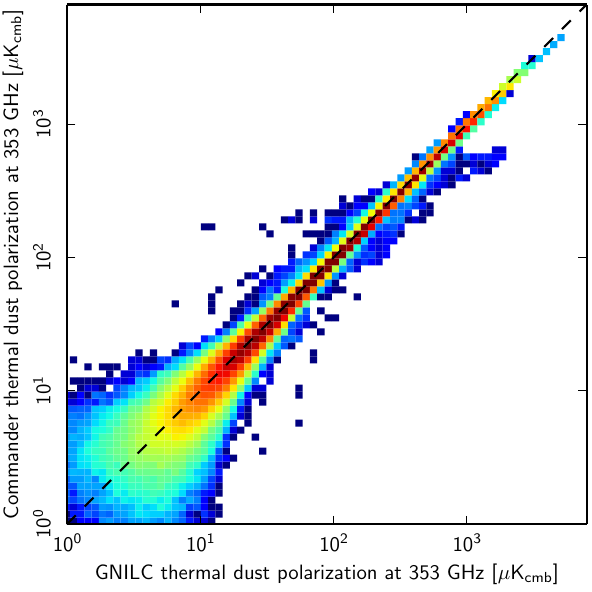}
\end{center}
\caption{$P$--$P$ scatter plot between the thermal dust polarization amplitude at 353\,GHz, as estimated with \gnilc\ and \commander. Colours indicate the density of points on a logarithmic scale.}
\label{fig:gnilc_dust_P}
\end{figure}

Before considering astrophysical components, it is instructive to consider the internal consistency between the \Planck\ 2018
polarization frequency maps. For this purpose, we employ the \commander\ model described in Sect.~\ref{sec:commander}, fitting a minimal three-component signal model (CMB, synchrotron, and thermal dust emission) to the seven polarized \Planck\ frequencies between 30 and 353\,GHz. The synchrotron component is modelled by a single power-law with a free spectral index, $\beta_{\mathrm{s}}$, in the frequency domain, while the thermal dust component is modelled as a modified blackbody with free spectral index, $\beta_{\mathrm{d}}$, and temperature, $T_{\mathrm{d}}$. In the
main analyses, the synchrotron spectral index is fixed spatially to $\beta_{\mathrm{s}}=-3.1$, matching the high-latitude temperature result found from the combination of \Planck\ 2015, WMAP, and Haslam data \citep{planck2014-a12}; as shown in Sect.~\ref{sec:pol_ind}, the \Planck\ measurements by themselves have little sensitivity to the synchrotron spectral index. For thermal dust, we fix $T_{\mathrm{d}}$ at the \commander\ result found from the \Planck\ 2018 temperature data in Appendix~\ref{app:foregrounds}; with a highest frequency of 353\,GHz, the \Planck\ polarization observations are insensitive to this parameter.

Additionally, we impose a spatial smoothness prior on both synchrotron and thermal dust emission to reduce noise-induced degeneracies between the various components.  This takes the form of a Gaussian smoothing kernel with $40\arcm$ FWHM for synchrotron emission and $10\arcm$ FWHM for thermal dust emission. The widths of these priors are chosen to match the resolution at which the data have a significant signal-to-noise ratio; see Appendix~\ref{app:commander} for further details.

Given this model, the top panels in Fig.~\ref{fig:comm_pol_residuals} show residual maps of the form \textcolor{black}{data minus model ($\vec d_{\nu} - \vec s_{\nu}$)} for each \Planck\ frequency map, all smoothed to a common resolution of $40\arcm$. The colour scales cover $\pm20\muK$ for the LFI channels, and $\pm5\muK$ for the HFI channels. Ideally, each of these maps should be consistent with instrumental noise alone, and for the three LFI channels this appears to be a reasonable approximation. The only clearly visible artefacts in these maps correspond to regions of high foreground amplitudes, which most likely are due to a low level of
residual temperature-to-polarization leakage, for instance from bandpass mismatch between individual detectors. \textcolor{black}{In particular, the sharp morphology of the Galactic plane residuals corresponds to the shape of temperature foregrounds, not polarization foregrounds.} 

In contrast, significant large-scale residuals may be seen at all four HFI frequencies, with patterns typically aligning with the
\Planck\ scanning strategy. Collectively, these features correspond to effective calibration uncertainties that couple the CMB dipole and foregrounds to the reconstructed CMB polarization signal.  Although these residuals are significant, their amplitudes are almost an order of magnitude smaller than in the 2015 data.  Moreover, the latest end-to-end simulations describe the residuals to a high level of precision \citep{planck2016-l03}.

The bottom panel in Fig.~\ref{fig:comm_pol_residuals} shows the reduced $\chi^2$ per pixel, as defined by
\textcolor{black}{
\begin{equation}
  \chi_{\mathrm{red}}^2(p) = \frac{1}{\nu_{\mathrm{dof}}}\sum_{\nu=1}^{N_{\mathrm{band}}} \left(\frac{d_{\nu}-s_{\nu}(p)}{\sigma_{\nu}(p)}\right)^2.
  \label{eq:chisq}
\end{equation}
This map is summed over Stokes $Q$ and $U$ parameters and evaluated at $\nside=1024$, corresponding to the resolution of the LFI frequency maps. The total number of degrees of freedom is therefore approximately $\nu_{\mathrm{dof}}=2\cdot(3+4\cdot4)- 2\cdot3 = 32$, accounting for three LFI maps at $\nside=1024$, four HFI maps at $\nside=2048$, and three fitted component maps, each with an angular resolution comparable to the size of an $\nside=1024$ pixel. The colour range corresponds to $\pm3\sigma$ in terms of expected statistical variation for 32 degrees of freedom.} Note that $\sigma_{\nu}(p)$ only accounts for white noise. The smoothness of this $\chi^2$ map clearly suggests that the \Planck\ 2018 polarization observations are dominated by instrumental white noise on intermediate and small angular scales, not by systematic effects or foreground artefacts.

\subsection{Polarization amplitude}

Next, we consider the polarization amplitude of synchrotron emission at 30\,GHz and thermal dust emission at 353\,GHz, naively defined as $P^{\mathrm{s}} = \sqrt{Q_{\mathrm{s}}^2 + U_{\mathrm{s}}^2}$.  As discussed by \citet{plaszczynski2014}, this estimator is intrinsically noise-biased; however, since we are only interested in it for comparison and consistency purposes, the noise bias is not critical for this paper. The resulting maps are shown in Figs.~\ref{fig:comm_pol_dust}--\ref{fig:gnilc_pol_dust}, as estimated by \commander, \gnilc, and \smica. For \commander, the synchrotron map is smoothed to $40\arcm$ FWHM and the thermal dust emission map is smoothed to $5\arcm$ FWHM.  For \smica, the corresponding smoothing scales are $40\arcm$ and $12\arcm$ FWHM.  For \gnilc\ the effective angular resolution varies over the sky, depending on the local signal-to-noise ratio. The \commander\ maps correspond to the amplitudes evaluated at monochromatic reference frequencies, while the \gnilc\ and \smica\ maps correspond to bandpass-integrated maps at 30 and 353\,GHz, respectively.

Two sets of \gnilc\ products are delivered for the \Planck\ 2018 release: (i)~the \gnilc\ Stokes $I,$ $Q$, and $U$ maps of thermal dust emission at uniform $80'$ resolution, with the associated \gnilc\ noise covariance matrix maps ($II$, $IQ$, $IU$, $QQ$, $QU$, and $UU$);  and (ii)~the \gnilc\ Stokes $I,$ $Q$, and $U$ maps of thermal dust emission at variable resolution ($80'$ to $5'$) over the sky, with the associated \gnilc\ noise-covariance-matrix maps, along with a beam FWHM map indicating the corresponding variable resolution of the dust over the sky regions. The \Planck\ 2018 \gnilc\ dust products are analysed in great detail in \citep{planck2016-l11B}.

Figure~\ref{fig:gnilc_dust_P} shows a scatter plot between the \commander\ and \gnilc\ thermal dust amplitudes, both evaluated for a common resolution of $80\arcm$ FWHM. Overall, the agreement is very good, and the Pearson's correlation coefficient between the two maps is $r=0.999$.  Similar good agreement is observed between the \smica\ and the \commander\ and \gnilc\ maps, except for very high values of $P$, for which \smica\ applies an inpainting mask during processing to avoid ringing.  The main notable difference between the \commander\ and \gnilc\ maps is an overall relative
scaling of around 5\,\%, corresponding to the fact that no colour corrections are applied to the \gnilc\ map, and it therefore
corresponds to the dust signal as observed through the \Planck\ 353-GHz bandpass. This distinction between the two maps is
important to bear in mind when subjecting either one to statistical analysis.

Based on these polarization amplitude maps, one can compute the corresponding polarization fraction, defined as $p = P/I$, where $P$ is the polarization amplitude, and $I$ is the corresponding total intensity. This quantity is useful for modelling and characterizing astrophysical emission processes, and is therefore of great interest to astrophysical theorists. However, it is also highly sensitive to systematic errors in the intensity component, and in particular to the zero level, which is difficult to constrain for the \Planck\ measurements. A careful analysis of the thermal dust polarization fraction derived from the \Planck\ 2018 measurements, including zero level uncertainties, is provided in \citet{planck2016-l11B}, and we refer the interested reader to that paper for full details.

\begin{figure}[htbp]  
\begin{center}
  \includegraphics[width=\linewidth]{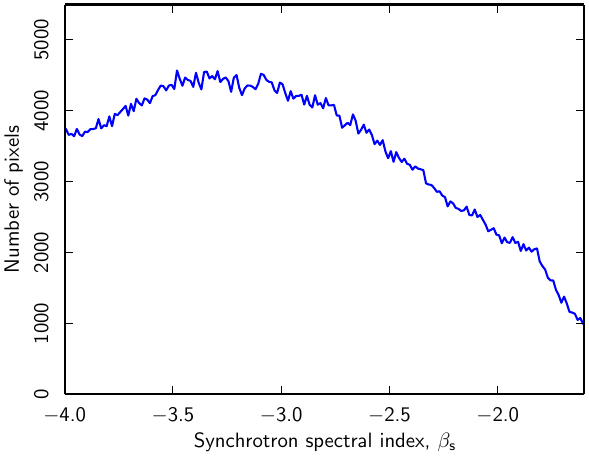}\\
  \includegraphics[width=\linewidth]{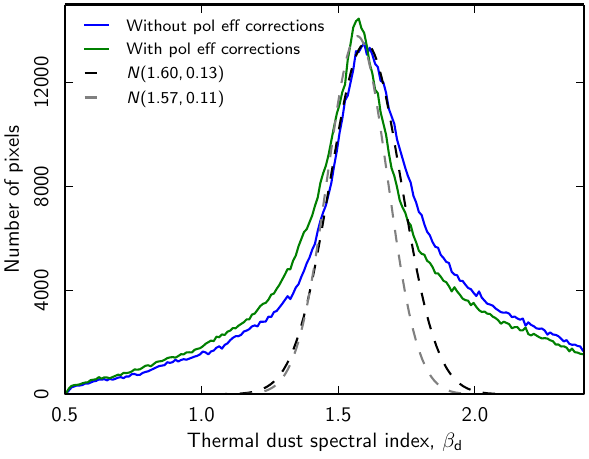}
\end{center}
\caption{Distribution of spectral indices for polarized synchrotron (top panel) and thermal dust (bottom panel) emission as estimated with \commander\ without applying any informative Gaussian prior. The synchrotron spectral index shown in this plot is estimated with a $5^{\circ}$ FWHM smoothing scale, and the thermal dust spectral index is estimated with a $3^{\circ}$ FWHM smoothing scale. For the thermal dust case, results are shown both with (green curve) and without (blue curve) applying polarization efficiency corrections at 100--217\,GHz. The dashed lines in this case indicate Gaussian fits to the central peak.}
\label{fig:comm_pol_beta}
\end{figure}

\subsection{Synchrotron and thermal dust spectral indices}
\label{sec:pol_ind}

Next, we consider the spectral energy distributions (SEDs) for polarized synchrotron and thermal dust emission. For simplicity, we focus primarily on the effective spectral index for either process, noting that \Planck\ has very limited sensitivity to estimate additional spectral parameters in polarization.

Starting with \commander, we note that the main analysis discussed above is performed with informative (delta function or Gaussian) priors on both $\beta_{\mathrm{s}}$ and $\beta_{\mathbf{d}}$. In order to quantify the intrinsic information content and statistical strength of the \Planck\ data to constrain these parameters at a more basic level, it is useful also to perform \emph{prior-free} runs. The results from such analyses are summarized in Fig.~\ref{fig:comm_pol_beta}, for synchrotron emission in the top panel and thermal dust emission in the bottom panel. In either case, the Gaussian prior is removed only on the component in question, not both simultaneously. In all cases, however, a broad uniform prior is imposed in order to exclude completely unphysical values. The synchrotron analysis is performed at a smoothing scale of $5^{\circ}$ FWHM, while the thermal dust analysis is performed at a smoothing scale of $3^{\circ}$ FWHM. \textcolor{black}{This scale was determined by considering a series of scales (1, 2, 3, 5, 10 deg), and identifying the largest scale that did not result in leakage artifacts.}

For synchrotron emission, we find a very broad distribution between $\beta_{\mathrm{s}}=-4$ and $-1.5$, with both ends being defined by the uniform prior. There is a weak preference for values between $\beta_{\mathrm{s}}=-3.5$ and $-3.0$, consistent with the value of $\beta_{\mathrm{s}}=-3.1$ found by combining \Planck, \WMAP, and Haslam temperature data in \citet{planck2014-a12}, but overall, it is clear that the \Planck\ polarization data by themselves do not significantly constrain the spectral index of synchrotron emission at scales smaller than $5^{\circ}$. For the main analysis, we therefore fix the spectral index for polarized synchrotron emission at the best-fit value derived from the 2015 temperature data, corresponding to $\beta_{\mathrm{s}}=-3.1$. This value is also consistent within the uncertainties with corresponding results derived by
\citet{kogut2007}, \citet{dunkley2009b}, \citet{bennett2012}, \citet{fuskeland2014}, \citet{vidal2014}, and \citet{krachmalnicoff2018}.

For thermal dust emission, the situation is more informative, since the HFI data constrain thermal dust emission more strongly than the LFI data constrain synchrotron emission. Focusing for the moment on the blue curve in Fig.~\ref{fig:comm_pol_beta}, corresponding to the nominal data set considered in this paper, we observe a clear peak centred around $\beta_{\mathrm{d}} \approx 1.60$, and with a width of $0.10$--$0.15$. The distribution exhibits heavy tails toward both steep and shallow spectral indices, which is typical for noise-dominated data; these pixels are mostly located at high Galactic latitudes, where the dust amplitude is low. Motivated by these results, we adopt a Gaussian prior for the \commander\ analysis of $\beta_{\mathrm{d}}=1.60\pm0.10$ for the main analysis, acknowledging that the standard deviation quoted above over-estimates the intrinsic scatter in the dust population because of instrumental noise. Note that the uncertainty in this prior refers to the standard
deviation of the map, not the error in the mean of the central value.

As mentioned in Sect.~\ref{sec:inputs}, the \Planck\ 2018 HFI polarization measurements are associated with small but non-negligible uncertainties in terms of polarization efficiencies, $\epsilon$. By default, polarization efficiency corrections are not included in the analyses presented in this paper, but instead we assess their impact by comparing results with and without these corrections. The green curve in the bottom panel Fig.~\ref{fig:comm_pol_beta} shows the distribution of $\beta_{\mathrm{d}}$ with application of these corrections at frequencies between 100 and 217\,GHz. Overall, we see that these polarization efficiencies shift the distribution by $\Delta\beta_{\mathrm{d}} = -0.03$.

\begin{figure}[htbp]  
\begin{center}
  \includegraphics[width=\linewidth]{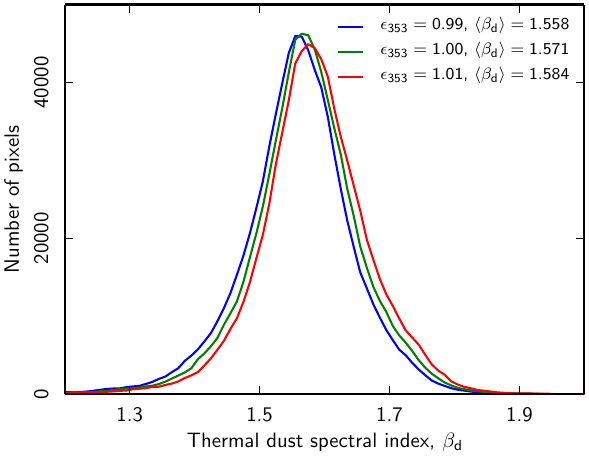}
\end{center}
\caption{Effect on the spectral index of polarized thermal dust emission, $\beta_{\mathrm{d}}$, when changing the polarization
efficiency correction at 353\,GHz, $\epsilon_{353}$. A shift of $\epsilon_{353}$ by 1\,\% translates into a change in
$\beta_{\mathrm{d}}$ of 0.013.}
\label{fig:comm_pol_beta_353}
\end{figure}

The nominal polarization-efficiency corrections described in \citet{planck2016-l03} and \citet{planck2016-l05} do not include any robust estimates for the 353-GHz channel, since the CMB signal that is used to estimate these corrections is faint at this frequency. However, it is reasonable to assume that it is associated with similar uncertainties as the other HFI channels. In Fig.~\ref{fig:comm_pol_beta_353}, we show the $\beta_{\mathrm{d}}$ posterior distributions resulting from changing $\epsilon_{353}$ by 1\,\% in either direction from its nominal value. In this case, we find that a shift of $\epsilon_{353}$ by 1\,\% translates into a change in $\beta_{\mathrm{d}}$ of 0.013. Combined with the uncertainties arising from the 100- to 217-GHz frequencies, we therefore consider the total systematic uncertainty on $\beta_{\mathrm{d}}$ due to polarization efficiency corrections to be 0.04.

\begin{figure}[htbp]  
\begin{center}
  \includegraphics[width=\linewidth]{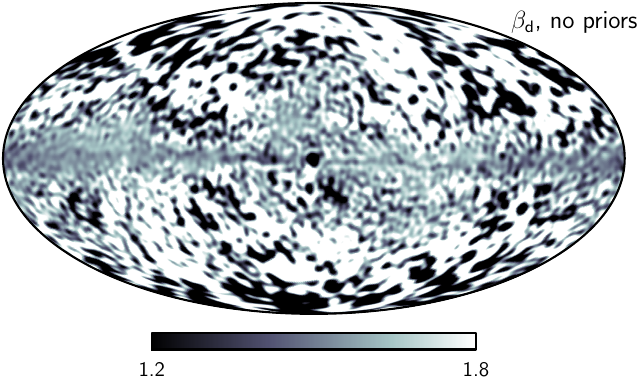}\\
  \includegraphics[width=\linewidth]{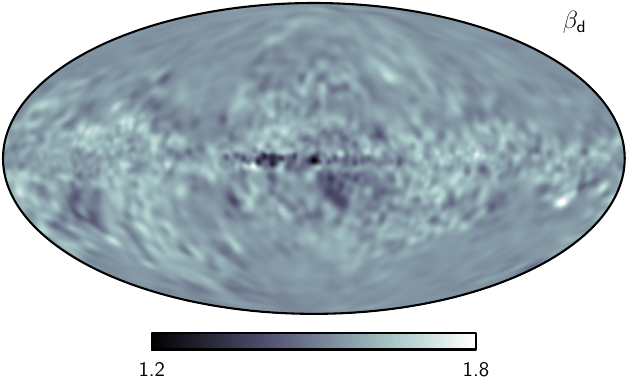}
\end{center}
\caption{Spatial distribution of the spectral index of polarized thermal dust emission, $\beta_{\mathrm{d}}$, as estimated with
\commander\, adopting a smoothing scale of $3^\circ$ FWHM. In the top panel no Gaussian prior is applied.  In the bottom panel a Gaussian prior of $\beta_{\mathrm{d}} = 1.60\pm0.10$ is applied. In both cases, the spectral index of synchrotron emission is fixed to $\beta_{\mathrm{s}}=-3.1$.}
\label{fig:comm_pol_beta_map}
\end{figure}

The top panel in Fig.~\ref{fig:comm_pol_beta_map} shows the spatial distribution of $\beta_{\mathrm{d}}$ from the prior-free analysis without polarization efficiency corrections. In this plot the statistical power of the \Planck\ observations to constrain the
spectral index is seen very clearly from position to position, depending on the local dust polarization amplitude. Near the Galactic plane, the data are sufficiently strong to determine the spectral index well per resolution element, while at high latitudes the measurements are fully dominated by instrumental noise. The bottom panel shows the corresponding result when applying the supporting Gaussian prior. From this figure, it is clear that the $\beta_{\mathrm{d}}$ distribution and prior presented above are dominated by measurements in the Galactic plane, where the signal-to-noise ratio is substantially larger than at high Galactic latitudes.

Next, we perform a blind analysis of polarization spectral indices with \smica. This analysis is performed by running \smica\ with a foreground dimension of $\nfg=2$ (that is, with a two-column foreground emissivity matrix $\tens F$), as defined in
Eq.~(\ref{eq:smica:model}), corresponding to synchrotron and thermal dust emission. Spectral priors are  imposed during
the multi-frequency fit so that synchrotron emission vanishes at 353\,GHz and thermal dust emission vanishes at 30\,GHz.

\begin{figure}[htbp]  
\centering
\includegraphics[width=\columnwidth]{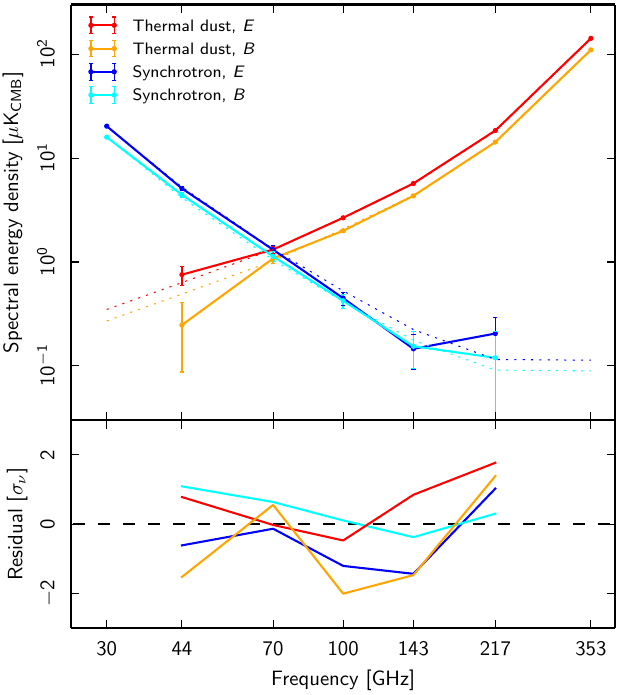}
\caption{\emph{Top}: Synchrotron and thermal dust \textcolor{black}{full-sky-averaged} SEDs as estimated blindly by \smica. Red and blue curves indicate thermal dust and synchrotron $E$ modes, respectively, and orange and cyan curves indicate corresponding $B$ modes. Dotted lines indicate the best-fit spectra for a power-law fit with $\beta_{\mathrm{s}}=-3.10\pm0.06$ for synchrotron, and a modified blackbody fit with $\beta_{\mathrm{d}}=1.53\pm0.01$ and $T_{\mathrm{d}}=19.6\mathrm{K}$ for thermal dust
emission.  \emph{Bottom:} Residual spectral energy densities relative to best-fit models, measured in units of the data
uncertainty, $\sigma_{\nu}$.}
  \label{fig:polsedsmica}
\end{figure}

The results from these calculations are summarized in Fig.~\ref{fig:polsedsmica} for both $E$-mode and $B$-mode  polarization. Parametric best-fits are indicated by dotted lines.  These are, however, only the products of post-processing the raw \smica\ results \textcolor{black}{by fitting a modified blackbody spectrum to the measured data points with $\chi^2$ minimization.  They} do not correspond to active priors as they do in the Bayesian analysis discussed above. In these particular fits, polarization efficiency corrections are applied to the 100, 143, and 217\,GHz data, and colour corrections are applied in post-analysis.

The best-fit spectral parameters derived in this blind manner are $\beta_{\mathrm{s}}=-3.10\pm0.06$ and
$\beta_{\mathrm{d}}=1.53\pm0.01$, both corresponding to full-sky averages. Furthermore, these fits provide a statistically
sufficient model across the full frequency range, as indicated by the residual spectra shown in the bottom panel of
Fig.~\ref{fig:polsedsmica}.  All residuals are within $2\sigma$ of their statistical errors. 

The \smica\ measurements of the polarized thermal dust spectral index are in excellent agreement with the corresponding results presented in \citet{planck2016-l11A}, based on both frequency cross-correlation power spectra at high Galactic latitudes and simple colour ratios between the 217- and 353-GHz channels at low Galactic latitudes. At the same time, $\beta_{\mathrm{d}}$ is lower by 0.07 or $3\sigma$ compared to the \commander\ results presented above. To understand the origin of these differences, it is instructive to take a closer look at the 217/353 colour ratio, which is the fastest, simplest and most transparent estimator available.

The results from this estimator may be summarized as follows.  We subtract one of the cleaned CMB maps from the \Planck\ 217- and 353-GHz polarization HM split maps to form two statistically independent foreground-plus-noise maps.  We smooth these maps to $3^{\circ}$ FWHM to increase the effective signal-to-noise ratio per pixel.  We then compute the cross-polarization amplitude between the two halves of the split, and we finally form the CMB-corrected colour ratio between the 217 and 353\,GHz maps. Given some estimate of the thermal dust temperature, this ratio may then be easily translated into estimates of the thermal dust spectral index.  We adopt a constant temperature of 19.6\,K in the following.

First, we consider the impact of different CMB estimates produced by each of the four analysis pipelines. With the above procedure, we find median estimates of $\beta_{\mathrm{d}}=1.57$, 1.54, 1.55, and 1.54, when subtracting the \commander, \nilc, \sevem, and \smica\ CMB polarization maps, respectively. Different noise-weighting and foreground-modelling assumptions thus account for $\Delta\beta_{\mathrm{d}}\approx0.03$.

Second, the effect of polarization efficiencies has already been addressed above in the context of \commander.  We find similar sensitivities to the polarization efficiencies on the 217/353 colour ratio, as the median estimates for each of the four codes when applying these corrections are $\beta_{\mathrm{d}}=1.54$, 1.52, 1.52, and 1.52, corresponding to an effective shift of
$\Delta\beta_{\mathrm{d}}\approx0.02$--0.03.

Third and finally, a small effect is due to different bandpass treatments. Specifically, in \citet{planck2016-l11A}, bandpass
integration effects are taken into account by the so-called colour correction technique, in which a multiplicative correction based on some fiducial spectral parameters is applied to the nominal thermal dust SED at a given reference frequency. The same approach is adopted for the \smica\ results. In contrast, \commander\ performs a full integral over the product of the bandpass and the SED for each set of spectral parameters. These two different approaches agree to 0.07\,\% at 143\,GHz, 0.7\,\% at 217\,GHz, and 1.3\,\% at 353\,GHz. In sum, these small differences translate into a net shift of $\Delta\beta_{\mathrm{d}}=0.015$ in terms of the thermal dust spectral index.

Recognizing the significant systematic uncertainties on the thermal dust spectral index from both modelling aspects and polarization efficiencies, we adopt a total systematic uncertainty of 0.05, defined by the above shifts added in quadrature with a statistical uncertainty of 0.02 \citep{planck2016-l11A}. As a single point estimate, we adopt the average value of the colour-ratio-derived estimates without polarization efficiency corrections, for a total final estimate of $\beta_{\mathrm{d}}=1.55\pm0.05$. This estimate is conservative, and corresponds to marginalizing over all analysis methods and known
uncertainties.

\subsection{Synchrotron and thermal dust angular power spectra}

\begin{figure}[htbp]  
  \centering
  \includegraphics[width=0.5\textwidth]{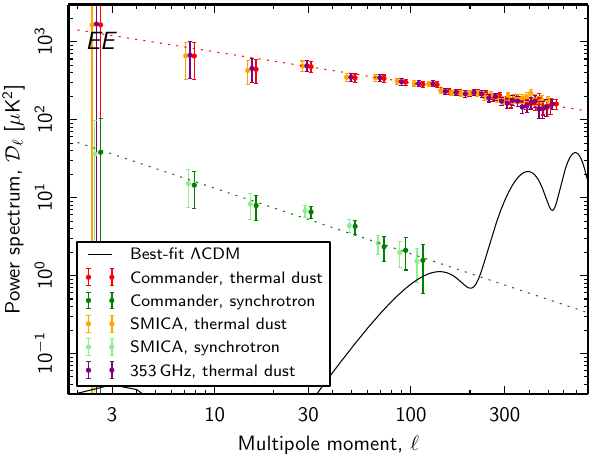} \\
  \includegraphics[width=0.5\textwidth]{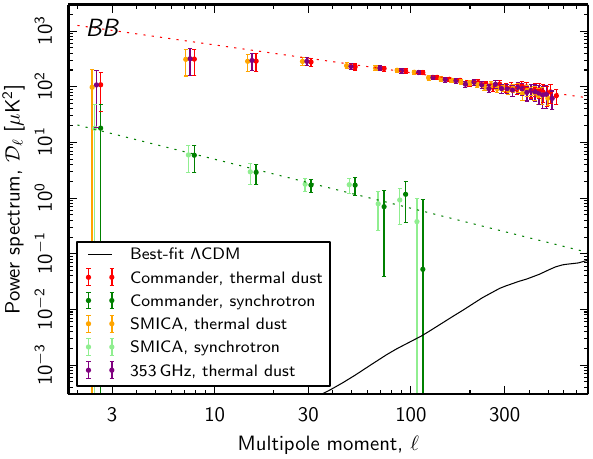}
  \caption{$EE$ (top) and $BB$ (bottom) power spectra for synchrotron and thermal dust as computed from the
\commander, \smica, and 353-GHz frequency maps; see \citet{planck2016-l11A} for algorithmic details. All spectra are
evaluated outside the common polarization mask, over 78\,\% of the sky.  Dashed lines indicate the best-fit power-law fits for the \commander\ case, as reported in Table~\ref{tab:fg_powspec}.  Error bars indicate $3\sigma$ uncertainties. All spectra have been colour corrected to monochromatic reference frequencies of 30\,GHz for synchrotron and 353\,GHz for thermal dust emission, respectively.}
  \label{fig:fg_powspec}
\end{figure}

Finally we consider the angular power spectra of polarized synchrotron and thermal dust emission as estimated by \commander\ and \smica. We estimate the $EE$ and $BB$ angular cross-spectra outside the common CMB mask for the half-mission split with {\tt XPol} (see \citealt{tristram2005} and \citealt{planck2016-l11A} for details). The results from these calculations are summarized in Fig.~\ref{fig:fg_powspec}. \commander\ results are shown in red (for thermal dust emission) and green (for synchrotron emission);  \smica\ results are shown in orange and light green.  For comparison, direct 353-GHz cross-correlation results are shown in purple, derived using the same methodology as in \citet{planck2016-l11A}. \textcolor{black}{As in that analysis, the best-fit \Planck\ 2018 $\Lambda$CDM CMB spectrum, shown as a black solid line, has been subtracted from the raw estimate.} Dotted coloured lines indicate best-fit power law fits to the \commander\ spectra, as defined by
\begin{equation}
D_{\ell} = q\,\left(\frac{\ell}{80}\right)^\alpha.
\end{equation}
Overall, we find excellent agreement between the \commander, \smica, and 353-GHz results, demonstrating that the derived component maps are robust with respect to specific algorithmic details for the particular angular ranges and sky coverage considered here.

Table~\ref{tab:fg_powspec} summarizes the angular power spectra in terms of best-fit power-law models for \commander\ and \smica, and in terms of the $EE/BB$ ratio, all derived using the same machinery as in \citet{planck2016-l11A}. For thermal dust, corresponding results are also given for the direct 353-GHz cross-correlation approach.  Power-spectrum amplitudes have been colour corrected to monochromatic reference frequencies of 30 and 353\,GHz for synchrotron and thermal dust emission, respectively. \textcolor{black}{The two masks considered in Table~\ref{tab:fg_powspec} are defined in \citet{planck2016-l11A}.  Note, however, that only thermal dust emission results are shown for the mask with a sky fraction of $f_{\mathrm{sky}}=0.42$. The signal-to-noise ratio for synchrotron emission is too low to support robust power spectrum estimates in the same region.}

\begin{table}[htbp]                                                                                                                                                
\begingroup                                                                                                                                     
\newdimen\tblskip \tblskip=5pt
\caption{Best-fit power-law parameters to the angular power spectra of synchrotron (30\,GHz) and thermal dust emission (353\,GHz), evaluated with the XPol power spectrum estimator \citep{tristram2005} as detailed in \citet{planck2016-l11A}. Frequency cross-correlation results are derived using precisely the same methodology as in \citet{planck2016-l11A}, while \commander\ and \smica\ results are derived using the same power spectrum estimation tools, but applied to the half-mission maps presented in this paper. Note that all uncertainties are statistical, and do not account for systematic or modelling uncertainties.  Power spectrum amplitudes refer to monochromatic reference frequencies of 30 and 353\,GHz for synchrotron and thermal dust emission, respectively. } 
\label{tab:fg_powspec}
\nointerlineskip                                                                                                                                                                                     
\vskip -3mm 
\footnotesize 
\setbox\tablebox=\vbox{ %
  \newdimen\digitwidth 
  \setbox0=\hbox{\rm 0}
  \digitwidth=\wd0 \catcode`*=\active \def*{\kern\digitwidth}
\newdimen\signwidth
\setbox0=\hbox{+}
\signwidth=\wd0
\catcode`!=\active
\def!{\kern\signwidth}
\newdimen\decimalwidth
\setbox0=\hbox{.}
\decimalwidth=\wd0
\catcode`@=\active
\def@{\kern\signwidth}
\def\s#1{\ifmmode $\rlap{$^{\rm #1}$}$ \else \rlap{$^{\rm #1}$}\fi}
\halign{ \hbox to 1in{#\leaderfil}\tabskip=1.5em&
    \hfil#\hfil\tabskip=1em&
    \hfil#\hfil\tabskip=0pt\cr
\noalign{\doubleline}
\omit\hfil \hfil& $q$ [$\mu\textrm{K}_{\textrm{CMB}}^2$]& $\alpha$\cr
\noalign{\vskip 3pt\hrule\vskip 4pt}
\noalign{\vskip 3pt}
{\bf Thermal dust, \boldmath$f_{\mathrm{sky}}=0.42$, $\ell=40$\bf--600}\hfil\cr
\noalign{\vskip 6pt}
\multispan3 {\commander}\hfil\cr
\hglue 1em$EE$& $60\pm2$& $-0.39\pm0.03$\cr
\hglue 1em$BB$& $32\pm1$& $-0.49\pm0.05$\cr
\noalign{\vskip 2pt}
\hglue 1em$BB$/$EE$& \multispan2 \hfil$0.52$\hfil\cr
\noalign{\vskip 3pt}
\multispan3 {\smica}\hfil\cr
\noalign{\vskip 3pt}
\hglue 1em$EE$& $62\pm2$& $-0.18\pm0.04$\cr
\hglue 1em$BB$& $32\pm1$& $-0.45\pm0.05$\cr
\noalign{\vskip 2pt}
\hglue 1em$BB$/$EE$& \multispan2 \hfil$0.48$\hfil\cr
\noalign{\vskip 3pt}
\multispan3 {Frequency map cross-correlation}\hfil\cr
\noalign{\vskip 3pt}
\hglue 1em$EE$& $59\pm2$& $-0.28\pm0.04$\cr
\hglue 1em$BB$& $32\pm1$& $-0.48\pm0.06$\cr
\noalign{\vskip 2pt}
\hglue 1em$BB$/$EE$& \multispan2 \hfil$0.50$\hfil\cr
\noalign{\vskip 3pt}
\noalign{\vskip 8pt}
{\bf Thermal dust, \boldmath$f_{\mathrm{sky}}=0.78$, $\ell=40$\bf--600}\hfil\cr
\noalign{\vskip 6pt}
\multispan3 {\commander}\hfil\cr
\noalign{\vskip 3pt}
\hglue 1em$EE$& $323\pm4$& $-0.40\pm0.01$\cr
\hglue 1em$BB$& $199\pm3$& $-0.50\pm0.02$\cr
\noalign{\vskip 2pt}
\hglue 1em$BB$/$EE$& \multispan2 \hfil$0.57$\hfil\cr
\noalign{\vskip 3pt}
\multispan3 {\smica}\hfil\cr
\noalign{\vskip 3pt}
\hglue 1em$EE$& $318\pm4$& $-0.34\pm0.01$\cr
\hglue 1em$BB$& $205\pm3$& $-0.55\pm0.02$\cr
\noalign{\vskip 2pt}
\hglue 1em$BB$/$EE$& \multispan2 \hfil$0.54$\hfil\cr
\noalign{\vskip 3pt}
\multispan3 {Frequency map cross-correlation}\hfil\cr
\noalign{\vskip 3pt}
\hglue 1em$EE$& $313\pm4$& $-0.41\pm0.01$\cr
\hglue 1em$BB$& $187\pm3$& $-0.50\pm0.02$\cr
\noalign{\vskip 2pt}
\hglue 1em$BB$/$EE$& \multispan2 \hfil$0.57$\hfil\cr
\noalign{\vskip 3pt}
\noalign{\vskip 8pt}
{\bf Synchrotron, \boldmath$f_{\mathrm{sky}}=0.78$, $\ell=4$\bf--140}\hfil\cr
\noalign{\vskip 6pt}
\multispan3 {\commander}\hfil\cr
\hglue 1em$EE$& $2.3\pm0.1$& $-0.84\pm0.05$\cr
\hglue 1em$BB$& $0.8\pm0.1$& $-0.76\pm0.09$\cr
\noalign{\vskip 2pt}
\hglue 1em$BB$/$EE$& \multispan2 \hfil$0.34$\hfil\cr
\noalign{\vskip 3pt}
\multispan3 {\smica}\hfil\cr
\noalign{\vskip 3pt}
\hglue 1em$EE$& $2.4\pm0.2$& $-0.88\pm0.04$\cr
\hglue 1em$BB$& $0.9\pm0.2$& $-0.75\pm0.07$\cr
\noalign{\vskip 2pt}
\hglue 1em$BB$/$EE$& \multispan2 \hfil$0.34$\hfil\cr
\noalign{\vskip 4pt\hrule}
}}
\endPlancktable
\endgroup
\end{table}

Overall, in terms of angular power spectra for polarized thermal dust emission, we find excellent agreement \textcolor{black}{over 78\% of the sky}  between the frequency cross-correlation technique and the \commander\ and \smica\ component-separation techniques. The only statistically significant discrepancy is seen for the spatial power-law index parameter, $\alpha$, for which formally a $6\sigma$ difference is observed between \commander\ and \smica. However, we note that in terms of absolute values the difference is only $\Delta\alpha=0.06$, and no systematic uncertainties are included in these numbers. Finally, it is worth noting that the two analyses are carried out with different angular resolutions, corresponding to $5\arcm$ and $12\arcm$ FWHM respectively. Due to its lower resolution, the \smica\ analysis is somewhat more sensitive to high-multipole systematics than the \commander\ and 353\,GHz analyses.

We also observe excellent agreement between the \commander\ and \smica\ maps in terms of polarized synchrotron emission. In addition, we note that the $BB/EE$ ratio measured from the \Planck\ 2018 data is 0.34, which is very similar to the corresponding value of 0.36 estimated from the \Planck\ 2015 data.

\textcolor{black}{The thermal dust $BB$ spectrum is in general lower than the $EE$ spectrum. For intermediate values of $\ell$, this has been interpreted as the result of statistical alignment of filamentary structure in the interstellar medium with the local direction of the Galactic magnetic field \citep[][and references therein]{planck2016-l11A}.  Empirically, this asymmetry of $BB$ relative to $EE$ extends to the lowest multipoles, as seen in Fig.~\ref{fig:fg_powspec} by the decrease of $BB$ below the power law and, interestingly here, no such strong decrease for $EE$.  The amount of asymmetry appears to depend on both multipole and sky fraction \citep{planck2016-l11A}.  For $BB$ in particular, the departures from the power law at lower multipoles show a dependence on Galactic hemisphere (see the comparison on Northern and Southern cuts of the sky in \citealp{planck2016-l11A}), which in turn suggests some relationship to the large-scale structure of the magnetic field.  Further discussion is beyond the scope of this paper.}

\begin{figure}[htbp]  
  \centering
  \includegraphics[width=\columnwidth]{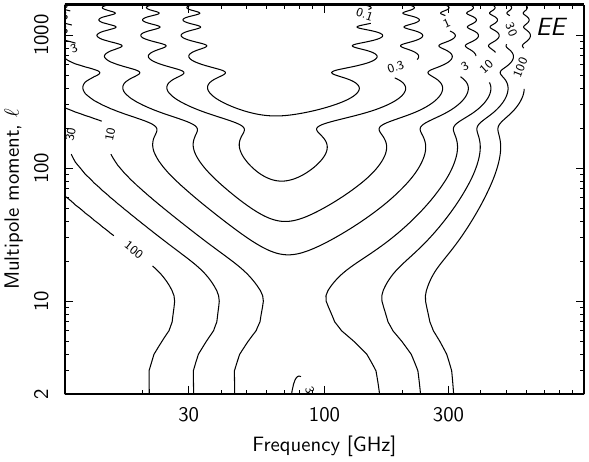}\\
  \includegraphics[width=\columnwidth]{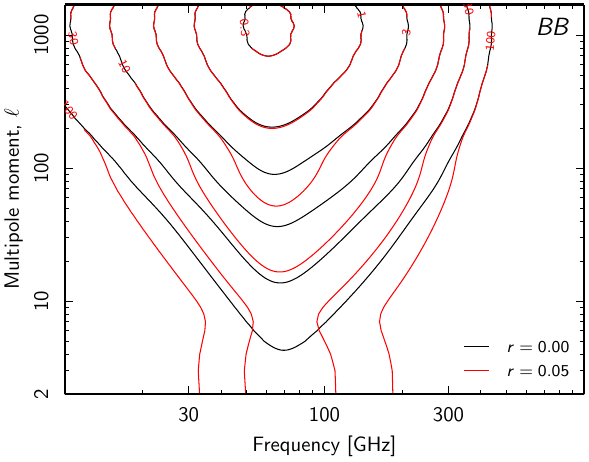}
  \caption{Amplitude ratio between total polarized foregrounds and CMB
    as a function of both multipole moment and frequency, as defined
    by $f(\ell,\nu) =
    [C_{\ell}^{\mathrm{fg}}(\nu)/C_{\ell}^{\mathrm{CMB}}]^{1/2}$, with
    parameters derived from 78\,\% of the sky as estimated by
    \commander. The top and bottom panels show $EE$ and $BB$ spectra,
    and the black and red contours in the latter corresponds to
    tensor-to-scalar ratios of $r = 0.0$ and 0.05, respectively.}
  \label{fig:fg_cl_ratio}
\end{figure}

Based on the best-fit power spectrum and SED parameters reported above, Fig.~\ref{fig:fg_cl_ratio} summarizes the
foreground-to-CMB ratio in terms of the quantity $f(\ell,\nu) = [C_{\ell}^{\mathrm{fg}}(\nu)/C_{\ell}^{\mathrm{CMB}}]^{1/2}$ as a
function of both frequency and angular scale. As expected, the overall picture is very similar to that presented from the \Planck\ 2015 data in \citet{planck2014-a12}, with one small but notable exception: Because the best-fit value of the optical depth of reionization is lower in the \Planck\ 2018 $\Lambda$CDM model than in the corresponding 2015 model, the relative foregrounds-to-CMB ratio is higher at low $EE$ multipoles, further emphasizing the importance of accurate foreground modelling for large-scale polarization CMB analysis.

\section{Conclusions}
\label{sec:conclusions}

In this paper we have presented cleaned CMB temperature and polarization maps derived from the \Planck\ 2018 data set, as well as new polarized synchrotron and thermal dust emission maps. These maps represent a new state-of-the-art characterization of the microwave sky.

The main scientific motivation underlying the work between the \Planck\ 2015 and 2018 data releases has been reduced instrumental systematics, \textcolor{black}{in particular for the polarization measurements}. As demonstrated in this and companion papers, the work has been successful, as the updated \Planck\ frequency maps exhibit significantly lower contamination on all angular scales. For polarization, we find that the lower systematics in frequency maps translates directly into lower systematics in CMB and foreground maps. Additionally, new  end-to-end CMB-plus-noise simulations have been constructed that more accurately reproduce residual systematics observed in the real data. For full-mission data, these simulations are accurate to $\lesssim3\,\%$ for $\ell\lesssim1500$ in both temperature and polarization. On smaller scales, non-negligible biases are observed, and caution is warranted when subjecting the maps to detailed statistical analysis on scales smaller than $\ell\gtrsim1500$.

It is important to note that the 2018 data release does not represent a globally optimal reduction of the \Planck\ time-ordered data that is ideal for all purposes. In particular, the updated data set does not include single detector maps, and the new frequency maps have complicated bandpass properties \citep{planck2016-l03}. As a result, accurate reconstruction of astrophysical temperature foreground properties is non-trivial. Thus, while the \Planck\ 2018 release represents a significant step forward in our understanding of the polarized microwave sky compared to the 2015 release, the associated temperature
results, for which the astrophysics are richer, do not represent a similar improvement. Indeed, for several intensity applications we anticipate that external users may find the 2015 products more useful than the corresponding 2018 products. One concrete example of this is the \Planck\ astrophysical sky model as presented in \citet{planck2014-a12}, which includes intensity estimates of both CO line emission and thermal dust emission. The same considerations apply both to \gnilc\ and \commander; while chronologically formally superseded by the current results, we believe that the 2015 temperature astrophysical foreground models represent more accurate approximations to the true sky than the ones presented in the 2018 data release. To avoid confusion, we therefore do not release the corresponding 2018 foreground temperature products.

Fortunately, these issues are largely unimportant for CMB reconstruction purposes.  The analyses presented in this paper and in \citet{planck2016-l06} reach the same conclusion regarding the CMB temperature results, namely that the \Planck\ 2018 CMB temperature data are for all practical purposes statistically consistent with the corresponding 2015 rendition. Of course, this is the direct result of the very high signal-to-noise ratio of the \Planck\ measurements, in that small variations in the processing procedure make very little difference in the final maps compared to the intrinsic sample variance of the true CMB sky.

For large-scale CMB polarization at $\ell\lesssim50$, we find that the \Planck\ 2018 data are compatible with end-to-end
simulations. However, it is critical to note that the observations are \emph{not} consistent with uncorrelated white noise at any angular scales. Any statistical analysis of the \Planck\ 2018 polarization data must therefore always be accompanied by a corresponding analysis of the associated end-to-end simulations. In addition, analysis of half-data split sky maps is strongly encouraged in order to probe stability with respect to both noise and residual instrumental systematics.

In addition to improving the large-scale CMB polarization map, the new data processing also results in improved astrophysical polarization results. One concrete example of this is the fact that the \Planck\ 2018 data for the first time allow a pixel-by-pixel
estimation of the spectral index of thermal dust emission over the full sky. Corresponding analyses based on previous data sets
invariably led to clearly nonphysical results obviously driven by instrumental systematics. With this new data set, we obtain a typical spectral index of polarized thermal dust emission of $\beta_{\mathrm{d}}=1.55\pm0.05$, where the uncertainty accounts both for systematic uncertainties and different analysis techniques. This estimate is largely consistent with comparable results derived from temperature measurements. Also, for polarized synchrotron emission, we are for the first time able to fit the spectral index pixel-by-pixel, and obtain physically meaningful values, even if the signal-to-noise ratio is low; the full-sky averaged synchrotron spectral index for polarized emission is $\beta_{\mathrm{s}}=-3.1\pm0.1$. For thermal dust emission we find a $BB/EE$ angular power spectrum ratio of 0.5, largely independent of sky fraction, while for synchrotron emission we find a lower ratio of 0.34.

\begin{figure}[t]  
\begin{center}
  \includegraphics[width=\columnwidth]{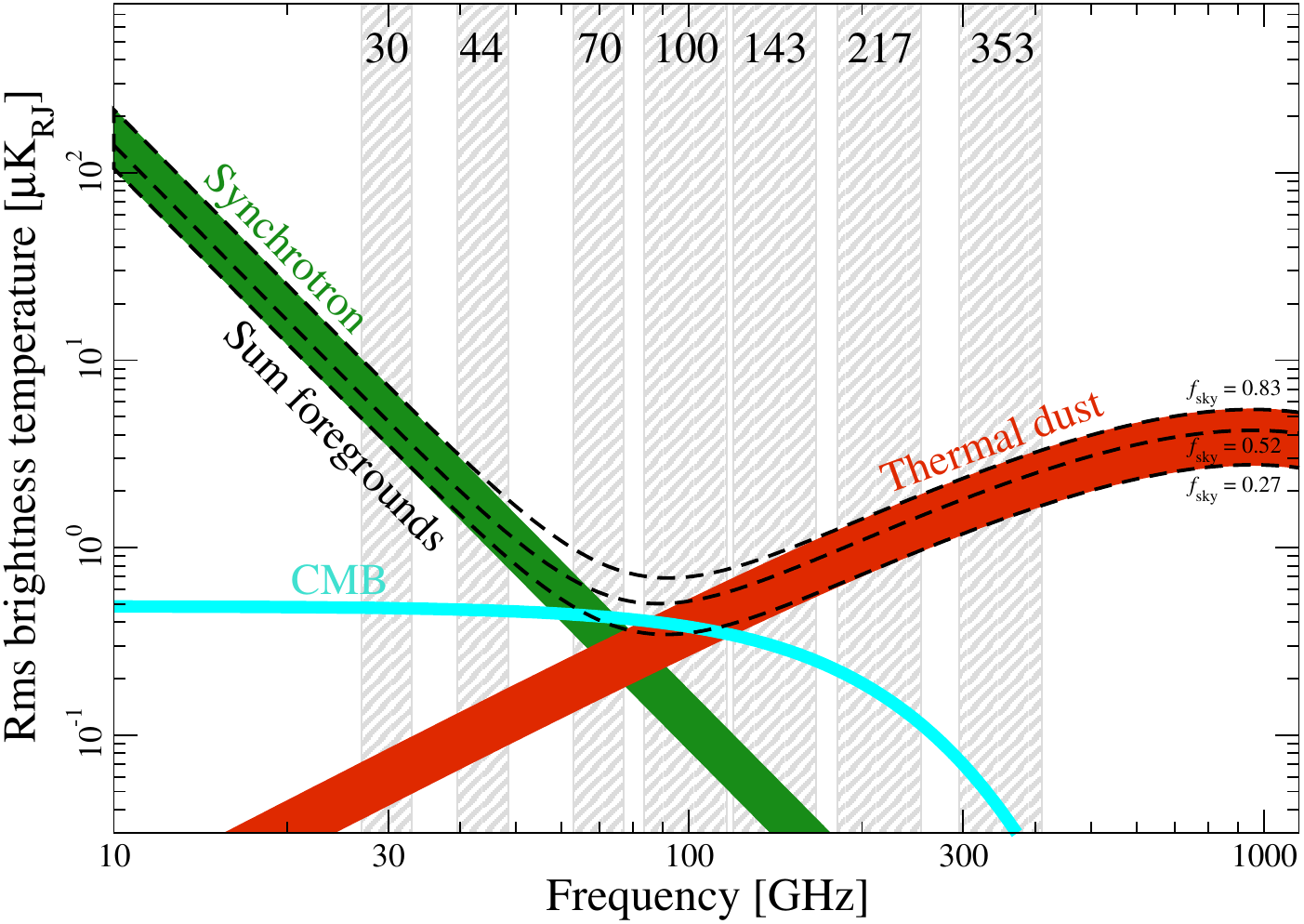}
\end{center}
\caption{Polarization amplitude rms as a function of frequency and astrophysical components, evaluated at a smoothing scale of $40\arcm$ FWHM. The green band indicates polarized synchrotron emission, and the red band indicates polarized thermal dust emission. The cyan curve shows the CMB rms for a $\Lambda$CDM model with $\tau=0.05$, and is strongly dominated by $E$-mode polarization. The dashed black lines indicate the sum of foregrounds evaluated over three different masks with $f_{\mathrm{sky}}=0.83$, 0.52, and 0.27. The widths of the synchrotron and thermal dust bands are defined by the largest and smallest sky coverages.}
\label{fig:overview_pol}
\end{figure}

In Fig.~\ref{fig:overview_pol} we plot the rms of the polarization amplitude as a function of frequency for polarized CMB, synchrotron, and thermal dust emission, evaluated with an angular resolution of $40\arcm$ FWHM. The CMB component is estimated from a simulation drawn from the best-fit \Planck\ 2018 $\Lambda$CDM spectrum, and is dominated by $E$-mode polarization. The synchrotron and thermal dust emission components are based on the \commander\ sky model, by
cross-correlating half-mission sky maps. The dotted lines indicate the sum of the foreground components for three different masks, defined by thresholding the total \commander\ foreground model evaluated at 70\,GHz, near the foreground minimum. Three masks are shown, corresponding to 27, 52, and 82\,\% of the sky. The widths of the foreground bands are defined by the two extreme masks. This figure provides a convenient summary of the properties of the polarized sky in the CMB frequencies measured by \Planck, and it updates the corresponding polarization panel of figure~51 in \citet{planck2014-a12}.
  
Before concluding we briefly summarize some important points regarding limitations and recommended usage of the various \Planck\ component separation products presented in this paper.

\begin{itemize}

\item For polarization analysis, the \Planck\ 2018 data products are superior to the 2015 products in all respects, and the new maps entirely supercede the previous release.

\item For CMB temperature analysis, we consider the 2015 and 2018 data products as equivalent in terms of overall data quality. Most differences between the two generations of cleaned CMB maps are due to different processing choices, rather than fundamental data quality. For instance, for \commander\ the 2018 CMB temperature maps are more constrained by data selection issues than the 2015 maps, and as a result the new maps are more contaminated by CO emission. In contrast, for \smica\ some minor glitches regarding inter-frequency calibration have been resolved in the 2018 maps, and the new maps are therefore somewhat more reliable. For \nilc\ and \sevem, only small changes are observed between the two releases. In
all cases, the differences are small, typically less than $2\muK$ at high Galactic latitudes with a smoothing scale of $80\arcm$ FWHM.

\item For temperature foreground analysis, the 2015 release provides a number of distinct advantages compared to the 2018 release, including no pixelization issues near bright sources in the Galactic plane, more transparent bandpass definitions, and, most importantly, the availability of robust single-bolometer and detector-set maps. For these reasons, we consider the 2015 temperature foreground products from both \commander\ and \gnilc\ to be more reliable than the 2018 products. For the same reason, we anticipate the 2015 temperature data set to continue to play an important role for astrophysical component-separation purposes.

\item The noise properties of the \Planck\ observations are complicated both in temperature and polarization, and usage of
end-to-end simulations is essential to capture all uncertainties. However, even the best currently available simulations are only accurate to a few percent in power. When employing these simulations for quantitative scientific analysis, it is essential to check that the statistic of choice is not sensitive to this level of uncertainty.

\end{itemize}

With these caveats in mind, we end our discussion by recalling the original motivation and goal of the \Planck\ mission, namely ``\ldots \emph{to measure the fluctuations of the CMB with an accuracy set by fundamental astrophysical limits}'' \citep{planck2005-bluebook}. For temperature, this goal was achieved already with the \Planck\ 2015 release. With the 2018 data release, \Planck\ provides a new state-of-the-art for the field also in terms of polarization.

\begin{acknowledgements}
The Planck Collaboration acknowledges the support of: ESA; CNES, and CNRS/INSU-IN2P3-INP (France); ASI, CNR, and INAF (Italy); NASA and DoE (USA); STFC and UKSA (UK); CSIC, MINECO, JA, and RES (Spain); Tekes, AoF, and CSC (Finland); DLR and MPG (Germany); CSA (Canada); DTU Space (Denmark); SER/SSO (Switzerland); RCN (Norway); SFI (Ireland); FCT/MCTES (Portugal); ERC and PRACE (EU). A description of the Planck Collaboration and a list of its members, indicating which technical or scientific activities they have been involved in, can be found at \href{url}{http://www.cosmos.esa.int/web/planck/planck-collaboration}.  This work has received funding from the European Union’s Horizon 2020 research and innovation programme under grant agreement numbers 687312, 776282 and 772253.
\end{acknowledgements}


\bibliographystyle{aa}

\bibliography{Planck_bib,diffuse_compsep}

\appendix

\section{Commander}
\label{app:commander}

The \commander\ analysis framework as applied to previous \Planck\ releases is described in detail by
\citet{eriksen2004,eriksen2008} and \citet{planck2014-a12}. This approach implements a standard Bayesian fitting procedure based on Monte Carlo and Gibbs sampling, in which an explicit parametric model including cosmological, astrophysical, and instrumental parameters is fitted to the observations through the posterior distribution.

Due to its very general approach to CMB analysis, it is more appropriate to refer to \commander\ as a framework rather than
as a specific and well-defined algorithm. For instance, the implementation that is employed for the \Planck\ 2017 analysis has
been re-written from scratch compared to the 2015 version, and the current version is sometimes referred to as
\commandertwo\ \citep{seljebotn2017}. The main difference between the old and the new implementations is their different choice of basis functions for the amplitude degrees of freedom, and their different treatment of instrumental beams. While \commanderone\ adopted real-space pixels as its fundamental basis set and required uniform angular resolution across frequencies, \commandertwo\ adopts spherical harmonics as its fundamental basis set and supports different angular
resolutions at different frequencies. As a result, the new implementation supports signal reconstruction at the full angular
resolution of the \Planck\ observations.

\subsection{Amplitude sampling algorithm}

For full algorithmic specifics regarding the new implementation, see \citet{seljebotn2017};  here we review only the main equations. First, we adopt a general signal model on the following form,
\begin{align}
  \s_{\nu}(\theta) &= s_{\nu}(\a_i, \beta_i, g_{\nu}, \m_{\nu}) \\ & =
  g_{\nu} \sum_{i=1}^{N_{\textrm{comp}}} \F_\nu^i(\beta_i)\a_{i}
\end{align}
where $\a_i$ is an amplitude vector for component $i$ at a given reference frequency, $\beta_i$ is a general set of spectral parameters for the same component, $g_\nu$ is a multiplicative calibration factor for frequency $\nu$, and $\m_{\nu}$ are monopole and dipole amplitudes.  The quantity $\F_\nu^i(\beta_i)$ is a general projection operator that translates from the reference amplitude vector to the basis of the observed data at a given frequency.  As such, it accounts for both the choice of basis functions, and for spectral effects such as the frequency dependence of the component in question and unit conversions.

\begin{table}[t]
\begingroup
\newdimen\tblskip \tblskip=5pt
\caption{Overview of spectral and spatial priors adopted in the \commander\ analysis. Parameters denoted $A$ and $\theta$ correspond to the spatial angular power spectrum prior, as defined in Eq.~(\ref{eq:prior}), where $A$ is defined relative to the reference frequency of the component in question in units of $\muK_{\mathrm{RJ}}^2$. \label{tab:commander_priors}}
\nointerlineskip
\vskip -4mm
\footnotesize
\setbox\tablebox=\vbox{ %
\newdimen\digitwidth
\setbox0=\hbox{\rm 0}
\digitwidth=\wd0
\catcode`*=\active
\def*{\kern\digitwidth}
\newdimen\signwidth
\setbox0=\hbox{+}
\signwidth=\wd0
\catcode`!=\active
\def!{\kern\signwidth}
\newdimen\decimalwidth
\setbox0=\hbox{.}
\decimalwidth=\wd0
\catcode`@=\active
\def@{\kern\signwidth}
\halign{ \hbox to 1.8in{#\leaderfil}\tabskip=2em&
  #\hfil\tabskip=0em\cr
\noalign{\doubleline}
\omit\hfil Component\hfil&\omit\hfil Prior\hfil\cr
\noalign{\vskip 5pt\hrule\vskip 3pt}
\noalign{\vskip 3pt}
\multispan{2}{\bf Temperature} \hfil\cr
\noalign{\vskip 5pt}
\hglue 1em CMB                     & $T_{\mathrm{CMB}}=2.755\mathrm{K}$\cr
\noalign{\vskip 3pt}
\hglue 1em Low-frequency component & $\beta_{\mathrm{lf}} = -3.1\pm0.5$\cr
\omit                              & $A_{\mathrm{lf}} = 10^5\muK_{\mathrm{RJ}}^2$\cr
\omit                              & $\theta_{\mathrm{lf}} =
30\arcm$ FWHM\cr
\noalign{\vskip 3pt}
\hglue 1em Thermal dust emission   & $\beta_{\mathrm{d}} = 1.55\pm0.1$\cr
\omit                              & $T_{\mathrm{d}} = (19.5\pm3)\,\mathrm{K}$\cr
\omit                              & Cosine apodization,\cr
\omit&                               \hglue 2em$5000\le\ell\le6000$\cr
\noalign{\vskip 3pt}
\hglue 1em CO emission             & Spatially uniform line ratios\cr
\omit                              & $A_{\mathrm{CO}} = 10^4\muK_{\mathrm{RJ}}^2$\cr
\omit                              & $\theta_{\mathrm{CO}} = 15\arcm$ FWHM\cr
\noalign{\vskip 3pt}
\hglue 1em Radio source component  & $a_{\mathrm{cs}} \ge 0$\cr
\noalign{\vskip 8pt}
\multispan{2}{\bf Polarization} \hfil\cr
\noalign{\vskip 5pt}
\hglue 1em CMB & $T_{\mathrm{CMB}}=2.755\,\mathrm{K}$\cr
\noalign{\vskip 3pt}
\hglue 1em Synchrotron emission    & $\beta_{\mathrm{s}}$ spatially uniform\cr
\omit                              & $A_{\mathrm{s}} = 10^2\muK^2$\cr
\omit                              & $\theta_{\mathrm{s}} = 40\arcm$ FWHM\cr
\noalign{\vskip 3pt}
\hglue 1em Thermal dust emission   & $\beta_{\mathrm{d}} = 1.6\pm0.1$\cr
\omit                              & $T_{\mathrm{d,pol}} = T_{\mathrm{d,int}}$\cr
\omit                              & $A_{\mathrm{d}} = 50\muK_{\mathrm{RJ}}^2$\cr
\omit                              & $\theta_{\mathrm{d}} = 10\arcm$ FWHM\cr
\noalign{\vskip 5pt\hrule\vskip 3pt}
}}
\endPlancktablewide
\endgroup
\end{table} 

As mentioned above, \commanderone\ adopted pixels as its basis set for all diffuse components, requiring identical angular resolution at all frequencies. In this case, the projection operator reduces to the so-called mixing matrix, $\F = \M$, which translates signal amplitudes from a reference frequency to any other observed frequency. In contrast, \commandertwo\ employs different types of basis functions for different components. For diffuse components, it adopts spherical harmonics, and the projection operator therefore becomes the product of the mixing matrix, which is defined in pixel space, and a spherical
harmonics transform, $\F = \M\Y$. For compact objects (radio sources in the current analysis), the map projection is performed
through a local real-space {\tt FEBeCoP} template per source, $\B_{\mathrm{F}}$, and therefore $\F = \M\B_{\mathrm{F}}$. Finally, fixed template corrections such as monopole, dipole, or zodiacal light corrections, summarized by some overall real-space template matrix per frequency, $T_{\nu}$, are implemented directly as $\F = \T$, and the fitted parameters are thus defined directly as the template amplitude at the respective frequency.

Computationally speaking, by far the most expensive part in \commander\ is to fit for the linear amplitudes, which corresponds to sampling from the conditional distribution $P(\a|\d,\ldots)$.  As shown by, e.g., \citet{jewell2004} and \citet{wandelt2004}, this can be done by solving the so-called Wiener filter equation by conjugate gradients,
\begin{equation}
\left(\S^{-1} + \P^{\rm T} \N^{-1} \P\right) \a = \P^{\rm T} \N^{-1} \d + \P^{\rm T} \N^{-1/2} \omega_1 + \S^{-1} \omega_2.
\label{eq:wiener}
\end{equation}
Here $\S$ is the (prior) covariance matrix of the signal amplitudes, $\P$ is the end-to-end projection operator from amplitude space to data space, $\N$ is the data noise covariance matrix, and $\omega_i$ are Gaussian random vectors with zero mean and unit variance. If the maximum posterior solution is desired rather than a sample drawn from the posterior, one simply sets $\omega_i$ to zero.

The computational expense for solving this equation depends directly on the complexity of the projection operator, $\P$. In most \commanderone-type analyses, which employ a pixel basis for all components and impose no spatial priors, i.e., $\S=0$, all matrix multiplications are given by diagonal matrices. In contrast, as implemented in \commandertwo, $\P$ involves one spherical harmonic transform per frequency channel, and the computational scaling of the left-hand side increases from $\mathcal{O}(N_{\mathrm{pix}})$ to $\mathcal{O}(N_{\mathrm{pix}}^{3/2})$. Accordingly, the associated CPU time required per sample increases from minutes to tens of hours. This additional cost, however, is very well justified by the new flexibility in terms of beam treatment, which now supports arbitrary resolution at each frequency.

By virtue of being a Gibbs-sampling procedure, \commander\ requires one sampling step for each parameter under consideration, such as spectral or calibration parameters.  However, these parameters are  sampled with exactly the same methods in \commandertwo\ as in \commanderone, and the details will not be repeated here; see \citet{eriksen2008} and \citet{planck2014-a12}.

\subsection{\commander\ 2018 signal model and priors}
\label{app:commprior}

As discussed in Sects.~\ref{sec:methods} and \ref{sec:inputs}, the maps provided in the \Planck\ 2018 release include only full frequency maps, not individual detector or detector-set maps. This has significant consequences for our ability to reconstruct some important parameters, in particular CO line emission. For this reason, we adopt a simpler signal model in the 2018 analysis than in the 2015 analysis. Explicitly, the basic model considered in the current analysis, as defined in Rayleigh-Jeans temperature units, reads\footnote{For simplicity, bandpass integration and unit conversion effects are omitted from this  expression. Such effects are handled as in earlier implementations, through construction of fast, splined look-up tables based on direct bandpass convolution for the relevant parameters; see \citet{planck2013-p03d} for an overview of the basic equations.}
\begin{align}
  s_{\nu} =\   g_{\nu} \biggl[& \Y\a_{\mathrm{cmb}} \gamma(\nu) \\
     &+ \Y\a_{\mathrm{lf}}
     \left(\frac{\nu}{\nu_{\mathrm{lf}}}\right)^{\beta_{\mathrm{lf}}(p)} \\
     &+ \Y\a_{\mathrm{d}}
     \left(\frac{\nu}{\nu_d}\right)^{\beta_{\mathrm{d}}(p)+1}
     \left(\frac{e^{h\nu_{\mathrm{d}}/kT_{\mathrm{d}}(p)}-1}{e^{h\nu/kT_{\mathrm{d}}(p)}-1}
     \right) \\
     &+ \Y\a_{\mathrm{co}} h_{\nu} \\
     &+ \sum_{i=1}^{N_{\mathrm{src}}} \B_{\mathrm{F},\nu,i} a_{\mathrm{cs},i}
     \left(\frac{\nu}{\nu_{\mathrm{cs}}}\right)^{\alpha_{\mathrm{cs}}(p)}
     \\
     &+ \sum_{i=1}^{N_{\mathrm{temp}}} \T_{\nu} a_{\mathrm{temp},i}
     \biggr],
\end{align}
where the various components correspond to, from top to bottom, CMB, low-frequency/synchrotron emission, thermal dust emission, CO line emission, compact objects, and template corrections. The full model applies only to temperature analysis, since CO line emission, point sources, and template corrections are all omitted from the polarization analysis. For temperature, we refer to the second term as a ``low frequency component,'' since it includes both synchrotron, free-free, and anomalous microwave emission, while for polarization we refer to it as ``synchrotron'', since that is the only component that is significantly detected at low frequencies in polarization.

In the above expression, $\gamma(\nu)$ is the conversion factor between thermodynamic and Rayleigh-Jeans units,
$\nu_{\mathrm{lf}} = 30$\,GHz is the reference frequency for the low-frequency component, $\nu_{\mathrm{d}} = 857$\,GHz is the thermal dust reference frequency for temperature (353\,GHz for polarization), $h_{\nu}$ is the CO line ratio between 100 and 217 or 353\,GHz, respectively, and all other quantities are defined above.

To complete the specification of a model used for Bayesian analysis, we also have to choose priors for the various parameters. Starting with the spectral parameters, we adopt the same types of priors as in previous analyses. Technically speaking, for each parameters these are given as the product of three different priors, each serving a different purpose. First, we impose a uniform prior between two hard limits for numerical reasons; this makes it easier to precompute look-up tables for all unit conversion and bandpass integration quantities. Second, we impose a Jeffreys' ignorance prior, which effectively normalizes posterior volume effects due to the specific choice of parametrization. Third, and by far most importantly, we adopt Gaussian informative priors with physically motivated means and standard deviations for all spectral parameters. The values of these are listed in Table~\ref{tab:commander_priors}.

Next, we need to specify spatial priors on the amplitude degrees of freedom. With the new \commandertwo\ implementation -- which models all diffuse components, not just the CMB, in spherical harmonic space -- we are now able to impose informative spatial priors on the foregrounds through the signal covariance matrix, $\S$, in Eq.~(\ref{eq:wiener}). In this paper, we define this matrix in harmonic space in terms of a standard angular power spectrum, ${\cal D}_{\ell}$, per component. In principle, this could be used to enforce physically motivated power spectra for each component, for instance a $\Lambda$CDM spectrum for the CMB, or a power-law spectrum for synchrotron or thermal dust emission. However, in the present analysis, we choose to be minimally constraining, and simply use this new feature to enforce smoothness of the foreground components on small scales. For all components except thermal dust intensity, we implement this by defining a reference prior spectrum given by the shape of a Gaussian smoothing kernel multiplied by an overall amplitude that is larger than the actual sky signal in the high signal-to-noise regime. Thus, the prior takes the form
\begin{equation}
 {\cal D}_{\ell}^{i} = A^i e^{-\ell(\ell+1)\sigma^2},
  \label{eq:prior}
\end{equation}
where $\sigma^2 = \theta_{\mathrm{FWHM}}^2 / (8\ln 2)$, $\theta_{\mathrm{FWHM}}$ is the FWHM of the desired Gaussian smoothing kernel in radians, and $A^i$ is the uniform power spectrum amplitude. This type of prior simply acts as a smooth apodization of the high-$\ell$ spectra, and its main function is to prevent the ringing that would otherwise occur around objects  with a sharp cutoff in harmonic space, given by some $\ell_{\mathrm{max}}$. For the special case of thermal dust emission in intensity, we employ a simple cosine apodization between $\ell=5000$ and 6000, in order to retain as much signal as possible. The spatial prior values adopted for the various components are summarized in Table~\ref{tab:commander_priors}.

Finally, we need to impose priors on the zero levels and dipoles for each map. For HFI zero levels, we adopt the CIB offsets defined in table~6 of \citet{planck2014-a09}, while for the LFI we adopt a vanishing monopole at 30\,GHz. At 44 and 70\,GHz, we impose no priors on the zero levels, but rather fit them freely, obtaining best-fit values of 17 and 21\muK, respectively. For the HFI channels between 100 and 545\,GHz, the best-fit zero levels are 12.4\muK, 22.0\muK, 71.0\muK, 431\muK, and 0.346\,MJy\,$\textrm{sr}^{-1}$, respectively, while the 857-GHz zero level is fixed \textcolor{black}{at 0.64\,MJy\,$\textrm{sr}^{-1}$ from \citet{planck2014-a09}. For comparison, the nominal CIB offsets listed in the same reference} correspond to 12\muK, 21\muK, 68\muK, 451\muK, and 0.35\,MJy\,$\textrm{sr}^{-1}$\,MJy\,$\textrm{sr}^{-1}$.

We only fit for dipoles in the 70- and 100-GHz channels, a choice determined by inspection of the residual maps resulting from an initial analysis in which no dipoles are fitted. The best-fit \commander-derived amplitudes of the 70- and 100-GHz dipoles are 2.0 and 2.3\muK, respectively.

\subsection{Sampling compact objects}

A significant new feature of \commandertwo\ is its ability to fit compact sources with multi-resolution frequency maps, while at the same time accounting for the full asymmetric beam structure at each frequency. As described above, for a single source this is done through the following parametric model,
\begin{equation}
s_{\nu}(p) = \B_{\mathrm{F},\nu} a_{\mathrm{cs}}
\left(\frac{\nu}{\nu_{\mathrm{cs}}}\right)^{\alpha_{\mathrm{cs}}(p)},
\end{equation}
where $\B_{\mathrm{F},\nu,i}$ is a full {\tt FEBeCoP} template evaluated at the pixel closest to the point source in question, $\a$ is the source amplitude in units of mJy, and we assume a simple power-law frequency scaling with a spectral index of $\alpha_{\mathrm{cs}}$ in flux density units (in the current analysis we only consider this component for radio sources, for which a power-law model is a reasonable spectrum). Thus, each source is associated with only two free parameters, the amplitude $\a$ and the spectral index $\alpha$, across all frequencies. This simple two-parameter model, however, is not likely to be adequate for a full fit between 30 and 857\,GHz for many sources. As a result, when fitting the free parameters, we only include frequencies between 30 and 143\,GHz in the actual fit; however, the resulting parameters are used to extrapolate to the higher frequencies when fitting other parameters.

Source locations are not identified internally in \commander, but rather defined by external catalogues. Unfortunately, no full-sky, deep, and high-resolution catalogue of radio sources exists for the microwave frequencies, and we therefore construct a hybrid catalogue by combining four different catalogues. First, we include all sources in the AT20G catalogue for declinations below $-15^{\deg}$, for a total of 4499 sources.  By virtue of being closest to our frequency range, this catalogue is adopted as an overall reference. Thus, we compute an effective source number density per area of AT20G sources, and adopt this as a threshold density. This threshold is then applied to the GB6 catalogue, including all sources above a flux density defined by requiring that the area number density is the same as for AT20G. This results in 5814 GB6 sources. Next, for sky regions not
covered by either AT20G or GB6, we employ the same algorithm to the NVSS catalogue, resulting in 1527 NVSS sources. Finally, we also include all sources found in the PCCS2 catalogue, except for excluding duplicates in the already considered catalogues; this results in 352 unique sources. Thus, at this stage, the full sky has been populated by sources with a nearly uniform number density, for a total of 12\,192 sources. 

It is important to note that the catalogue positions defined above are only used as candidates for source positions. Including a non-existing source will not bias any other parameter, since its relevant parameters are fitted jointly with all other parameters; the only detrimental effect of including too many sources is a slight increase in the overall noise level.

\begin{figure}[t]
\begin{center}
  \includegraphics[width=1\linewidth]{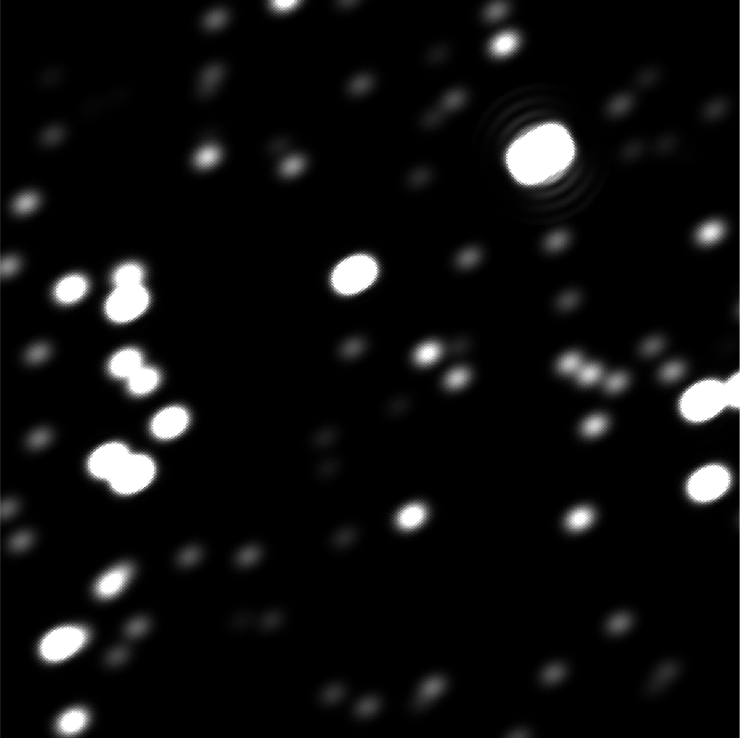}\\
  \includegraphics[width=1\linewidth]{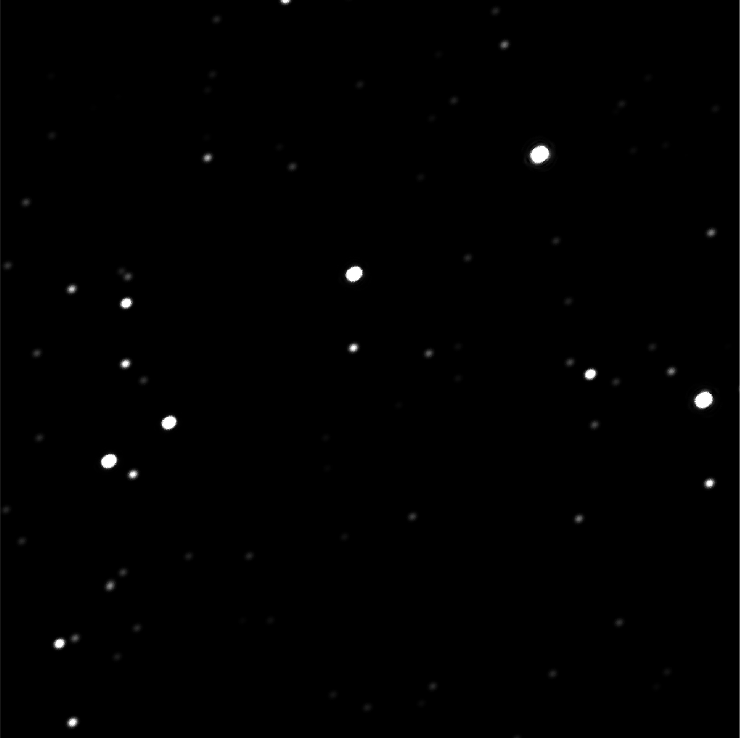}
\end{center}
\caption{Enlargement of the compact source map fitted with \commander\ using real-space spatial {\tt FEBeCoP} templates and a power-law spectral model. Shown here is a $10^\circ \times 10^\circ$ region centreed on the South Galactic Pole (SGP), and the top and bottom panels showing the effective point source maps at 30 and 100\,GHz, respectively. Note the significantly asymmetric beam structures in the 30-GHz map.}
\label{fig:comm_sources}
\end{figure}

Figure \ref{fig:comm_sources} shows an enlargement of the final \commander\ compact source map for 30 and 100\,GHz, generated as described above. The plot shows a $10^\circ \times 10^\circ$ region centred on the South Galactic Pole, for which the \Planck\ scanning strategy provides relatively poor cross-linking. As a result, the asymmetric properties of the 30-GHz beams are clearly visible. Another feature seen in these plots is the large number of overlapping sources in the 30-GHz map. If we included only this single frequency while fitting the spectral properties of the sources, there would be significant degeneracies between such overlapping sources. However, when we include higher frequencies, for which the beams are
smaller and neighboring sources overlap less, these degeneracies are effectively broken.

\subsection{Confidence masks}
\label{app:comm_mask}

For \commander, we establish the following prescription for defining a temperature confidence mask. First, the base temperature mask is defined by smoothing the \commander\ $\chi^2$ map with a $30\arcm$ FWHM Gaussian beam, suppressing instrumental noise fluctuations, and then thresholding the smoothed map at a value of 50, which corresponds to a roughly $4\sigma$ confidence level at high Galactic latitudes. This mask removes any pixel for which the \commander\ model obviously breaks down in terms of total $\chi^2$. However, based on frequency residual maps, one does observe residuals corresponding to specific components that are not easily picked up by the total $\chi^2$. To capture these, we augment the base mask with three specifically targeted masks. First, we remove any pixels brighter than $10\,\textrm{mK}$, to eliminate particularly bright radio sources. Second, we exclude by hand the Virgo and Coma clusters and the Crab Nebula. Third, noting that CO emission represents a particularly difficult problem with the current data set, we smooth the \commander\ 2018 CO emission map shown in Fig.~\ref{fig:comm_fg_temp} with a $30\arcm$ FWHM beam, and exclude any pixels for which the CO amplitude is larger than $50\muK_{\mathrm{RJ}}$ at 100\,GHz.

\begin{figure}[t]
\begin{center}
\includegraphics[width=\columnwidth]{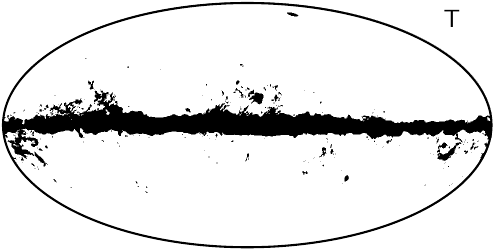}
\includegraphics[width=\columnwidth]{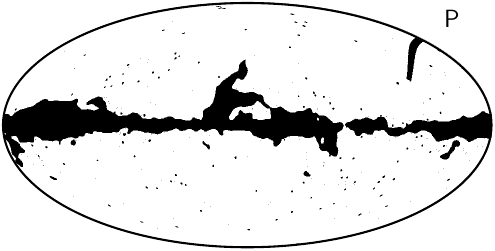}
\end{center}
\caption{\commander\ masks in temperature (top) and polarization (bottom).}
\label{fig:dx12_masks_commander}
\end{figure}

The polarization confidence mask is constructed in a similar way, with a few specific modifications. First, a base mask is produced by smoothing and thresholding the $\chi^2$ map shown in Fig.~\ref{fig:comm_pol_residuals}. Second, we remove all pixels for which the polarized thermal dust amplitude smoothed to $3^{\circ}$ FWHM is larger than $20\muK_{\mathrm{RJ}}$ at 353\,GHz (see Fig.~\ref{fig:comm_pol_dust}); this mask excludes pixels for which large values are observed in the frequency residual maps shown in Fig.~\ref{fig:comm_pol_residuals}, but which are not robustly picked up by the $\chi^2$ values. Third, we remove all pixels corresponding to the cosmic ray contaminated ring discussed in Sect.~\ref{sec:fullmission}. Fourth, we remove particularly bright point sources based on the PCCS2 source catalogue. The resulting masks for both temperature and
polarization are shown in Fig.~\ref{fig:dx12_masks_commander}.

\begin{figure}[htbp]
\begin{center}
  \includegraphics[width=\columnwidth]{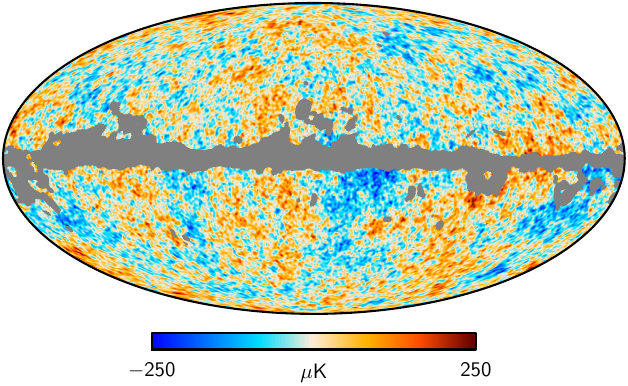}
  \includegraphics[width=\columnwidth]{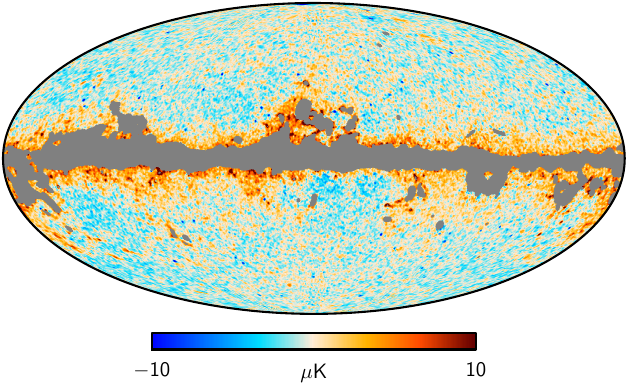}
\end{center}
\caption{\emph{Top:} \commander\ CMB temperature map used for the   \Planck\ low-$\ell$ temperature likelihood analysis, smoothed to $60\arcm$ FWHM resolution. The grey region indicates the mask adopted for the likelihood analysis, which retains 86\% of the sky. \emph{Bottom:} Difference between the low-$\ell$ likelihood and full-resolution \commander\ CMB temperature maps, smoothed to $60\arcm$ FWHM resolution.}
\label{fig:comm_lowl_vs_fullres}
\end{figure}

\begin{figure}[htbp]
\begin{center}
  \includegraphics[width=\columnwidth]{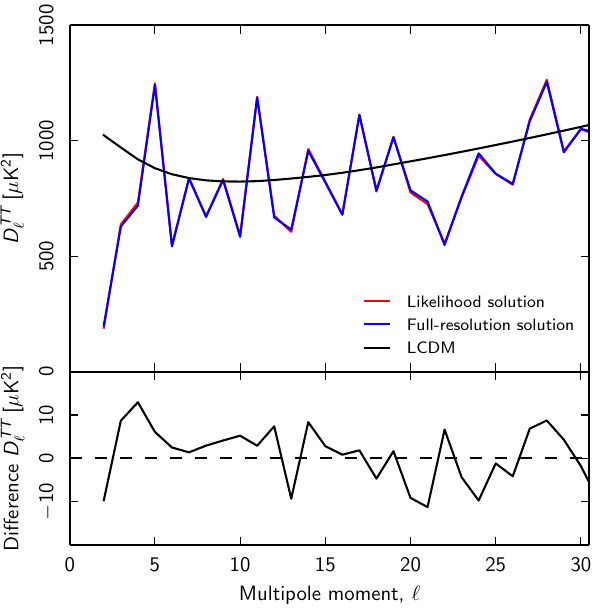}
\end{center}
\caption{\emph{Top:} Low-$\ell$ temperature power spectra derived from the low-$\ell$ likelihood \commander\ map (red) and from the full-resolution \commander\ map (blue), both evaluated over the low-$\ell$ likelihood mask shown in Fig.~\ref{fig:comm_lowl_vs_fullres}. \emph{Bottom:} Difference between the two spectra shown in the top panel. }
\label{fig:comm_cls_lowl_vs_fullres}
\end{figure}

\subsection{Comparison between low-$\ell$ likelihood and full-resolution \commander\ maps }

As described in \citet{planck2016-l05}, the \commander\ algorithm is used to generate the low-$\ell$ temperature sky map that feeds the \Planck\ 2018 CMB likelihood, as it was in previous \Planck\ releases. The set-up adopted for that analysis is, however, somewhat different than the one adopted for the main analysis presented in this paper, primarily due to the different angular scales in question. Specifically, since the likelihood map is only used for large angular scales, covering primarily only $\ell\le30$, the full analysis is carried out with \commanderone, and all input frequency maps are smoothed to a common angular resolution of $40\arcm$ FWHM, similar to the \Planck\ 2015 processing. Finally, the \commanderone\ low-$\ell$ analysis internally estimates the CMB power spectrum as one of the parameters in the Bayesian parametric model, and the corresponding Gaussian constrained realization samples \citep{eriksen2008} provide the necessary inputs for the Blackwell-Rao likelihood estimator employed by the \Planck\ temperature-only likelihood \citep{chu2005}. For the combined \Planck\ temperature and polarization likelihood, which is map-based rather than power-spectrum-based, a single constrained-realization sample is adopted as the low-$\ell$ likelihood temperature component.   We have verified that the choice of the particular sample used has no significant effect on the actual power spectrum outside the analysis mask.

The top panel of Fig.~\ref{fig:comm_lowl_vs_fullres} shows the low-$\ell$ likelihood temperature map with the corresponding likelihood mask marked in grey. The bottom panel shows the difference map with respect to the full-resolution \commander\ map, after the latter is smoothed to $60\arcm$ FWHM resolution. Over most of the sky, the absolute difference between the two maps is $\lesssim2\muK$, increasing to $5\muK$ near the Galactic plane. A few bright spots exhibit differences at the 10-$\mu$K level. These differences are dominated by thermal dust emission (see, e.g., Fig.~\ref{fig:comm_fg_temp}), and are in effect due to the different angular resolutions adopted for the two analyses. Since the likelihood analysis is performed at an angular resolution of $40\arcm$ FWHM, the thermal dust spectral index and temperature are also estimated at an angular resolution of $40\arcm$ FWHM. In contrast, the high-resolution analysis estimates the thermal dust spectral index at $10\arcm$ FWHM, and the corresponding temperature at $5\arcm$ resolution. As a consequence, the assumed spectral priors have a relatively larger impact in the high-resolution analysis than in the low-$\ell$ analysis.

Nevertheless, these differences are small in terms of absolute numbers, and have a negligible impact on the derived angular power spectrum. This is explicitly shown in Fig.~\ref{fig:comm_cls_lowl_vs_fullres}, in which the top panel shows the individual spectra computed from each of the two maps, and the bottom panel shows their difference. Overall, the absolute differences
are smaller multipole-by-multipole than $10\muK^2$, corresponding to $\lesssim1\,\%$ in absolute power and $\lesssim0.05\,\sigma$ in terms of cosmic variance. There is also no overall trend in the difference spectrum that might pull systematically on cosmological parameters. The two maps are statistically equivalent in terms of temperature power spectra.

\section{Needlet Internal Linear Combination}
\label{app:nilc}

The goal of \nilc\ is to estimate the CMB from multi-frequency observations while minimizing the contamination from Galactic and extragalactic foregrounds, and instrumental noise.  The method makes a linear combination of the data from the input maps with minimum variance on a frame of spherical wavelets called needlets \citep{narcowich06localizedtight}.  Due to their unique properties, needlets enable localized filtering in both pixel space and harmonic space.  Localization in pixel space allows the weights of the linear combination to adapt to local conditions of foreground contamination and noise, whereas localization in harmonic space allows the method to favour foreground rejection on large scales and noise rejection on small scales.  Needlets permit the weights to vary smoothly on large scales and rapidly on small scales, which is not possible by cutting the sky into zones prior to processing \citep{2009A&A...493..835D}.

The \nilc\ pipeline \citep{2012MNRAS.419.1163B, 2013MNRAS.435...18B} is applicable to scalar fields on the sphere, hence we work separately on maps of temperature and the $E$ and $B$ modes of polarization.  The decomposition of input polarization maps into $E$ and $B$ is performed on the full sky.  At the end of the processing, the CMB $Q$ and $U$ maps
are reconstructed from the $E$ and $B$ maps.

\begin{figure}[htbp]
  \centering
  \includegraphics[width=\linewidth]{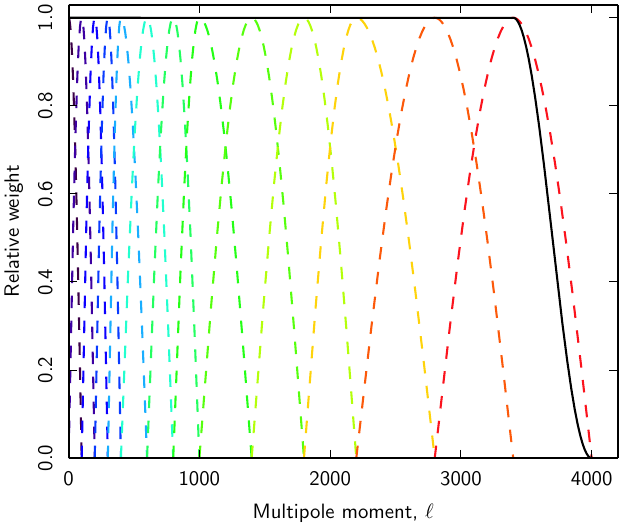}
\caption{Needlet bands used in the analysis.  The solid black line shows the normalization of the needlet bands, i,e., the total
filter applied to the original map after needlet decomposition and synthesis of the output map from needlet coefficients.}
 \label{fig:needlet-bands}
\end{figure}

Prior to applying \nilc, all of the input maps are convolved or deconvolved in harmonic space to a common resolution corresponding to a Gaussian beam of 5\arcm\ FWHM.  Each map is then decomposed into a set of needlet coefficients. For each scale $j$, needlet coefficients of a given map are stored in the form of a single \healpix\ map. The filters $h^{j}_{l}$ used to compute filtered maps are given by
\begin{eqnarray*} 
h^{j}_{l} = \left\{
\begin{array}{rl} 
\cos\left[\left(\frac{\ell^{j}_{\mathrm{peak}}-\ell}{\ell^{j}_{\mathrm{peak}}-\ell^{j}_{\mathrm{min}}}\right)
\frac{\pi}{2}\right]& \mathrm{for }\, \ell^{j}_{\mathrm{min}} \le \ell < \ell^{j}_{\mathrm{peak}},\\ 
\\
1\hspace{0.5in} & \mathrm{for }\, \ell = \ell_{\mathrm{peak}},\\
\\
\cos\left[\left(\frac{\ell-\ell^{j}_{\mathrm{peak}}}{\ell^{j}_{\mathrm{max}}-\ell^{j}_{\mathrm{peak}}}\right)
\frac{\pi}{2}\right]& \mathrm{for }\, \ell^{j}_{\mathrm{peak}} < \ell \le \ell^{j}_{\mathrm{max}}. 
\end{array} \right. 
\end{eqnarray*}
For each scale $j$, the filter has compact support between the multipoles $\ell^{j}_{\mathrm{min}}$ and $\ell^{j}_{\mathrm{max}}$ with a peak at $\ell^{j}_{\mathrm{peak}}$ (see Table~\ref{tab:needlet-bands} and Figure~\ref{fig:needlet-bands}). The needlet coefficients are computed from these filtered maps on \healpix\ pixels with $\nside$ equal to the smallest power of $2$ larger than $\ell^{j}_{\mathrm{max}}/2$.

\begin{table}[t]
\begingroup \newdimen\tblskip \tblskip=5pt
  \caption{List of needlet bands used in the \nilc\ analysis.}
  \label{tab:needlet-bands} 
\vskip -4mm
\footnotesize
\setbox\tablebox=\vbox{
\newdimen\digitwidth
\setbox0=\hbox{\rm 0}
\digitwidth=\wd0
\catcode`*=\active
\def*{\kern\digitwidth}
\newdimen\signwidth
\setbox0=\hbox{+}
\signwidth=\wd0
\catcode`!=\active
\def!{\kern\signwidth}
\newdimen\decimalwidth
\setbox0=\hbox{.}
\decimalwidth=\wd0
\catcode`@=\active
\def@{\kern\signwidth}
\halign{ \hbox to 0.9in{#\leaderfil}\tabskip=2em&
    \hfil#\hfil&
    \hfil#\hfil&
    \hfil#\hfil&
    \hfil#\hfil\tabskip=0pt\cr
\noalign{\doubleline}
\omit\hfil Band\hfil&$\ell_{\rm min}$&$\ell_{\rm peak}$&$\ell_{\rm max}$&$N_{\rm side}$\cr
\noalign{\vskip 4pt\hrule\vskip 5pt}
$j=1$&            ***0& ***0& *100& **64\cr
\hglue 1.65em 2&  ***0& *100& *200& *128\cr
\hglue 1.65em 3&  *100& *200& *300& *256\cr
\hglue 1.65em 4&  *200& *300& *400& *256\cr
\hglue 1.65em 5&  *300& *400& *600& *512\cr
\hglue 1.65em 6&  *400& *600& *800& *512\cr
\hglue 1.65em 7&  *600& *800& 1000& *512\cr
\hglue 1.65em 8&  *800& 1000& 1400& 1024\cr
\hglue 1.65em 9&  1000& 1400& 1800& 1024\cr
\hglue 1.22em 10& 1400& 1800& 2200& 2048\cr
\hglue 1.22em 11& 1800& 2200& 2800& 2048\cr
\hglue 1.22em 12& 2200& 2800& 3400& 2048\cr
\hglue 1.22em 13& 2800& 3400& 4000& 2048\cr
\noalign{\vskip 5pt\hrule\vskip 4pt}
}}
\endPlancktable
\endgroup
\end{table}

\textcolor{black}{Due to the deconvolution of sky maps to an effective smoothing scale of 5\arcm\ FWHM, noise levels in lower-resolution frequency maps are boosted. To limit this effect, we include only those multipoles for which the ratio between the corresponding beam transfer function and that of a corresponding 5\arcm\ Gaussian is larger than 0.01 for each frequency map. This cut is made separately for each needlet scale, such that only those frequency maps containing valid harmonic content that spans the entire bandwidth of a given needlet scale contribute to that particular scale.}

In order to improve the measurement of CMB temperature anisotropy near the Galactic plane, we have used a very small preprocessing mask with a sky fraction of 99.8\,\%. The procedure to generate the preprocessing mask is as follows. First we implement the \nilc\ pipeline on the full-mission data sets. Then we identify the pixels where the CMB is more than $500\,\mu \mathrm{K}_{\mathrm{CMB}}$, and assign a value of 0 to all the pixels that are within $6\arcm$ and a value of~1 to other pixels.  We implement this preprocessing mask on the sky maps in the next run of the \nilc\ pipeline.  Prior to implementing the pipeline on the sky maps, the mask regions are filled using the inpainting procedure adopted by the Planck Sky Model.

Estimates of the covariance matrices of needlet coefficients for each scale are computed by smoothing all possible products of needlet coefficients with Gaussian beams.  In this way, an estimate of needlet covariances at each point is obtained as a local, weighted average of needlet coefficient products. The FWHMs of the Gaussian windows used for the analysis are chosen to support the computation of statistics; 4225 samples or more samples are averaged. Choosing a smaller FWHM results in excessive error in the covariance estimates, and hence excessive bias. Choosing a larger FWHM results in less localization, and hence some loss of efficiency of the needlet approach.

A patch of angular radius $\theta$ and area $2\pi(1-\cos(\theta))$ contains $N/(4\pi) \times 2\pi\{1-\cos(\theta)\}$ modes. If we wish to have $M$ modes in that patch, we simply solve for the corresponding $\theta$.  We chose FWHM$ = 2 \times \theta$ for the Gaussian beam that we use to smooth the covariance matrix.  Hence in order to determine the covariance matrix at a particular point, we have given more weight to those pixels that are close to that point, and less weight to those pixels that are far away.  However, this strategy is not optimal for the largest scales.

\begin{figure}[htbp]
  \centering
  \includegraphics[width=\columnwidth]{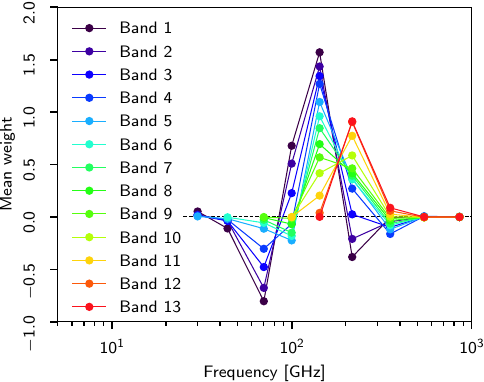}
  \includegraphics[width=\columnwidth]{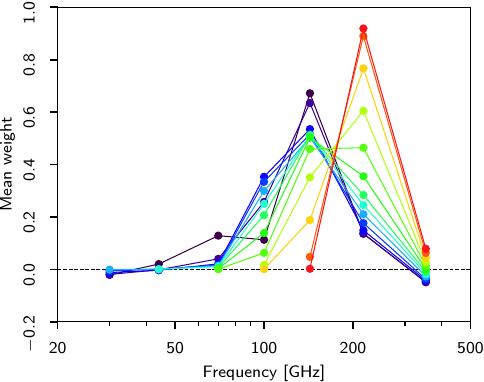}
  \includegraphics[width=\columnwidth]{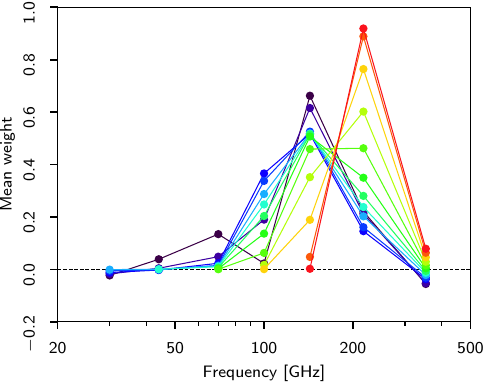}
  \caption{Full-sky average of needlet weights for different frequency channels and needlet bands. From top to bottom, the panels show results for temperature, $E$ modes, and $B$ modes.}
  \label{fig:needlet-weights}
\end{figure}

Figure \ref{fig:needlet-weights} shows that the $70$, $100$, $143$, and $217$ GHz channels have contributed most to the final
reconstruction of the NILC CMB map. However, other channels are also important because these channels are tracers of the foreground signals, and help us to find optimal weights for the final solution.

\begin{figure}[htbp]
  \centering
  \includegraphics[width=\linewidth]{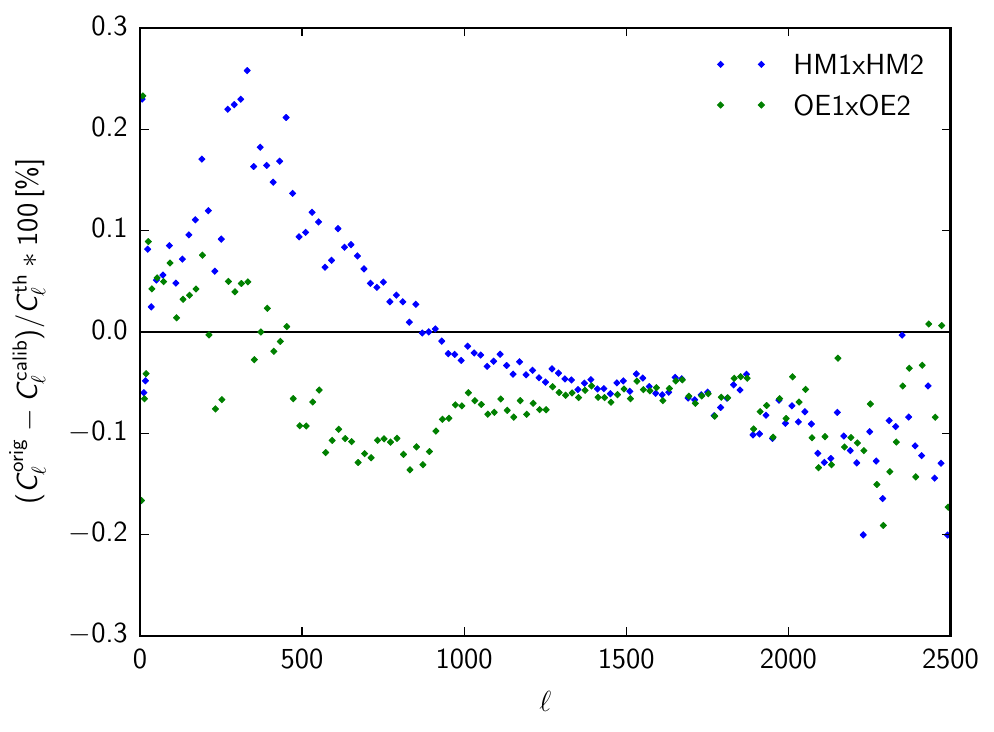}
  \caption{Difference of angular power spectra obtained with and without considering the correction to calibration coefficients estimated by \smica.}
  \label{fig:nilc-calib}
\end{figure}

Calibration errors are a serious issue for precise measurement of the CMB, as they conspire with the ILC filter to cancel out the
CMB. This effect is particularly strong in the high signal-to-noise ratio regime, on large scales in particular. We investigate the impact of this calibration bias by redoing the analysis with slightly modified calibration coefficients, and computing the difference between the CMB spectra estimated in the two cases. We adopt the calibration coefficients determined by \smica\ in Appendix~\ref{app:smica}. Figure~\ref{fig:nilc-calib} shows the impact of calibration on the angular power spectrum of the CMB temperature. Comparison of angular power spectra for two data splits shows that the impact is less than 0.5\,\%.

\section{SEVEM}
\label{app:sevem}

The \sevem\ method \citep{2008A&A...491..597L,2012MNRAS.420.2162F} produces cleaned CMB maps at different frequencies by subtracting a linear combination of templates constructed internally from the data. In particular, the templates are typically obtained as the subtraction of two close \Planck\ frequency channels, filtered to the same resolution to remove the CMB signal.   The cleaning is achieved simply by subtracting a linear combination of the templates $t_j(\vec{x})$ from the data, with coefficients $\alpha_j$ obtained by minimizing the variance outside a given mask:
\begin{equation}
T_{\rm c}(\vec{x},\nu)=d(\vec{x},\nu)- \sum_{j=1}^{n_{\rm t}} \alpha_j t_j(\vec{x}).
\label{eq:sevem_basic_formula}
\end{equation}
Here $n_{\rm }t$ is the number of templates used, while $T_{\rm c}(\vec{x},\nu)$ and $d(\vec{x},\nu)$ correspond to the cleaned and raw maps at frequency $\nu$, respectively.  The same expression applies for $T$, $Q$, or $U$. Note that we estimate  the linear coefficients  $\alpha_j$ independently for $Q$ and $U$ maps, following what was done for the previous release\footnote{In principle, it would be desirable to estimate the linear coefficients taking into account the spinorial character of the $Q$ and $U$ components, since this allows us to keep the physical coherence of the foreground residuals, following, for instance, the method proposed by \citep{fernandez-cobos2016}. However, in practice, this does not seem to have any significant impact on the results from \Planck\ data and, therefore, for simplicity, we work with independent coefficients for $Q$ and $U$ maps.}.
  
The cleaned frequency maps are then combined in harmonic space, taking into account the noise level, resolution, and (optionally) an estimate of the foreground or systematic residuals of each cleaned channel, to produce a final CMB map at the required resolution.

\subsection{Implementation for temperature}

For temperature, we have followed the same procedure as for the \Planck\ 2015 release (see \citealt{planck2014-a11} for further details). As before, we clean the 100-, 143-, and 217-GHz frequency channels with four templates, three of them constructed as the difference between two nearby \Planck\ channels ($30-44$, $44-70$, $545-353$), such that the first channel is convolved with the beam of the second one and vice versa, and a fourth template given by the 857-GHz channel, convolved with the 545-GHz beam. The cleaned frequency maps have the same resolution as the corresponding original raw data map. To reduce the contamination from point sources in the templates, we follow the same approach as in the previous release. First, point sources are detected in each frequency map using the Mexican-Hat-Wavelet algorithm  \citep{lopezcaniego2006,planck2014-a35}. The upper part of Table~\ref{tab:ps_sevem} gives the number of point sources detected in intensity and polarization for all the \Planck\ frequency channels over the full-sky, at Galactic latitudes $\left| b\right| > 20\deg $.  We then inpaint the holes corresponding to the positions of those point sources in the frequency maps (at their original resolution) involved in the construction of the templates. Note that the size of the hole depends on the amplitude of the detected source and the resolution of the considered channel. The filling is done with a simple diffusive inpainting scheme, which replaces one pixel with the mean value of the neighbouring pixels in an iterative way. To avoid possible inconsistencies when performing the subtraction of two maps to construct a template, the diffusive inpainting is performed for all of the sources detected in both channels. For instance, when constructing the ($30-44$)\,GHz template, all sources detected at 30 and 44\,GHz are inpainted in the two frequency maps before subtraction\footnote{Note that if a map is used to construct more than one template, several inpainted versions of that map will be constructed in the appropriate way in order to match the pair.}.

\begin{table*}[t]                                                                                                                                                   
\begingroup                                                                            
\newdimen\tblskip \tblskip=5pt
\caption{Number of detected sources in intensity and polarization. The upper part of the table refers to sources detected in the raw frequency maps, while the lower part gives the number of point sources detected in the cleaned \sevem\ frequency maps after inpainting the originally detected sources. A list with the positions of all the sources and the corresponding masks used in the \sevem\ pipeline are available in the Planck Legacy Archive. \label{tab:ps_sevem}}
\nointerlineskip                                                                                                                                                                                     
\vskip -4mm
\footnotesize                                                                                                                                      
\setbox\tablebox=\vbox{ %
\newdimen\digitwidth                                                                                                                          
\setbox0=\hbox{\rm 0}
\digitwidth=\wd0
\catcode`*=\active
\def*{\kern\digitwidth}
\newdimen\signwidth
\setbox0=\hbox{+}
\signwidth=\wd0
\catcode`!=\active
\def!{\kern\signwidth}
\newdimen\decimalwidth
\setbox0=\hbox{.}
\decimalwidth=\wd0
\catcode`@=\active
\def@{\kern\signwidth}
\halign{ \hbox to 1.5in{#\leaderfil}\tabskip=1em&
  \hfil#\hfil\tabskip=1em&
  \hfil#\hfil\tabskip=1em&
  \hfil#\hfil\tabskip=1em&
  \hfil#\hfil\tabskip=1em&
  \hfil#\hfil\tabskip=1em&
  \hfil#\hfil\tabskip=1em&
  \hfil#\hfil\tabskip=1em&
  \hfil#\hfil\tabskip=1em&
  \hfil#\hfil\tabskip=0em\cr
\noalign{\doubleline}
 \omit\hfil Map\hfil&30\,GHz&44\,GHz&70\,GHz&100\,GHz&143\,GHz&217\,GHz&353\,GHz&545\,GHz&857\,GHz\cr
\noalign{\vskip 5pt\hrule\vskip 3pt}
\omit {\bf Raw}\hfil\cr
\noalign{\vskip 4pt}
\hglue 2em $T$ (full-sky)&                            1593& 923&1307&2162& 3479& 4955& 5794& 8145& 11876\cr
\hglue 2em $T$ ($\left| b \right| >20 \deg $)&*977& 470&*648& *809& 1093& 1289& 1588& 2898& *6117\cr
\hglue 2em $P$ (full-sky)&                            *195&  *64& **74& *237&  *349& *632& 1075\cr
\hglue 2em $P$ ($\left| b \right| >20 \deg $)&**65&  *19& **15&  **56&  **63&  **87& *134\cr
\noalign{\vskip 8pt}
\omit {\bf Cleaned}\hfil\cr
\noalign{\vskip 4pt}
\hglue 2em $T$ (full-sky)&                             \dots&\dots& 420&1475&2117&3675&\dots&\dots&\dots\cr
\hglue 2em $T$ ($\left| b \right| >20 \deg $)&\dots&\dots& *37& **93&*230&*553&\dots&\dots&\dots\cr
\hglue 2em $P$ (full-sky)&                            \dots&\dots& *10& **48&**73&*199&\dots&\dots&\dots\cr
\hglue 2em $P$ ($\left| b \right| >20 \deg $)&\dots&\dots& **1&  ***1&***4&**16&\dots&\dots&\dots\cr
\noalign{\vskip 5pt\hrule\vskip 3pt}
}}
\endPlancktablewide                                                                                                                                            
\endgroup
\end{table*}

In addition, for this release, we also provide a cleaned CMB map for the 70-GHz channel. This map is constructed at its original resolution and $N_{\rm side}=1024$, and has been cleaned with two templates, one constructed as the 30-GHz channel (convolved with the 44-GHz beam) minus the 44-GHz one (convolved with the 30-GHz beam), and a second template obtained as the difference between the 353 and 143 channels, constructed at a resolution equal to that of the 70-GHz channel. This second template has been chosen to trace the emission of the thermal dust, but avoding, as far as possible, the CO contamination (which is mostly present in the 100- and 217-GHz maps). Point source emission in the templates has also been reduced with the inpainting mechanism already described.

\begin{table*}[t]
\begingroup \newdimen\tblskip \tblskip=5pt
 \caption{Linear coefficients, $\alpha_j$, of the templates used to clean individual frequency maps with \sevem\ for temperature. The  353--143 template has been produced at the same resolution as the 70-GHz frequency channel, while the 857-GHz map has been convolved with the 545-GHz beam. The rest of the templates are constructed such that the first map in the subtraction is convolved with the beam of the second map and vice versa.}
\label{table:sevem_coef_T}
\nointerlineskip
\vskip -3mm
\footnotesize
\setbox\tablebox=\vbox{
\newdimen\digitwidth
\setbox0=\hbox{\rm 0}
\digitwidth=\wd0
\catcode`*=\active
\def*{\kern\digitwidth}
\newdimen\signwidth
\setbox0=\hbox{+}
\signwidth=\wd0
\catcode`!=\active
\def!{\kern\signwidth}
\newdimen\expsignwidth
\setbox0=\hbox{$^{-}$}
\expsignwidth=\wd0
\catcode`@=\active
\def@{\kern\expsignwidth}
\halign{\hbox to 1.3in{#\leaderfil}\tabskip=2em&
     \hfil#\hfil\tabskip=1em&
     \hfil#\hfil&
     \hfil#\hfil&
     \hfil#\hfil\tabskip=0pt\cr
\noalign{\doubleline}
\omit&\multispan4\hfil Coefficients $\alpha_j$\hfil\cr
\noalign{\vskip -3pt}
\omit&\multispan4\hrulefill\cr
\noalign{\vskip 3pt}
\omit\hfil Template\hfil& 70\GHz&100\GHz& 143\GHz& 217\GHz\cr
\noalign{\vskip 3pt\hrule\vskip 5pt}
   *$30-44$& $1.68 \times 10^{-1}$&$-9.15 \times 10^{-2}$& !$3.47\times 10^{-3}$&$-1.57\times 10^{-1}$\cr  
   *$44-70*$&\ldots&!$4.19 \times 10^{-1}$& !$1.97\times 10^{-1}$&!$4.22 \times 10^{-1}$\cr 
   $353-143$& $6.68 \times 10^{-3}$&\ldots& \ldots&\ldots\cr  
   $545-353$&\ldots&!$4.21\times 10^{-3}$&$!6.32\times 10^{-3}$&!$1.72\times 10^{-2}$\cr 
      !*857&\ldots&$-3.23\times 10^{-5}$&$-5.04\times 10^{-5}$&$-1.04\times 10^{-4}$\cr 
\noalign{\vskip 5pt\hrule\vskip 3pt}}}
\endPlancktablewide
\endgroup
\end{table*}

The coefficients of the linear combination used for cleaning the frequency maps are given in Table~\ref{table:sevem_coef_T}. They have been calculated by minimizing the variance of the corresponding cleaned map outside a mask that excludes the brightest 1\,\% of the sky and all the point sources detected in intensity.  The cleaning procedure introduces a certain level of correlation between the 100-, 143-, and 217-GHz cleaned frequency maps, due to the use of the same templates, but one frequency map is not used to clean the others. The cleaned 70-GHz channel is, however, more correlated, since it is part of one of the templates used to clean the higher frequency channels. Moreover, the 143-GHz map is also used to clean the 70-GHz channel. Therefore, one should bear in mind these correlations when carrying out analyses with the cleaned single
frequency maps. A possible way to reduce the correlations introduced by the cleaning process would be, when possible, to use pairs of cleaned frequency maps constructed with different splits, although this would be at the expense of decreasing the signal-to-noise ratio (e.g., to work with the cleaned 143-GHz even-ring and with the 217-GHz odd-ring maps, since the templates are constructed with the corresponding split).

\begin{figure*}[htpb]
  \begin{center}
    \begin{tabular}{c @{\hspace{0.5\tabcolsep}} c @{\hspace{0.5\tabcolsep}} c}
      \includegraphics[width=0.32\linewidth]
      {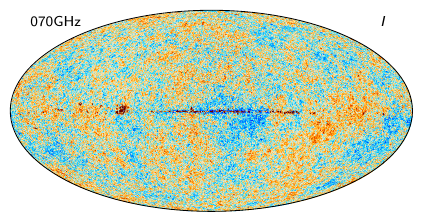}&
      \includegraphics[width=0.32\linewidth]
      {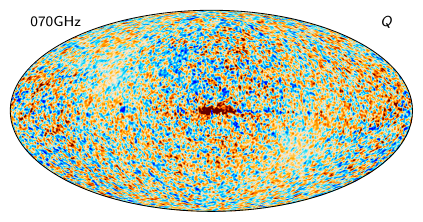}&
      \includegraphics[width=0.32\linewidth]
      {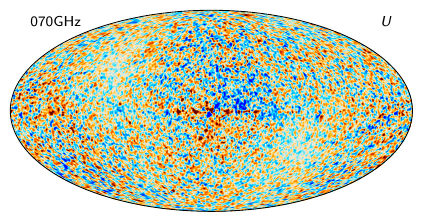}\\
      \includegraphics[width=0.25\linewidth]{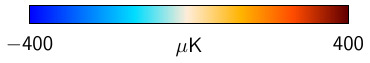}&
      \multicolumn{2}{c}{
        \includegraphics[width=0.25\linewidth]{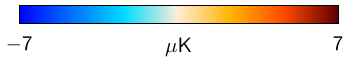}
      }\\
      \includegraphics[width=0.32\linewidth]
      {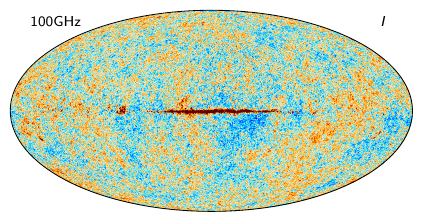}&
      \includegraphics[width=0.32\linewidth]
      {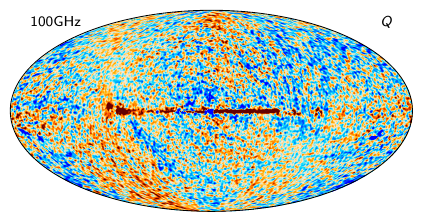}&
      \includegraphics[width=0.32\linewidth]
      {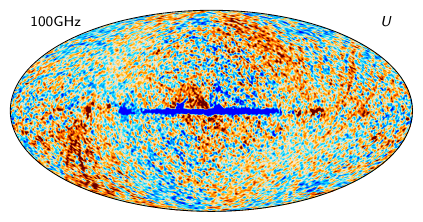}\\
      \includegraphics[width=0.32\linewidth]
      {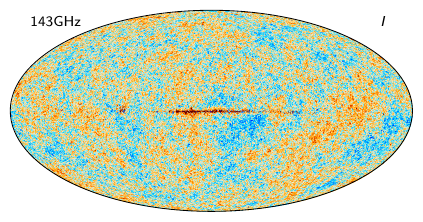}&
      \includegraphics[width=0.32\linewidth]
      {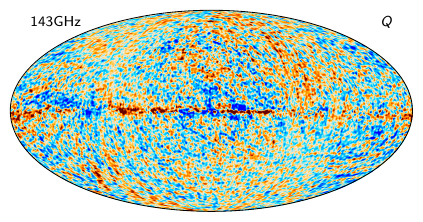}&
      \includegraphics[width=0.32\linewidth]
      {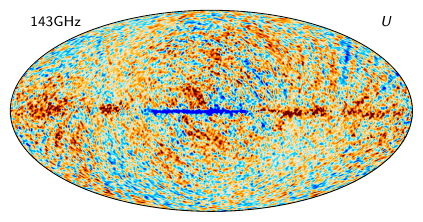}\\
      \includegraphics[width=0.25\linewidth]{figs/colourbar_400uK}&
      \multicolumn{2}{c}{
        \includegraphics[width=0.25\linewidth]{figs/colourbar_2p5uK}
      }\\
      \includegraphics[width=0.32\linewidth]
      {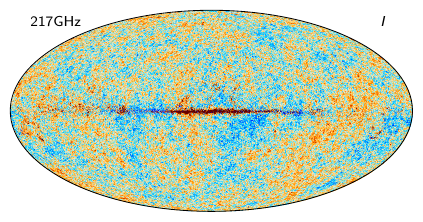}&
      \includegraphics[width=0.32\linewidth]
      {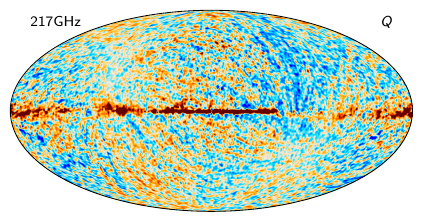}&
      \includegraphics[width=0.32\linewidth]
      {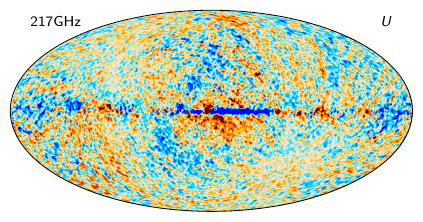}\\
      \includegraphics[width=0.25\linewidth]{figs/colourbar_400uK}&
      \multicolumn{2}{c}{
        \includegraphics[width=0.25\linewidth]{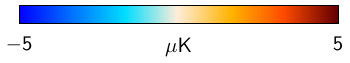}
      }
      \end{tabular}
        \end{center}
\caption{Cleaned single-frequency CMB maps from the \sevem\ pipeline.   The cleaned maps in intensity (left column) are 
given at their original resolution, while the polarization maps ($Q$ and $U$, middle and right columns) have been smoothed with a Gaussian beam of 80\arcmin\ FWHM resolution for better visualization. Rows show results for different frequencies (70, 100, 143, and 217\,GHz).}
\label{fig:sevem_freqmaps}
\end{figure*}

Following the same approach as in the previous release, after the frequency maps are cleaned, they are inpainted, in a first step, at the positions of the point sources identified in the corresponding raw maps. In a second step, the Mexican-Hat-Wavelet algorithm is again run on the cleaned frequency maps, and the newly detected sources (see lower part of Table~\ref{tab:ps_sevem}) are further inpainted. The combined area inpainted outside the \sevem\ confidence mask for the 143- and 217-GHz channels (those used to construct the final CMB map) corresponds to around $0.4\%$ of the sky, while it is fully covered by the common confidence mask. Note that the same inpainting strategy is applied to the simulations processed through the \sevem\ pipeline, to make sure that any possible effect introduced by this procedure is statistically taken into account. Finally, the monopole and dipole are removed from the full-sky cleaned maps (note that this is different from the previous release, where monopole and dipole were removed outside the \sevem\ confidence mask). The cleaned intensity maps for the 70-, 100-, 143-, and 217-GHz channels are shown in Fig.~\ref{fig:sevem_freqmaps}.

\begin{figure}[htbp]
\begin{center}
\includegraphics[width=\columnwidth]{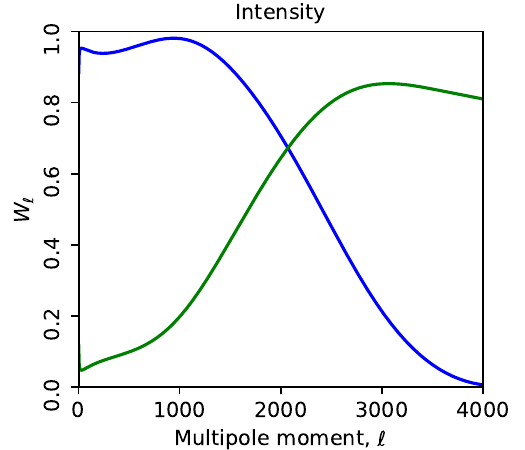}
\end{center}
\caption{Weights used to combine the cleaned single-frequency maps into the final \sevem\ CMB map for temperature, corresponding to 143\,GHz (blue line) and to 217\,GHz (green line).  The weights do not sum to unity because they include the effect of deconvolving the beams of the frequency maps and convolving with the 5\arcmin\ Gaussian beam of the final map. }
\label{fig:weights_sevem_I}
\end{figure}

The final \sevem\ CMB map is constructed by combining the cleaned 143- and 217-GHz\ maps in harmonic space.\footnote{In principle one could also include the cleaned 70- and 100-GHz maps in the combined solution. However, given the lower resolution and higher noise level of these channels, the improvement in the signal-to-noise ratio of the combined map is modest. Taking into account also that the addition of these channels could potentially  introduce contamination from low-frequency foregrounds or CO emission, we decided to combine only the 143- and 217-GHz cleaned channels in the final map.} In particular, the maps are weighted at each multipole, taking into account the noise level and resolution of the maps, as well as a rough estimation of the expected foreground residuals. This estimation has been updated with respect to the previous release by using the FFP8 simulations. The total weights are shown in Fig.~\ref{fig:weights_sevem_I}. The resolution of the combined map corresponds to a Gaussian beam of 5\arcmin\ FWHM and \healpix\ resolution $\nside = 2048$, with maximum multipole $\ell_{\mathrm{max}} = 4000$.  A monopole and a dipole are also removed from the full-sky map.

\subsection{Implementation for polarization}

A similar procedure is applied independently to the frequency maps of the Stokes $Q$ and $U$ parameters to obtain cleaned polarization CMB maps, which are aftewards combined in harmonic space to produce the final \sevem\ maps. Given the narrower frequency coverage available for polarization, a different choice of templates needs to be defined in this case. In the previous release, only two cleaned channels (100 and 143\,GHz) were combined to produce the final polarization map. However, due to the significant improvement of the \Planck\ data in polarization for the current release, we are now able to clean the 217-GHz channel and to include this map in the final combination. This produces a significant improvement in the signal-to-noise ratio of the cleaned \sevem\ CMB polarization maps with respect to the previous version. In addition, in the updated pipeline, the cleaned maps are produced at full resolution ($\nside=2048$ instead of $\nside=1024$ for the 100-, 143-, and 217-GHz channels, as well as the combined map). As for the previous release, a cleaned 70-GHz map is also provided at its native resolution.  To reduce point source contamination, inpainting similar to that in the previous release is performed.

The first step of the pipeline is to inpaint the positions of the sources detected in those channels that will be used to construct
templates. These point sources are detected using a non-blind approach, among the intensity candidates, using the filtered fusion technique \citep{argueso2009}. The upper part of Table~\ref{tab:ps_sevem} shows the number of sources detected in polarization in each of the frequency channels. The size of the holes to be inpainted takes into account both the amplitude of the source and the beam of the channel. As in the intensity case, when performing the subtraction of two maps to construct a template, the diffuse inpainting is performed for all of the sources detected in both channels. Note that the inpainting is always done at the native resolution of the channel and independently for $Q$ and $U$ maps.

Once the maps have been inpainted, each template is constructed as the subtraction of two frequency channels processed to a common resolution. Given the smaller number of channels in polarization, the maps to be cleaned are also used to construct templates. In this sense, the cleaned maps at different frequencies are, in general, less independent than in the intensity case (the exception is the 70-GHz channel, which is not used as part of the templates for polarization).  Six templates (one of them at two different resolutions) are generated to produce cleaned maps at 70, 100, 143, and 217\,GHz. In particular, to trace the synchrotron emission, the ($30-44$)\,GHz template is constructed, where the 30-GHz map is smoothed with the 44-GHz beam and vice versa. To trace the thermal dust, templates are produced at ($217-143$),  ($217-100$), and ($143-100$) with 1\deg\ resolution (this smoothing is included in order to increase the signal-to-noise ratio of the template), and at ($353-217$) and ($353-143$) with 10\arcm\ resolution. In addition, this last template is also constructed at the resolution of the 70-GHz beam, in order to clean that channel.  The produced templates are then subtracted from the (non-inpainted) raw data at their native resolution. Table~\ref{table:sevem_coef_P} shows the list of templates used to clean each map, as well as the corresponding coefficients of the linear combination. These coefficients have been obtained by minimizing the variance of each cleaned map outside a mask excluding the brightest 3\,\% of the sky and all the point sources detected in polarization.  Note that to clean the 100- and 143-GHz maps, the same combination of templates as in the previous release is used, although now templates and cleaned maps are produced at $\nside=2048$.

Once the frequency maps are cleaned, inpainting at the position of the point sources detected at each of those channels is carried out. Moreover, once these cleaned inpainted maps are ready, the non-blind point source detection algorithm is run again on them and additional point sources detected (see lower part of Table~\ref{tab:ps_sevem}). These newly identified sources are also inpainted.  This second iteration of the algorithm was performed for intensity in the previous release but not for polarization; in this version it is done for both cases. The joint area inpainted outside the \sevem\ mask in the three cleaned channels used to produce the combined maps corresponds to around a $0.04\%$ of the sky, and is fully covered by the common confidence mask.  As for intensity, exactly the same inpainting procedure is applied to the simulations processed through the \sevem\ pipeline, to account for any possible effects introduced by this step.  The cleaned $Q$ and $U$ maps for the 70-, 100-, 143-, and 217-GHz channels are shown in Fig.~\ref{fig:sevem_freqmaps}. The maps have been smoothed with a Gaussian beam with 80\arcmin\ FWHM resolution to allow for better visualization.

\begin{table*}[t]
\begingroup
\newdimen\tblskip \tblskip=5pt
  \caption{Linear coefficients $\alpha_j$ for each of the templates used to clean individual frequency maps with \sevem\ for polarization.}
\label{table:sevem_coef_P}
\nointerlineskip
\vskip -3mm
\footnotesize
\setbox\tablebox=\vbox{
\newdimen\digitwidth
\setbox0=\hbox{\rm 0}
\digitwidth=\wd0
\catcode`*=\active
\def*{\kern\digitwidth}
\newdimen\signwidth
\setbox0=\hbox{+}
\signwidth=\wd0
\catcode`!=\active
\def!{\kern\signwidth}
\newdimen\expsignwidth
\setbox0=\hbox{$^{-}$}
\expsignwidth=\wd0
\catcode`@=\active
\def@{\kern\expsignwidth}
\halign{\hbox to 1.2in{#\leaderfil}\tabskip=1.2em&
     \hfil#\hfil\tabskip=0.7em&
     \hfil#\hfil&
     \hfil#\hfil&
     \hfil#\hfil&
     \hfil#\hfil&
     \hfil#\hfil&
     \hfil#\hfil&
     \hfil#\hfil\tabskip=0pt\cr
\noalign{\doubleline}
\omit&\multispan8\hfil Coefficients $\alpha_j$\hfil\cr
\noalign{\vskip -3pt}
\omit&\multispan8\hrulefill\cr
\noalign{\vskip 3pt}
\omit\hfil Template\hfil& 70\GHz\ $Q$& 70\GHz\ $U$& 100\GHz\ $Q$& 100\GHz\ $U$
& 143\GHz\ $Q$& 143\GHz\ $U$& 217\GHz\ $Q$& 217\GHz\ $U$\cr
\noalign{\vskip 3pt\hrule\vskip 5pt}
   *30$-$44*&$2.72\times 10^{-2}$&$3.53\times 10^{-2}$&$0.96\times
   10^{-2}$& $1.29 \times 10^{-2}$& $3.43 \times 10^{-3}$& $6.81
   \times 10^{-3}$& $1.21 \times 10^{-2}$& $1.79 \times 10^{-2}$\cr  
143$-$100& \dots& \ldots& \ldots& \ldots& \ldots& \ldots& 8.63 $\times 10^{-1}$& 7.19 $\times 10^{-1}$\cr
217$-$100& \ldots& \ldots& \ldots& \ldots& 1.52 $\times 10^{-1}$& 1.47
$\times 10^{-1}$& \ldots& \ldots\cr
217$-$143& \ldots& \ldots&  9.38 $\times 10^{-2}$& 8.27
$\times 10^{-2}$& \ldots& \ldots& \ldots& \ldots\cr
353$-$143& 1.13 $\times 10^{-2}$& 0.98 $\times 10^{-2}$& \ldots& \ldots& \ldots& \ldots& 1.17 $\times 10^{-1}$& 1.13 $\times 10^{-1}$\cr
353$-$217& \ldots& \ldots&  1.32 $\times 10^{-2}$& 1.30
$\times 10^{-2}$& 2.83 $\times 10^{-2}$& 2.72 $\times 10^{-2}$& \ldots& \ldots\cr
\noalign{\vskip 5pt\hrule\vskip 3pt}}}
\endPlancktablewide
\endgroup
\end{table*} 

\begin{figure}[t]
\begin{center}
\includegraphics[width=\columnwidth]{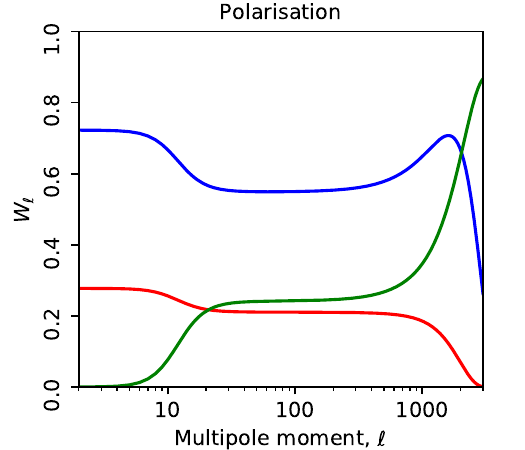}
\end{center}
\caption{Weights used to combine the cleaned single-frequency maps into the final \sevem\ CMB maps for polarization. The different lines correspond to 100- (red), 143- (blue), and 217-GHz (green) channels. The weights do not sum to unity because they include the effect of deconvolution by the beams of the frequency maps and convolving with the 5\arcmin\ Gaussian beam of the final map. }
\label{fig:weights_sevem_P}
\end{figure}

The last step is to combine the cleaned single-frequency maps in order to produce the final $Q$ and $U$ cleaned CMB maps. This is done by combining in harmonic space the cleaned 100, 143, and 217-GHz maps. The weights take into account the noise of each channel and its resolution. In addition, recognizing the fact that the 217-GHz channel is likely to be somewhat more susceptible to large-scale systematic residuals than the other two channels, we also introduce a relative down-weighting of the 217-GHz channel on the largest scales. This can be seen in Fig.~\ref{fig:weights_sevem_P}, where the harmonic weights
are given for the 100- (red), 143- (blue), and 217-GHz (green) channels. The same weights are applied for $E$ and $B$.  The resolution of the combined map corresponds to a Gaussian beam of FWHM 5\arcm\ and \healpix\ resolution $\nside = 2048$,
with a maximum multipole $\ell_{\mathrm{max}} = 3000$. \textcolor{black}{We consider a lower $\ell_{\mathrm{max}}$ for polarization than for intensity due to the lower signal-to-noise ratio of the polarization data.}

\subsection{Masks}
\label{sec:sevem_mask}

\begin{figure}[t]
\begin{center}
\includegraphics[width=\columnwidth]{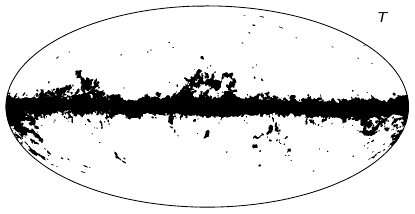}
\includegraphics[width=\columnwidth]{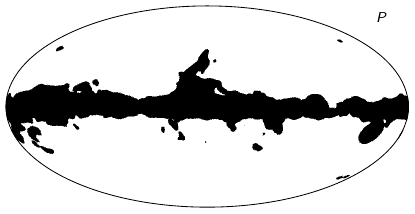}
\end{center}
\caption{\sevem\ masks in temperature (top) and polarization (bottom).}
\label{fig:dx12_masks_sevem}
\end{figure}

In temperature, the \sevem\ confidence mask is generated following a similar procedure to that of the previous release. Specifically, we define the mask by thresholding maps constructed as the difference between two different CMB reconstructions. As in 2015, we construct these differences at $\nside=256$, with resolution given by a Gaussian beam with $\hbox{FWHM} =30\arcm$. In particular, three combinations are considered: the cleaned ($217-143$)\,GHz and ($143-100$)\,GHz maps and the difference between two cleaned, combined CMB maps, whose linear coefficients have been obtained by minimizing the variance outside two different masks. From each of these maps, one mask is constructed by removing the brightest pixels (and its direct neighbours) down to a certain threshold. The three masks are multiplied to produce the final
confidence mask, which is then smoothed with a Gaussian beam of 1$^\circ$ to avoid sharp edges and upgraded to full resolution. The thresholds that define the masks are chosen by looking at the amplitude of the extrema and the dispersion of the cleaned 100-, 143-, and 217-GHz channels and the combined map after applying the considered mask, trying to find a compromise between reducing the values of these quantities while keeping a reasonable sky fraction. In particular, thresholds removing between 8 and 10\,\% of the sky were found to be adequate for the differences considered. In addition, a small region
near the Galactic plane with a relatively high contamination, but that was not captured with these values of the thresholds, was manually masked by applying a circle of 0\pdeg3 radius. This removed around 350 additional pixels, without modification of the thresholds, which would lead to a larger reduction of the area allowed by the mask.  The final \sevem\ confidence mask in intensity leaves a suitable sky fraction of 83.8\,\%, and is shown in the top panel of Fig.~\ref{fig:dx12_masks_sevem}.

\begin{figure}[t]
\begin{center}
\includegraphics[width=\columnwidth]{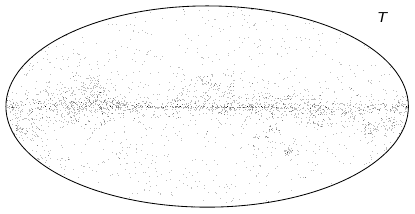}
\includegraphics[width=\columnwidth]{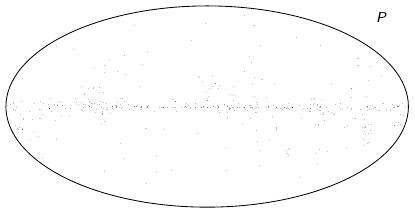}
\end{center}
\caption{\sevem\ masks in temperature (top) and polarization (bottom) for inpainted point sources.}
\label{fig:dx12_masks_inpainting_sevem}
\end{figure}

In polarization, given the lower signal-to-noise ratio of the reconstructed CMB maps, a different approach from that of intensity needs to be considered to identify the reliable regions of the sky. Several aspects of the approach to construct the polarization confidence mask have been modified with respect to the previous release, and the new method is described here in detail.  In particular, we have defined the confidence mask as the product of two individual masks: one specific mask based on the achieved CMB reconstruction, and a second one customized to avoid the regions more contaminated by thermal dust. 

For the specific mask, the first step is to downgrade the CMB reconstructed maps ($Q$ and $U$) to a resolution equivalent to a Gaussian beam with $\hbox{FWHM} = 90\arcm$ and $\nside=128$.  From these maps, we estimate locally the rms of $P$ (i.e., $\sqrt{Q^2+U^2}$) at each position by caculating the rms of the pixels included in a circle with a given radius centred on the considered pixel. We then estimate the expected rms of $P$ for a map containing only CMB and noise. For the noise,
this is obtained by estimating this quantity locally for the odd-even half-difference map, processed through the \sevem\
pipeline, at the resolution being considered, using the same procedure as for the cleaned maps. For the CMB, we simply obtained the rms of $P$, averaging over simulations. Since the CMB and noise are independent, their rms values are added quadratically. The ratio between the rms of the cleaned maps over that expected for a CMB-plus-noise map is then constructed. Pixels with larger ratios are expected to be more contaminated; the specific mask is defined by those pixels above a given threshold. This mask is then smoothed with a Gaussian beam of $\hbox{FWHM}=90\arcm$ to avoid sharp boundaries, and upgraded to $\nside=2048$. We explored several values for the radius of the circle (to locally estimate the rms) and for the amplitude of the threshold, finding that a value of 15 pixels (at $\nside=128$) for the radius and a threshold of 1.5 produced good results.

To construct the dust mask, we use the raw 353-GHz channel, smoothed at a resolution of 90\arcm\ and $\nside=128$. The rms of $P$ is obtained at each pixel as explained above, and a fixed fraction of pixels with the largest rms values is included in the mask. This mask is again smoothed with a Gaussian beam of 90\arcm\ and upgraded to $\nside=2048$. To construct this mask, we have chosen a radius for estimating the rms of four pixels and excluded 15\,\% of the sky. Finally, the specific mask and the dust mask are multiplied together, passing 80.3\,\% of the sky. The \sevem\ confidence mask in polarization is shown in the bottom panel of Fig.~\ref{fig:dx12_masks_sevem}. 

We conclude with some additional comments about the best way to deal with inpainted pixels. The most conservative approach is to explicitly exclude all of the inpainted areas from the analysis.  This implies the inclusion of a large number of holes in the confidence mask, which can be damaging for certain analyses, especially those performed in harmonic space. The diffusive inpainting strategy considered above seems to effectively reduce the emission from detected point sources while, at the same time, not introducing evident artefacts in the cleaned maps (recall that we are filling small holes, corresponding to scales where the background is usually smooth). Therefore, we have only masked those inpainted pixels which are directly excluded by the general algorithm used to construct the confidence mask. For intensity, this leaves a joint inpainted area outside the \sevem\ mask in the two cleaned channels (143 and 217\,GHz) used to construct the final CMB map of around 0.4\,\% of the sky.
For polarization, the corresponding joint area (from the cleaned 100-, 143-, and 217-GHz channels) also covers around  0.04\,\% of the sky.   Moreover, for both intensity and polarization, the exact same procedure is applied to the simulations processed through the \sevem\ pipeline, to ensure that any unexpected spurious effects are statistically taken into account. We believe that this is a good way to proceed in order to find a compromise between reducing point-source contamination in the cleaned maps and providing a well-behaved confidence mask for CMB analysis.  This is the same approach used in the previous release.  Nonetheless, for certain types of analysis, as for example the local study of compact objects, it may be necessary to discard, or at least to be aware of, the inpainted regions. For these cases, we also provide masks of the pixels inpainted in each of the cleaned frequencies, as well as the joint mask for those channels that are used to construct the final
CMB maps. The joint masks of inpainted pixels are given in Fig.~\ref{fig:dx12_masks_inpainting_sevem} for intensity (top) and
polarization (bottom). Note that additional inpainting is also performed during the template construction, but those positions are not included in these masks since those pixels are not directly inpainted on the cleaned maps. Finally, we point out that the masks for inpainted point sources given in Fig.~\ref{fig:dx12_masks_inpainting_sevem} have been included in the
confidence common masks (Fig.~\ref{fig:commonmask}), to reduce possible point source contamination in all the CMB maps. If it is desired to carry out an analysis of the \sevem\ CMB maps without explicitly including point source holes in the mask, the \sevem\ confidence masks should be considered.

\section{Spectral Matching Independent Component Analysis (\smica)}
\label{app:smica}

\def\bC{\tens{C}}
\def\bG{\tens{G}}
\def\ba{\vec{a}}
\def\bb{\vec{b}}
\def\ncha{N_\mathrm{cha}}
\def\nfg{N_\mathrm{fg}}

The general operation of \smica\ (Spectral Matching Independent Component Analysis;
\citealp{delabrouille2003,cardoso2008}) and the main changes with respect to the 2015 release are summarized in Sect.~\ref{sec:smica}.  In this appendix, we provide additional implementation details.  There are several masking and pre-processing operations whose specifics vary depending on the target map (CMB or foregrounds, temperature or polarization), but the general methodology is the same, following these steps:

\begin{enumerate}

\item Preprocessing of the input maps by point source subtraction and masking/inpainting.  This step also includes additional masking (Galactic plane, etc.).

\item Estimation of the spectral statistics $\widehat\bC_\ell$ via Eq.~(\ref{eq:smica:scm}) from the spherical harmonic coefficients computed from the preprocessed maps, possibly with some additional masking to remove particularly bright
objects.

\item Fitting of a \smica\ model to beam-corrected $\widehat\bC_\ell$, from which the \smica\ harmonic weights $\vec{w}_\ell$ are computed.

\item Computation of the spherical harmonic coefficients from the preprocessed maps and linear combination as per Eq.~(\ref{eq:smica:ilc}), to synthesize a map with a specified effective Gaussian beam.

\item Determination of a ``confidence mask''.
\end{enumerate}
The specifics of the production of each \smica\ map are given below.

\subsection{Temperature analysis}
\label{app:smica:temp}

The \smica\ 2018 temperature map is a hybrid of two complementary CMB renderings, namely $X_\textrm{high}$, which  includes only HFI observations, and is specialized for high Galactic latitudes, and intermediate and small angular scales, and $X_\textrm{full}$, which includes all \Planck\ channels, and provides us with additional content.  The final \smica\ temperature map is then constructed as a weighted sum of these two maps, following Eq.~\ref{eq:smica:merge}.  The two sky areas to be hybridized are defined by a smooth mask shown at Fig.~(\ref{fig:smica:tmask}).  In polarization, we do not resort to such a hybrid scheme.

\paragraph {Recalibration}

As in previous releases, a preliminary \smica\ fit (calibration run) is conducted, with calibration coefficients left unconstrained at 100 and 217\,GHz.  This fit involves only HFI channels, is limited to the first peak ($30\leq\ell\leq 300$), and involves spectral matrices estimated over a clean part of the sky.  It yields relative calibration coefficients 1.0004 at 100\,GHz and 1.0005 at 217\,GHz.   These values are consistent with the results reported in \citet{planck2016-l03} and \citet{planck2016-l05}.

\begin{figure*}
  \begin{center}
    \begin{tabular}{cc}
      \includegraphics[width=0.5\linewidth]{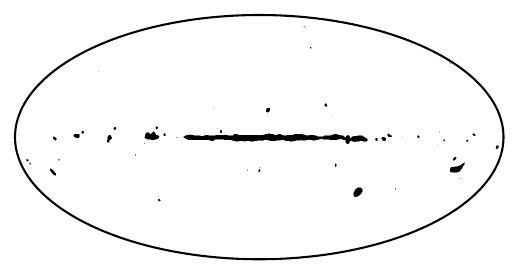}&
      \includegraphics[width=0.5\linewidth]{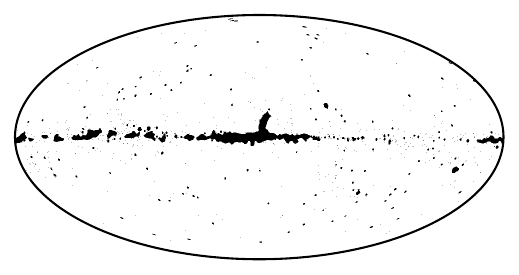}\\
    \end{tabular}
  \end{center}
  \caption{\smica\ pre-processing masks.  {\it Left:} for intensity analysis, covering $f_\textrm{sky}=98.5\,\%$.   {\it Right:} for polarization analysis, covering $f_\textrm{sky}=97.3\,\%$.}
  \label{fig:smica_processing_masks}
\end{figure*}

\begin{figure}[t]
  \begin{center}
    \includegraphics[width=\columnwidth]{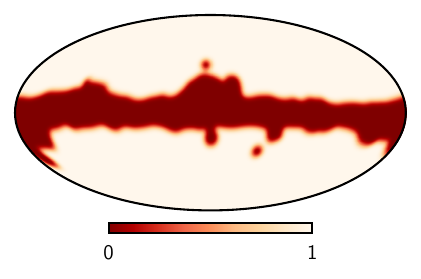}
  \end{center}
  \caption{The \smica\ transition mask used to combine the $X_\textrm{high}$ and the $X_\textrm{full}$ CMB renderings.}
    \label{fig:smica:tmask}
\end{figure}

\paragraph {Preprocessing}
The input maps are preprocessed for point sources as follows.  In the maps from 30\,GHz to 353\,GHz, we try to fit and subtract the strongest point sources detected at the 5\,$\sigma$ level in the PCCS2 catalogue (\citealt{planck2014-a35}).  Any point source with an unsatisfactory fit is left ``as is'' in the map.  In a second step, in each of the input maps from 44\,GHz to 353\,GHz, we mask all the point sources detected at more than 50\,$\sigma$ (unless they have already been subtracted in the previous step).  The masked areas at all frequencies are then combined to form a common point-source mask.  In addition to that point-source mask, we include a small mask, hereafter ``the Galactic mask'', blocking the Galactic plane, plus a small number of selected regions (such as the LMC).  The resulting ``preprocessing mask'' is shown in Figure~\ref{fig:smica_processing_masks}.  In order to minimize leakage in the subsequent computation of spherical harmonic coefficients, the masked areas under this common mask are filled in by a simple diffusive inpainting procedure.

\paragraph{Spectral statistics}

The computation of the spherical harmonic coefficients entering in the spectral statistics $\widehat\bC_\ell$ differs between $X_\textrm{high}$ and $X_\textrm{full}$.  For $X_\textrm{high}$, we apply an apodized version of the transition mask, while for
$X_\textrm{full}$, we use the full sky.  In both cases, we use the preprocessed maps with additional masking of bright objects or regions.  For $X_\textrm{full}$, which invloves all \Planck\ frequency channels, the point source mask is augmented with all the sources detected at more than 50\,$\sigma$ at frequencies 30\,GHz, 545\,GHz, and 857\,GHz, and the new holes are again filled in by diffusive inpainting.  We also mask part of Galactic region using an apodized version of the Galactic mask.  For $X_\textrm{high}$, which invloves only HFI channels, we mask all the point sources detected at 5\,$\sigma$ at frequencies 100\,GHz, 143\,GHz, and 217\,GHz, even if already subtracted.  The resulting holes are then apodized over 30\arcm.

\paragraph{Spectral fits}

For producing the $X_\textrm{high}$ map, \smica\ processing is conducted, fitting the spectral covariance matrices $\widehat\bC_\ell$ over the multipole range $25\leq\ell\leq 1000$.  For this fit, the calibration is kept fixed at the values found in the calibration run.  The free parameters are the (binned) CMB spectrum ${\bC}_\ell^{\rm cmb}$, the positive matrices $\tens P_\ell$, and the $6\times \nfg$ foreground emissivity matrix $\tens F$.

For producing the $X_\textrm{full}$ map, a first run is devoted to estimating the foreground emissivity matrix $\tens F$, and a recalibration factor for the 70-GHz channel (this factor is found to be 1.0019).  This fit is conducted over the multipole range $2\leq\ell\leq 150$.  In a second run, we fit (binned versions of) $\bC_\ell^{\rm cmb}$ and $\tens P _\ell$ over the multipole range $10\leq\ell\leq 1000$, keeping fixed the calibration (vector $\vec a$) and the foreground emissivity matrix $\tens F$.

\paragraph{Map synthesis}

The \smica\ fits produce parametric estimates of $\bC_\ell$, from which spectral weights $\vec{w}_\ell$ are readily obtained.  They are shown in Figure~\ref{fig:smica_T_weights} for $X_\textrm{high}$ (top panel) and $X_\textrm{full}$ (middle panel).   Those weights are applied to spherical harmonic coefficients computed from the preprocessed input maps.  The spatial transition weights used to hybridize $X_\textrm{high}$ and $X_\textrm{full}$ are shown in Fig.~\ref{fig:smica:tmask}.

\paragraph{Confidence mask}

The confidence mask combines a point source mask and a Galactic mask determined by a procedure similar to the one used for the 2015 release.  It is documented in the Explanatory Supplement.

\paragraph{Inpainting} Final inpainting of the CMB maps is no longer performed in the \smica\ pipeline, but is carried out through a procedure common to all methods, as described in Sect.~\ref{sec:masks}.

\paragraph{SZ-free CMB map}
A CMB map free of SZ contamination is produced by a simple adaptation of Eq.~(\ref{eq:smica:ilc}) as follows.  That expression yields weights $\vec{w}_\ell$, which, at each multipole $\ell$, mimimize the output power while enforcing unit gain towards the CMB signal.  In other words, it is the minimizer of $\vec{w}_\ell\adj \bC_\ell \vec{w}_\ell$ subject to $\vec{w}_\ell\adj \ba=1$.  One can solve the same problem with the additional constraint that the weights should also cancel the SZ signal, that is, enforcing the additional  constraint  $\vec{w}_\ell\adj \bb=0$  where  $\bb$ denotes the SZ emission law.  The minimizer of $\vec{w}_\ell\adj \bC_\ell \vec{w}_\ell$ subject to $\vec{w}_\ell\adj \ba=1$ and $\vec{w}_\ell\adj \bb=0$ is easily found in closed form (see \citealt{Remazeilles2011a}) as
\begin{equation}\label{eq:smica:ilcnosz}
  \vec{w}_\ell = \bC_\ell\inv  \bG (\bG\adj \bC_\ell\inv \bG)\inv \vec{c}
\end{equation}
where $\bG = [ \ba\ \bb ]$ and $\vec{c} = [ 1\ 0]\adj$.

\begin{figure}[t]
  \begin{center}
     \includegraphics[width=\columnwidth]{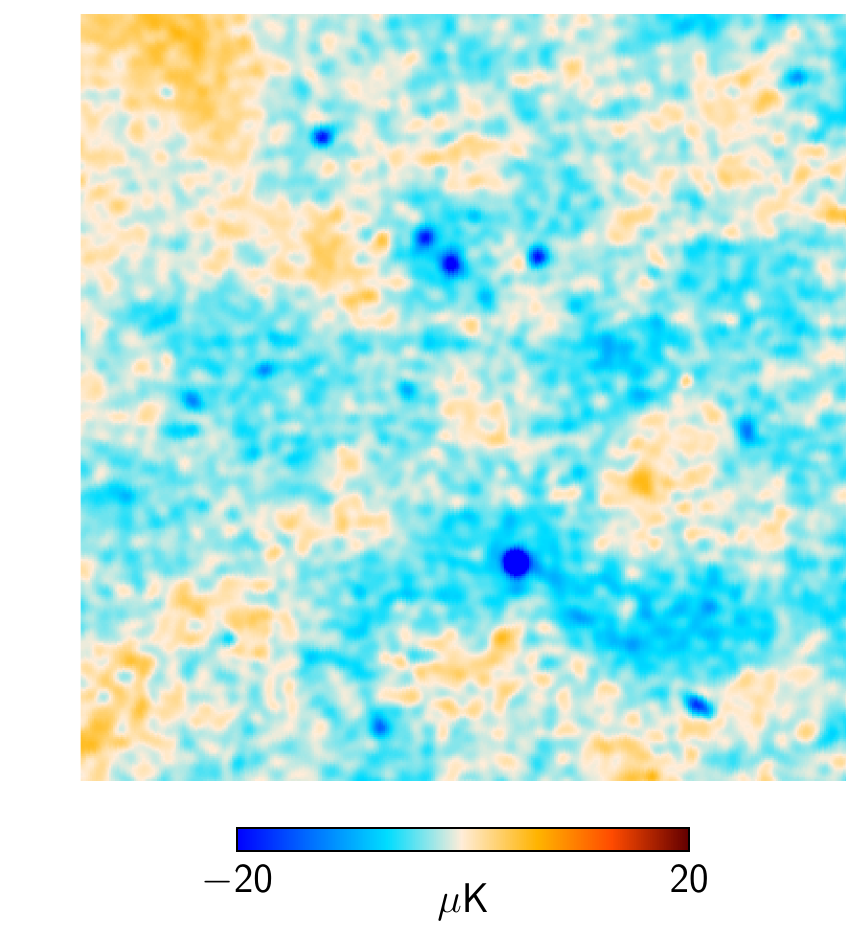}
  \end{center}
  \caption{Difference between the \smica\ CMB map and its SZ-free version. The patch shown is $20 \deg\times 20\deg$ centered on $(l, b) = (46\pdeg3, 53\deg)$.}
  \label{fig:diffszmap}
\end{figure}

\begin{figure}[htbp]
  \begin{center}
    \includegraphics[width=\columnwidth]{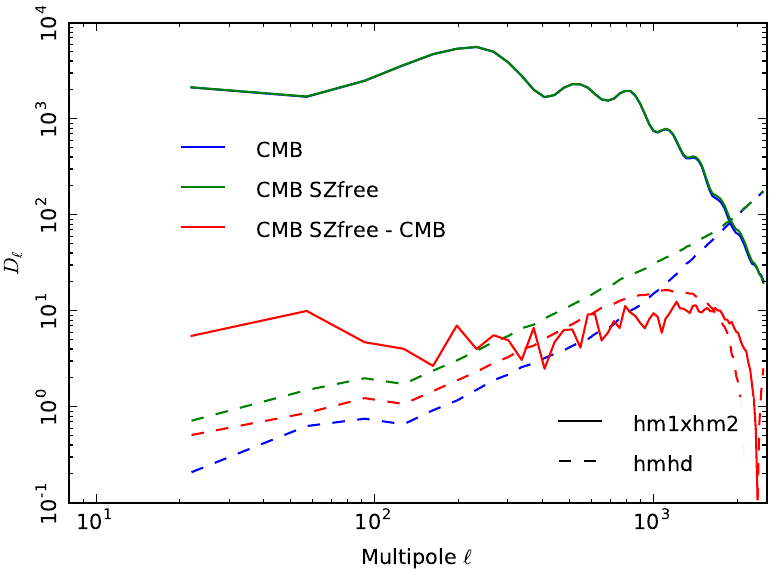}
  \end{center}
  \caption{Angular spectra for the CMB (blue lines), the CMB SZ-free version (green lines), and their difference spectra (red lines), computed on the \smica\ confidence mask. Half-mission cross-spectra (solid line) and half-mission difference spectra (dashed line) are shown to assess the signal and noise differences between the two CMB maps. }
  \label{fig:diffszspec}
\end{figure}

Figure~\ref{fig:diffszmap} shows an enlargement of the difference between the \smica\ CMB maps derived with and without SZ projection. Figure~\ref{fig:diffszspec} compares the angular power spectra of these two maps.  \textcolor{black}{Both versions of the \smica\ CMB maps are considered in the lensing study \citep{planck2016-l08}.}

\begin{figure}[t]
  \begin{center}
    \includegraphics[width=\columnwidth]{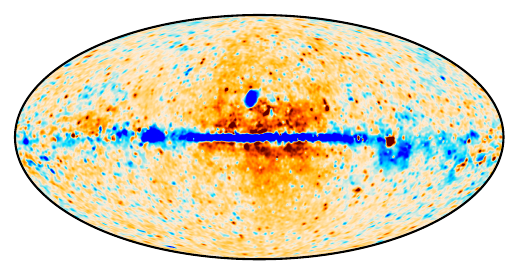}
    \includegraphics[width=\columnwidth]{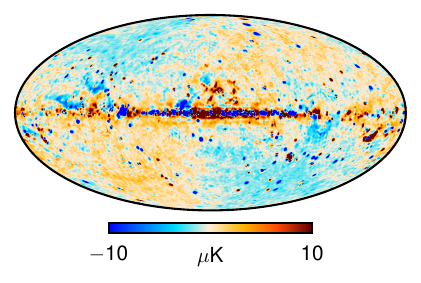}
  \end{center}
  \caption{CMB difference maps in temperature at 80\arcmin\ resolution. {\it Top:} Difference between the \smica\ 2015 released map and the 2015b map (without recalibration of the 44-GHz channel). {\it Bottom:} Difference between the 2015b and the 2018 map. }
  \label{fig:smica_diff2015}
\end{figure}

\paragraph{Changes with respect to the 2015 release}
Figure~\ref{fig:cmb_diff_release_maps} shows, for all pipelines, the differences in CMB temperature maps from 2015 to 2018.  In the \smica\ case, the difference could have three origins: changes in the input data, changes in the \smica\ pipeline, and changes in recalibration procedure.  We show here that the difference is mostly due to recalibration by producing a CMB map, referred to as the ``2015b map'', obtained from the 2015 data by running the 2015 pipeline with the sole exception that, as for the 2018 release, the 30\,GHz and 44\,GHz channels are \emph{not} recalibrated.  Figure~\ref{fig:smica_diff2015} shows the differences from the 2015 map to this 2015b map (top panel) and from this 2015b map to the 2018 map, while the bottom panel of Fig.~\ref{fig:cmb_diff_release_maps} shows the difference from the 2015 map to the 2018 map.  These three pairwise comparisons make it clear that, in temperature, most of the differences between 2015 and 2018 should be attributed to changes in calibration, rather than to changes in the \smica\ pipeline.

\subsection{Polarization analysis}

\paragraph{CMB reconstruction.} We now turn to the construction of \smica\ polarization maps, and start with the CMB map. First, a significant modification to the \smica\ 2018 pipeline is the fact that $E$ and $B$ modes are now processed independently; in contrast, the 2015 analysis fitted and filtered these modes jointly.  \smica\ uses all seven \Planck\ polarized
channels.  When producing either $E$-mode or $B$-mode CMB maps, the foreground emission is taken to have maximal dimension: $\nfg=7-1=6$.

The input maps are  preprocessed as follows.  First, in each of the input frequency maps, all point sources detected at the 5\,$\sigma$ level are masked and the holes are filled by diffusive inpainting.  Second, the bright pixels (with amplitude ten times larger than the standard deviation of the map) are similarly masked and inpainted.  Finally, a small Galactic mask -- obtained by thresholding a combination of the 30-GHz and 353-GHz maps -- is applied.  The resulting mask, shown in Figure~\ref{fig:smica_processing_masks}, covers 97\,\%\ of the sky.  In the 2015 release, the same processing mask was used for polarization and intensity.

As for temperature, we proceed in two steps.  A first \smica\ fit is performed to estimate the foreground emissivity matrix $\tens F$ over the range $5\leq\ell\leq 150$.   A second \smica\ fit is then performed in the range $2\leq\ell\leq 1000$ over parameters $C_\ell^{\rm cmb}$ and $\tens P_\ell$, while $\tens F$ is kept fixed at the value found in the first run.  The right panel of Fig.~\ref{fig:smica_T_weights} shows the resulting harmonic weights.  These result from spectral statistics $\hat\bC_\ell$ computed from the preprocessed maps without additional masking, unlike in the temperature case.

\paragraph{Confidence mask.}

A \smica\ polarization confidence mask has been produced and released, but appears not to be conservative
enough.  For that reason, we recommend using the common confidence mask to analyse \smica\ polarized CMB maps.

\subsection{Polarized foreground reconstruction.} 
The results presented in Sect.~\ref{sec:pol_ind}, regarding the polarized dust and synchrotron emission, are based on a dedicated, blind \smica\ fit with a foreground emissivity matrix $\tens F$ composed only of $\nfg=2$ columns.  The total foreground contribution to a spectral covariance matrix $\bC_\ell$ being $\tens F \tens P_\ell \tens F \adj$, a blind fit can only determine the factors $\tens F$ and $\tens P_\ell$ up to multiplication by an invertible $2\times2$ matrix~$\tens T$.  Indeed, for any such matrix $\tens T$, one can define $ \tilde{\tens P}_\ell=\tens T \tens P_\ell \tens T \adj$ and $\tilde{\tens F} =\tens F \tens T\inv$ and see that the transfomed pair $(\tilde{\tens F}, \tilde{\tens P}_\ell)$ contributes as much as the original pair $({\tens F}, {\tens P}_\ell)$ to the spectral covariance matrix, since, by construction $ \tens F \tens P_\ell \tens F \adj = \tilde {\tens F} \tilde{\tens P}_\ell \tilde{\tens   F} \adj $.  Therefore the likelihood is insensitive to the value of $\tens T$.  Since a blind fit is (by definition) conducted without constraining either $\tens F$ nor $\tens P_\ell$, the matrix $\tens T$ cannot be determined from the data without imposing extra constraints.  This degeneracy could be fixed by constraining ${\tens P}_\ell$ to be diagonal, but this would be equivalent to fitting a (wrong) model of uncorrelated synchrotron and dust emissions.  We choose instead to fix the degeneracy as follows.  We conduct a blind \smica\ fit and, in a post processing step, we select (without affecting the quality of the \smica\ fit) a matrix $T$ given by
\begin{displaymath}
  \tens T
  = \left[
    \begin{array}{cc}
      \tens F_{30, 1} & \tens F_{30, 2} \\ \tens F_{353, 1} & \tens F_{353, 2}
    \end{array}
  \right]\,,
\end{displaymath}
so that the first row of $\tilde{\tens F} = \tens F\tens T\inv$ becomes $[ 0 , 1 ]$ and its last row becomes $[1, 0]$.  In other words, we fix the indeterminacy in the blind fit of a two-template foreground model by assuming that the entire foreground signal at 30\,GHz is only synchrotron, and that the entire foreground signal at 353\,GHz is only thermal dust.  We checked that performing a second fit where the synchrotron contribution at 353\,GHz is not zero but an extrapolated value (and similarly for dust at 30\,GHz), has no significant effect on fitted values and, unsurprisingly, that it does not affect either of the reconstructed maps.

Maps of polarized dust and synchrotron emission are synthesized from harmonic coefficients computed over the full sky, except for point sources detected at 5\,$\sigma$, which are masked and inpainted.   This is carried out independently for each input map.  The $Q$ and $U$ maps are synthesized with an effective Gaussian beam of 3\deg\ (FWHM) for synchroton and 12\arcm\ for dust.  The SEDs of dust and synchrotron emission shown in Fig.~\ref{fig:polsedsmica} are determined from a dedicated \smica\ fit based on spectral covariance matrices computed from about 70\,\% of the sky.

\section{\gnilc}
\label{app:gnilc}

The formalism of \gnilc\ has been described in detail in \cite{Remazeilles2011b} and \cite{planck2016-XLVIII}. The main
characteristics can be summarized as follows: (i) a \gnilc\ map at given frequency is a weighted linear combination (ILC) of the
\Planck\ frequency maps having minimimum variance; (ii) \gnilc\ performs localized analysis in both harmonic space and pixel
space via needlet (spherical wavelet) decomposition \citep{Narcowich2006}, and as such it adapts component separation to local conditions of contamination both over the sky and over angular scale; and (iii) \gnilc\ uses not only spectral information,   but also spatial information (angular power spectra) of the non-Galactic components (CIB, CMB, and noise) in order to disentangle the Galactic signal from the CIB, CMB, and noise contamination. Therefore, \gnilc\ is a blind, model-independent, data-driven component-separation method, in the sense that there is no prior assumption/parametrization of the Galactic foreground properties.

There are, however, a few differences in the  \gnilc\ processing steps between intensity and polarization. For intensity, the processing is identical to that of \cite{planck2016-XLVIII}, i.e., the prior information is both spectral and spatial, and consists of the \Planck\ best-fit CMB temperature power spectrum, $C_\ell^{\rm \Lambda CDM}$ \citep{planck2014-a13}, the \Planck\ CIB best-fit auto/cross power spectra across frequency pairs,  $C_\ell^{\rm CIB}(\nu_1,\nu_2)$ \citep{planck2013-pip56}, and the \Planck\ noise power spectra across frequencies, $C_\ell^{\rm noise}(\nu)$. For polarization, prior information is only spectral for the CMB, consisting of the CMB SED,\footnote{Given that the amplitude of the CMB  $B$-mode power spectrum is unknown, we cannot use spatial information on the CMB as a prior when performing \gnilc\ on polarization data.} while the noise prior is still spectral and spatial, comprising the \Planck\ noise power spectra at each frequency. In practice, \Planck\ noise power spectra are derived from the half-difference of the first and second halves of each stable pointing period (``rings'') of \Planck, in which the sky emission cancels out and leaves an estimate of the full-survey noise.

From those prior power spectra, we simulate Gaussian realizations of the CMB map, $y^{\rm CMB}(p)$, the correlated CIB maps, $y_\nu^{\rm CIB}(p)$, and the noise maps, $y_\nu^{\rm noise}(p)$, where $\nu$ denotes the frequencies and $p$ the pixels. The simulated total "nuisance" map is defined as
\begin{equation}
y_\nu(p)
 \equiv g_\nu\, y^{\rm CMB}(p) + y_\nu^{\rm CIB}(p) + y_\nu^{\rm noise}(p)
\end{equation}
 for intensity, where $g_\nu$ is the derivative of a blackbody with respect to temperature, and 
 \begin{equation}
 y_\nu(p) 
 \equiv y_\nu^{\rm noise}(p)
 \end{equation}
 for polarization, since the CIB is assumed to be unpolarized and we have no spatial information on the CMB polarization. 
  
We perform a needlet decomposition of both the simulated nuisance maps and the \Planck\ frequency maps. We thus define ten needlet windows, $\{h^{(j)}_\ell\}_{1\leq j \leq 10}$, as Gaussian bandpass filters in harmonic space to perform component separation on different ranges of multipoles independently.\footnote{The needlet windows satisfy the relation $\sum_{j=1}^{10} (h^{(j)}_\ell)^2 = 1$ to ensure the conservation of the total power when synthesizing all the needlet maps to reconstruct the complete map.} The spherical harmonic coefficients of the simulated maps, $y_\nu(p)$, are bandpass-filtered as $h^{(j)}_\ell\,a_{\ell m }(\nu)$. The inverse spherical harmonic transform of the filtered coefficients produces ten needlet maps,  $y_\nu^{(j)}(p)$ (one for each needlet scale), for each frequency. Each needlet map, $y_\nu^{(j)}(p)$, contains temperature fluctuations at the specific range of angular scales probed by the associated needlet window, with statistical properties determined by the prior power spectra at these scales. 
  
For each needlet scale $(j)$, we compute the covariance matrix of the nuisance map (noise for polarization; CMB plus CIB plus noise for intensity) in each pixel $p$ for all pairs of frequencies $a$ and $b$ as:
\begin{equation}
  \left[ {\rm R_n}^{(j)} (p) \right]_{a\,b}
  = \sum_{p' \in \mathcal{D}^{(j)}(p)}\, y_a^{(j)}(p')\, y_b^{(j)}(p')\,,
\end{equation}
where in practice the pixel domain, $\mathcal{D}^{(j)}(p)$, is defined by the convolution in real space of the product of needlet maps, $y_a^{(j)}(p)\, y_b^{(j)}(p)$, with a Gaussian kernel whose the width is a function of the needlet scale considered. Note that the prior covariance matrix of the nuisance map, ${\rm R_n}(p)$, is blind about the particular realization of CMB, CIB, and noise that is found in the observed \Planck\ data.

Similarly, the data (\Planck\ frequency maps), $d_\nu(p)$, are decomposed onto the same needlet frame, and the frequency-frequency covariance matrix of the data is computed in each pixel for each needlet scale as:
\begin{equation}
  \left[ {\rm \widehat{R}_d}^{(j)} (p) \right]_{a\,b}
  = \sum_{p' \in \mathcal{D}^{(j)}(p)}\, d_a^{(j)}(p')\, d_b^{(j)}(p')\,.
\end{equation}

As described in \cite{planck2016-XLVIII}, the prior power spectra are thus used  to obtain a model of the frequency-frequency covariance matrix, ${\rm R_n}$, of the nuisance contribution (CIB, CMB, and noise) to the total data covariance matrix, $\widehat{\rm R}_{\rm d}$ ($9\times 9$ matrices for intensity, $7\times 7$ for polarization). The signal-to-nuisance ratio, where signal stands for Galactic emission, is obtained via the matrix ${\rm R_n}^{-1/2}\widehat{\rm R}_{\rm d}{\rm R_n}^{-1/2}$, which is estimated locally over the sky and over different ranges of angular scales via needlet decomposition of the maps. The eigenstructure of the matrix ${\rm R_n}^{-1/2}\widehat{\rm R}_{\rm d}{\rm R_n}^{-1/2}$ allows us to discriminate those eigenvalues that are close to unity (therefore corresponding to nuisance) from those that correspond to the contribution of Galactic emission.\footnote{In practice, the distinction between the two sets of eigenvalues is performed via the Akaike Information Criterion, which prevents the method from overfitting the foreground subspace.} This allows us to estimate the local dimension, $m$, of the Galactic signal subspace over the sky and over scales, i.e., the finite number of independent (not physical) components\footnote{Those independent components are related to the subset of eigenvectors, or principal components, of the matrix ${\rm R_n}^{-1/2}\widehat{\rm R}_{\rm d}{\rm R_n}^{-1/2}$ for which the associated eigenvalues depart from unity.} onto which the correlated Galactic emission can be decomposed.   We note ${\rm U_S}$ the matrix collecting the selected subset of $m$ eigenvectors of the matrix ${\rm R_n}^{-1/2}\widehat{\rm R}_{\rm d}{\rm R_n}^{-1/2}$ that form an orthogonal basis of the Galactic signal subspace.

The data\footnote{Needlet coefficients of \Planck\ frequency maps.} are then projected onto the identified Galactic signal subspace, and an $m$-dimensional ILC is performed on the projected data in order to further minimize any part of the nuisance that did not project orthogonally to the Galactic subspace:
\begin{equation}\label{eq:gnilc_dust}
  \widehat{s}^{\,\,\rm dust\, (j)}_\nu (p) 
  = \sum_{\nu'} {\rm W}_{\nu\nu'}^{(j)}(p) \, d_{\nu'}^{(j)} (p)\,.
\end{equation}
The matrix of \gnilc\ weights can be written in compact form as \citep{Remazeilles2011b}:
\begin{equation}\label{eq:gnilc_weights}
 {\rm W}
  = {\rm F}\,\left( {\rm F}^t \, {\rm \widehat{R}_d}^{-1} \, {\rm F}\right)^{-1}\, {\rm F}^t \, {\rm \widehat{R}_d}^{-1}  \,,
\end{equation}
with the estimated foreground mixing matrix, ${\rm F}$, given by
\begin{equation}
 {\rm F} =   {\rm R_n}^{1/2}\, {\rm U_S} \,.
\end{equation}
For polarization, where there is no prior on the CMB power spectra, the ILC is replaced by a constrained ILC
\citep{Remazeilles2011a}, for which the vector of weights in frequency is constrained to be orthogonal to the CMB SED. In practice, this is done through a Gram-Schmidt orthogonalization of the set of eigenvectors collected in matrix ${\rm U_S}$ with respect to the CMB SED vector $g_{\nu}$. This constraint ensures that the GNILC weights (Eq.~\ref{eq:gnilc_weights}) project out any CMB polarization signal in the reconstructed dust polarization map. 

The \gnilc\ filters (Eq.~\ref{eq:gnilc_weights}) are invariant if $\rm F$ is replaced by $\rm F\,T$ for any invertible matrix $\rm T$. Therefore, the true foreground mixing matrix does not need to be known by \gnilc; the only useful information is a set of independent components onto which the correlated Galactic emission can be decomposed.

The estimated needlet maps of dust emission (Eq.~\ref{eq:gnilc_dust}) are then synthesised to reconstruct the complete \gnilc\ dust maps, as follows.  The spherical harmonic coefficients, $\widehat{a}_{\ell m}^{(j)}(\nu)$, of the needlet dust maps, $\widehat{s}^{\,\,\rm dust\, (j)}_\nu (p)$, are again bandpass-filtered by the needlet windows as $h_\ell^{(j)} \widehat{a}_{\ell m}^{(j)}(\nu)$. The filtered coefficients are then inverse-spherical-harmonic transformed into maps, and coadded across needlet scales to form the complete \gnilc\ dust map, accounting for all the angular scales.

\gnilc\ has many advantages over template subtraction, parametric methods, or smoothing procedures.  First, it is a one-shot component-separation method that does not rely on subtraction of any template, such as a CMB template map, coming from another component-separation process. This prevents the propagation of CMB foreground residuals (e.g., dust and CIB residuals in the CMB map) to the reconstructed Galactic map.

The second advantage is related to noise filtering in \Planck\ polarization maps, where \gnilc\ performs better than a simple smoothing. Given that \gnilc\ is a minimum-variance linear combination of frequency maps, the overall noise level in the \gnilc\ maps will always be lower than the noise level in smoothed \Planck\ maps at the same frequency and equal resolution:
\begin{eqnarray}
{1\over \sigma^2_{\rm \gnilc}(353\,\hbox{GHz})} = {1\over \sigma^2_{\Planck}(30\,\hbox{GHz})} + ... + {1\over \sigma^2_{\Planck}(353\,\hbox{GHz})},
\end{eqnarray}
where $\sigma_{\rm \gnilc}(353\,\hbox{GHz})$ is the noise rms in the \gnilc\ 353-GHz map, and $\sigma_{\Planck}(353\,\hbox{GHz})$ is the noise rms in the \Planck\ 353-GHz map.  Moreover, a simple smoothing of the \Planck\ 353-GHz $Q$ and $U$ maps will mitigate CMB $E$ and $B$ modes but not cancel them on large scales, and there is no reliable CMB $B$-mode template to be subtracted. Conversely, \gnilc\ is an orthogonal projection to the flat CMB SED, and therefore cancels out any CMB $E$- and $B$-mode polarization at all angular scales.

Third, \gnilc\ filtering is performed locally over the sky and over scales via wavelet decomposition. This enables optimization
of the component-separation process given local variations of contamination over the sky and over scales.

Finally, the \gnilc\ method is blind, since it does not rely on any assumption about Galactic foregrounds. Most important, \gnilc\
allows for outputting Galactic foreground maps at all frequencies, e.g., at 100--143\,GHz, without relying on the extrapolation of
high-frequency templates with arbitrary emission laws. This is particularly useful in the context of decorrelation effects and
searches for primordial $B$ modes \citep{Tassis2015, planck2016-L}, where we can no longer rely on simple emission laws to extrapolate dust foregrounds to CMB frequencies.

\section{Intensity foregrounds}
\label{app:foregrounds}

In this appendix, we review the temperature foreground products derived by \commander\ and \gnilc\ from the \Planck\ 2018 frequency maps. As discussed in Sect.~\ref{sec:inputs} and elsewhere, these results are not intended for scientific analysis, but are included here for reference and completeness purposes.

\subsection{\commander\ analysis} 

We start our discussion with a review of the \commander\ intensity analysis. For a summary of the methodology and model definitions used in this work, see Sect.~\ref{sec:commander} and Appendix~\ref{app:commander}. In short we fit a parametric
five-component model to the \Planck\ 2018 data by maximizing the standard Bayesian posterior. The 2018 model includes the following components: (1)~CMB; (2)~a single power-law foreground model with a free spectral index per pixel to describe the sum of low-frequency foregrounds (synchrotron, free-free, and anomalous microwave emission); (3)~a modified blackbody with a free emissivity and temperature to describe thermal dust; (4)~a line-emission component at 100, 217, and 353\,GHz, with fixed line ratios between channels to describe CO emission; and (5)~a catalogue of 12\,192 known point source positions,
each source being fitted with a free flux density and spectral index.

We first consider the parameters of the derived astrophysical model in intensity, starting with the point source component, which represents one of the most novel aspects of the \commander\ 2018 model compared to previous versions.

Starting with the amplitude maps, the most notable difference with earlier results is caused by the explicit inclusion of a radio point source component in the latest model. Each object in this component is associated with an overall flux density and spectral index across all frequencies, while the spatial projection into each frequency component is performed through a full {\tt FEBeCoP} calculation, accounting for the asymmetric beam profile in the respective frequency channel. Only frequencies up to and including 143\,GHz are included when fitting the flux densities and spectral indices, to avoid biases from modelling errors at high frequencies. However, the resulting model is also extrapolated to higher frequencies when fitting other components. Infrared and sub-mm sources are not explicitly modelled in this approach, since they are well
described for the \Planck\ frequencies within the diffuse thermal dust component, which has 5\arcm\ FWHM resolution.

As described in Appendix~\ref{app:commander}, the total catalogue used in this work represents a combination of four separate source catalogs, three of which (AT20G, GB6, and NVSS; \citealp{murphy2010,gregory1996,condon1998}) are selected to cover disjoint regions of the sky, and the fourth (PCSS2; \citealp{planck2014-a35}) includes microwave sources that are not detected by any of the former three. In Table~\ref{tab:ptsrc_summary}, we provide summary statistics for the fits produced in the current analysis, broken down according to reference catalogue. From left to right, columns show: (1)~catalogue name; (2)~catalogue reference frequency; (3)~total number of sources used in our combined catalogue; (4)~number of sources statistically detected by \commander\ in the \Planck\ 2018 data; (5)~average flux density recalibration factor relative to the reference catalogue (no colour corrections are applied); and (6)~Pearson's $r$ correlation coefficient evaluated between the reference catalogue and \commander-estimated flux densities.

\begin{figure*}
\begin{center}
  \includegraphics[width=0.98\columnwidth]{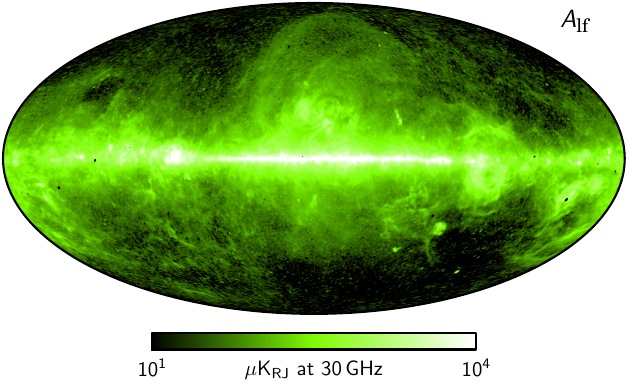}
  \includegraphics[width=0.98\columnwidth]{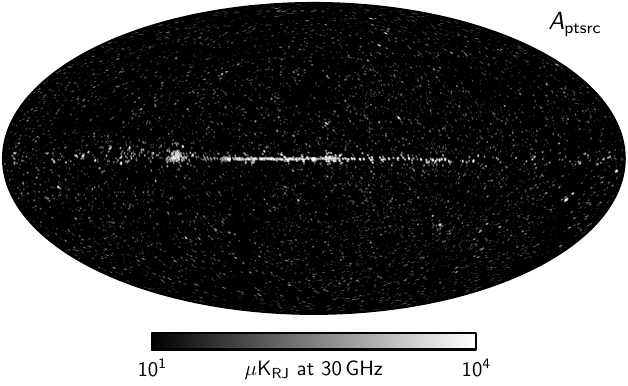}\\
  \includegraphics[width=0.98\columnwidth]{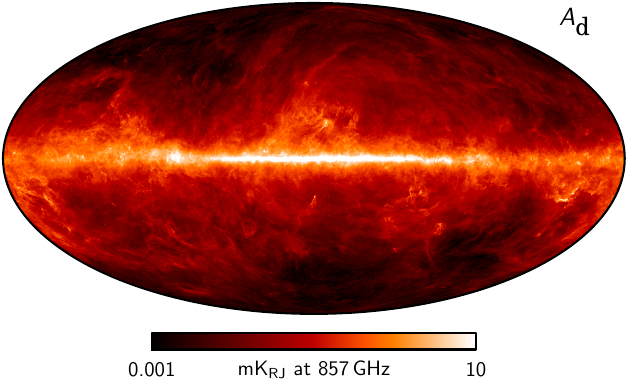}
  \includegraphics[width=0.98\columnwidth]{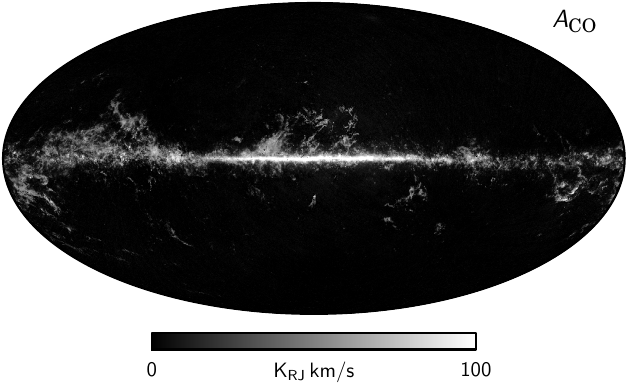}
\end{center}
\caption{\commander\ foreground amplitude maps, derived from the \Planck\ 2018 data set in intensity. The top-left panel shows the combined low-frequency foreground map at $40\arcm$ FWHM resolution, evaluated at 30\,GHz, and accounts for synchrotron, free-free, and anomalous microwave emission. The top-right panel shows the derived radio point source map, as observed in the 30-GHz frequency channel. The bottom-left panel shows thermal dust emission at $10\arcm$ FWHM resolution, evaluated at 857\,GHz. Neither the CIB nor high-frequency point sources are fitted explicitly in the \commander\ 2018 temperature model, and these are therefore in effect included in this thermal dust emission map. The bottom-right panel shows the CO line-emission map, evaluated for the 100-GHz channel.}
\label{fig:comm_fg_temp}
\end{figure*}

\begin{table*}[t]                                                                                                                                                   
\begingroup                                                                            
\newdimen\tblskip \tblskip=5pt
\caption{Summary of \commander\ point-source fits. Each row corresponds to one reference catalogue, as described in the text. Columns indicate, from left to right: (1)~catalogue name; (2)~catalogue reference frequency; (3)~total number of catalogue sources selected for the current analysis; (4)~number of statistically detected sources in the current analysis; (5)~detection rate; (6)~relative average normalization factor between \commander-derived and reference flux densities;
(7)~Pearson's $r$ correlation coefficient between \commander-derived and reference flux densities; and (8)~reference publication. \label{tab:ptsrc_summary}}
\nointerlineskip                                                                                                                                                                                     
\vskip -4mm
\footnotesize                                                                                                                                      
\setbox\tablebox=\vbox{ %
\newdimen\digitwidth                                                                                                                          
\setbox0=\hbox{\rm 0}
\digitwidth=\wd0
\catcode`*=\active
\def*{\kern\digitwidth}
\newdimen\signwidth
\setbox0=\hbox{+}
\signwidth=\wd0
\catcode`!=\active
\def!{\kern\signwidth}
\newdimen\decimalwidth
\setbox0=\hbox{.}
\decimalwidth=\wd0
\catcode`@=\active
\def@{\kern\signwidth}
\halign{ \hbox to 1.2in{#\leaderfil}\tabskip=0.5em&
  \hfil#\hfil\tabskip=1em&
  \hfil#\hfil\tabskip=1em&
  \hfil#\hfil\tabskip=1em&
  \hfil#\hfil\tabskip=1em&
  \hfil#\hfil\tabskip=1em&
  \hfil#\hfil\tabskip=1em&
  \hfil#\hfil\tabskip=0em\cr
\noalign{\doubleline}
\omit\hfil Catalog\hfil&$\nu_{\mathrm{ref}}$ [GHz]&$N_{\mathrm{tot}}$&$N_{\mathrm{det}}$&$f_{\mathrm{det}}$&$a$&Pearson's
 $r$&Reference\cr
\noalign{\vskip 5pt\hrule\vskip 3pt}
AT20G&20@*&  4499&  4096&  0.91&  0.977&  0.74& \citet{murphy2010}\cr
GB6&      *4.85&  5814&  3415&  0.59&  0.560&  0.69& \citet{gregory1996}\cr
NVSS&    *1.4*&  1527&  1094&  0.72&  0.163&  0.10& \citet{condon1998}\cr
PCCS2& 28.5*& *352& *313&  0.89&  0.867&  0.99& \citet{planck2014-a35}\cr
\noalign{\vskip 5pt\hrule\vskip 3pt}
}}
\endPlancktablewide                                                                                                                                            
\endgroup
\end{table*}

Several interesting features may be seen in Table~\ref{tab:ptsrc_summary}.  Starting with the PCCS2 sources \citep{planck2014-a35}, the correlation between the \commander\ and PCCS2 flux densities at 30\,GHz is very high, with a Pearson's correlation coefficient of 0.99. However, the best-fit relative amplitude between the two catalogues is $a=0.867$. Part of this is due to the fact that the \commander\ flux densities are intrinsically colour corrected, and therefore correspond to a monochromatic reference frequency of 30\,GHz, whereas the PCCS2 values correspond to flux densities directly observed in the 30-GHz map without colour correction. Considering that the effective frequency of the 30-GHz channel for a flat-spectrum source with a spectral energy distribution proportional to $\nu^{-2}$ is 28.4\,GHz, the difference in amplitude is expected to be roughly $(28.4/30)^{2}\approx0.90$. In addition, the \commander\ analysis takes into account the full asymmetry of the \Planck\ beams, and also exploits all frequencies between 30 and 143\,GHz in the fit, while the PCCS2 catalogue only considers a symmetric Gaussian beam model, and employs the LFI 30-GHz observations alone.

The \commander\ fits exhibit a slightly lower correlation coefficient relative to the AT20G source catalogue at 20\,GHz, with a numerical value of $r=0.74$ and a detection rate of 91\,\%. However, the flux-density calibration is very good, with a relative normalization factor of $a=0.977$. At 4.85\,GHz, the correlation with the GB6 catalogue flux densities is again very slightly weaker at $r=0.69$, and this time the detection rate is 59\,\%, with a relative normalization of $a=0.56$. Finally, this general trend of weakening correlations becomes even stronger at lower frequencies, with the NVVS catalogue at 1.4\,GHz only having a correlation coefficient of $r=0.10$ and a relative normalization of $a=0.163$. However, the detection rate remains fairly
high, at 72\,\%. NVSS and \commander\ thus agree on the existence of the set of sources, but disagree significantly on their amplitudes. This is, of course, not unexpected, when extrapolating all the way from 1.4\,GHz to 30--143\,GHz. The point source component as evaluated for the 30\,GHz channel is plotted in the top right panel of Fig.~\ref{fig:comm_fg_temp}.

Next, we consider the amplitude parameter maps of the diffuse foreground components, as shown in Fig.~\ref{fig:comm_fg_temp}. Starting with the top left panel, this figure shows the joint low-frequency foreground component, which includes synchrotron, free-free, and anomalous microwave emission as evaluated at 30\,GHz and smoothed to a resolution of $40\arcm$ FWHM.\footnote{Although all components are formally estimated without   internal smoothing during the \commander\ analysis, the resulting maps are completely noise dominated on small scales. In practice, each component map therefore needs to be smoothed to the resolution corresponding to the most relevant frequency map for visualization  purposes.}  A similar low-frequency foreground map was presented in \citet{planck2013-p06}, derived from the \Planck\ 2013 data, and the most visually striking difference between these two maps is the absence of small-scale compact objects in the updated map. This is of course due to the fact that these sources are explicitly fitted out in the new model.  The resulting source amplitude map at 30\,GHz is shown in the top right panel.

The bottom left panel of Fig.~\ref{fig:comm_fg_temp} shows the thermal dust amplitude map evaluated at 857\,GHz and smoothed to $10\arcm$ FWHM. Visually speaking, this map is nearly identical to the corresponding 2015 map, since the thermal dust component is strongly dominated by the 545- and 857-GHz HFI frequency maps, and these have only changed by one or two percent since the last release (see Fig.~\ref{fig:dx11_vs_dx12_hfi}).

\begin{figure}[t]
\begin{center}
  \includegraphics[width=\columnwidth]{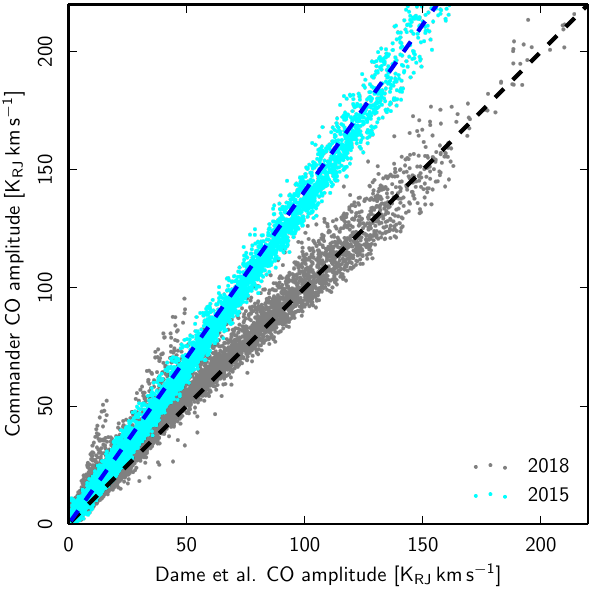}\\
\end{center}
\caption{$T$--$T$ scatter plots between the \citet{dame2001} $J$=1$\rightarrow$0 map and the \commander\ 2015 (blue dots) and 2018 (grey dots) CO maps. Note that the 2018 map has been directly calibrated to the Dame et al.\ map, and is therefore expected to have unity slope by construction, while the 2015 map was calibrated using the \Planck\ bandpasses; this difference explains the overall shift in slopes. The lower level of scatter around the best-fit slope in the 2015 map is due to including single-bolometer and detector-set maps, as opposed to the 2018 map, which exclusively uses co-added frequency maps.}
\label{fig:comm_co}
\end{figure}

At a strictly visual level, the same holds true for the CO component, shown in the bottom-right panel of Fig.~\ref{fig:comm_fg_temp}. However, in this case the reconstruction quality of the new map is notably worse than in the corresponding 2015 map, as shown in the top panel of Fig.~\ref{fig:comm_co}. This figure shows scatter plots between the Dame et al.\ CO survey map \citep{dame2001} and the \commander\ CO 2015 (cyan dots) and 2018 (grey dots) maps. Two effects are notable. First, we note that the slopes are different between the two maps, corresponding simply to the different overall normalization conventions adopted for the two maps. In particular, for the 2015 analysis we employed conversion factors between $\muK_{\mathrm{CMB}}$ and $\mathrm{K}_{\mathrm{RJ}}\, \mathrm{km}\,\mathrm{s}^{-1}$ derived directly from the \Planck\ bandpasses measured on the ground \citep{planck2013-p03d}. This is significantly more complicated with the single-CO line model employed in the current analysis, and with the 2018 co-added frequency maps. The scale of the current CO amplitude map is therefore instead directly set by regressing against the Dame et al.\ map, and the resulting scatter plot therefore by definition has a slope of unity.

More important than this choice of normalization, however, is the width and shape of the two scatter plots. Specifically, while the 2015 scatter plot exhibits a very tight overall correlation and no visually notable outliers, the 2018 scatter plot is broader overall and exhibits several outliers in the \commander\ map.  The reasons for this weaker correlation have already been discussed in Sect.~\ref{sec:inputs} and \citet{planck2016-l03}, and can be summarized as being due to the  lack of single-bolometer HFI maps and inaccuracies in the CO template corrections used during mapmaking.  As described in
 Appendix~\ref{app:commander}, the \commander\ CO map is used as a tracer for CO emission in the \commander\ confidence mask.

\begin{figure}[t]
\begin{center}
\includegraphics[width=\columnwidth]{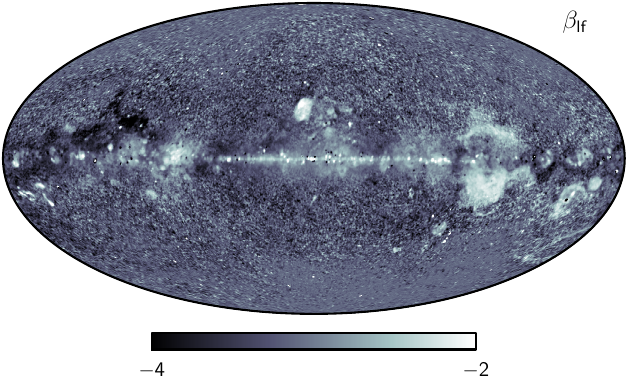}\\
\includegraphics[width=\columnwidth]{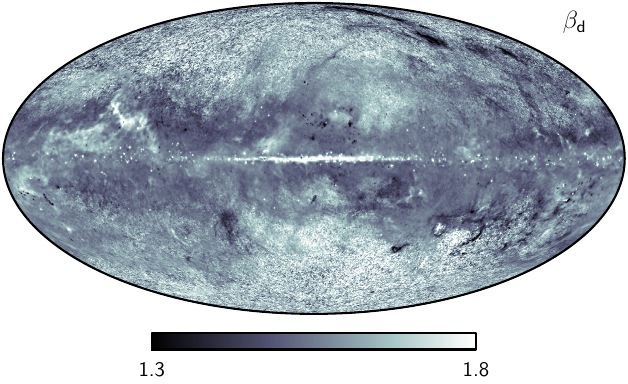}\\
\includegraphics[width=\columnwidth]{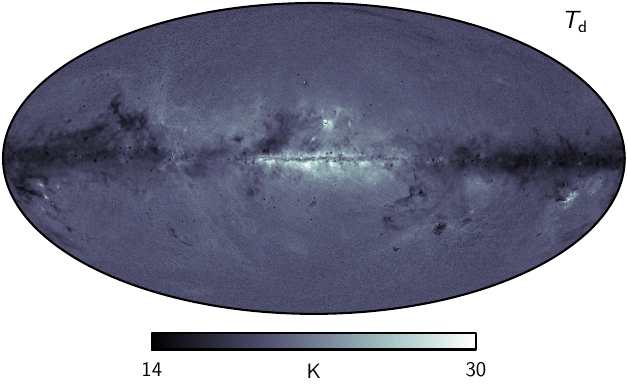}
\end{center}
\caption{\commander\ 2018 foreground spectral parameters. Rows show, from top to bottom, the low-frequency spectral index at a $40\arcm$ FWHM smoothing scale, the thermal dust spectral index at $10\arcm$ FWHM, and the thermal dust temperature at $5\arcm$ FWHM, respectively.}
\label{fig:spectral_params}
\end{figure}

Finally, we consider the spectral parameters for various components, shown in the left column of
Fig.~\ref{fig:spectral_params} for the low-frequency and thermal dust components. These can be compared to similar maps presented in the 2013 and 2015 \Planck\ releases \citep{planck2013-p06,planck2014-a12}.  Starting with the low-frequency spectral-index map, the two most notable changes with respect to the corresponding 2013 products are different priors on spectral index  ($\beta_{\mathrm{lf}}=-2.9\pm0.3$ in 2013 versus $\beta_{\mathrm{lf}}=-3.1\pm0.5$ in 2018), resulting in a darker map at high latitudes, and an overall higher signal-to-noise ratio resulting from the inclusion of four-years of LFI observations in these new maps, as opposed to only 14 months in 2013, resulting in larger areas being data-driven. Otherwise, the two maps are largely consistent.

Relatively speaking, larger changes are seen for the thermal dust spectral parameters when compared to the 2015 model presented in \citet{planck2014-a12}. Starting with the emissivity or spectral index, $\beta_\mathrm{d}$, one can see bright CO-like structures in the 2018 version, for instance near the Fan region at $(l,b)=(140^{\circ},10^{\circ})$; this indicates a stronger degeneracy between CO and thermal dust in the 2018 map than in the 2015 map, and results most likely from the lack of single-bolometer maps in the 2018 analysis. Similarly, one can see a strong dark region extending from the North to the South Ecliptic Pole in the new map. This feature is well-known in \Planck\ mapmaking, and arises from bandpass mismatch between different bolometers used to create a single map. Although the most recent mapmaking process makes a great
effort to suppress this effect \citep{planck2016-l03}, the lack of single-bolometer and detector-set maps carries a significant price for subsequent component separation: while it was possible to remove single bolometers for which this effect was particularly pronounced in 2015 (see figure~2 of \citealp{planck2014-a12}), only full frequency maps are available in the 2018 analysis.

\begin{figure}[t]
\begin{center}
  \includegraphics[width=\columnwidth]{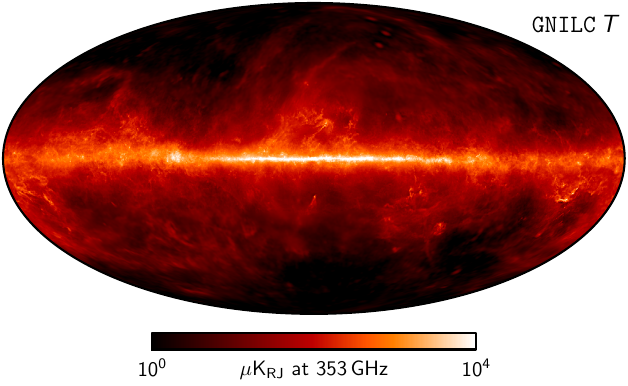}\\
  \includegraphics[width=\columnwidth]{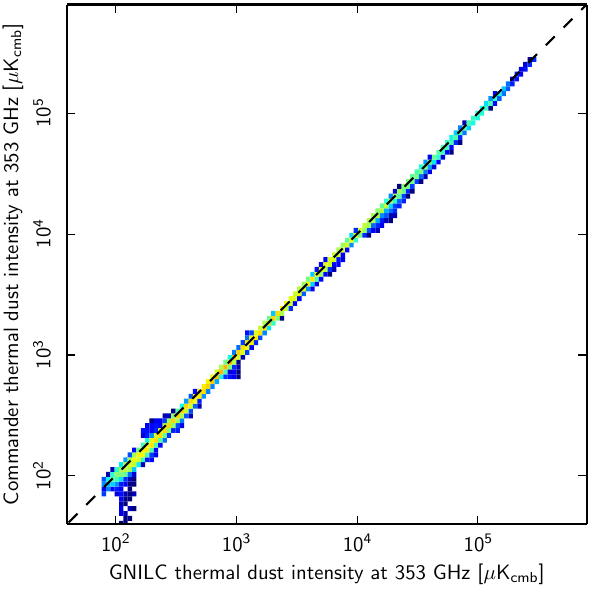}
\end{center}
\caption{(\emph{Top}): \gnilc\ thermal dust intensity map at 353\,GHz with spatially varying angular resolution.  (\emph{Bottom}): $T$--$T$ scatter plot between the thermal dust intensity \commander, both smoothed to a common angular resolution of $80\arcm$ FWHM. An offset of $421\muK$ has been subtracted from the \gnilc\ map in both panels (see main text for details).
}
\label{fig:gnilc_dust_T}
\end{figure}

At high latitudes, the most notable effect is a brighter overall distribution of small-scale fluctuations, which correspond to
small-scale cosmic infrared background (CIB) fluctuations. When interpreting these fluctuations, however, it is important to recall that the two-parameter $\beta$--$T$ modified blackbody model exhibits a strong degeneracy between the spectral index and temperature in the low signal-to-noise regime. The fluctuations seen in the 2018 $\beta$ map were thus also present in the 2015 rendition, but in that case were seen in the temperature map. The main reason for the apparent shift is the choice of thermal dust temperature prior, or, to be more precise, the angular resolution at which it is fitted.  In 2015 the thermal dust temperature was fitted at $40\arcm$ FWHM, while in the updated analysis it is fitted at $5\arcm$ FWHM.  As a result, the 2015 temperature map had higher effective signal-to-noise per resolution element, and therefore less dependence on the prior and
more structure at high latitudes. In contrast, the 2018 temperature map has less signal-to-noise per resolution element, stronger prior dependency, and accordingly also shows less structure at high latitudes, as the temperature is driven to the prior mean, and fluctuations are instead captured in the spectral index map. In general, we caution against over-interpreting the individual parameters of the modified blackbody model in the low signal-to-noise regime, since small changes in the input can lead to relatively large variations in parameter values. In contrast, the resulting SED arising from the parameters is robust. 

For completeness, we note that the best-fit CO line ratio between 100 and 217\,GHz (353\,GHz) is $h_{2018}=0.58$ ($h_{353}=0.20$), as estimated by \commander\ from the \Planck\ 2018 data set. For comparison, the corresponding 2013 values for these two parameters were $h_{2018}=0.595$ and $h_{353}=0.295$. However, for the reasons discussed above, we do not attach physical significance to the lower value found in the new data set, but rather recommend continued usage of the previous values when using \Planck\ results for astrophysical analysis and forecasts.

Before concluding our discussion, we emphasize that while we do not consider the \commander\ 2018 intensity foreground analysis to be as robust as the corresponding 2015 analysis, this has only a very small effect on the corresponding CMB reconstruction after accounting explicitly for CO emission in the \commander\ confidence mask (see Sec.~\ref{app:comm_mask}). As far as CMB reconstruction is concerned, the only important factor is whether the sum of the apparent foregrounds may be modelled within the parameter space of the Bayesian model; whether or not those best-fit values represents the physically true sky is irrelevant. This is of course also precisely why blind CMB reconstruction methods, such as \nilc, \sevem, and \smica, perform very well. Nevertheless, the \textcolor{black}{fact that the \commander\ 2018 intensity products appear reasonable, and that shortcomings are understood, is reassuring}.

\setcounter{section}{7}
\setcounter{figure}{0}

\begin{figure*}[h]
  \begin{center}
    \begin{tabular}{ccc}
      \includegraphics[width=0.3\linewidth]{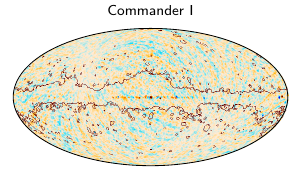}&
      \includegraphics[width=0.3\linewidth]{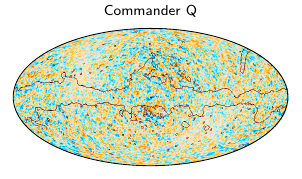}&
      \includegraphics[width=0.3\linewidth]{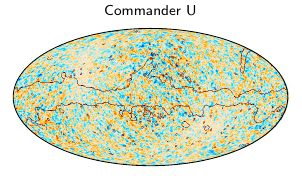}\\
      \includegraphics[width=0.3\linewidth]{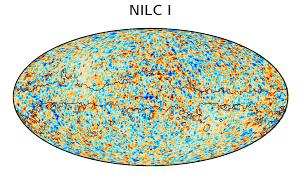}&
      \includegraphics[width=0.3\linewidth]{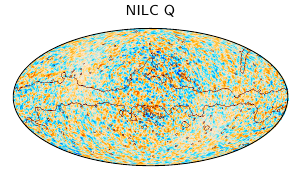}&
      \includegraphics[width=0.3\linewidth]{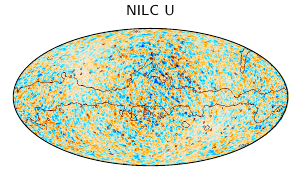}\\
      \includegraphics[width=0.3\linewidth]{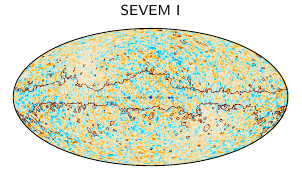}&
      \includegraphics[width=0.3\linewidth]{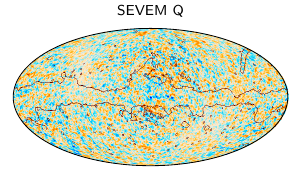}&
      \includegraphics[width=0.3\linewidth]{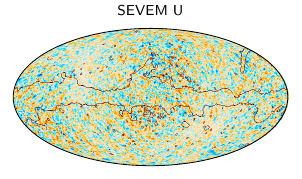}\\
      \includegraphics[width=0.3\linewidth]{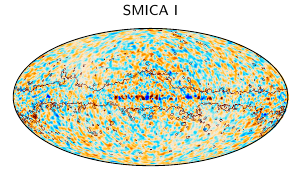}&
      \includegraphics[width=0.3\linewidth]{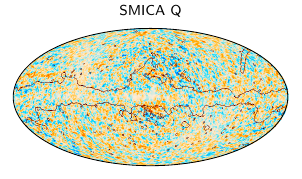}&
      \includegraphics[width=0.3\linewidth]{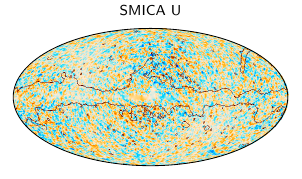}\\
      \includegraphics[width=0.25\linewidth]{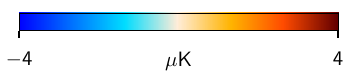}&
      \multicolumn{2}{c}{
        \includegraphics[width=0.25\linewidth]{figs/colourbar_2p5uK}
      }
    \end{tabular}
  \end{center}
  \caption{Odd-even half-difference CMB maps at 80\arcm\ resolution. Columns show Stokes $I$, $Q$, and $U$, while rows show results derived with different component-separation methods. The common mask is marked in red.}
  \label{fig:cmb_oehd_maps}
\end{figure*}

\begin{figure*}
  \begin{center}
    \begin{tabular}{ccc}
      \includegraphics[width=0.3\linewidth]{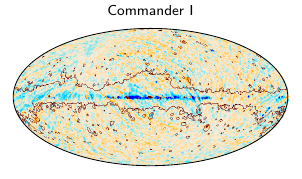}&
      \includegraphics[width=0.3\linewidth]{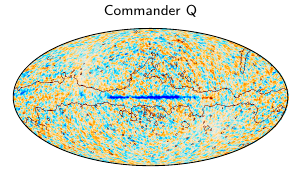}&
      \includegraphics[width=0.3\linewidth]{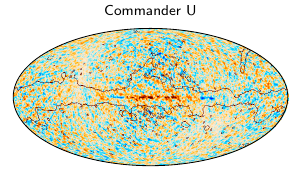}\\
      \includegraphics[width=0.3\linewidth]{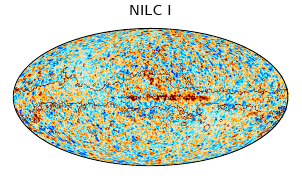}&
      \includegraphics[width=0.3\linewidth]{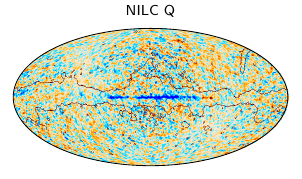}&
      \includegraphics[width=0.3\linewidth]{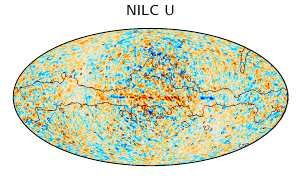}\\
      \includegraphics[width=0.3\linewidth]{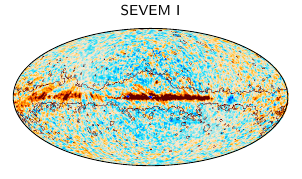}&
      \includegraphics[width=0.3\linewidth]{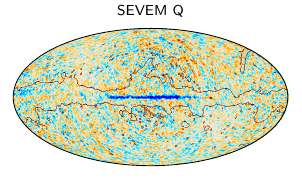}&
      \includegraphics[width=0.3\linewidth]{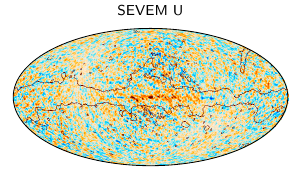}\\
      \includegraphics[width=0.3\linewidth]{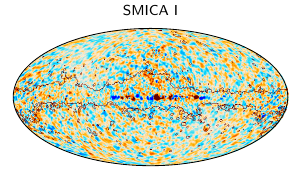}&
      \includegraphics[width=0.3\linewidth]{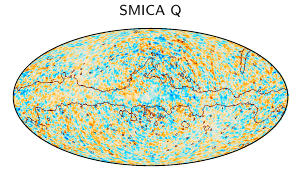}&
      \includegraphics[width=0.3\linewidth]{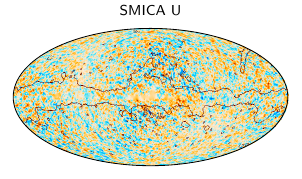}\\
      \includegraphics[width=0.25\linewidth]{figs/colourbar_4uK}&
      \multicolumn{2}{c}{
        \includegraphics[width=0.25\linewidth]{figs/colourbar_2p5uK}
      }
    \end{tabular}
  \end{center}
  \caption{Half-mission half-difference CMB maps at 80\arcmin\ resolution. Columns show Stokes $I$, $Q$, and $U$, while rows show results derived with different component-separation methods. The common mask is marked in red.}
  \label{fig:cmb_hmhd_maps}
\end{figure*}

\setcounter{section}{6}
\setcounter{figure}{0}

\subsection{Thermal dust intensity maps and their zero levels}

Finally, we compare the thermal dust intensity maps derived with \commander\ and \gnilc. Specifically, the top panel of
Fig.~\ref{fig:gnilc_dust_T} shows the \gnilc\ thermal dust intensity map evaluated at 353\,GHz, and the bottom panel shows a scatter plot between the \commander\ and \gnilc\ estimates, where the \commander\ model has been integrated over the 353-GHz channel bandpass. Overall, we observe good agreement between the two estimates.

The behaviour at low intensities is particularly interesting because it is sensitive to how the zero level of each map has been set.  By construction, the frequency maps delivered by the HFI DPC and used for component separation have a Galactic zero level consistent with an intensity of the dust foreground at high Galactic latitudes proportional to the column density of the ISM traced by the 21-cm emission of \hi\ at low column densities.  In the case of \gnilc, the processing does not adjust the monopoles contained in the input maps, the largest of which is the CIB monopole.  Therefore, the zero levels of the resulting \gnilc\ dust maps need to be adjusted prior to Galactic applications.  This has been accomplished here, just as in \citet{planck2016-XLVIII}, by correlation with the \hi\ map at high latitude, following the methodology set out in \citet{planck2013-p03f} and \citet{planck2013-p06b}.  At 353\,GHz, 421\muK\ is subtracted.  In the case of \commander, the zero level at each frequency is solved for explicitly within the component separation processing, with priors set equal to the value of the CIB monopole (see Appendix~\ref{app:commprior}).  The \commander\ offset found at 353\,GHz is 431\muK, separate from the thermal dust emission model.  Given these zero level adjustments, the agreement at low intensities
is satisfactory.

Especially for applications at low intensity, it critical to appreciate that there are significant uncertainties in the zero levels
of the \commander\ thermal dust intensity maps derived from the \Planck\ 2015 and 2018 frequency maps, as discussed in Section~6.1.1 of \citet{planck2014-a12}, and of \gnilc, as discussed in Section~2.2 of \citet{planck2016-l11B}, including the possibility of dust associated with ionized gas.
These uncertainties need to be evaluated and then propagated in any subsequent analyses using these thermal dust maps, \textcolor{black}{in particular when estimating modified blackbody parameters or the polarization fraction}.  Ideally, the uncertainties can be reduced through improved methods of zero level determination, such as exploitation of correlations with external data sets, including \hi\ and optical extinction \citep[e.g.,][]{planck2016-XLVIII}, or via spatial spectral variations \citep{wehus2014}.

\section{Extra CMB plots}
\label{app:splits}

In this Appendix, we present supporting plots relevant for the CMB discussion. These complement and elucidate the analyses and results presented in the main text, and are useful for reference purposes.

First, Figs.~\ref{fig:cmb_oehd_maps} and \ref{fig:cmb_hmhd_maps} show odd-even and half-mission half-difference maps, and as such, they represent our preferred tracers of noise and instrumental systematics, respectively. The former exhibit very few large-scale correlated features, whereas the latter show clear signatures of both the \Planck\ scanning strategy at high latitudes and Galactic contamination through calibration and leakage effects at low latitudes.

\setcounter{section}{7}
\setcounter{figure}{2}

\begin{figure*}
  \begin{center}
    \begin{tabular}{ccc}
      \includegraphics[width=0.29\linewidth]{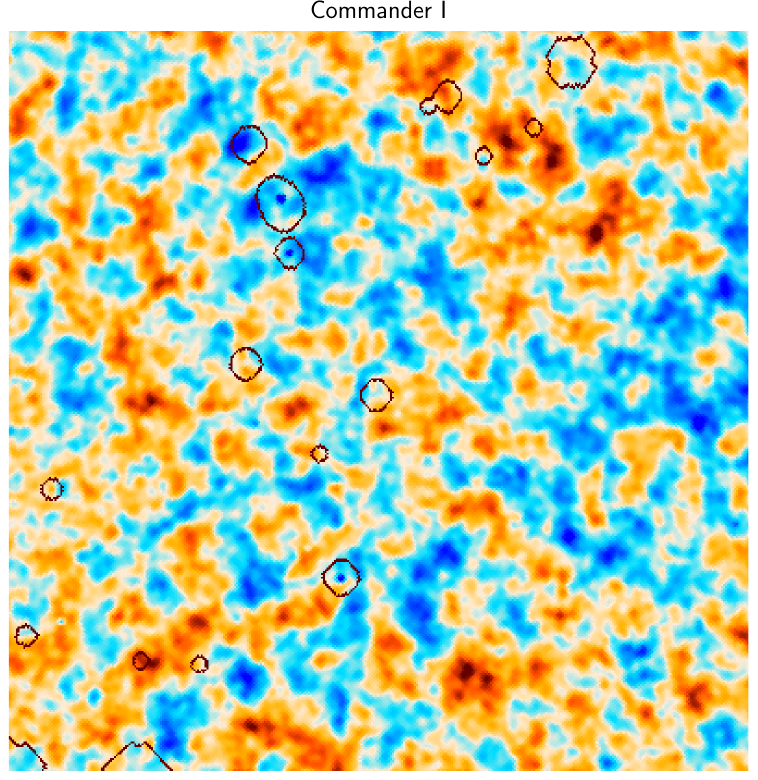}&
      \includegraphics[width=0.29\linewidth]{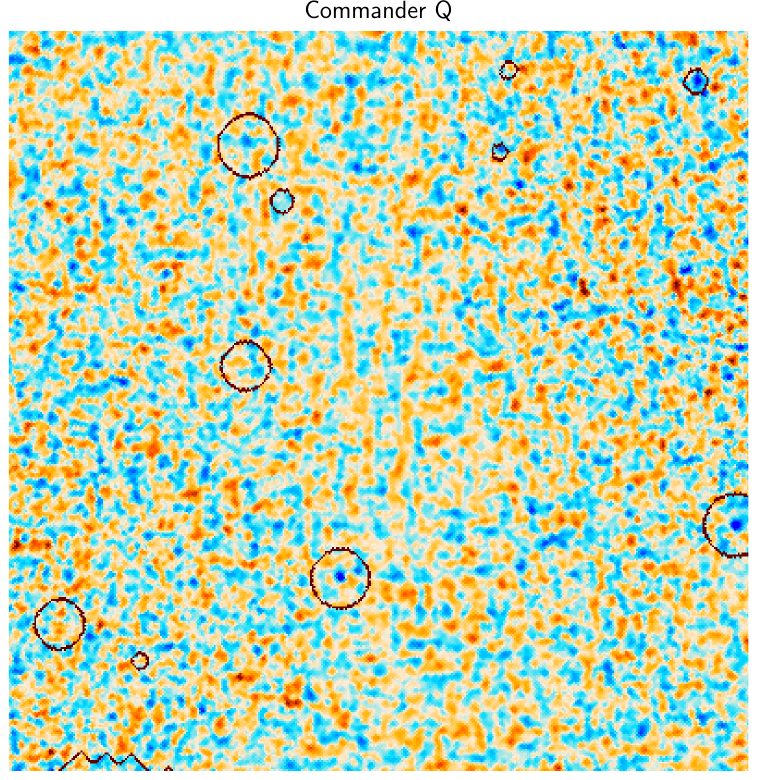}&
      \includegraphics[width=0.29\linewidth]{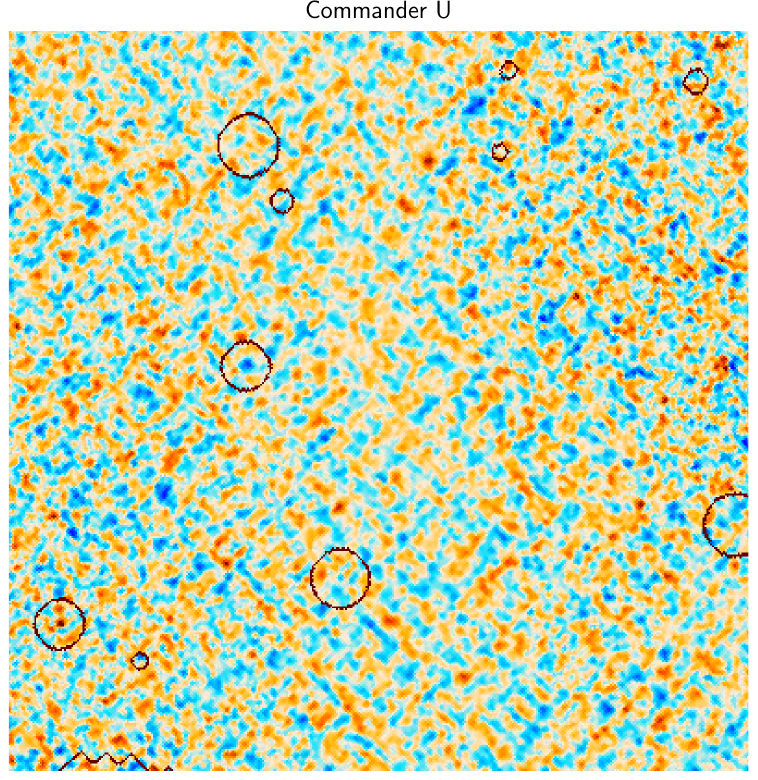}\\
      \includegraphics[width=0.29\linewidth]{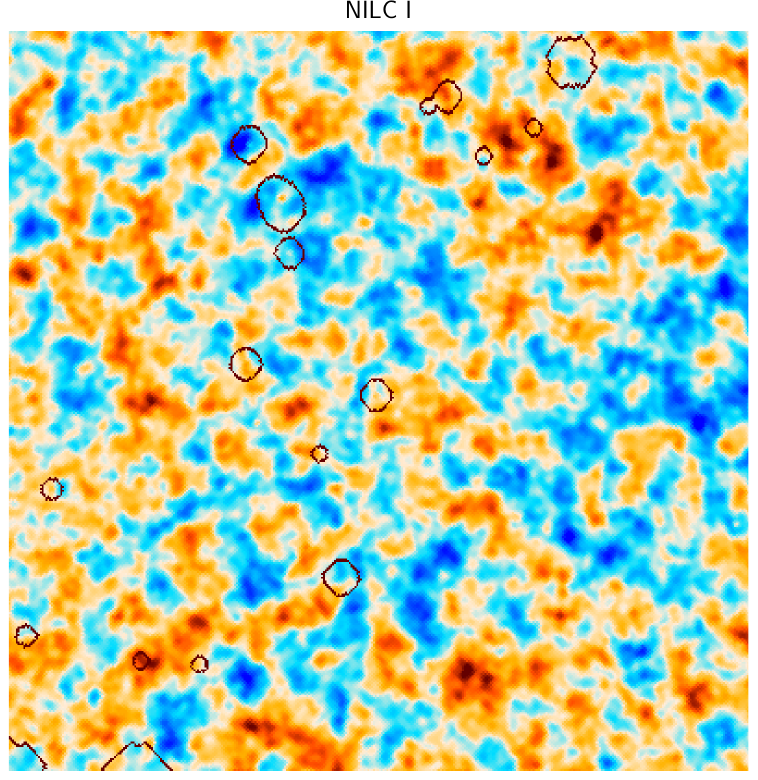}&
      \includegraphics[width=0.29\linewidth]{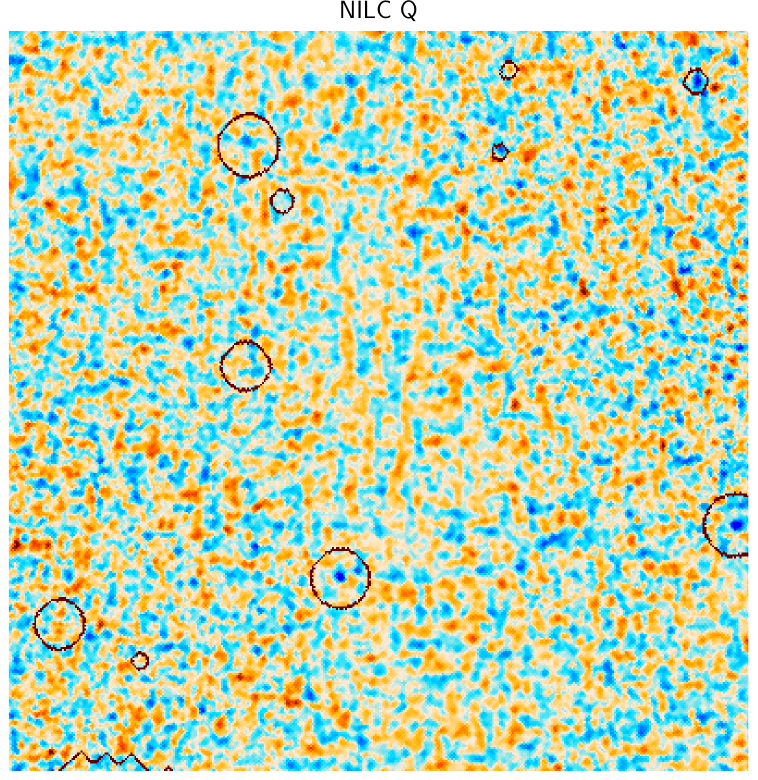}&
      \includegraphics[width=0.29\linewidth]{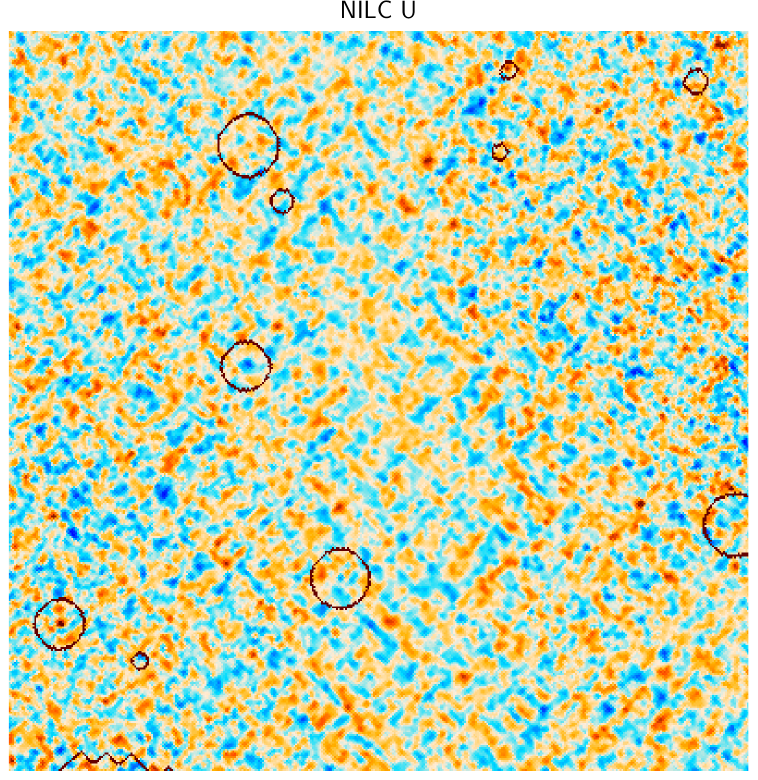}\\
      \includegraphics[width=0.29\linewidth]{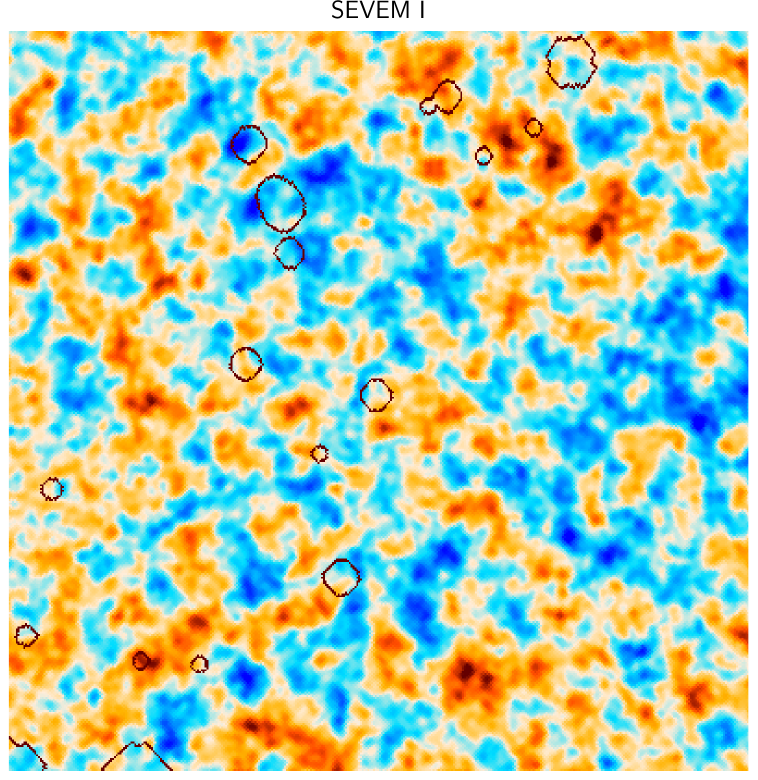}&
      \includegraphics[width=0.29\linewidth]{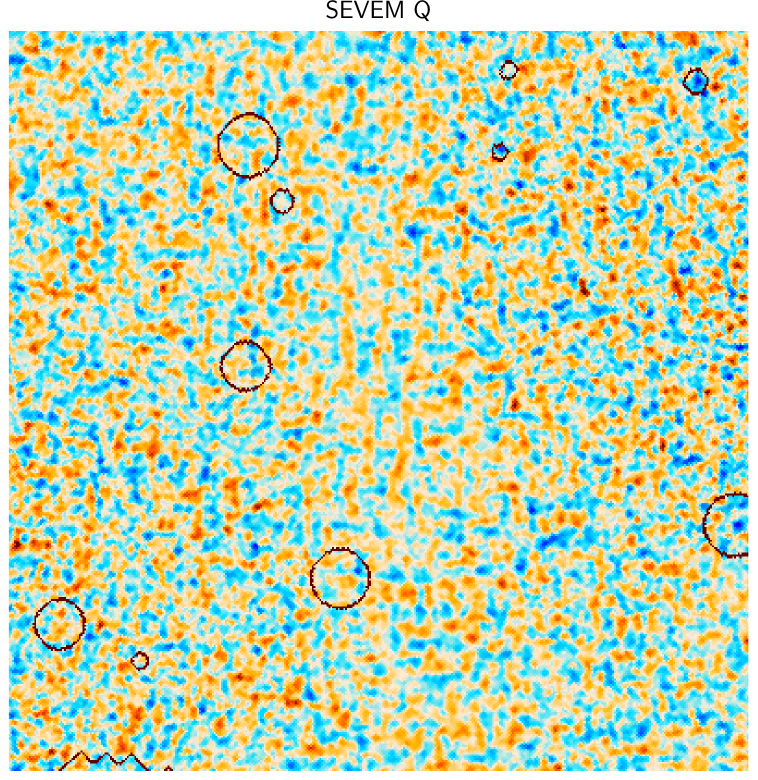}&
      \includegraphics[width=0.29\linewidth]{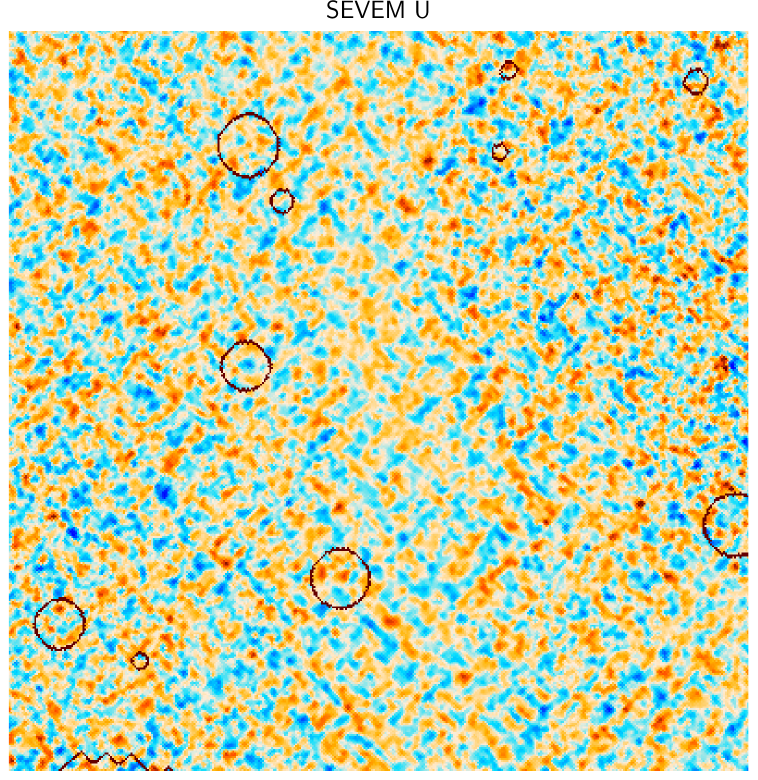}\\
      \includegraphics[width=0.29\linewidth]{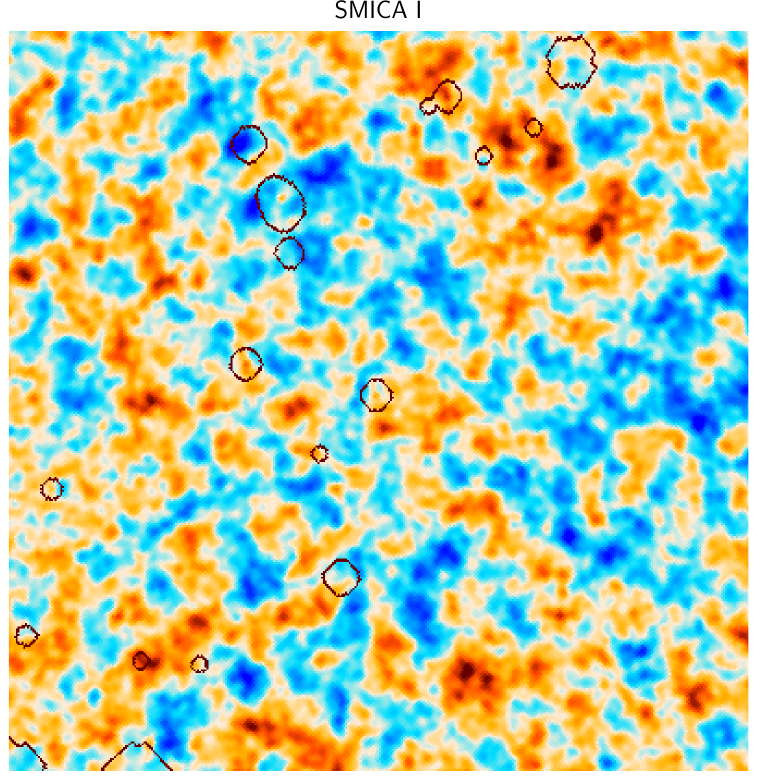}&
      \includegraphics[width=0.29\linewidth]{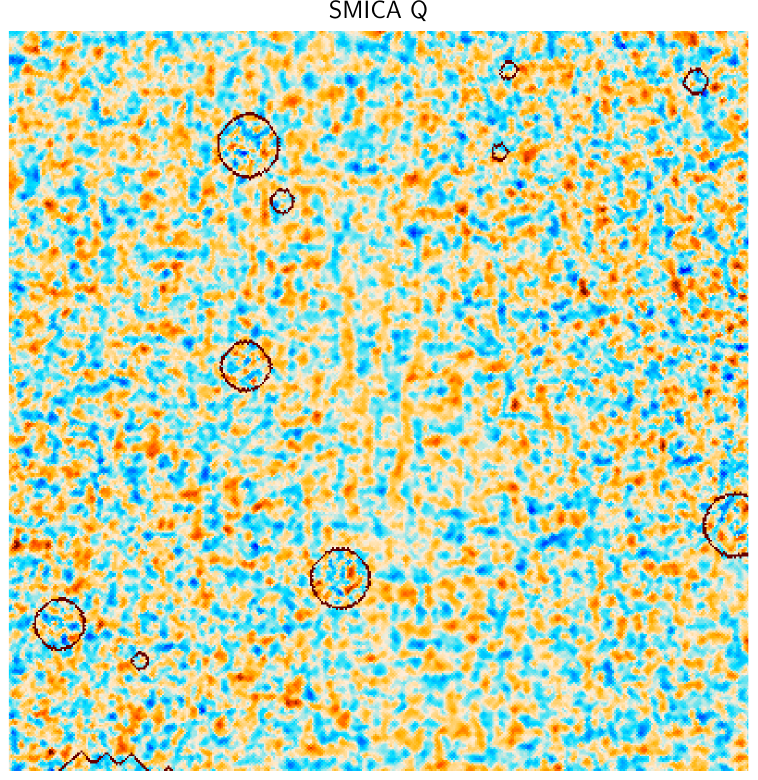}&
      \includegraphics[width=0.29\linewidth]{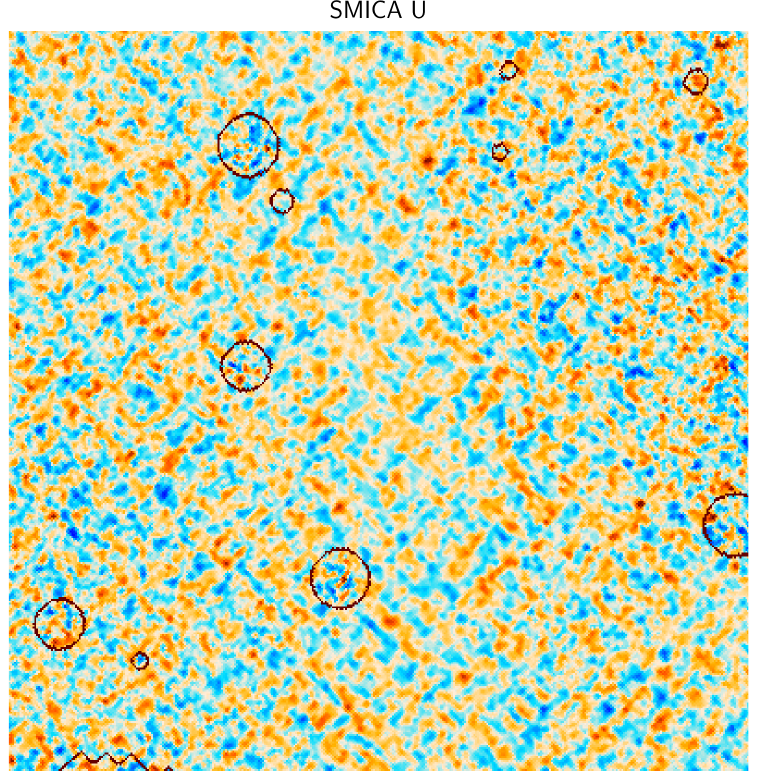}\\
      \includegraphics[width=0.29\linewidth]{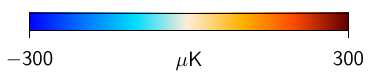}&
      \multicolumn{2}{c}{
        \includegraphics[width=0.29\linewidth]{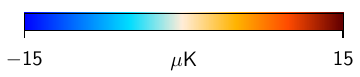}
      }
    \end{tabular}
  \end{center}
  \caption{CMB maps smoothed to a common resolution of 10\arcmin\ FWHM. The patch shown is $20 \deg\times 20\deg$ centred on the North Ecliptic Pole, $(l, b) = (96\pdeg38, 29\pdeg81)$. Columns show Stokes $I$, $Q$, and $U$, while rows show results derived with different component separation methods. The common mask is marked in red.}
  \label{fig:cmb_maps_zoom}
\end{figure*}

Next, Fig.~\ref{fig:cmb_maps_zoom} shows a $20^\circ \times 20^\circ$ zoom-in of the four cleaned CMB maps, centered on the North Ecliptic Pole. The polarization pattern expected from a typical $E$-mode signal ('+'-type in Stokes $Q$, and '$\times$'-type in Stokes $U$) is clearly visible.

\begin{figure*}
  \begin{center}
    \begin{tabular}{ccc}
      \includegraphics[width=0.29\linewidth]{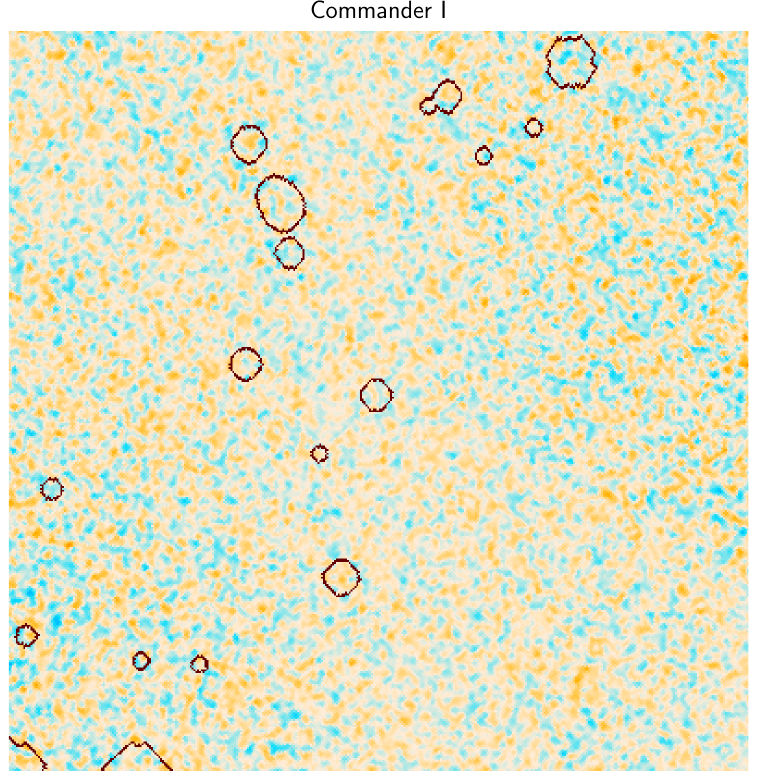}&
      \includegraphics[width=0.29\linewidth]{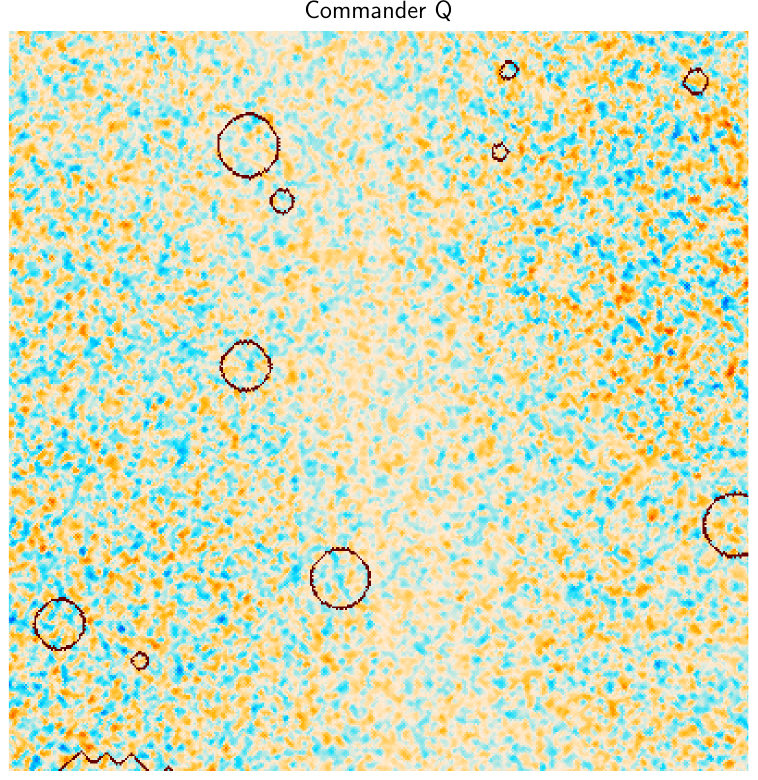}&
      \includegraphics[width=0.29\linewidth]{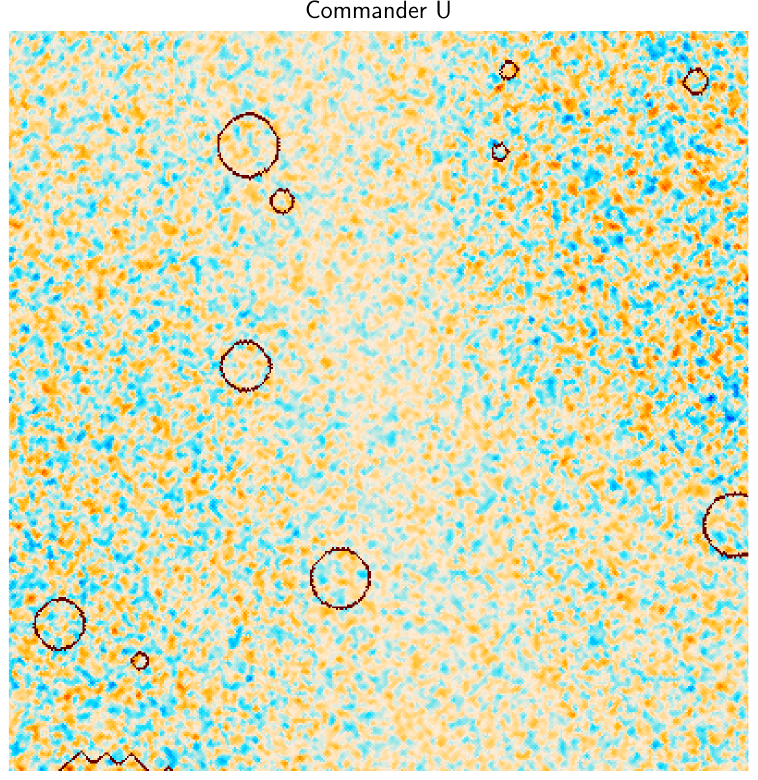}\\
      \includegraphics[width=0.29\linewidth]{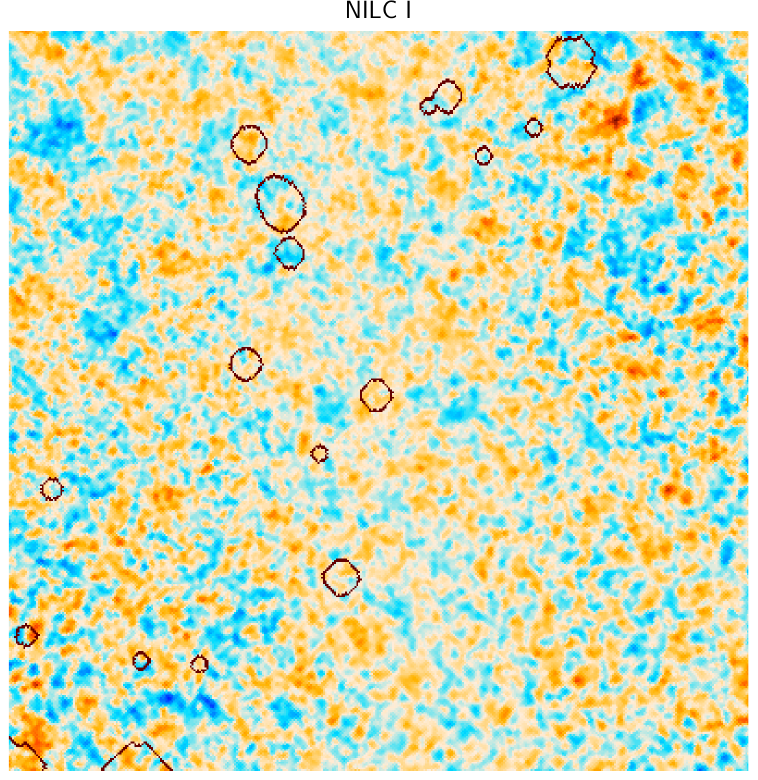}&
      \includegraphics[width=0.29\linewidth]{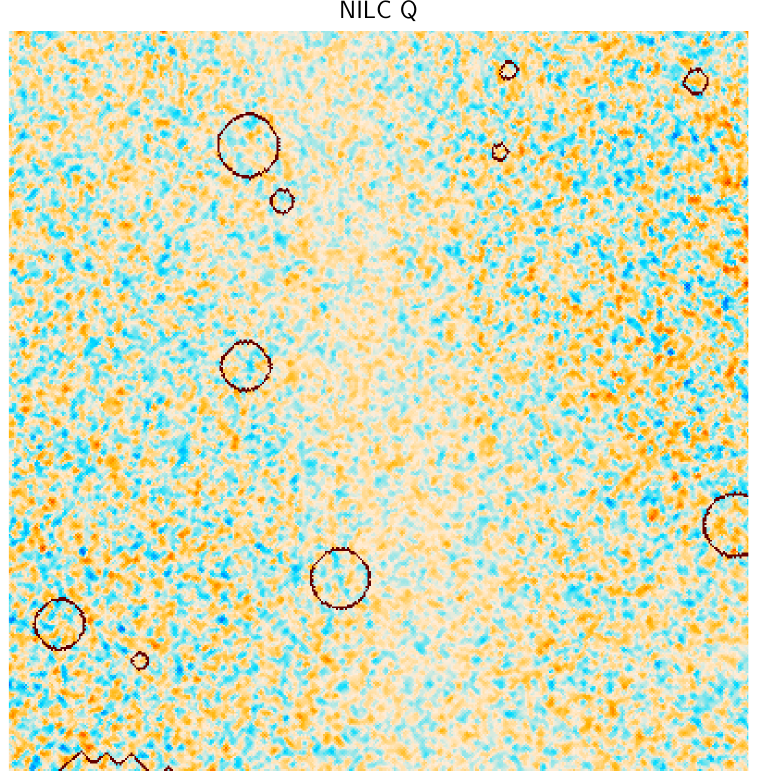}&
      \includegraphics[width=0.29\linewidth]{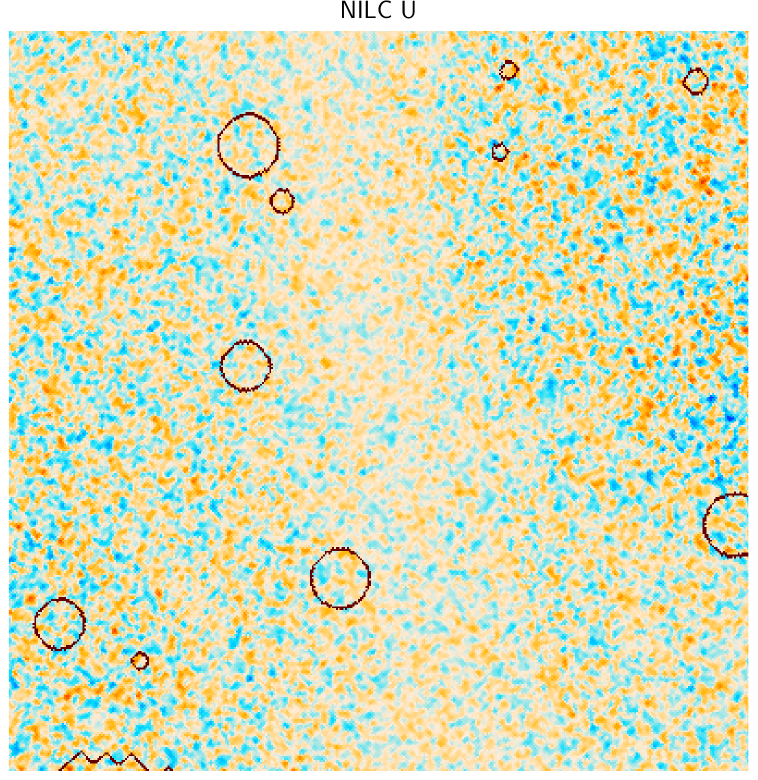}\\
      \includegraphics[width=0.29\linewidth]{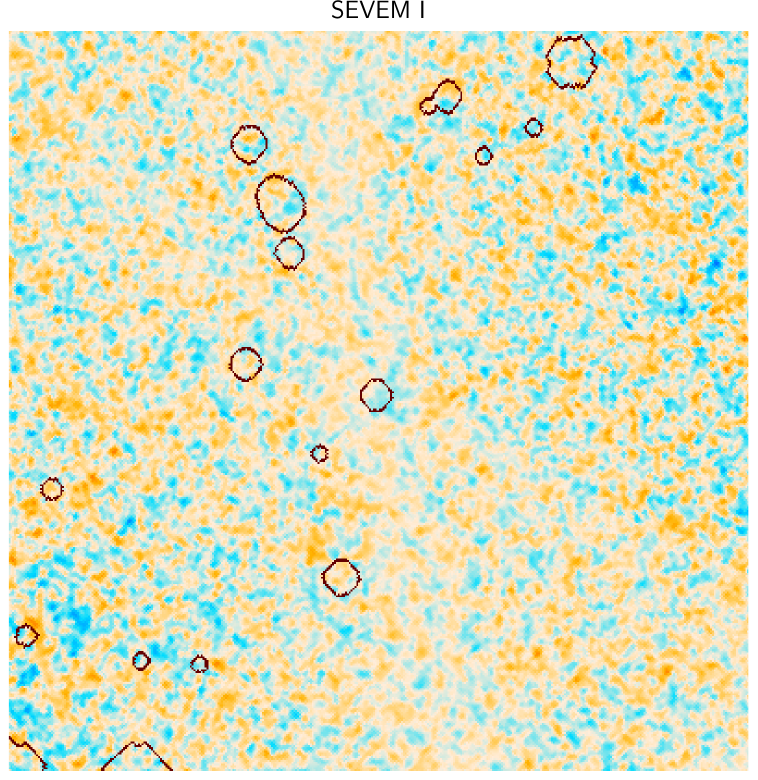}&
      \includegraphics[width=0.29\linewidth]{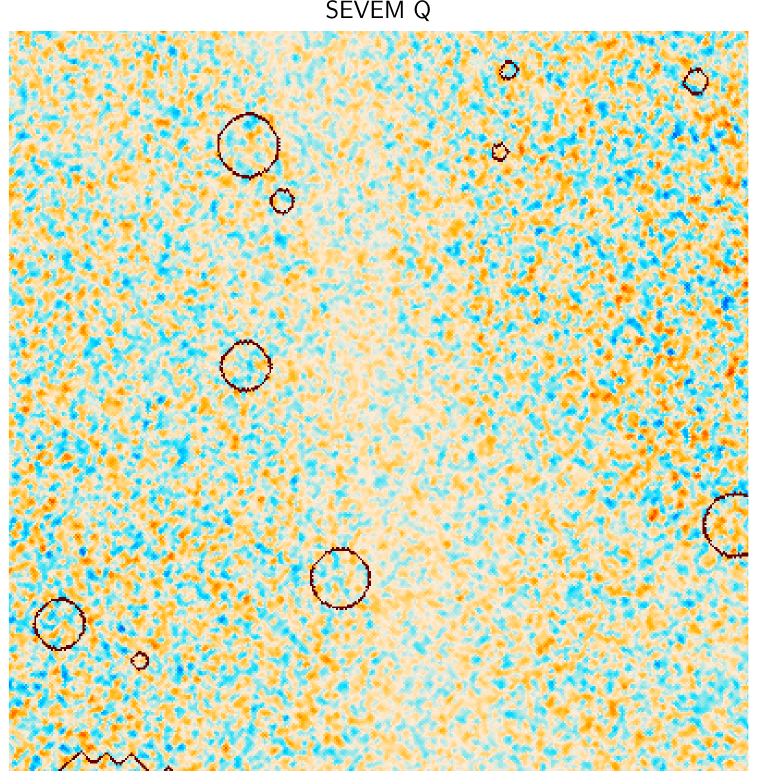}&
      \includegraphics[width=0.29\linewidth]{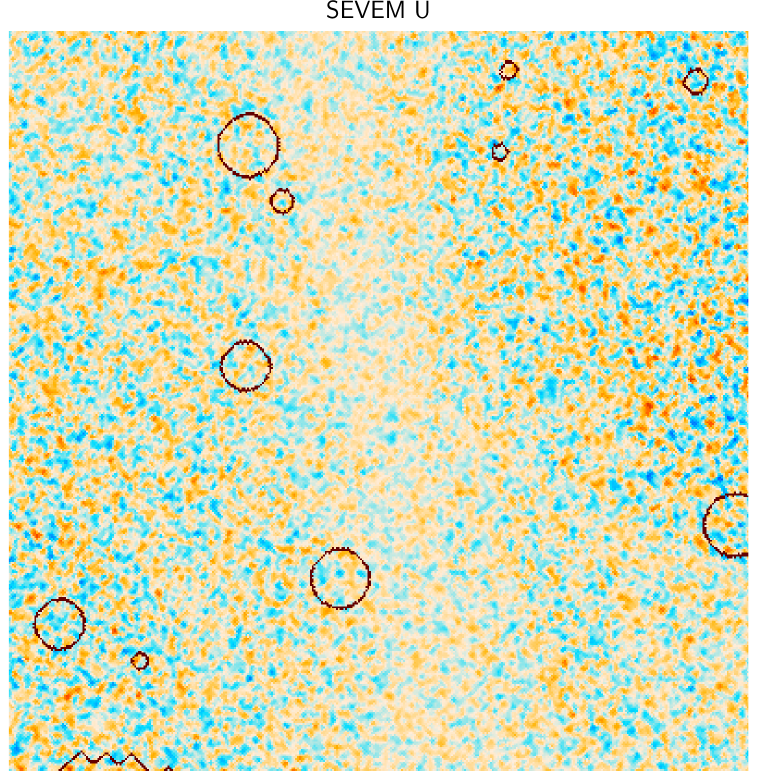}\\
      \includegraphics[width=0.29\linewidth]{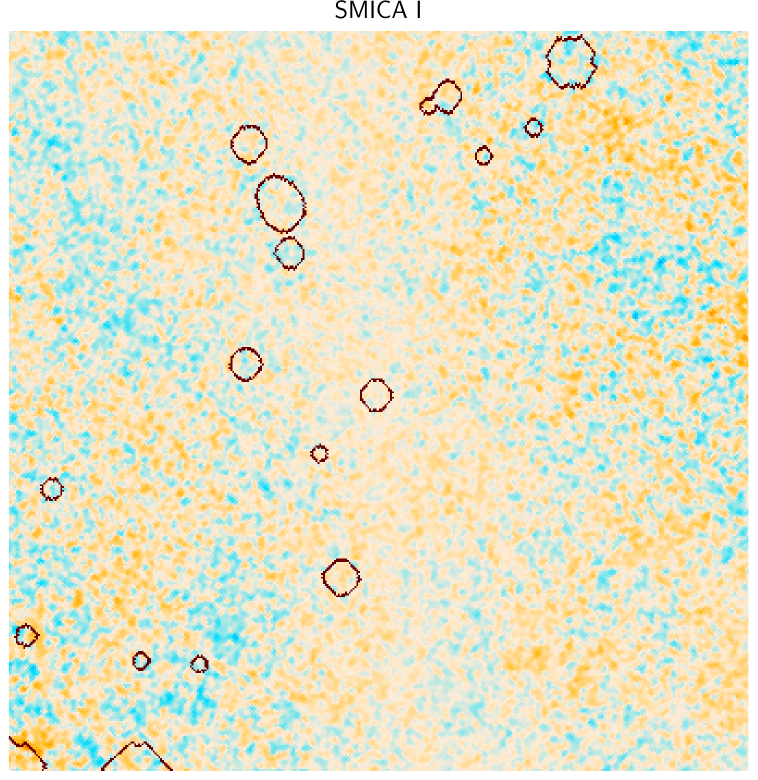}&
      \includegraphics[width=0.29\linewidth]{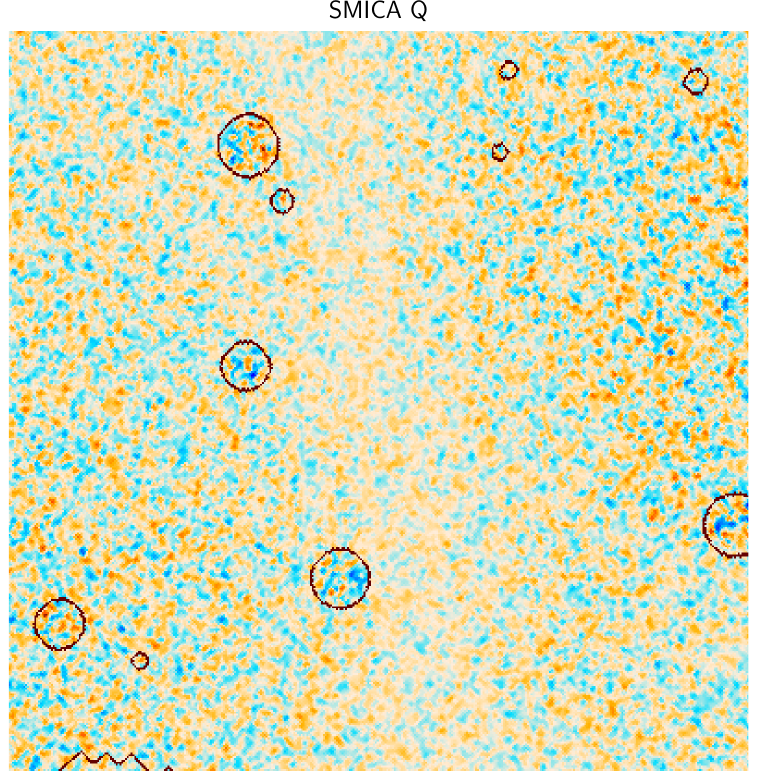}&
      \includegraphics[width=0.29\linewidth]{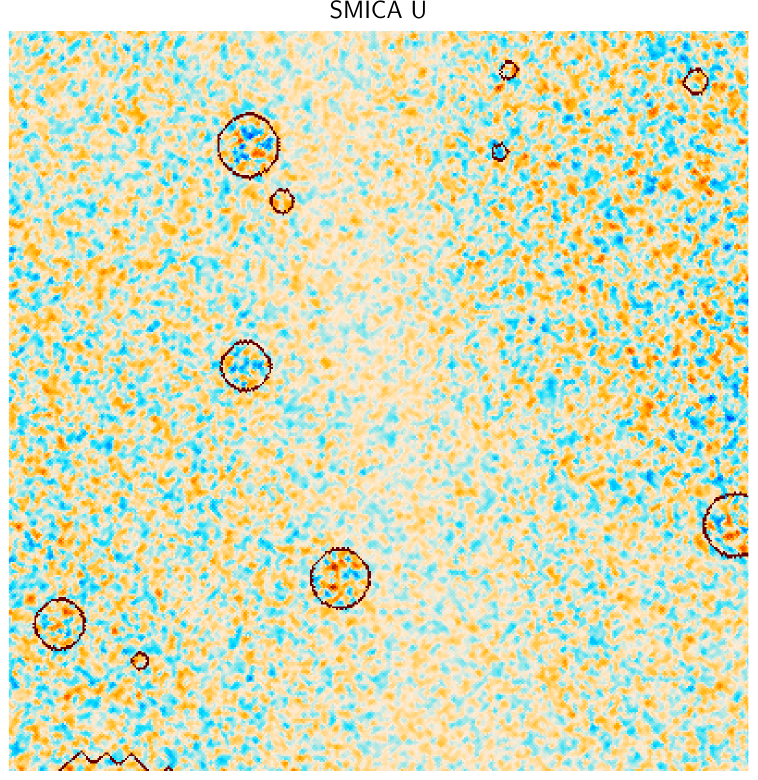}\\
      \multicolumn{3}{c}{
        \includegraphics[width=0.29\linewidth]{figs/colourbar_15uK}
      }
    \end{tabular}
  \end{center}
  \caption{Odd-even half-difference CMB maps smoothed to a common resolution of 10\arcmin\ FWHM. The patch shown is $20 \deg\times 20\deg$  centred on the North Ecliptic Pole, $(l, b) = (96\pdeg38, 29\pdeg81)$. Columns show Stokes $I$, $Q$, and $U$, while rows show results derived with different component-separation methods. The common mask is marked in red.}
  \label{fig:cmb_maps_zoom_oehd}
\end{figure*}

\begin{figure*}
  \begin{center}
    \begin{tabular}{ccc}
      \includegraphics[width=0.29\linewidth]{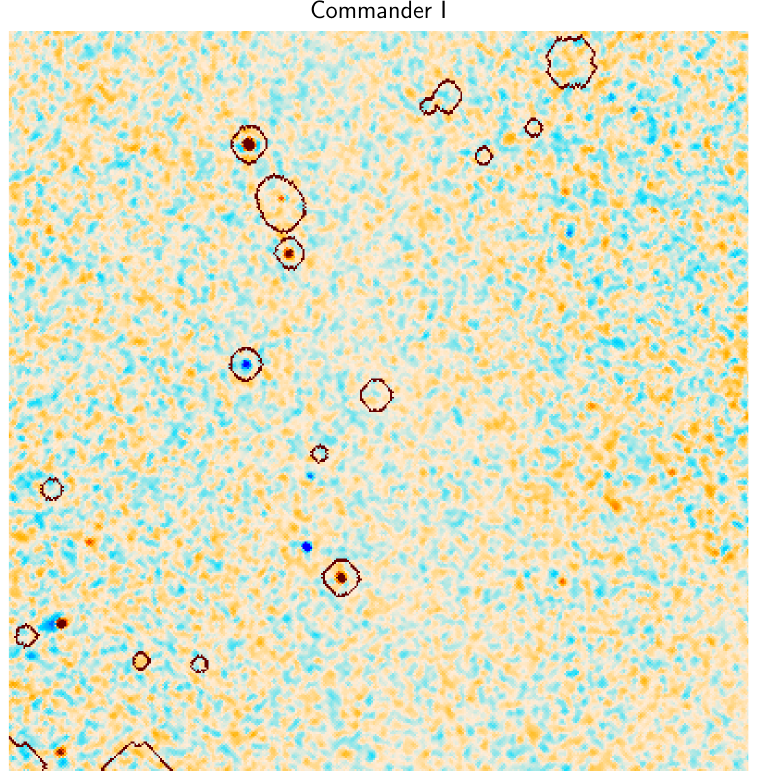}&
      \includegraphics[width=0.29\linewidth]{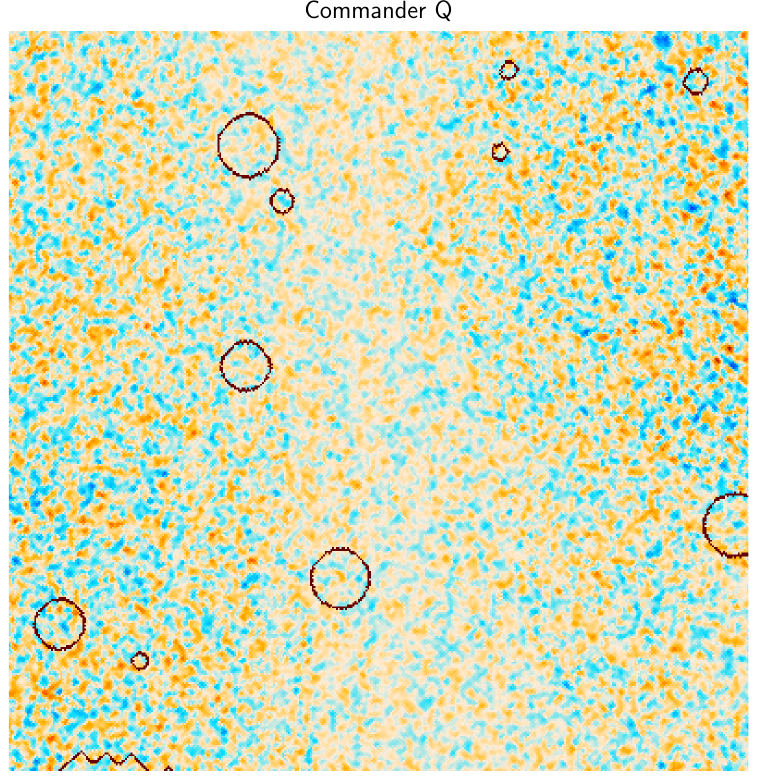}&
      \includegraphics[width=0.29\linewidth]{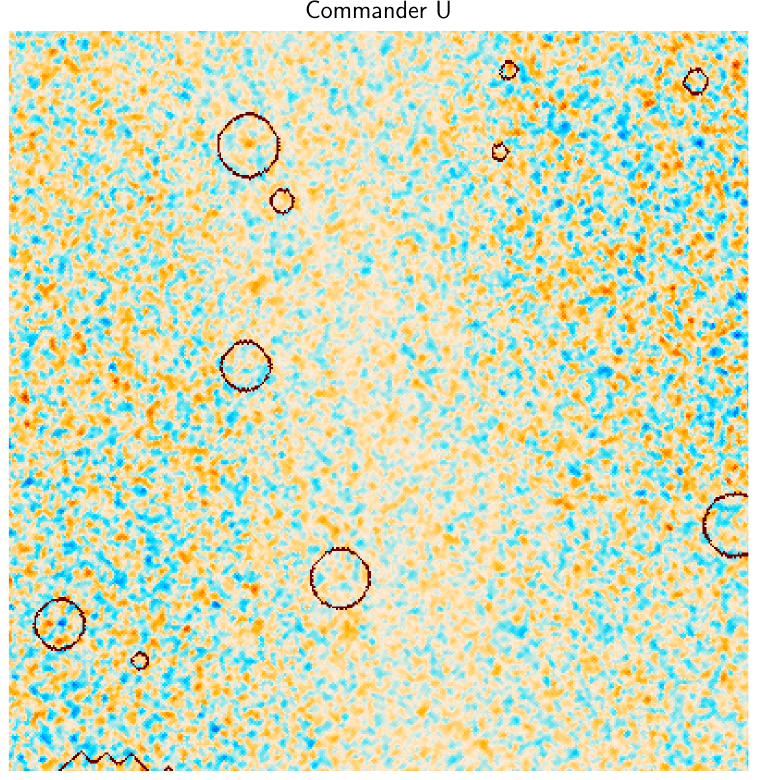}\\
      \includegraphics[width=0.29\linewidth]{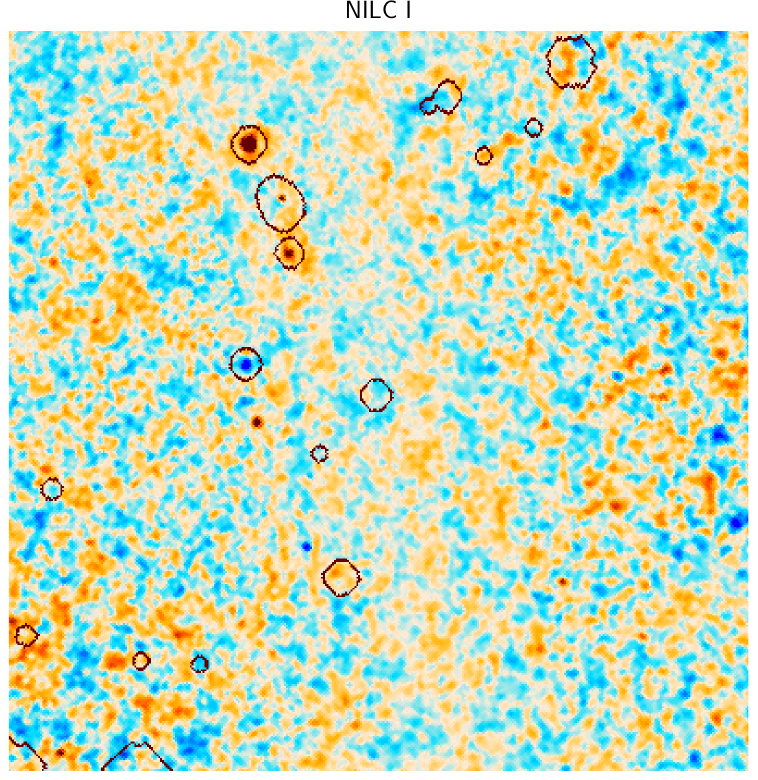}&
      \includegraphics[width=0.29\linewidth]{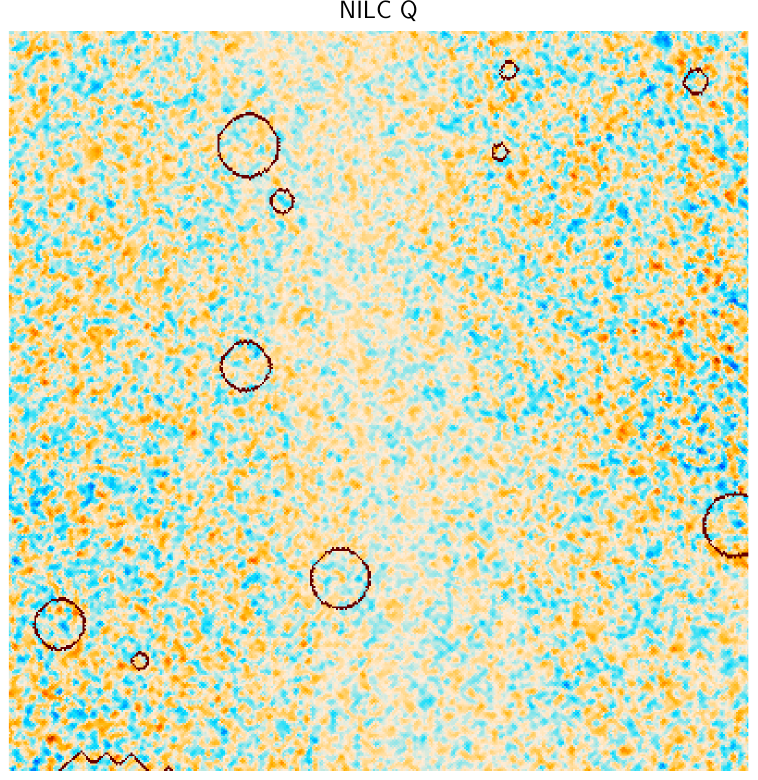}&
      \includegraphics[width=0.29\linewidth]{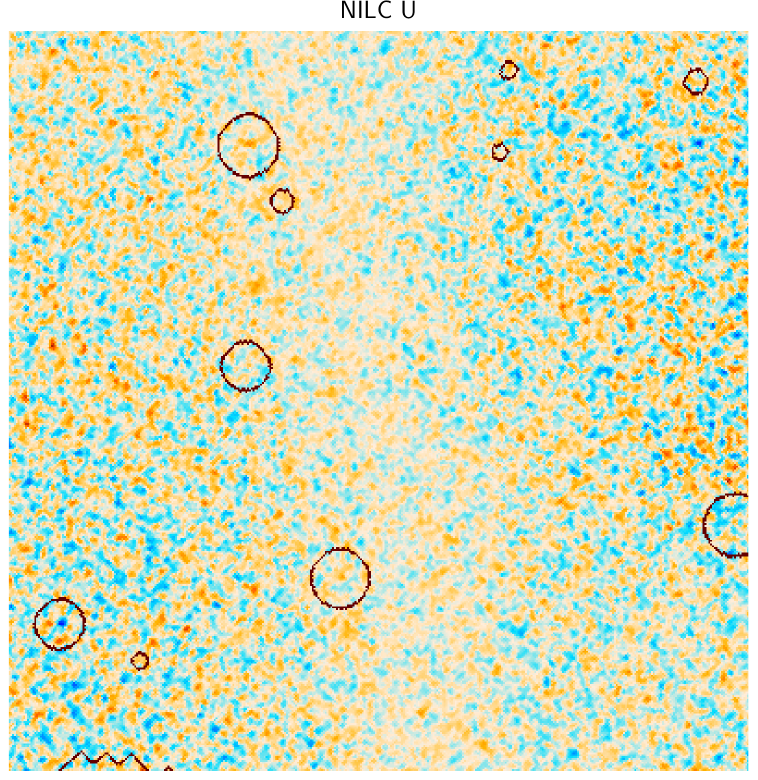}\\
      \includegraphics[width=0.29\linewidth]{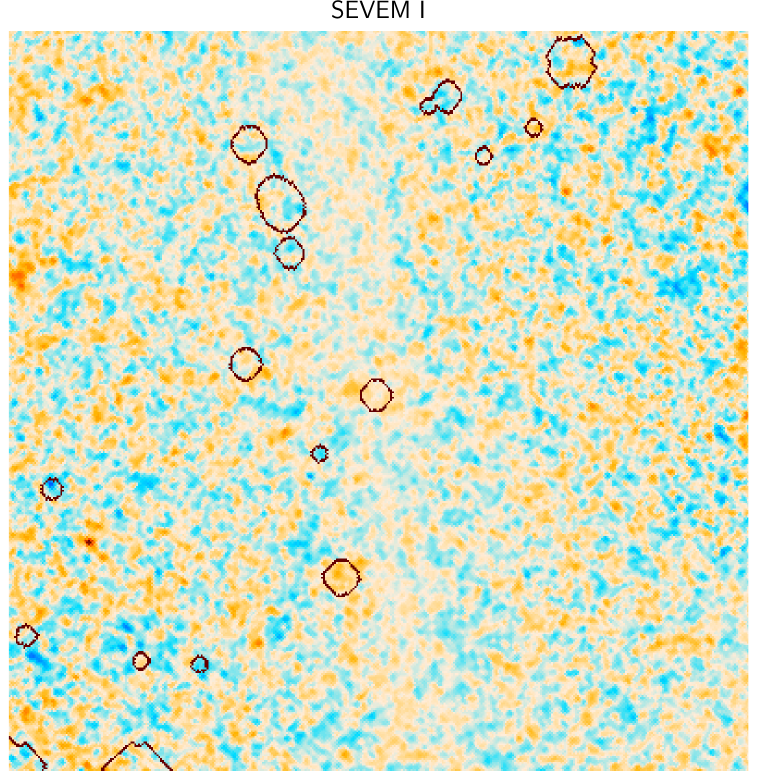}&
      \includegraphics[width=0.29\linewidth]{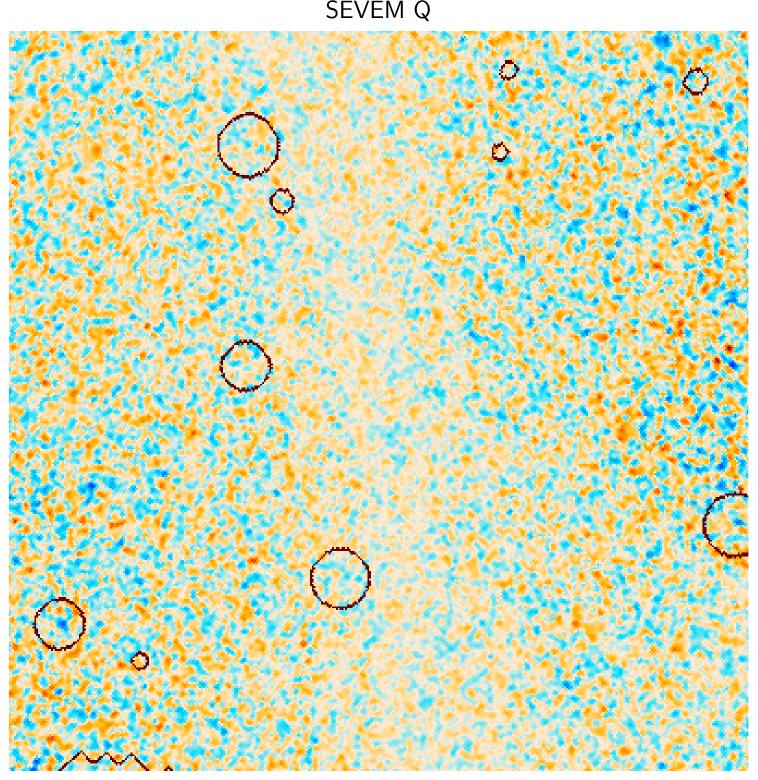}&
      \includegraphics[width=0.29\linewidth]{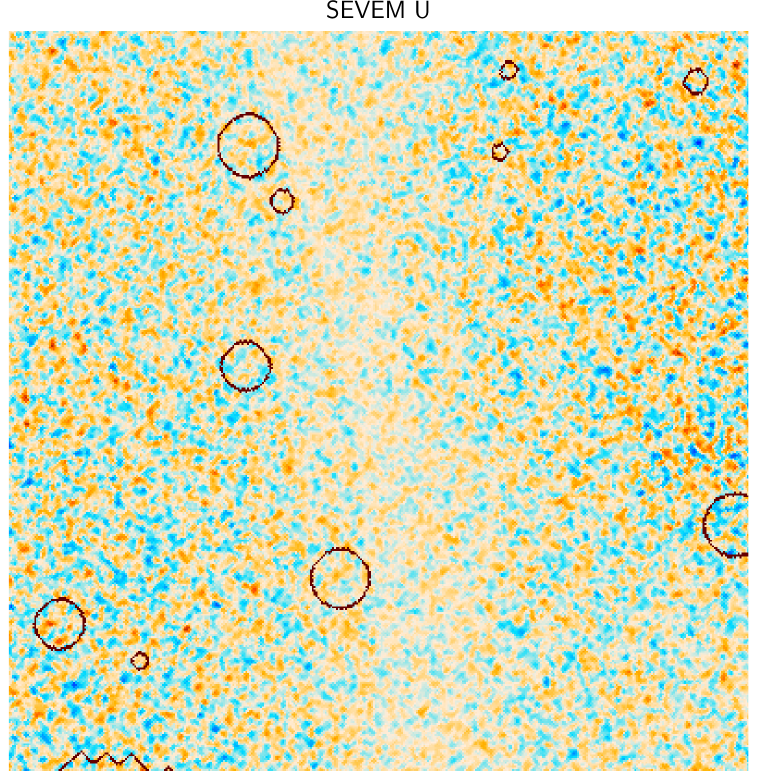}\\
      \includegraphics[width=0.29\linewidth]{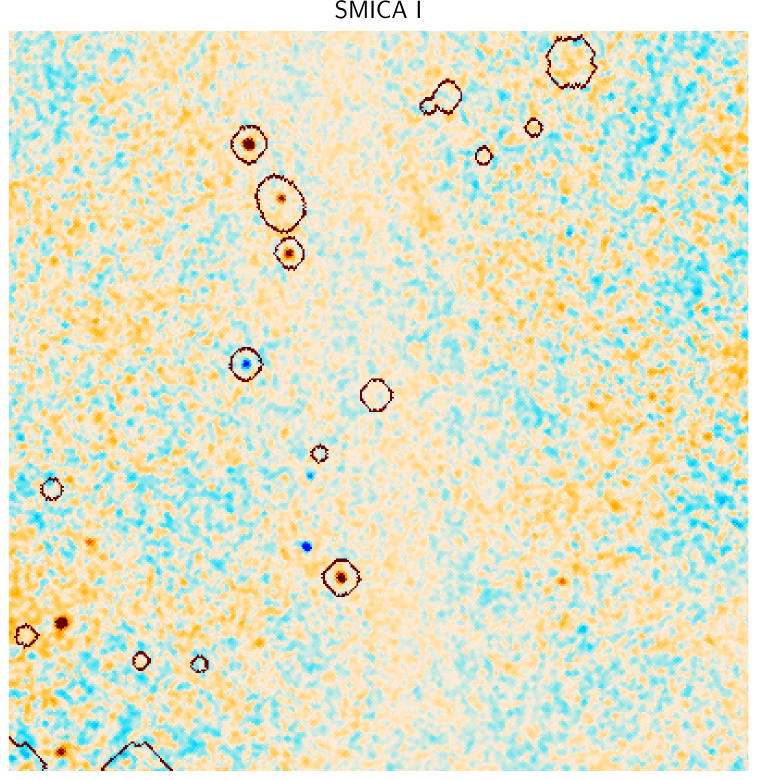}&
      \includegraphics[width=0.29\linewidth]{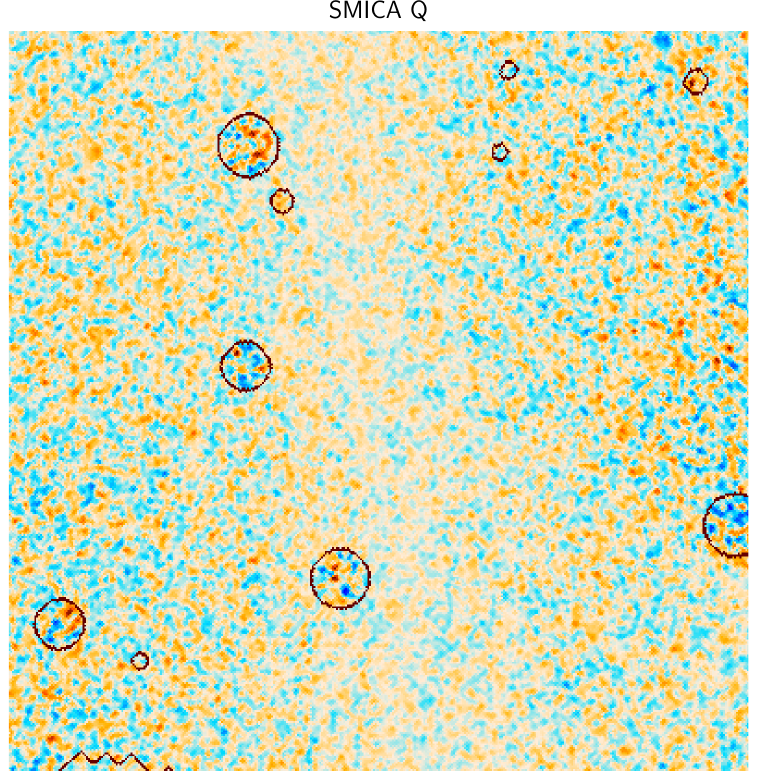}&
      \includegraphics[width=0.29\linewidth]{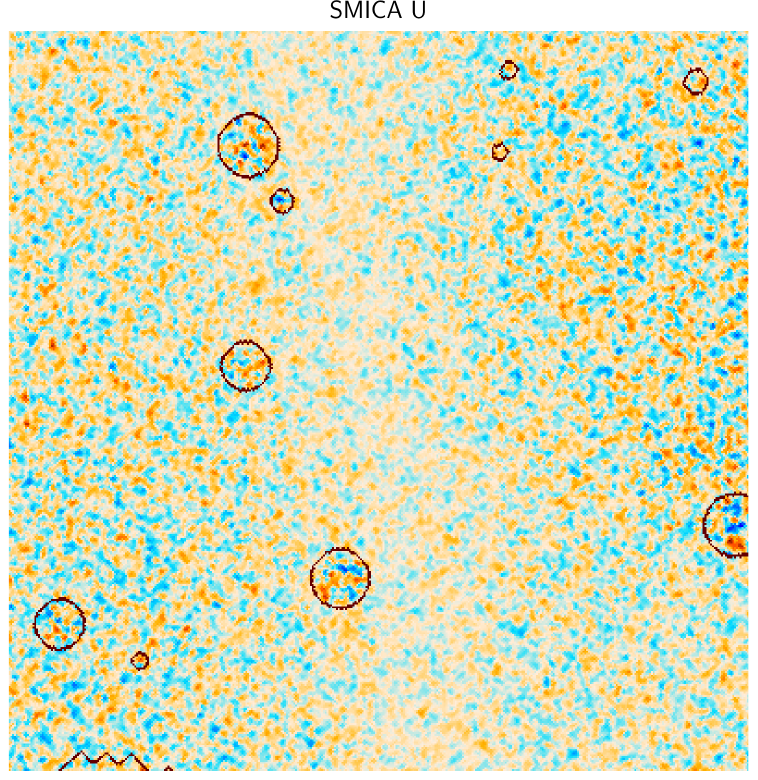}\\
      \multicolumn{3}{c}{
        \includegraphics[width=0.29\linewidth]{figs/colourbar_15uK}
      }
    \end{tabular}
  \end{center}
  \caption{Half-mission half-difference CMB maps at 20\arcmin\ resolution. The patch shown is $20 \deg\times 20\deg$ centred on the North Ecliptic Pole, $(l, b) = (96\pdeg38, 29\pdeg81)$. Columns show Stokes $I$, $Q$, and $U$, while rows show results derived with different component-separation methods. The common mask is marked in red.}
  \label{fig:cmb_maps_zoom_hmhd}
\end{figure*}

Figures~\ref{fig:cmb_maps_zoom_oehd} and \ref{fig:cmb_maps_zoom_hmhd} show enlargements of the odd-even and half-mission half-difference maps for the same region. In these maps, notable qualitative differences between the four CMB maps are observed, perhaps the most striking of which is the effect of different point source treatments adopted by the four pipelines. For instance, in the half-mission splits one can clearly see bright source residuals in the temperature maps for \commander, \nilc, and \smica, but not for \sevem.  These are due to changes in the amplitude of point sources between both periods of
observations, which show up when subtracting the half-mission splits. \sevem\ does not present these residuals because it explicitly inpaints known sources positions in each split, and therefore it reduces significantly this contaminant emission in the half-mission data before constructing the half-difference maps.  In the case of the \smica\ polarization maps, one can also see outlines of the processing mask adopted for inpainting in that case.

Another type of qualitative difference is seen between \commander\ on the one side, and the other three codes on the other
side.  \commander\ accounts explicitly for spatial variations in instrumental sensitivity at each frequency during Wiener
filtering, which corresponds to evaluating an exact inverse-noise-variance weighting pixel-by-pixel in the different channels. This procedure produces somewhat more uniform effective residual maps than the other three codes. 

\begin{figure*}
  \begin{center}
    \begin{tabular}{ccc}
      \includegraphics[width=0.7\linewidth]{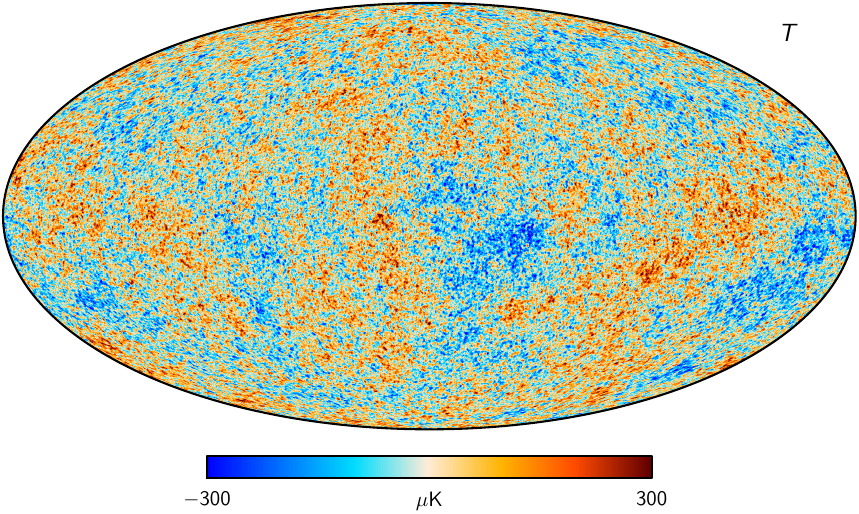}\\
      \includegraphics[width=0.7\linewidth]{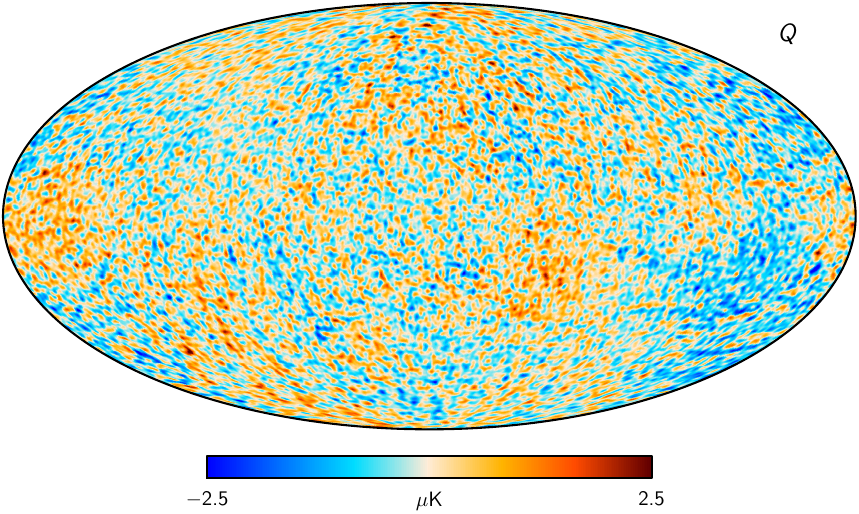}\\
      \includegraphics[width=0.7\linewidth]{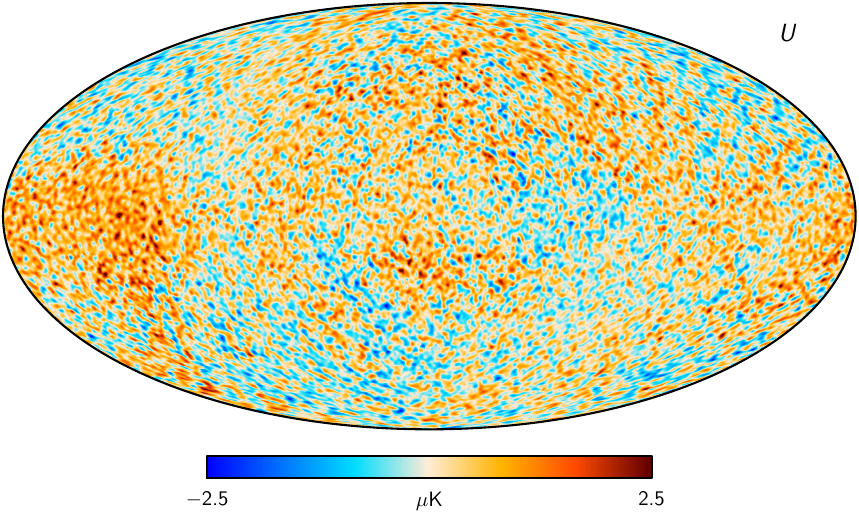}
    \end{tabular}
  \end{center}
  \caption{\commander\ constrained-realization CMB maps. The masked regions shown in Fig.~\ref{fig:inpaint} have been replaced with a Gaussian-constrained realization. Panels show, from top to bottom, Stokes parameters $I$, $Q$, and $U$. The temperature map is shown at $5\arcm$ FWHM angular resolution, while the polarization maps are shown at $80\arcm$ FWHM angular resolution. }
  \label{fig:commander_inpaint}
\end{figure*}

\begin{figure*}
  \begin{center}
    \begin{tabular}{ccc}
      \includegraphics[width=0.7\linewidth]{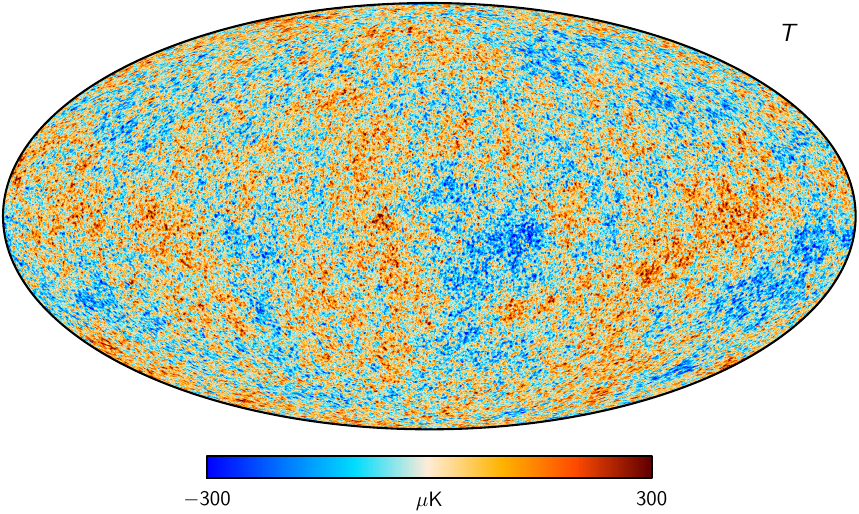}\\
      \includegraphics[width=0.7\linewidth]{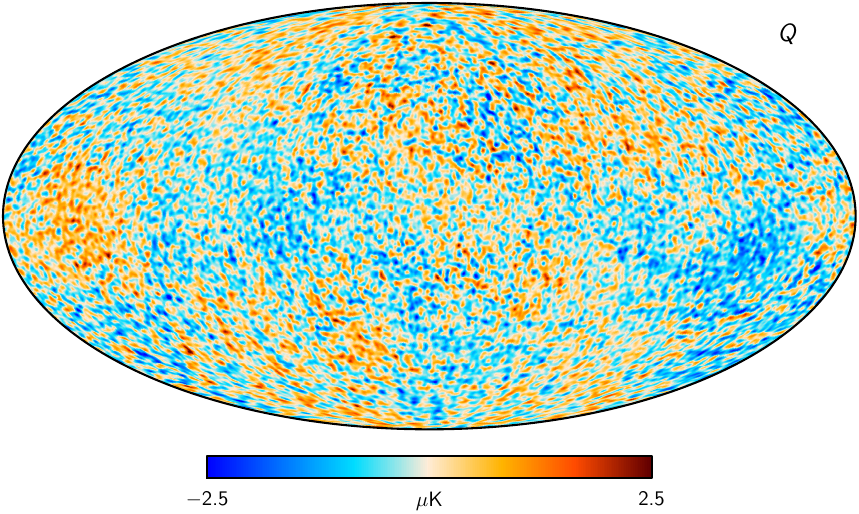}\\
      \includegraphics[width=0.7\linewidth]{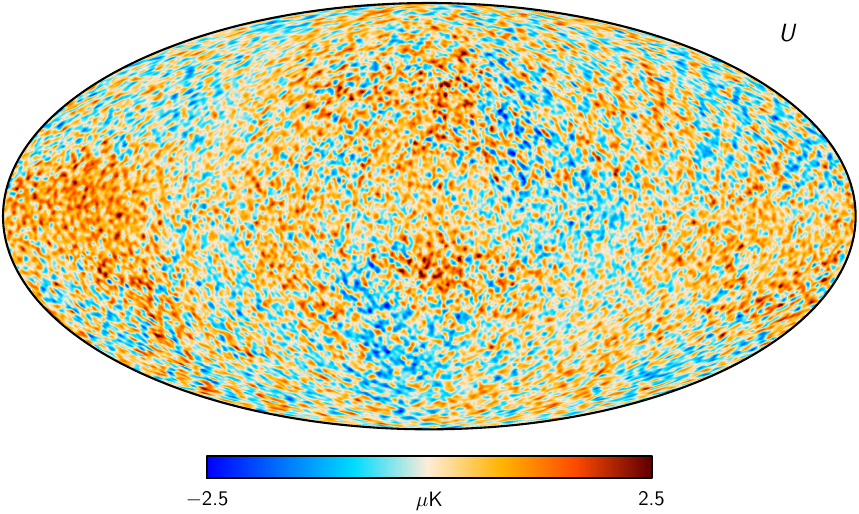}
    \end{tabular}
  \end{center}
  \caption{\nilc\ constrained-realization CMB maps. The masked regions shown in Fig.~\ref{fig:inpaint} has been replaced with a Gaussian-constrained realization. Panels show, from top to bottom, Stokes parameters $I$, $Q$, and $U$. The temperature map is shown at $5\arcm$ FWHM angular resolution, while the polarization maps are shown at $80\arcm$ FWHM angular resolution.}
  \label{fig:nilc_inpaint}
\end{figure*}

\begin{figure*}
  \begin{center}
    \begin{tabular}{ccc}
      \includegraphics[width=0.7\linewidth]{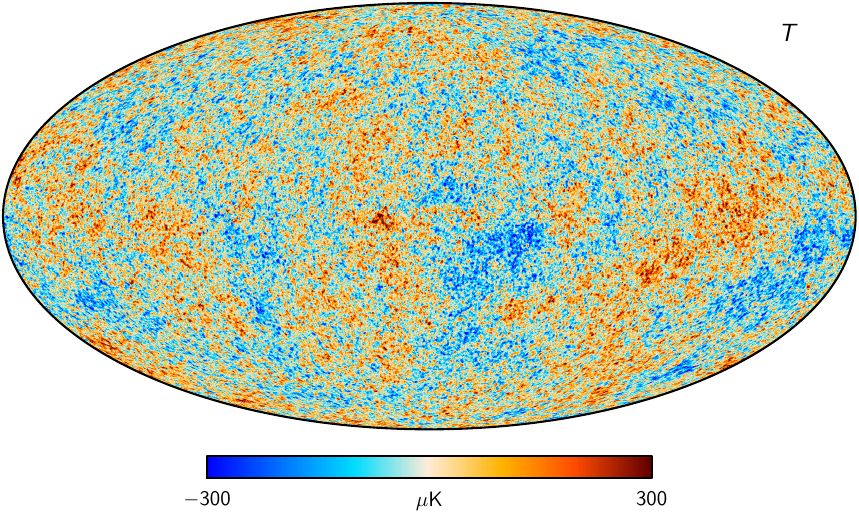}\\
      \includegraphics[width=0.7\linewidth]{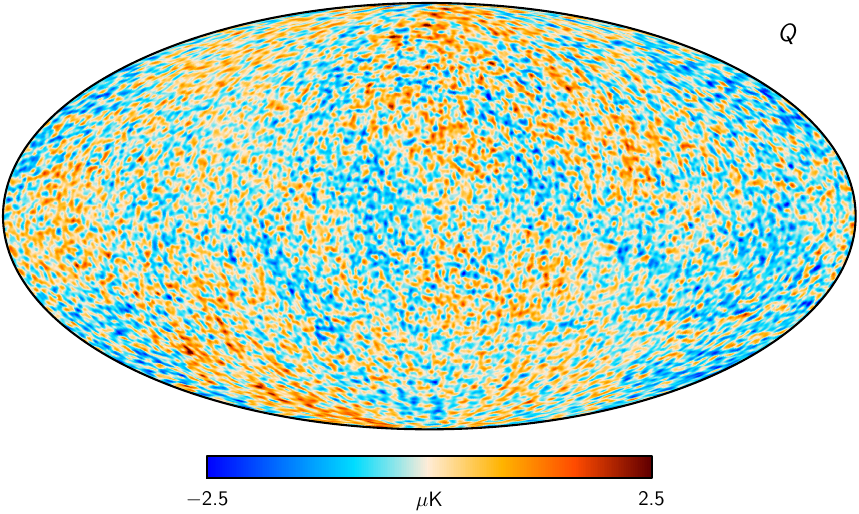}\\
      \includegraphics[width=0.7\linewidth]{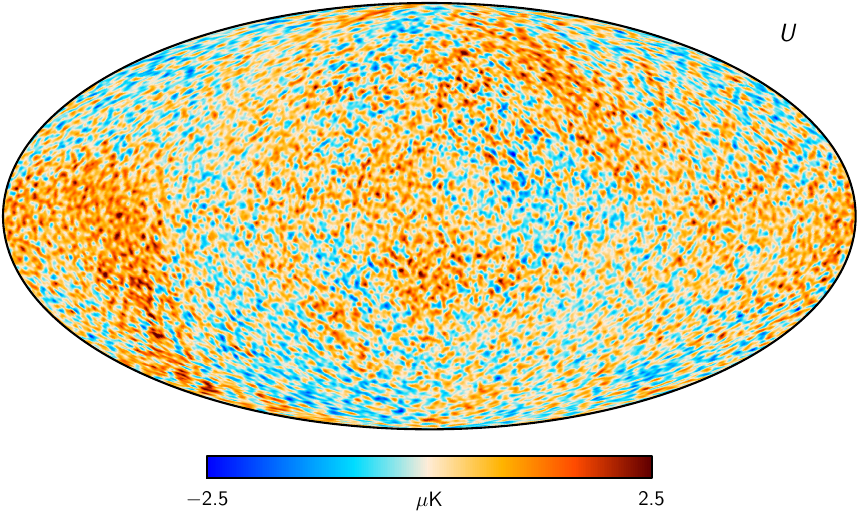}
    \end{tabular}
  \end{center}
  \caption{\sevem\ constrained-realization CMB maps. The masked regions shown in Fig.~\ref{fig:inpaint} has been replaced with a Gaussian-constrained realization. Panels show, from top to bottom, Stokes parameters $I$, $Q$, and $U$. The temperature map is shown at $5\arcm$ FWHM angular resolution, while the polarization maps are shown at $80\arcm$ FWHM angular resolution.}
  \label{fig:sevem_inpaint}
\end{figure*}

\begin{figure*}
  \begin{center}
    \begin{tabular}{ccc}
      \includegraphics[width=0.7\linewidth]{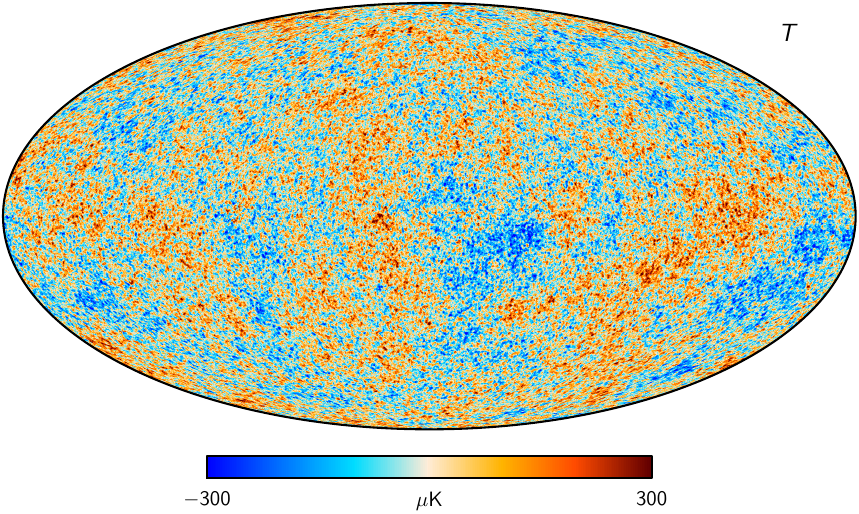}\\
      \includegraphics[width=0.7\linewidth]{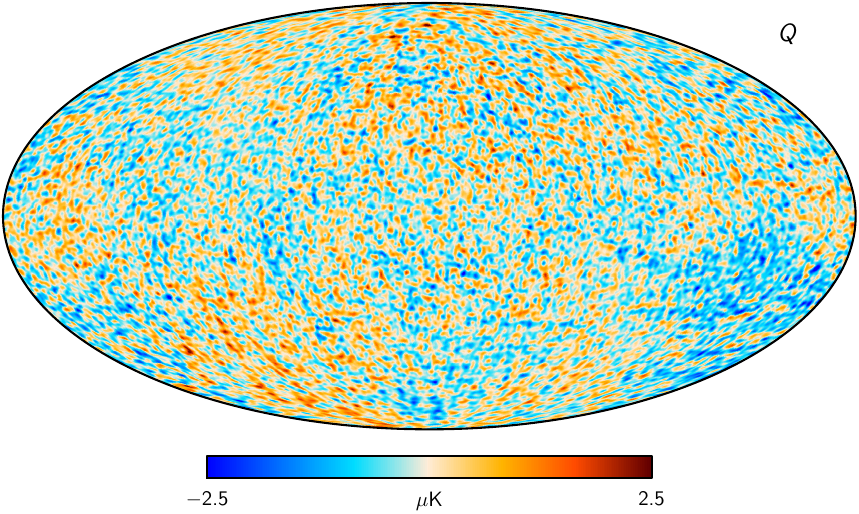}\\
      \includegraphics[width=0.7\linewidth]{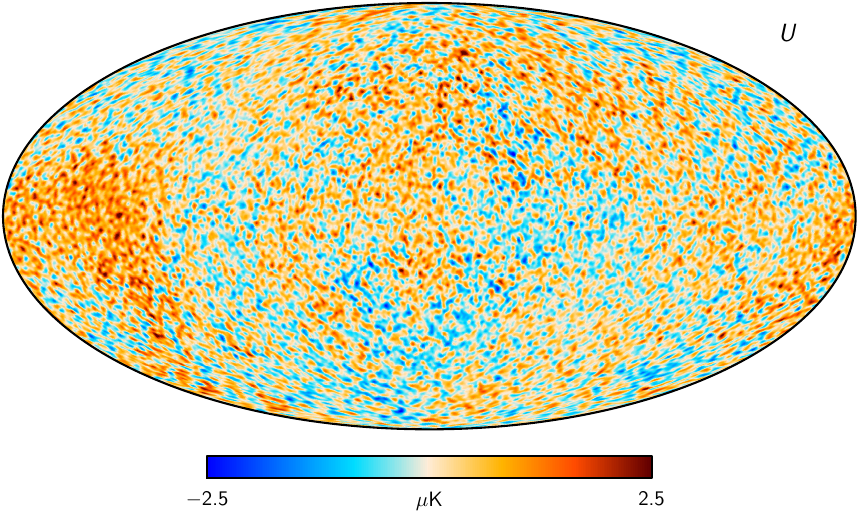}
    \end{tabular}
  \end{center}
  \caption{\smica\ constrained-realization CMB maps. The masked regions shown in Fig.~\ref{fig:inpaint} have been replaced with a Gaussian-constrained realization. Panels show, from top to bottom, Stokes parameters $I$, $Q$, and $U$. The temperature map is shown at $5\arcm$ FWHM angular resolution, while the polarization maps are shown at $80\arcm$ FWHM angular resolution.}
  \label{fig:smica_inpaint}
\end{figure*}

Next, Figs.~\ref{fig:commander_inpaint}--\ref{fig:smica_inpaint} show a single Gaussian-constrained realization evaluated for each of the cleaned CMB maps, with the inpainting mask shown in Fig.~\ref{fig:inpaint} applied. The temperature maps are shown at $5\arcm$ FWHM resolution, and the polarization maps are shown at $80\arcm$ FWHM resolution. These maps are primarily intended for presentation purposes, rather than scientific analysis, since their noise properties are complicated. If similar constrained realizations are required for quantitative analysis, we recommend users to employ a Gibbs sampler, for instance as implemented in \commander, to produce an ensemble of such realizations, which then collectively may be used to propagate
uncertainties.

\begin{figure*}
  \begin{center}
    \begin{tabular}{ccc}
      \includegraphics[width=0.3\linewidth]{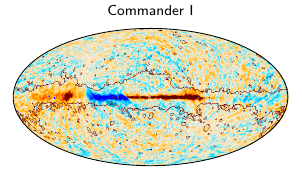}&
      \includegraphics[width=0.3\linewidth]{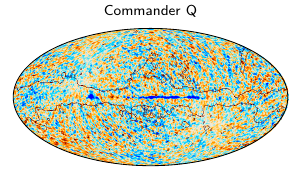}&
      \includegraphics[width=0.3\linewidth]{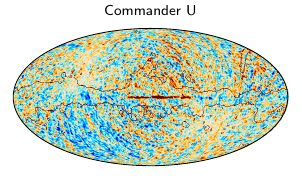}\\
      \includegraphics[width=0.3\linewidth]{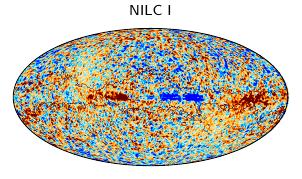}&
      \includegraphics[width=0.3\linewidth]{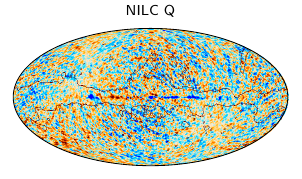}&
      \includegraphics[width=0.3\linewidth]{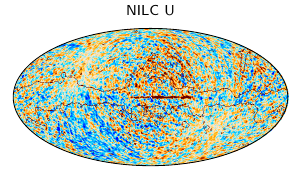}\\
      \includegraphics[width=0.3\linewidth]{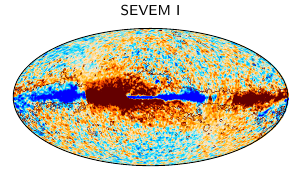}&
      \includegraphics[width=0.3\linewidth]{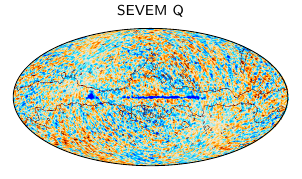}&
      \includegraphics[width=0.3\linewidth]{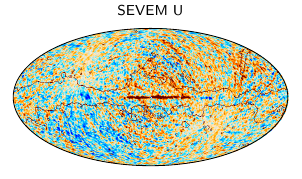}\\
      \includegraphics[width=0.3\linewidth]{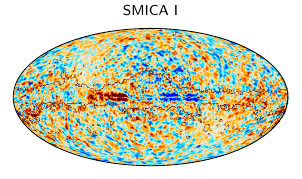}&
      \includegraphics[width=0.3\linewidth]{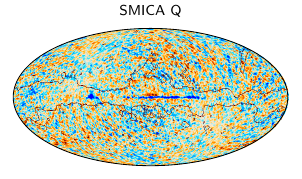}&
      \includegraphics[width=0.3\linewidth]{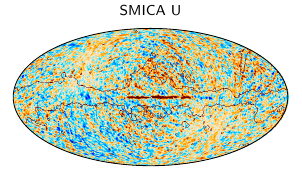}\\
      \includegraphics[width=0.25\linewidth]{figs/colourbar_4uK}&
      \multicolumn{2}{c}{
        \includegraphics[width=0.25\linewidth]{figs/colourbar_2p5uK}
      }
    \end{tabular}
  \end{center}
  \caption{First half-mission split-noise simulation maps at 80\arcmin\ resolution. Columns show Stokes $I$, $Q$, and $U$,
while rows show results derived with different component-separation methods. Monopoles and dipoles have been subtracted from the intensity maps, with parameters estimated outside a $|b|<30^{\circ}$ Galactic cut.}
  \label{fig:noise_hm1_maps}
\end{figure*}

\begin{figure*}
  \begin{center}
    \begin{tabular}{ccc}
      \includegraphics[width=0.3\linewidth]{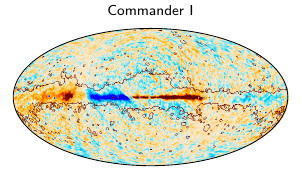}&
      \includegraphics[width=0.3\linewidth]{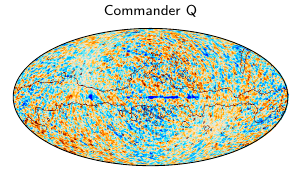}&
      \includegraphics[width=0.3\linewidth]{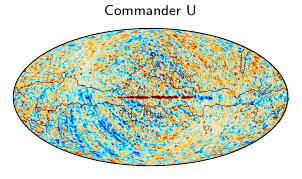}\\
      \includegraphics[width=0.3\linewidth]{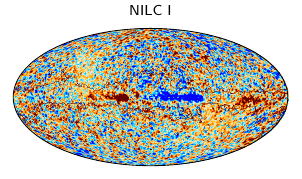}&
      \includegraphics[width=0.3\linewidth]{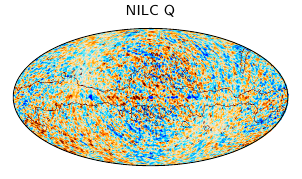}&
      \includegraphics[width=0.3\linewidth]{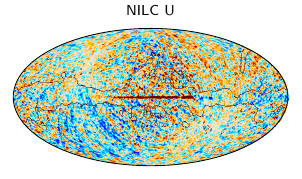}\\
      \includegraphics[width=0.3\linewidth]{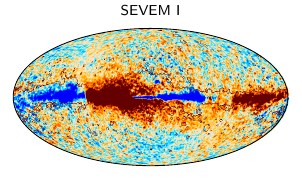}&
      \includegraphics[width=0.3\linewidth]{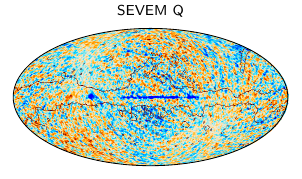}&
      \includegraphics[width=0.3\linewidth]{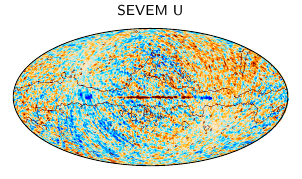}\\
      \includegraphics[width=0.3\linewidth]{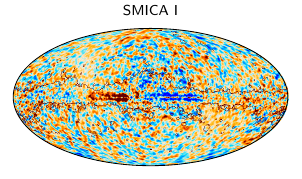}&
      \includegraphics[width=0.3\linewidth]{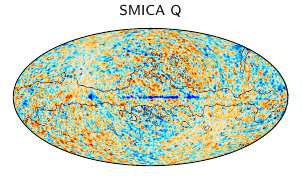}&
      \includegraphics[width=0.3\linewidth]{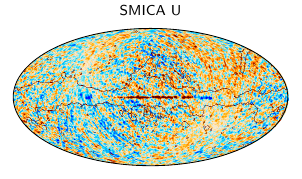}\\
      \includegraphics[width=0.25\linewidth]{figs/colourbar_4uK}&
      \multicolumn{2}{c}{
        \includegraphics[width=0.25\linewidth]{figs/colourbar_2p5uK}
      }
    \end{tabular}
  \end{center}
  \caption{Even ring split-noise simulation maps at 80\arcmin\ resolution. Columns show Stokes $I$, $Q$, and $U$, while rows show results derived with different component-separation methods. Monopoles and dipoles have been subtracted from the intensity maps, with parameters estimated outside a $|b|<30^{\circ}$ Galactic cut.}
  \label{fig:noise_oe1_maps}
\end{figure*}

\textcolor{black}{Finally, for illustration, Fig.~\ref{fig:noise_hm1_maps} shows one of the first half-mission noise simulations at 80\arcm\ FWHM resolution propagated through the four component separation methods. The simulation contains both instrumental noise and residual systematic effects. Some residual systematics can be seen in the Galactic plane, and are  especially apparent in the \sevem\ intensity map. These residuals come mainly from the 545\,GHz simulated map, which seems to have larger systematics than the other channels. This explains why this structure is not visible in polarization, and also why the greatest effect is on \sevem, which gives greater weight to this channel than the other methods. Since the maps have been smoothed to 80\arcm, the residuals extend beyond their original locations.  Nevertheless, the amplitude of these residuals is relatively small in absolute values, and moreover, they are mostly contained within the common confidence mask (marked in red), even without considering an extended version of the mask that should take into account the additional smoothing of the map. Therefore, we do not expect these residuals to affect significantly the analysis carried out with the simulations.  The NILC intensity map has higher noise at this resolution than the other pipelines, consistent with what has been seen in previous figures. Figure~\ref{fig:noise_oe1_maps} shows the same plot for one even-ring, split-noise simulation, from which similar conclusions can be derived.}

\section{N-point functions}
\label{app:npt_functions}

\begin{figure*}[htp!]
\begin{center}
\includegraphics[width=0.48\textwidth]{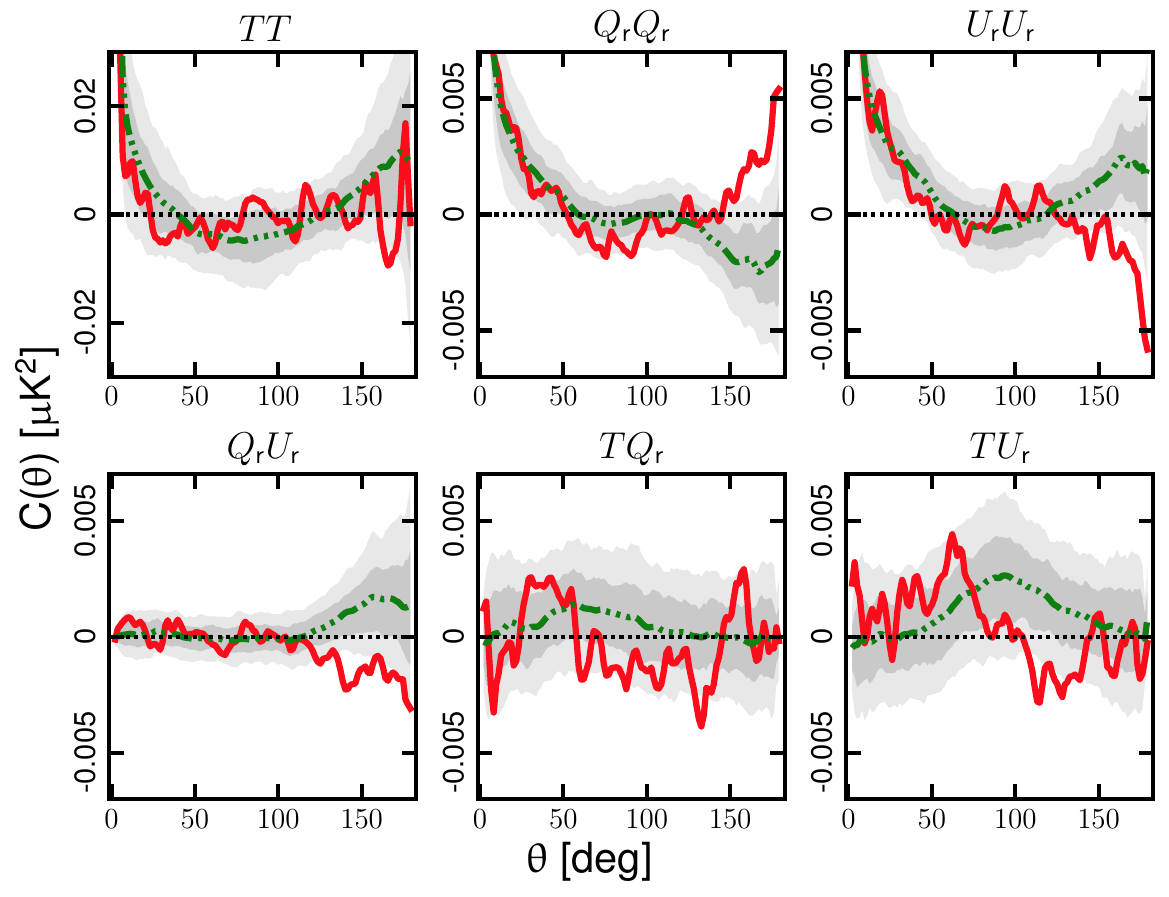}
\includegraphics[width=0.48\textwidth]{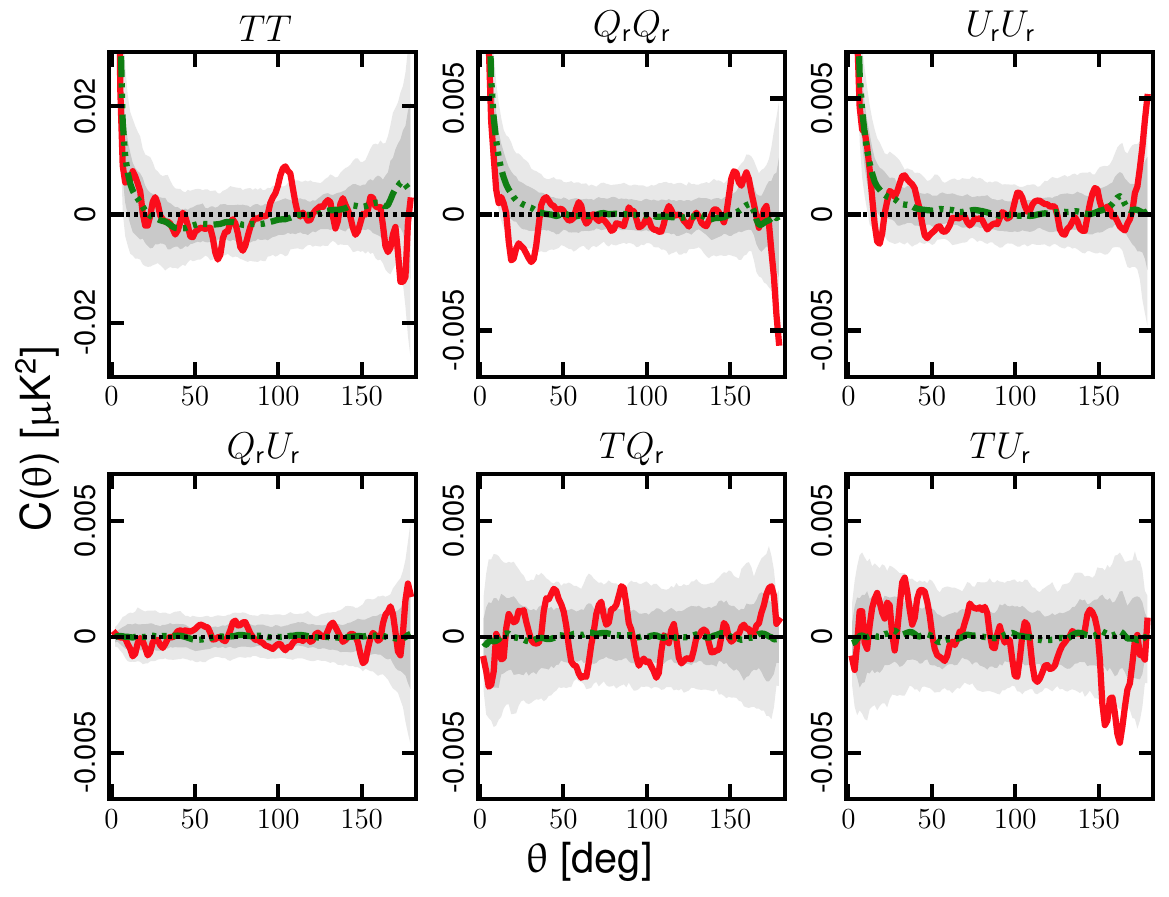} \\
\includegraphics[width=0.48\textwidth]{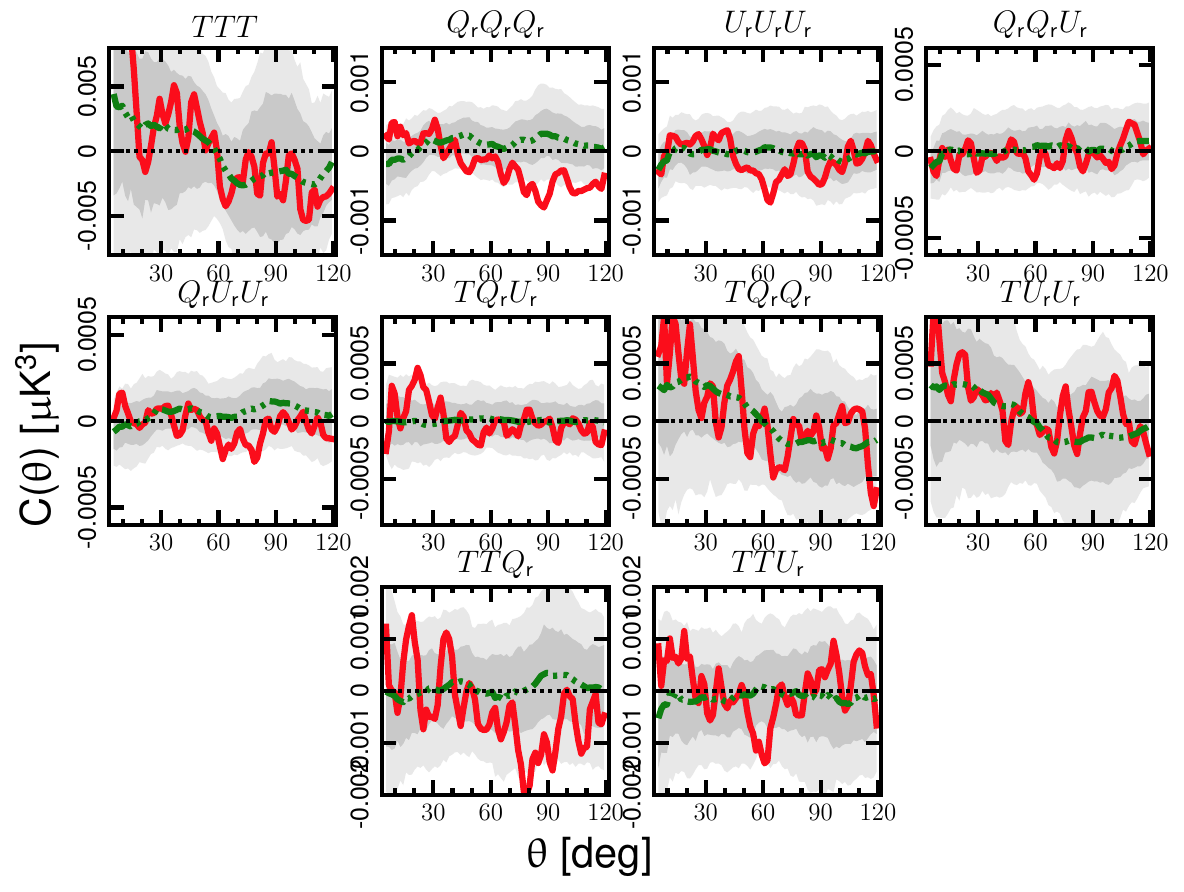}
\includegraphics[width=0.48\textwidth]{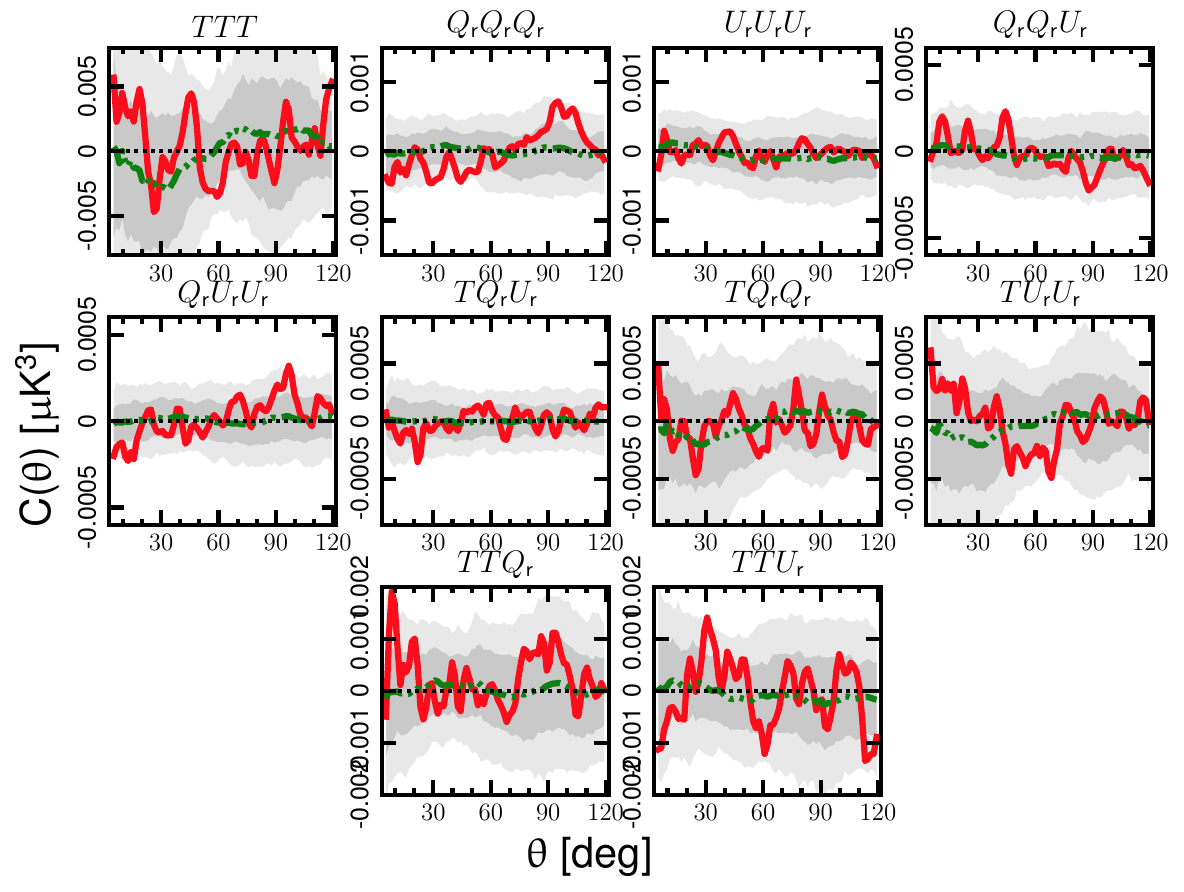}\\
\includegraphics[width=0.48\textwidth]{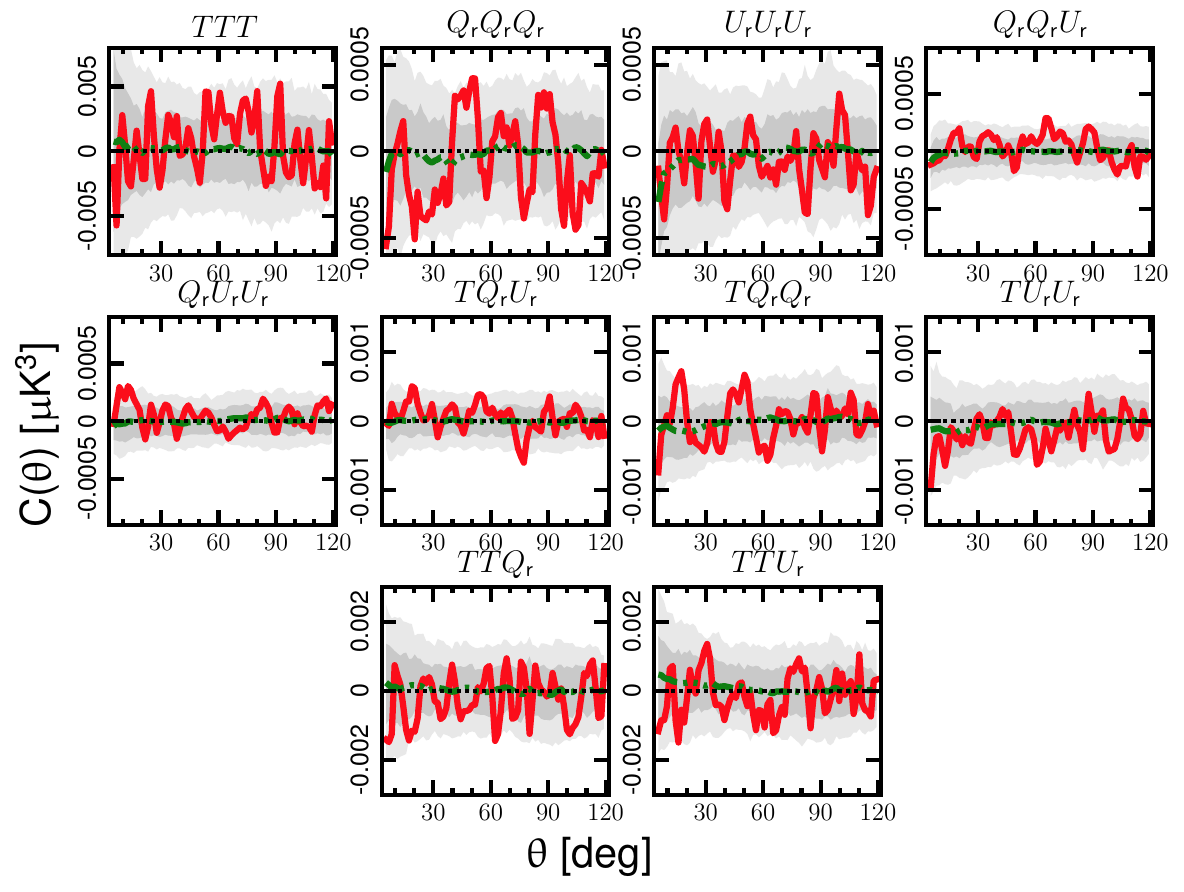}
\includegraphics[width=0.48\textwidth]{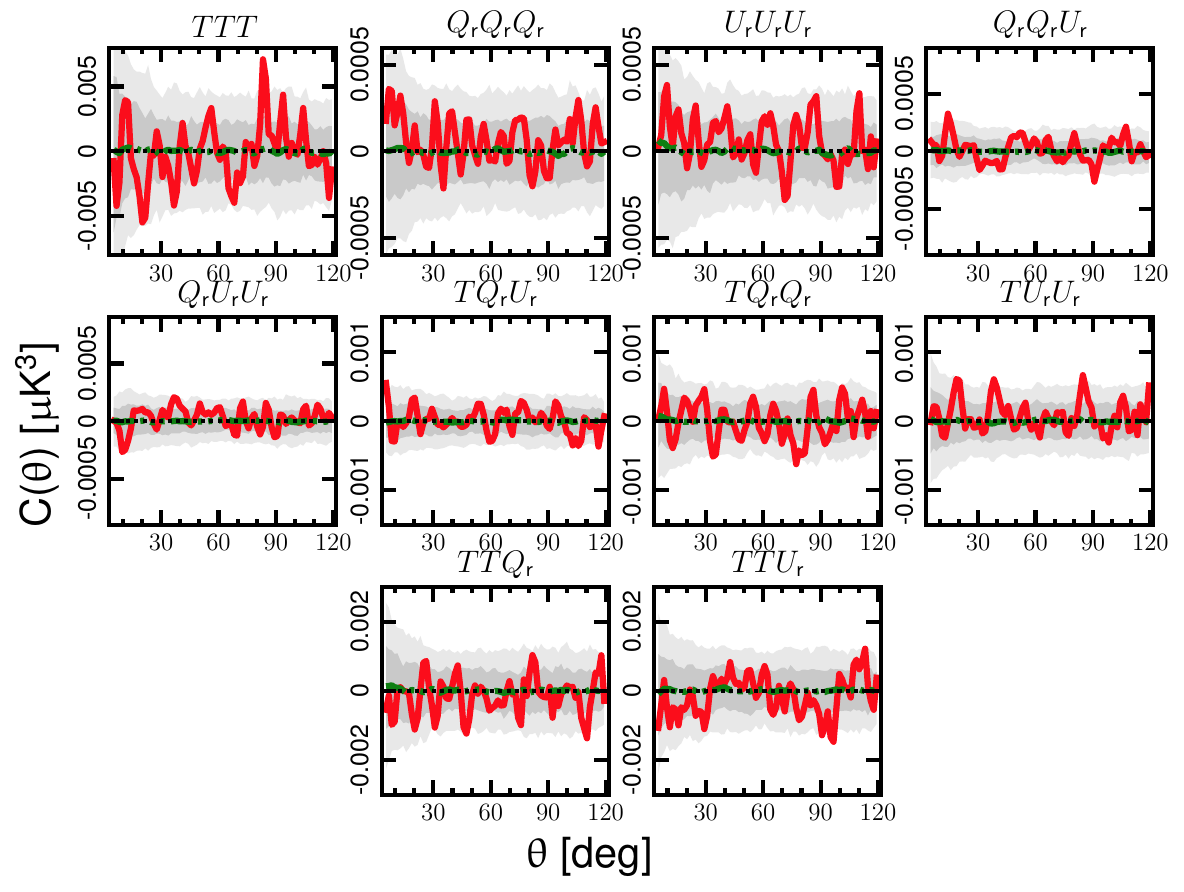}
\caption{The 2-point (upper panels), pseudo-collapsed (middle panels), and equilateral (lower panels) 3-point correlation functions determined from the $\nside=64$ \Planck\ \nilc\ HMHD (left panels) and OEHD (right panels) temperature and polarization map. The red solid line corresponds to the half-difference maps (HMHD or OEHD). The green triple-dot-dashed line indicates the mean determined from 300 FFP10 noise simulations. The shaded dark and light grey regions indicate the corresponding 68\,\% and 95\,\% confidence regions, respectively.}
\label{fig:npt_nilc}
\end{center}
\end{figure*}

\begin{figure*}[htp!]
\begin{center}
\includegraphics[width=0.48\textwidth]{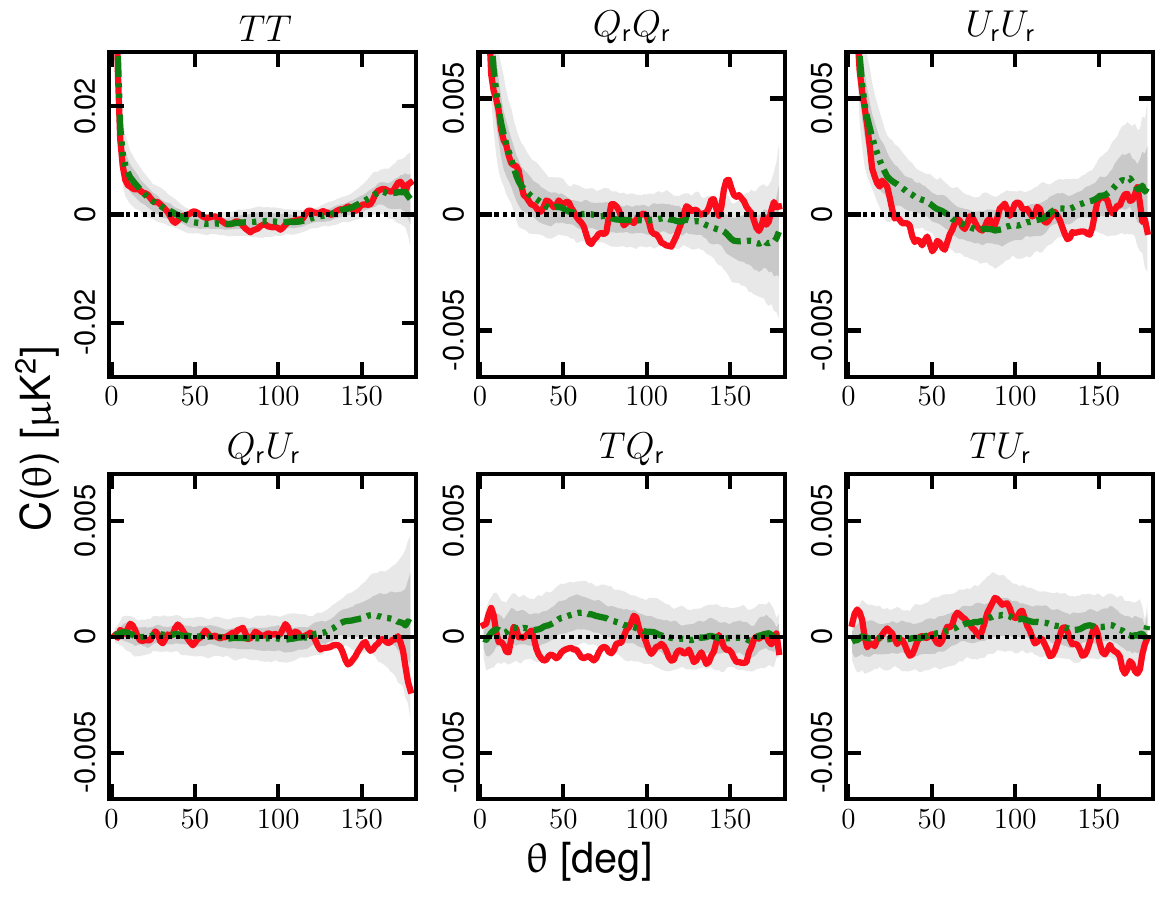}
\includegraphics[width=0.48\textwidth]{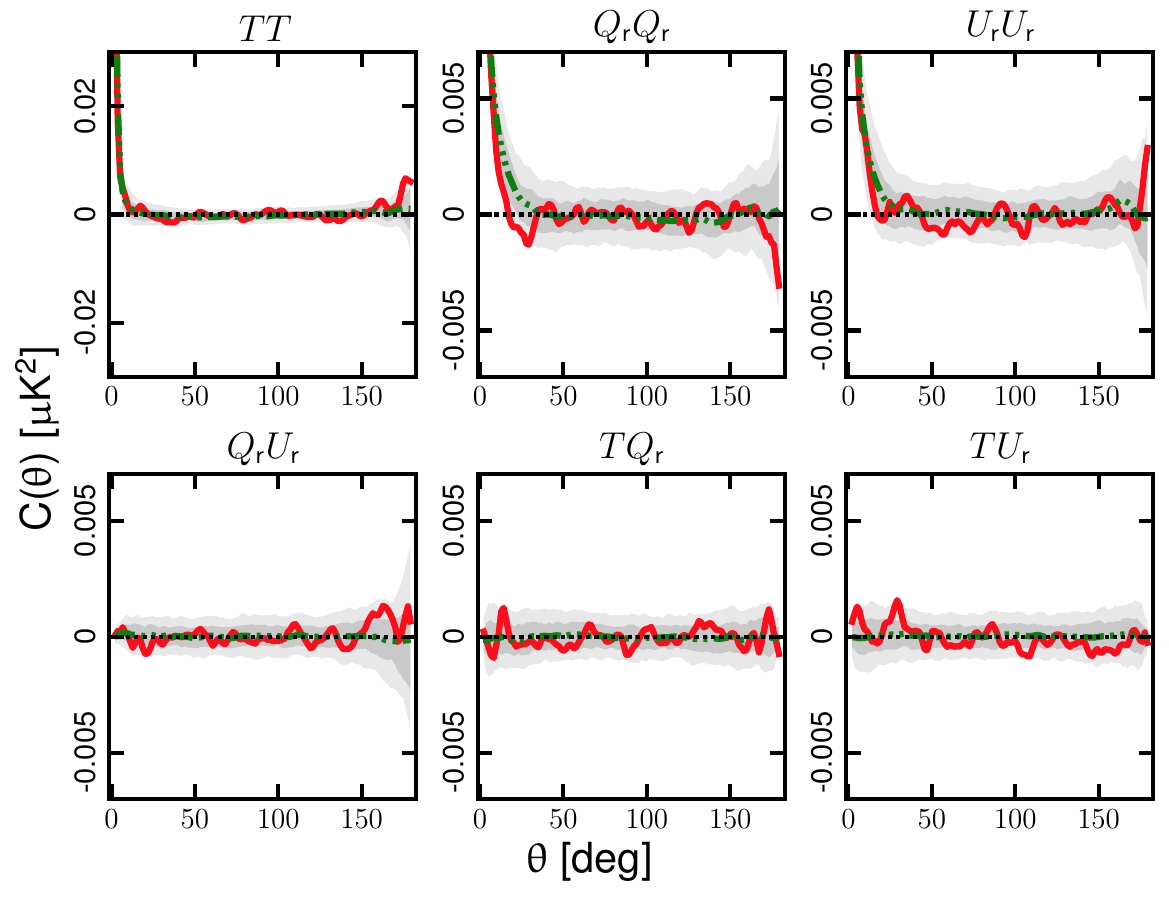} \\
\includegraphics[width=0.48\textwidth]{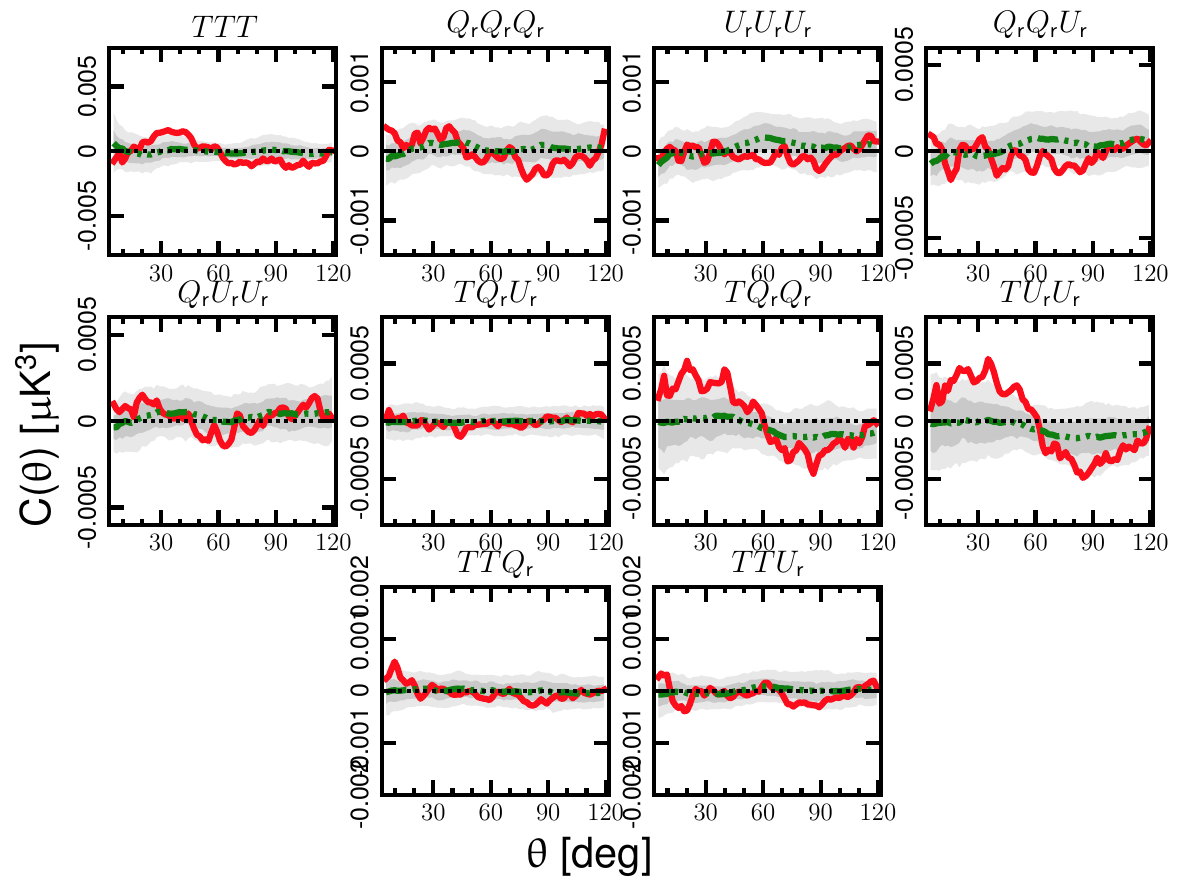}
\includegraphics[width=0.48\textwidth]{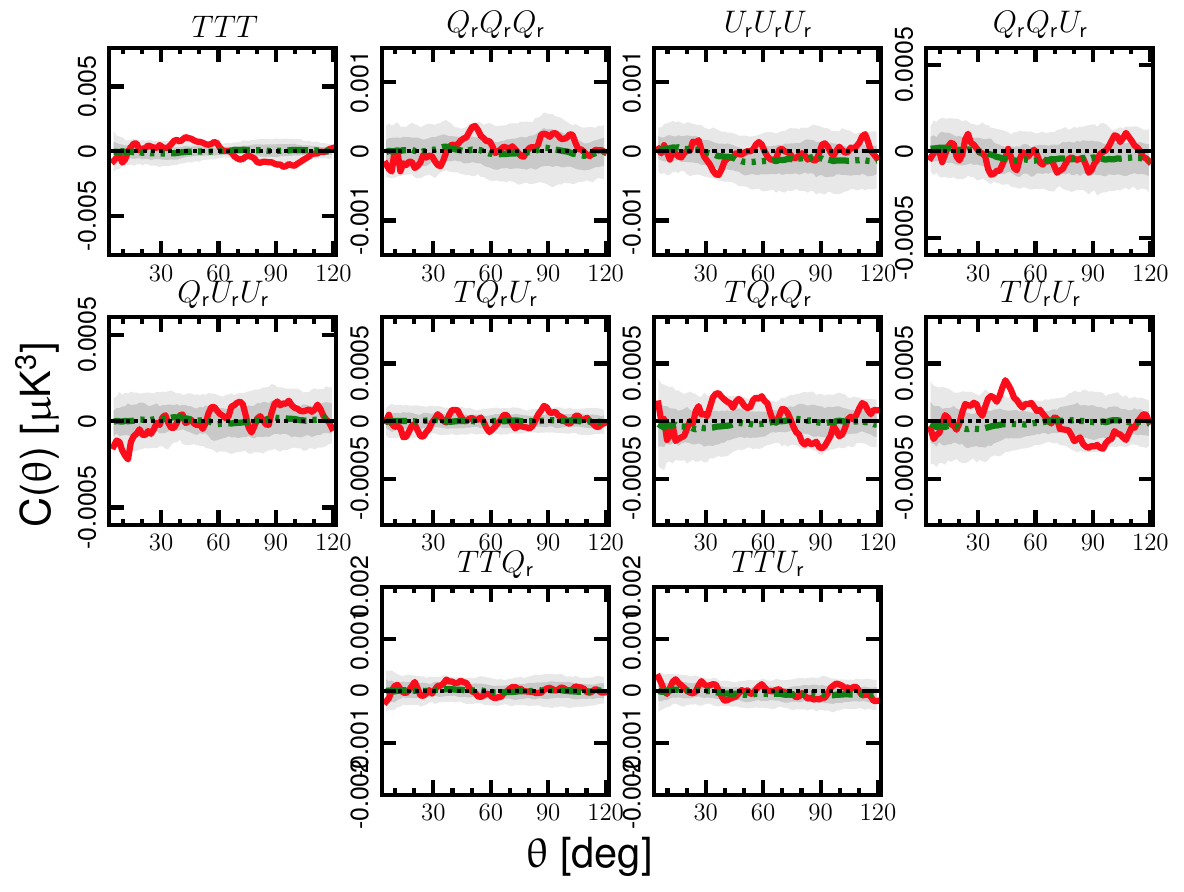}\\
\includegraphics[width=0.48\textwidth]{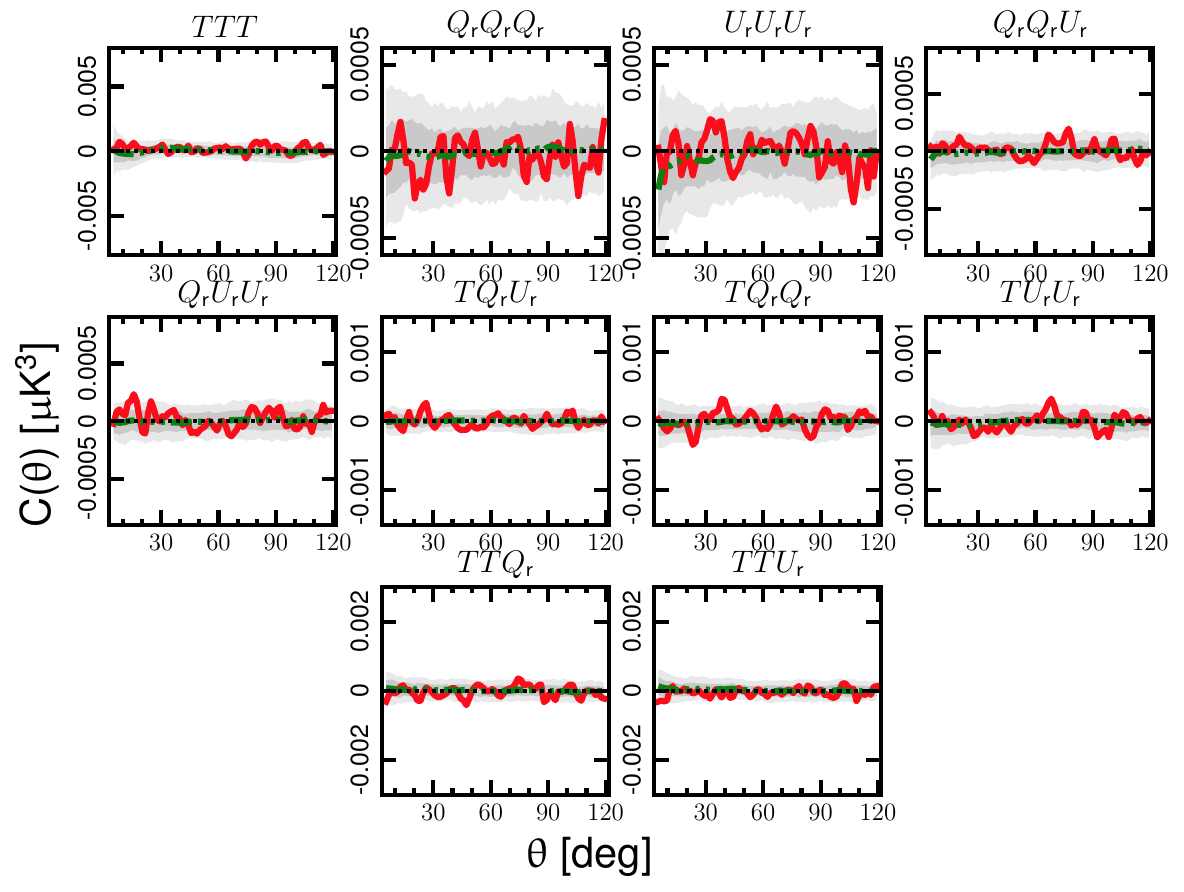}
\includegraphics[width=0.48\textwidth]{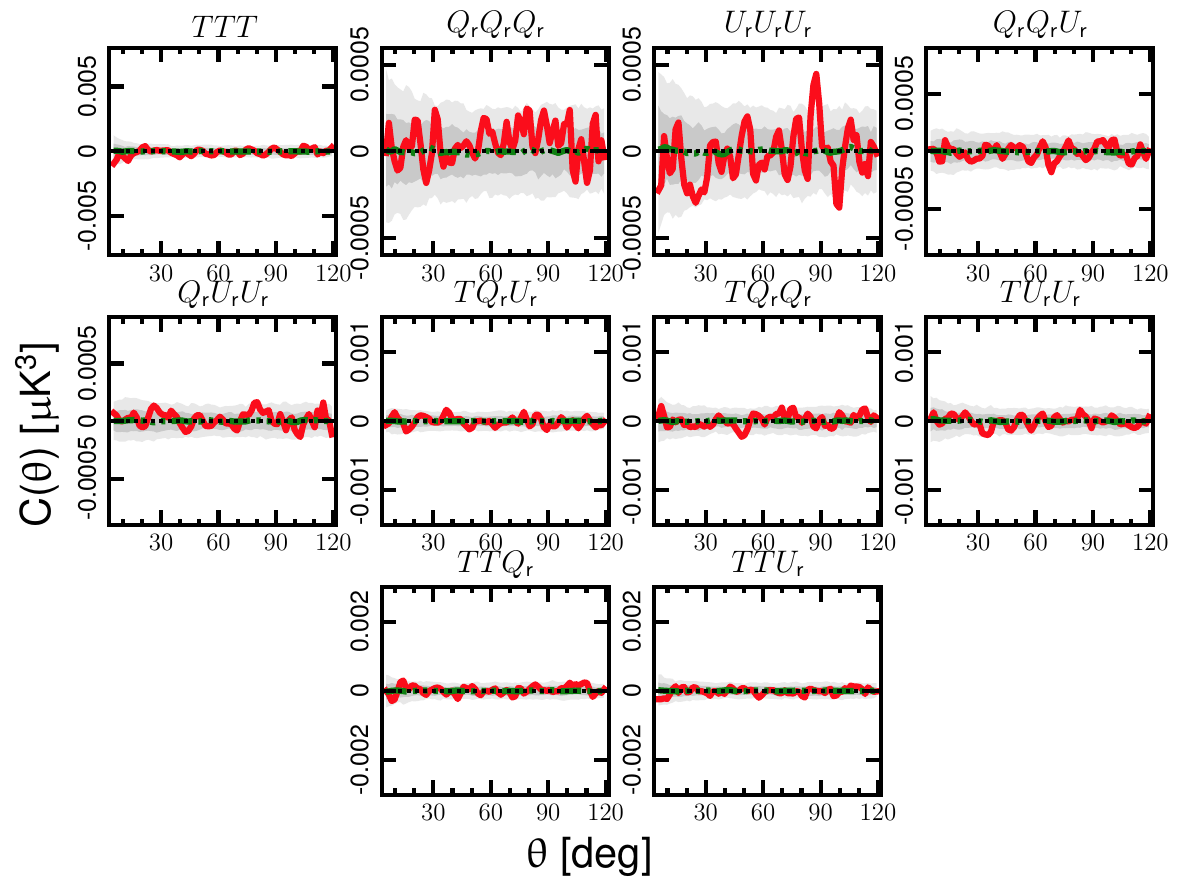}
\caption{The 2-point (upper panels), pseudo-collapsed (middle panels), and equilateral (lower panels) 3-point correlation functions determined from the $\nside=64$ \Planck\ \sevem\ HMHD (left panels) and OEHD (right panels) temperature and polarization map. The red solid line corresponds to the half-difference maps (HMHD or OEHD). The green triple-dot-dashed line indicates the mean determined from 300 FFP10 noise simulations. The shaded dark and light grey regions indicate the corresponding 68\,\% and 95\,\% confidence regions, respectively.}
\label{fig:npt_sevem}
\end{center}
\end{figure*}

\begin{figure*}[htp!]
\begin{center}
\includegraphics[width=0.48\textwidth]{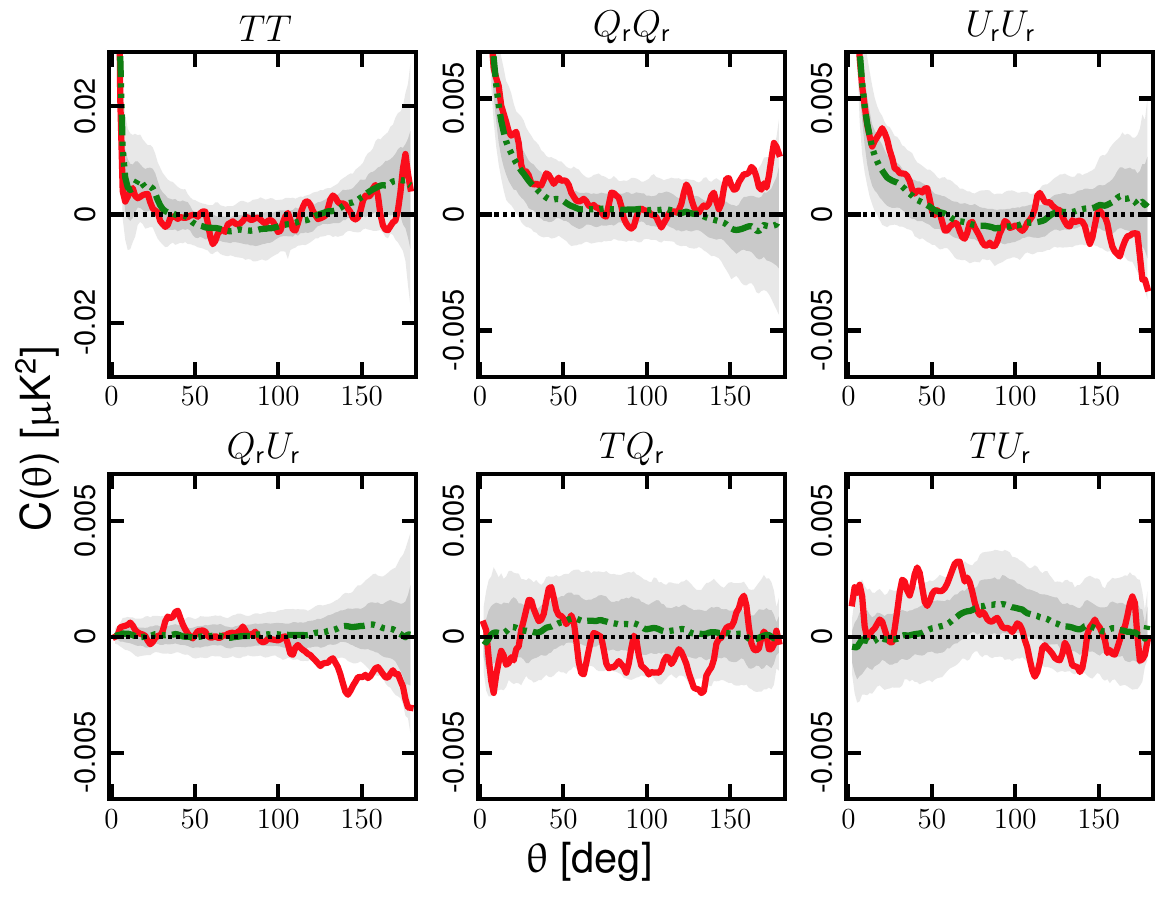}
\includegraphics[width=0.48\textwidth]{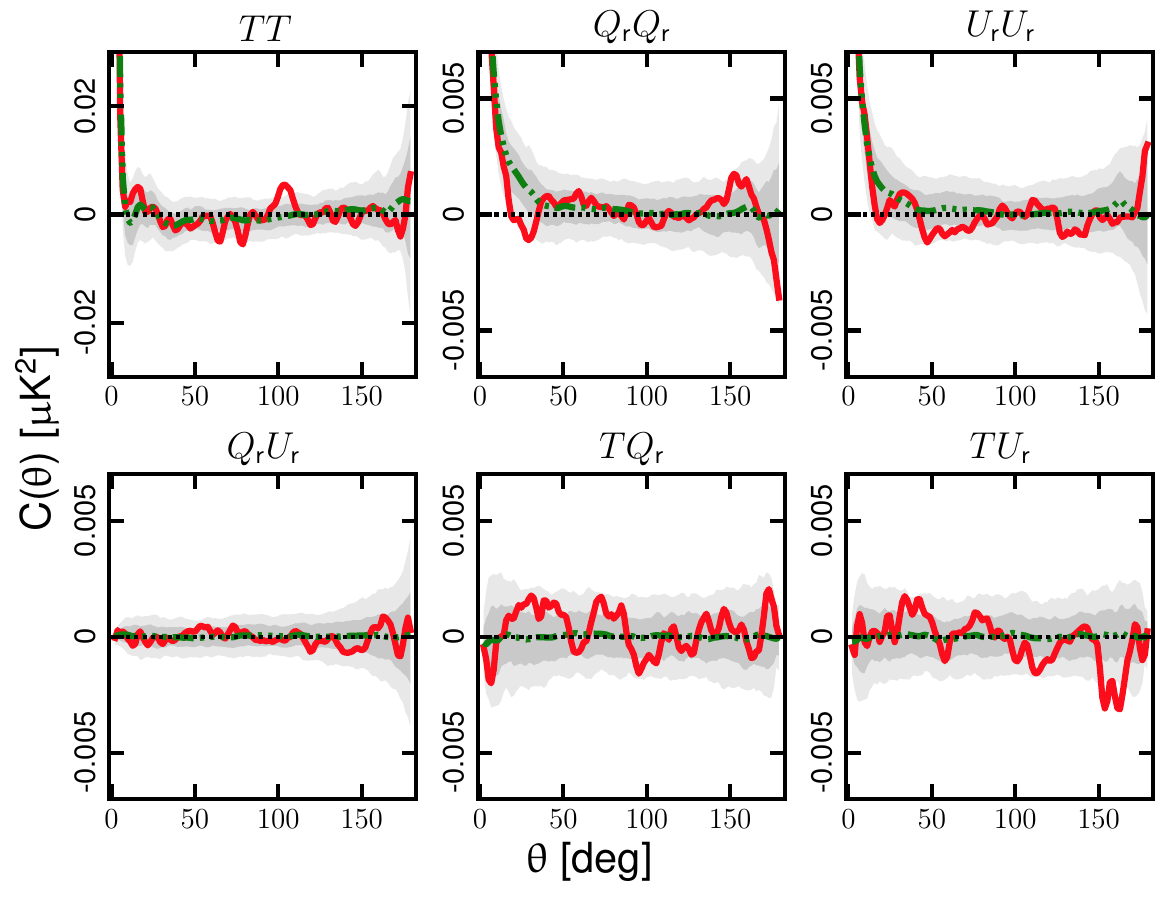} \\
\includegraphics[width=0.48\textwidth]{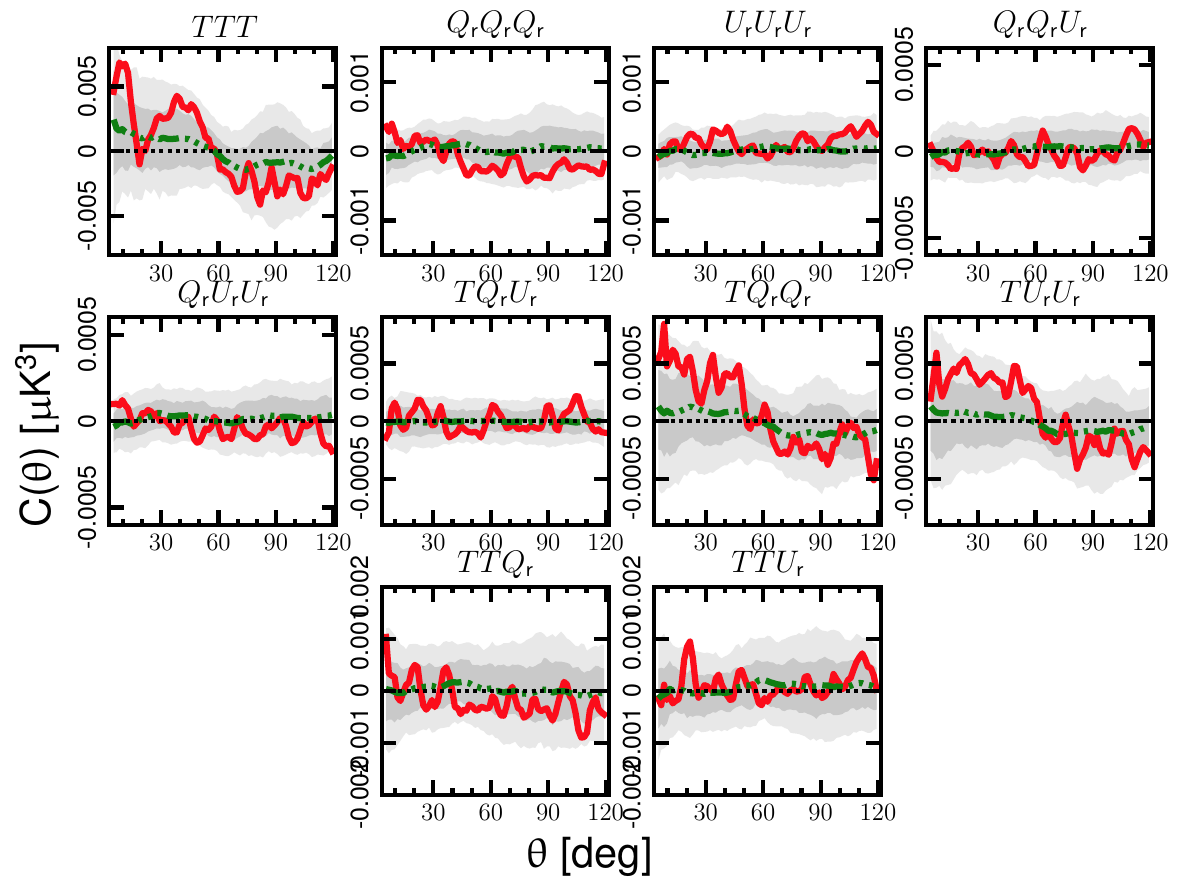}
\includegraphics[width=0.48\textwidth]{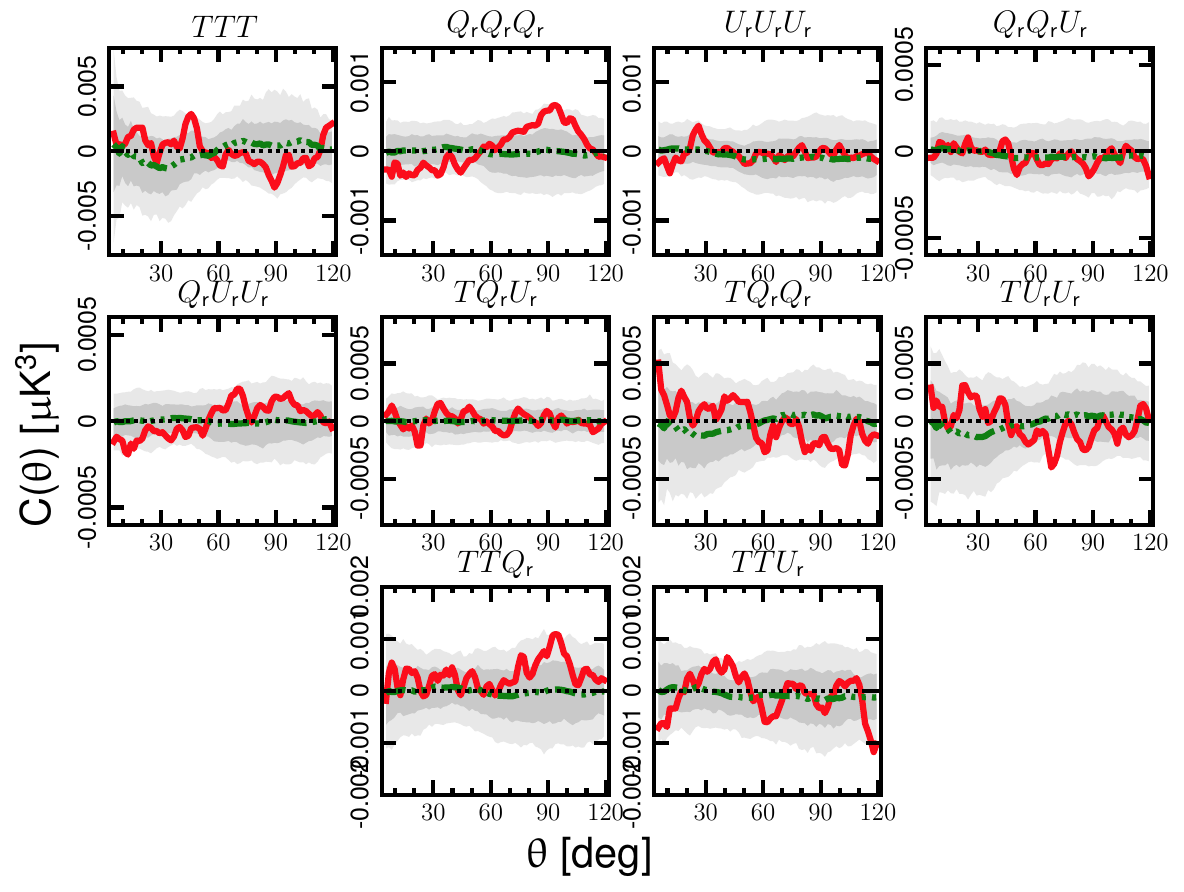}\\
\includegraphics[width=0.48\textwidth]{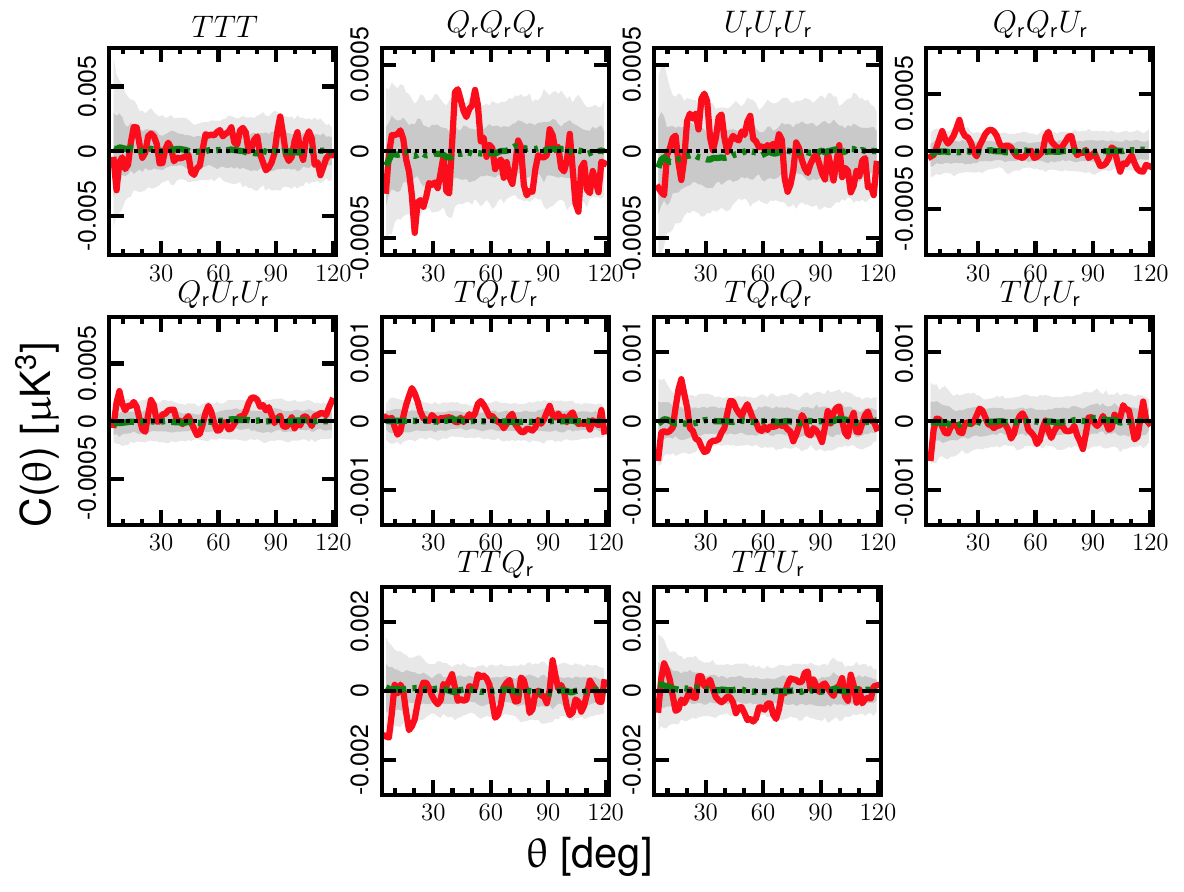}
\includegraphics[width=0.48\textwidth]{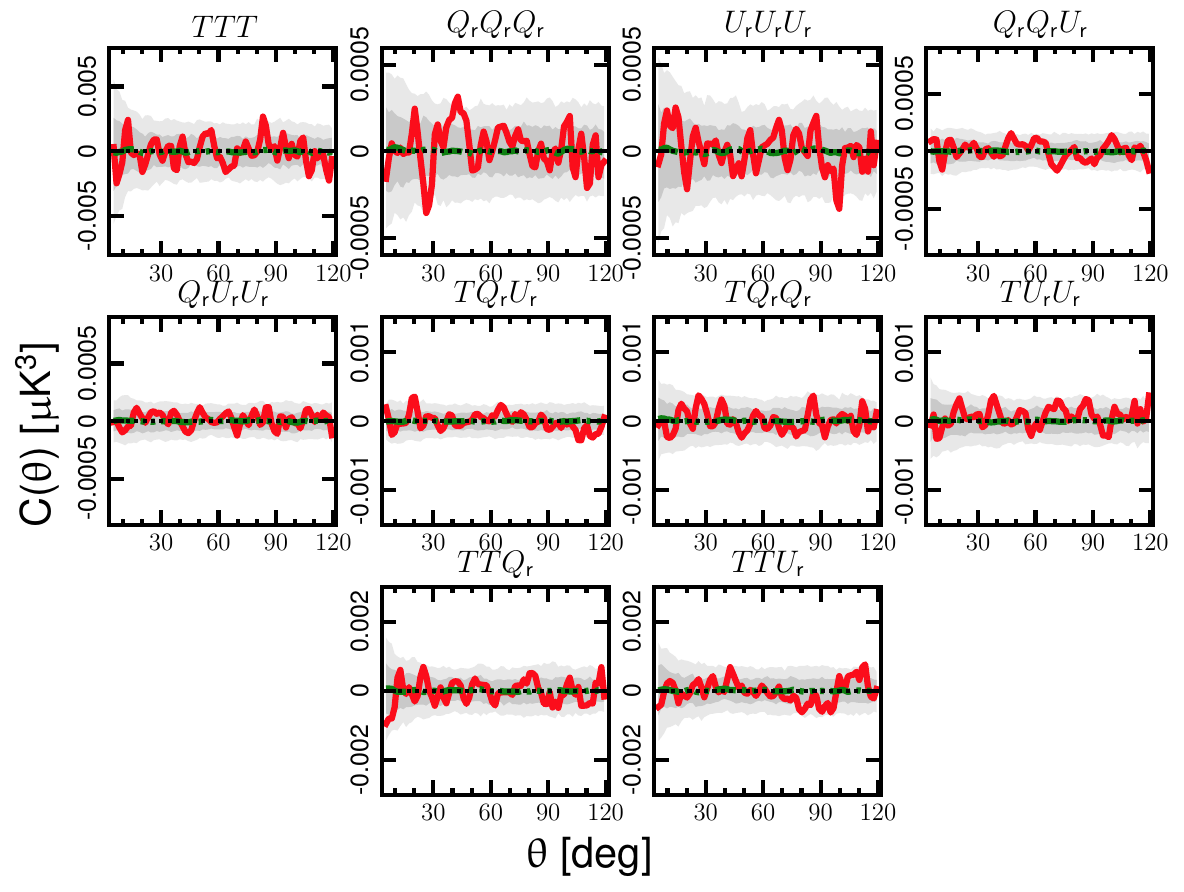}
\caption{The 2-point (upper panels), pseudo-collapsed (middle panels), and equilateral (lower panels) 3-point correlation functions determined from the $\nside=64$ \Planck\ \smica\ HMHD (left panels) and OEHD (right panels) temperature and polarization map. The red solid line corresponds to the half-difference maps (HMHD or OEHD). The green triple-dot-dashed line indicates the mean determined from 300 FFP10 noise simulations. The shaded dark and light grey regions indicate the corresponding 68\,\% and 95\,\% confidence regions, respectively.}
\label{fig:npt_smica}
\end{center}
\end{figure*}

Here we present 2-point and 3-point correlation functions for the HMHD and OEHD maps. These complement analyses and figures presented in the main text (Sect.~\ref{sec:2point_correlation}). Figures~\ref{fig:npt_nilc}, \ref{fig:npt_sevem}, and
\ref{fig:npt_smica} show the correlation functions for half-differences of the \nilc, \sevem, and \smica\ maps, respectively.

\end{document}